\documentclass[pdftex,twocolumn,3]{jour3}          

\usepackage[utf8]{inputenc}
\usepackage{epsf}
\usepackage{latexsym,amssymb,euscript}
\usepackage{widetext}
\usepackage[dvips]{graphicx}
\usepackage[numbers,sort&compress]{natbib}
\usepackage{amsmath}
\usepackage{nicefrac}
\usepackage{slashed}
\usepackage{booktabs}
\usepackage[linktocpage]{hyperref}
\usepackage{braket}
\usepackage{chngcntr}
\usepackage{bm}
\usepackage{bbold}
\usepackage{multicol}
\usepackage{graphics}
\usepackage{graphicx}
\usepackage{mciteplus}
\usepackage{caption}
\usepackage{subcaption}
\usepackage{pdfpages}
\usepackage[titletoc]{appendix}
\usepackage[normalem]{ulem}
\usepackage{microtype} 

\graphicspath{{./figures/}}
\hypersetup{
 linktocpage = false,
 urlcolor = citegreen,
 colorlinks = true,
 linkcolor = urlblue,
 anchorcolor = urlblue,
 citecolor = linkred,
 pdfstartview = {XYZ null null 1.25} 
           }
\usepackage[left=2cm, right=2cm]{geometry}
\usepackage{pstricks}
\usepackage{color}
\usepackage{xcolor}
\definecolor{urlblue}{rgb}{0.2,0.4,0.7}
\definecolor{citegreen}{rgb}{0,0.4,0.2}
\definecolor{linkred}{rgb}{0.9,0.2,0.1}
\usepackage{float}
\usepackage{academicons}
\definecolor{orcidlogocol}{HTML}{A6CE39}
\usepackage{fancyhdr}
\newcommand{\drv}{{\rm d}}

\newcommand{\as}{\alpha_s}
\newcommand{\LQCD}{\Lambda_{\rm QCD}}
\newcommand{\MSb}{\overline{\rm MS}}

\newcommand{\LL}{{\rm LL/LO}}

\newcommand{\NLLp}{{\rm NLL/NLO^+}}

\newcommand{\HENLOp}{{\rm HE}\mbox{-}{\rm NLO^+}}

\newcommand{\CmLL}{{\cal C}_m^\LL}

\newcommand{\CmNLLp}{{\cal C}_m^\NLLp}

\newcommand{\CmHENLOp}{{\cal C}_m^{{\rm HE}\text{-}{\rm NLO}^+}}

\newcommand{\DY}{\Delta Y}

\newcommand{\vqTTa}{\langle {\vec q}_T^{\;2} \rangle}

\newcommand{\E}{{\cal E}}

\newcommand{\Hb}{{\cal H}_b}
\newcommand{\Jpsi}{J/\psi}
\newcommand{\Yps}{\Upsilon}
\newcommand{\BCs}{B_c(^1S_0)}
\newcommand{\Bss}{B_c(^3S_1)}
\newcommand{\B}{{\cal B}}

\newcommand{\XQq}{X_{Qq\bar{Q}\bar{q}}}

\newcommand{\QXQq}{X_{Q\bar{Q}q\bar{q}}}

\newcommand{\QXbq}{X_{b\bar{b}q\bar{q}}}
\newcommand{\QXQu}{X_{Q\bar{Q}u\bar{u}}}
\newcommand{\QXQs}{X_{Q\bar{Q}s\bar{s}}}
\newcommand{\QXcu}{X_{c\bar{c}u\bar{u}}}
\newcommand{\QXcs}{X_{c\bar{c}s\bar{s}}}
\newcommand{\QXbu}{X_{b\bar{b}u\bar{u}}}
\newcommand{\QXbs}{X_{b\bar{b}s\bar{s}}}
\newcommand{\TQQ}{T_{4Q}}
\newcommand{\TQQZpp}{T_{4Q}(0^{++})}
\newcommand{\TQQTpp}{T_{4Q}(2^{++})}
\newcommand{\TQc}{T_{4c}}
\newcommand{\TQcZpp}{T_{4c}(0^{++})}
\newcommand{\TQcTpp}{T_{4c}(2^{++})}
\newcommand{\TQb}{T_{4b}}
\newcommand{\TQbZpp}{T_{4b}(0^{++})}
\newcommand{\TQbTpp}{T_{4b}(2^{++})}

\newcommand{{\HFNRevo}}{\textsc{HF-NRevo}}

\newcommand{{\Jethad}}{\tt JETHAD}
\newcommand{{\symJethad}}{\tt symJETHAD}
\newcommand{{\bsymJethad}}{\tt (sym)JETHAD}
\newcommand{{\Hell}}{\tt HELL}
\newcommand{{\RadISH}}{\tt RadISH}
\newcommand{{\Pegasus}}{\tt QCD-PEGASUS}
\newcommand{{\HOPPET}}{\tt HOPPET}
\newcommand{{\QCDNUM}}{\tt QCDNUM}
\newcommand{{\APFEL}}{\tt APFEL}
\newcommand{{\APFELpp}}{\tt APFEL++}
\newcommand{{\APFELppp}}{\tt APFEL(++)}
\newcommand{{\EKO}}{\tt EKO}

\newcommand{\orcidFGC}{\href{https://orcid.org/0000-0003-3299-2203}{\includegraphics[scale=0.1]{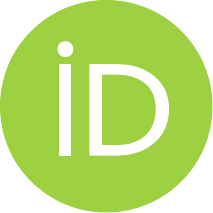}}}

\newcommand{\orcidGG}{\href{https://orcid.org/0009-0009-4270-4819}{\includegraphics[scale=0.1]{logo-orcid.pdf}}}

\smartqed  

\journalname{}

\begin{document}

\normalem
\title{\huge Bottomoniumlike states in proton collisions: \\[0.075cm] Fragmentation and resummation 
}

\subtitle{}

\author{
Francesco Giovanni Celiberto
\and
\thanksref{e1,addr1} \orcidFGC
\; 
Gabriele Gatto
\thanksref{e2,addr2,addr3} \orcidGG
}

\thankstext{e1}{{\it e-mail}:
\href{mailto:fceliberto@ectstar.eu}{francesco.celiberto@uah.es} (corresponding author)}
\thankstext{e2}{{\it e-mail}:
\href{mailto:gabriele.gatto@unical.it}{gabriele.gatto@unical.it}}

\institute{Universidad de Alcal\'a (UAH), E-28805 Alcal\'a de Henares, Madrid, Spain\label{addr1}
\and
Dipartimento di Fisica, Universit\`a della Calabria, I-87036 Arcavacata di Rende, Cosenza, Italy\label{addr2}
\and
Istituto Nazionale di Fisica Nucleare, Gruppo collegato di Cosenza, I-87036 Arcavacata di Rende, Cosenza, Italy\label{addr3}
}

\date{\today}

\maketitle


\section*{Abstract}
We study the semi-inclusive hadroproduction of doubly bottomed tetraquarks ($X_{b\bar{b}q\bar{q}}$) as well as fully bottomed ones ($T_{4b}$), to which we collectively refer as ``bottomoniumlike'' states.
We rely upon the variable-flavor number-scheme fragmentation at leading power, where a single parton perturbatively splits into the corresponding Fock state, which then hadronizes into the color-neutral, observed tetraquark.
To this end, we build new sets of DGLAP/\textsc{HF-NRevo} consistent, hadron-structure oriented collinear fragmentation functions, which we name {\tt TQHL1.1} and {\tt TQ4Q1.1} parametriz\-ations.
They extend and supersede the corresponding {\tt 1.0} versions recently derived in previous works.
The first family describes the fragmentation of doubly heavy tetraquarks and is based on an improved version of the Suzuki model for the heavy-quark channel.
The second family depicts the fragmentation of fully heavy tetraquarks and embodies initial-scale inputs for gluon and heavy-quark channels, both of them calculated by the hands of potential nonrelativistic QCD.
As a phenomenological application, we provide novel predictions for tetraquark-plus-jet high-energy distributions, computed within the NLL/NLO$^+$ hybrid factorization from {\tt (sym)JETHAD}, at 14~TeV and 100~TeV FCC.
\vspace{0.60cm} \hrule
\vspace{0.20cm}
{
 \setlength{\parindent}{0pt}
 \textsc{Keywords}: 
 High-energy QCD, Heavy flavor,~Bottom, 
 Tetraquarks, Exotics, Fragmentation, Resummation, {\tt TQHL1.1} \& {\tt TQ4Q1.1} FF release
}
\vspace{0.30cm} \hrule
\vspace{-0.55cm}

\setcounter{tocdepth}{3}
\renewcommand{\baselinestretch}{0.875}\normalsize
\tableofcontents
\renewcommand{\baselinestretch}{1.0}\normalsize
\clearpage


\section{Introduction}
\label{sec:intro}

The true nature of exotic hadrons, such as tetraquarks and pentaquarks, remains an open question in particle physics. 
These exotic states differ from conventional mesons and baryons in that they contain more than two or three valence quarks, which suggests more complex internal structures and interactions. 

Hadronic collisions at current and next-generation colliders, such as the Large Hadron Collider (LHC), and future facilities such as the Electron-Ion Collider (EIC)~\cite{AbdulKhalek:2021gbh,Khalek:2022bzd,Hentschinski:2022xnd,Amoroso:2022eow,Abir:2023fpo,Allaire:2023fgp} and the Future Circular Collider (FCC)~\cite{FCC:2018byv,FCC:2018evy,FCC:2018vvp,FCC:2018bvk}, offer
us a powerful avenue for studying these exotic states. By producing and analyzing these particles in high-energy collisions, one can probe their internal structure, including the core configuration of quarks and gluons, and their formation mechanisms.

Recent advancements in all-order perturbative techniques and QCD factorization provide us with new theoretical tools to investigate these processes. 
These approaches allow for precise computations of cross sections and distributions that can be compared with experimental data, leading to a better understanding of the dynamics underlying exotic-matter production.

An ambitious joint program, involving progress in exotic spectroscopy and complemented by the systematic employment of precision-QCD techniques, could reveal insights on the nature of strong-force interactions at play, the role of color confinement, and the basic properties of multiquark bound states. 
Ultimately, these studies could deepen our understanding of QCD and the fundamental nature of matter.

Exotic hadrons essentially fall into two primary categories: those with active gluon content, such as quark-gluon hybrids~\cite{Kou:2005gt,Braaten:2013boa,Berwein:2015vca} and glueballs~\cite{Minkowski:1998mf,Mathieu:2008me,D0:2020tig,Csorgo:2019ewn}, and those containing multiple quarks, like tetraquarks and pentaquarks~\cite{Gell-Mann:1964ewy,Jaffe:1976ig,Ader:1981db}. 
Hybrid states involve unconventional combinations of quarks and gluons, while glueballs are composed solely of gluons. 
On the other hand, tetraquarks and pentaquarks are thought to be described in terms of three and four valence-quark leading Fock states, respectively.

The observation of the $X(3872)$ particle by the Belle experiment at KEKB in 2003~\cite{Belle:2003nnu}, then confirmed by other Collaborations (see, \emph{e.g.} Refs.~\cite{CDF:2003cab,LHCb:2013kgk,CMS:2021znk,Swanson:2006st} and references therein), marked the turn of the so-called ``Exotic-matter Revolution'' or ``Second Quarkonium Revolution''.\footnote{It chronologically followed the (First) ``Quarkonium Revolution'', characterized by the remarkable discovery of the first doubly charmed hadron, the $\Jpsi$ quarkonium~\cite{SLAC-SP-017:1974ind,E598:1974sol,Bacci:1974za}.} 
This particle, characterized as a hidden-charm state containing a charm quark ($c$) and its corresponding antiquark ($\bar{c}$), represents the first observed hidden-charm tetraquark~\cite{Chen:2016qju,Liu:2019zoy}. 
More recently, in 2021, the LHCb experiment reported the detection of the $X(2900)$, marking the first observation of an exotic state with open-charm flavor, thus broadening our understanding of the exotic-matter domain~\cite{LHCb:2020bls}.

Although $X(3872)$ has conventional quantum numbers, its decay patterns violate isospin conservation, indicating a more intricate internal structure than that of traditional quarkonium states. 
This has led to the development of several alternative dynamical models, which extend beyond the standard quarkonium framework and align with the tetraquark hypothesis.
They include:
\begin{enumerate}
 \item \textbf{\textit{Compact diquarks}}: 
 This approach suggests that the $X(3872)$ consists of a tightly bound diquark-antidiquark pair, where two quarks are bound closely together in a configuration different from traditional mesons and baryons~\cite{Maiani:2004vq,tHooft:2008rus,Maiani:2013nmn,Maiani:2014aja,Maiani:2017kyi,Mutuk:2021hmi,Wang:2013vex,Wang:2013exa,Grinstein:2024rcu}. \\
 \item \textbf{\textit{Meson molecules}}: 
 This model proposes that the $X(3872)$ is a loosely bound state formed by two mesons, akin to a molecular bond, where the mesons are weakly interacting through the residual strong force~\cite{Tornqvist:1993ng,Braaten:2003he,Guo:2013sya,Mutuk:2022ckn,Wang:2013daa,Wang:2014gwa,Esposito:2023mxw,Grinstein:2024rcu}. \\
 \item \textbf{\textit{Hadroquarkonium states}}: 
 In this scenario, the $X(3872)$ is envisioned as a quarkonium core (a compact quark-antiquark state) surrounded by an orbiting light meson, thus resembling a hadronic analogue of an atom, where the quarkonium acts as a nucleus~\cite{Dubynskiy:2008mq,Voloshin:2013dpa,Guo:2017jvc,Ferretti:2018ojb,Ferretti:2018tco,Ferretti:2020ewe}.
\end{enumerate} 

These models aim to capture the complex nature of the $X(3872)$ and of similar exotic hadrons, providing a richer understanding of their structure and interactions.
Insights into the nature of the $X(3872)$ hadron could emerge from investigations of high-multiplicity proton collisions~\cite{Esposito:2020ywk}, as well as from the application of potential models to understand its hadronic thermal behavior~\cite{Armesto:2024zad}. 

In 2021, the first doubly charmed tetraquark, $T_{cc}^+$, was observed by the LHCb Collaboration~\cite{LHCb:2021vvq,LHCb:2021auc}. 
This state is described in Refs.~\cite{Fleming:2021wmk,Dai:2023mxm,Hodges:2024awq} as a nonrelativistic molecule composed of two $D$ mesons, modeled using the XEFT nonrelativistic effective field theory~\cite{Fleming:2007rp,Fleming:2008yn,Braaten:2010mg,Fleming:2011xa,Mehen:2015efa,Braaten:2020iye}.

Until recently, $X(3872)$ was the only exotic state identified in prompt proton collisions. 
This scenario drastically changed with the discovery of the doubly charmed $T_{cc}^+$ tetraquark and the observation of a new resonance in the double $\Jpsi$ invariant-mass spectrum~\cite{LHCb:2020bwg}. 
This newly observed resonance, named $X(6900)$, is widely considered a strong candidate for either the ground state $0^{++}$ or, more likely, the radial resonance $2^{++}$ of the fully charmed tetraquark $\TQc$~\cite{Chen:2022asf}.

From a theoretical perspective, singly heavy-flavor\-ed, $\QXQq$ tetraquarks, as well as fully heavy flavor\-ed, $\TQQ$ ones, may be among the most straightforward exotic states to investigate. 
On the one hand, exploring the core structure of $\QXQq$ states gives us a faultless opportunity to directly probe the strong force \emph{via} QCD interactions among heavy and light quarks, and heavy-light intermediate subystems, like diquarks.
Being their lowest Fock state $|Q\bar{Q}q\bar{q}\rangle$, the velocities of the $Q$ and $\bar{Q}$ constituent heavy quarks in the parent-tetraquark center-of-mass system are expected to be nonrelativistic.

On the other hand, given that the heavy-quark mass $m_Q$ lies above the perturbative-QCD threshold, a $\TQQ$ hadron can be seen as a composite system of two heavily nonrelativistic charm or bottom quarks and two heavily nonrelativistic anticharm or antibottom ones. 
Its leading Fock state, $|Q\bar{Q}Q\bar{Q}\rangle$, is not influenced by valence light quarks or dynamical gluons, which makes it analogous to quarkonia where the leading state is $|Q\bar{Q}\rangle$.

This suggests that theoretical techniques used for studying quarkonia are applicable to singly and doubly heavy tetraquarks as well. 
Consequently, just as charmonia are often likened to QCD ``hydrogen atoms''~\cite{Pineda:2011dg}, $\QXQq$ and $\TQQ$ particles might be viewed as QCD ``heavier nuclei'' or ``molecules'', depending on the theoretical framework adopted~\cite{Maiani:2019cwl}.
In the present work we propose another epithet for these exotic hadrons, collectively referring to them as ``quarkoniumlike'' particles.
This clearly follows from the fact that their leading Fock states contain one ($\QXQq$) or two ($\TQQ$) $|Q\bar{Q}\rangle$ subsystems.

Although impressive advancements have been made toward a deeper and more comprehensive understanding of mass spectra and decays of exotics since the discovery of the $X(3872)$ particle, understanding their dynamical production mechanisms remains elusive. 
To date, only a few model-dependent approaches, such as those based on color evaporation~\cite{Maciula:2020wri} and hadron-quark duality~\cite{Berezhnoy:2011xy,Karliner:2016zzc,Becchi:2020mjz}, have been proposed.

Further studies have addressed the impact of multiparticle interactions on heavy-tetraquark production at hadron colliders~\cite{Carvalho:2015nqf,Abreu:2023wwg} and unveiled possible signals of high-energy dynamics for tetraquark structures~\cite{Cisek:2022uqx}. 
Additionally, research has been conducted on exclusive radiative emissions of $\TQQ$ states at bottom factories~\cite{Feng:2020qee} and $\TQQ$ photoproduction at lepton-hadron colliding machines~\cite{Feng:2023ghc}.

Concerning the bottom-tetraquark sector, our level of knowledge is at an early stage.
The observation of two charged bottomoniumlike resonances in $Y(5S)$ decay events was first reported by the BELLE Collaboration~\cite{Belle:2011aa}, this strongly suggesting that exotic mechanisms are contributing to that channel.
Nonetheless, no tetraquark states containing bottom quarks, either $|b{\bar b}b{\bar b}\rangle$ or $|b{\bar b}q{\bar q}\rangle$, have been experimentally confirmed.

The observation of a resonance with a mass of 18.15 GeV in Cu$+$Au collisions was recently made by the ANDY Collaboration at the Relativistic Heavy-Ion Collider (RHIC)~\cite{ANDY:2019bfn}.
That signal was found to be compatible with predictions of $\TQb$ masses~\cite{Vogt:2021lei}.
On the lattice side, investigations on bottom-charmed and doubly bottomed tetraquarks were recently carried out in Refs.~\cite{Francis:2018jyb,Padmanath:2023rdu} and~\cite{Bicudo:2015vta,Leskovec:2019ioa,Alexandrou:2024iwi}, respectively.

The notably large cross sections for $X(3872)$ at high transverse momenta, as observed by LHC experiments~\cite{CMS:2013fpt,ATLAS:2016kwu,LHCb:2021ten}, have significant implications for unraveling its formation dynamics. 
These results provide a unique opportunity to refine theoretical models and decipher production mechanisms intrinsically linked to high-energy QCD, such as the leading-power \emph{fragmentation} of a single parton into the detected tetraquark.

The emerging complexity in the description of the production mechanism(s) of exotic tetraquarks calls for the use of a hadron-structure oriented approach.
To this end, we will derive two new families of fragmentation functions (FFs), named {\tt TQHL1.1} and {\tt TQ4Q1.1} sets, respectively depicting the collinear fragmentation of a single parton into the given tetraquark.

The {\tt TQHL1.1} determinations portray the fragment\-ation of doubly heavy tetraquarks: $\QXcu$, $\QXcs$, $\QXbu$, and $\QXbs$.
These functions take, as initial energy-scale inputs, calculations done by the hands of an enhanced version of the Suzuki model for the constituent heavy-quark channel~\cite{Suzuki:1977km,Suzuki:1985up,Amiri:1986zv,Nejad:2021mmp}.

They supersede the corresponding {\tt 1.0} version released in our previous work~\cite{Celiberto:2023rzw} (see also Ref.~\cite{Celiberto:2024mrq} for a review) by encoding a proper treatment of the normalization and other defining parameters.

The {\tt TQ4Q1.1} determinations portray the fragmentation of  fully heavy tetraquarks: $\TQcZpp$, $\TQbZpp$, and the corresponding radial excitations: $\TQcTpp$, $\TQbTpp$.
Building on a well-suited extension of modern quarkonium theory, these functions embody, as initial inputs, nonrelativistic QCD (NRQCD)~\cite{Caswell:1985ui,Thacker:1990bm,Bodwin:1994jh,Cho:1995vh,Cho:1995ce,Leibovich:1996pa,Bodwin:2005hm} treatments for gluon and heavy-quark channels~\cite{Feng:2020riv,Bai:2024ezn}.

They represent the successors of the corresponding {\tt 1.0} sets, recently released for $\TQc$ states~\cite{Celiberto:2024mab} and partially based on NRQCD (gluon channel) and Suzuki (heavy-quark channel).
Being the experimental information on tetraquark fragmentation still very limited, we believe that both the {\tt TQ4Q1.0} FFs and their {\tt 1.1} upgrade can serve as useful guidance for explorations at the LHC as well as at new-generation colliders.

Whereas the Suzuki picture and the NRQCD effective theory serve as building blocks to model the initial-scale inputs of our FFs, an essential ingredient is still missing to get a collinear factorization consistent description.
Indeed, our functions need to evolve in energy according to the Dokshitzer--Gribov--Lipatov--Altarelli--Parisi (DGLAP) equations~\cite{Gribov:1972ri,Gribov:1972rt,Lipatov:1974qm,Altarelli:1977zs,Dokshitzer:1977sg}.

Moreover, being our tetraquarks heavy-flavored particles, a correct description of their collinear-fragment\-ation production at moderate to large transverse momenta must rely on a zero-mass variable-flavor number-scheme (ZM-VFNS, or simply VFNS) treatment~\cite{Mele:1990cw,Cacciari:1993mq,Buza:1996wv}.
Within the VFNS, all quark flavors are treated as massless particles and take part into the DGLAP evolution.
The number of flavors grows by one every time a heavy-quark threshold is crossed.
The effect of finite heavy-quark masses is eventually retained in the initial-scale input of the heavy-hadron FFs.

To properly combine the heavy-flavor, hadron-struct\-ure oriented inputs of our FFs with a VFNS DGLAP, we will take advantage of key ingredients of the newly developed heavy-flavor nonrelativistic evolution ({\HFNRevo}) setup \cite{Celiberto:2024mex,Celiberto:2024bxu}.

Finally, for our phenomenological study, we will employ the $\NLLp$ hybrid factorization scheme, which integrates the resummation of energy leading logarithms (LL), next-to-leading ones (NLL) and beyond within the standard collinear framework. 
We will employ the {\Jethad} numerical interface and the {\symJethad} symbolic calculation plugin~\cite{Celiberto:2020wpk,Celiberto:2022rfj,Celiberto:2023fzz,Celiberto:2024mrq,Celiberto:2024swu} to generate predictions for high-energy observables related to tetraquark-plus-jet tags. 
Our analysis will cover center-of-mass energies from the 14~TeV LHC to the 100~TeV nominal energy of the FCC.

The structure of this article is the following. 
Section~\ref{sec:FFs} gives us technical details on the way the novel {\tt TQHL1.1} and {\tt TQ4Q1.1} heavy-tetraquark FFs are constructed.
Sec.~\ref{sec:hybrid_factorization} provides us insight on the hybrid-factorization framework within the $\NLLp$ accuracy.  
In Sec.~\ref{sec:results} we present and discuss our predictions for rapidity-interval and transverse-momentum differential rates sensitive to the emissions of a bottom\-oniumlike hadron ($\B$ state) accompanied by a jet. 
Finally, in Sec.~\ref{sec:conclusions} we present our conclusions and offer some perspectives for future studies.

\section[From quarkonia to tetraquarks]{Heavy-flavor fragmentation: From quarkonia to tetraquarks}
\label{sec:FFs}

In this section we present our strategy to build the two {\tt TQHL1.1} and {\tt TQ4Q1.1} FF families, respectively describing the production of singly and doubly heavy-flavored tetraquarks.
For the sake of completeness, in Sec.~\ref{ssec:FFs_quarkonia} we briefly review the NRQCD-inspired fragmentation production of vector quarkonia and charmed $B$ mesons, discussing general features of the corresponding {\tt ZCW19$^+$}~\cite{Celiberto:2022dyf,Celiberto:2023fzz} and {\tt ZCFW22}~\cite{Celiberto:2022keu,Celiberto:2024omj} VFNS FF determinations.
Then, initial energy-scale inputs and DGLAP/\textsc{HF-NRevo} evolution of {\tt TQHL1.1} and {\tt TQ4Q1.1} functions are respectively highlighted in Secs.~\ref{ssec:FFs_XQq} and~\ref{ssec:FFs_T4Q}. 

All the symbolic computations needed to obtain our functions were performed through {\symJethad}, the novel \textsc{Mathematica} plugin of {\Jethad}, suited to the symbolic manipulation of analytic expressions for ha\-dron\-ic structure and high-energy QCD~\cite{Celiberto:2020wpk,Celiberto:2022rfj,Celiberto:2023fzz,Celiberto:2024mrq,Celiberto:2024swu}.

Before starting our journey into the fragmentation production of quarkonium and quarkoniumlike states, we quickly introduce some general lore about the fragmentation of partons into hadrons with heavy flavor.

In contrast to light-flavored hadrons, the fragmentation mechanisms leading to the hadronization of heavy-flavored ones contain a further level of complexity. 
This follows from the fact that the masses of heavy quarks in their lowest Fock state fall into the perturbative-QCD region. 
As a result, while FFs for light hadrons are genuinely nonperturbative, the initial-scale inputs for heavy-hadrons' ones are believed to incorporate some perturbative elements.

For heavy-light hadrons, like $D$ or $B$ mesons or $\Lambda_Q$ baryons, the initial-scale fragmentation input can be envisioned as a two-step process~\cite{Cacciari:1996wr,Cacciari:1993mq,Kniehl:2005mk,Helenius:2018uul,Helenius:2023wkn}. 
In the first step, a parton $i$, produced in a hard scattering event with large transverse momentum $|\vec \kappa| \gg m_Q$, fragments into a heavy quark $Q$ with mass $m_Q$: charm or anticharm for $D$ mesons and $\Lambda_c$ baryons, bottom or antibottom for $B$ mesons and $\Lambda_b$ baryons. 
Given that $\alpha_s(m_Q) < 1$, we can compute this step perturbatively at an initial reference scale of ${\cal O}(m_Q)$. 
Since its timescale is shorter than that of hadronization, this part is often referred to as the short-distance coefficient (SDC) of the $(i \to Q)$ fragmentation process. 
The first next-to-leading-order (NLO) calculation of SDCs for singly heavy hadrons can be found in Ref.~\cite{Mele:1990yq,Mele:1990cw}.
Related studies at next-to-NLO were performed in Refs.~\cite{Rijken:1996vr,Mitov:2006wy,Blumlein:2006rr,Melnikov:2004bm,Mitov:2004du,Biello:2024zti}.

At larger times, the constituent heavy quark $Q$ hadronizes into the physical hadron.
This second step of the fragmentation is fully nonperturbative and can be obtained \emph{via} phenomenological models~\cite{Kartvelishvili:1977pi,Bowler:1981sb,Peterson:1982ak,Andersson:1983jt,Collins:1984ms,Colangelo:1992kh} or effective field theories~\cite{Georgi:1990um,Eichten:1989zv,Grinstein:1992ss,Neubert:1993mb,Jaffe:1993ie}.

The final step in constructing a comprehensive VFNS FF set for heavy-light hadrons involves accounting for the energy evolution. 
Starting from the initial-scale nonperturbative inputs discussed earlier, and assuming that these inputs are free from scaling-violation effects, numerical methods are typically used to solve the coupled DGLAP evolution equations at a specified perturbative accuracy.

\subsection{Vector quarkonia and charmed $B$ mesons}
\label{ssec:FFs_quarkonia}

\begin{figure*}[!t]
\centering

   \hspace{-0.25cm} \includegraphics[scale=0.41,clip]{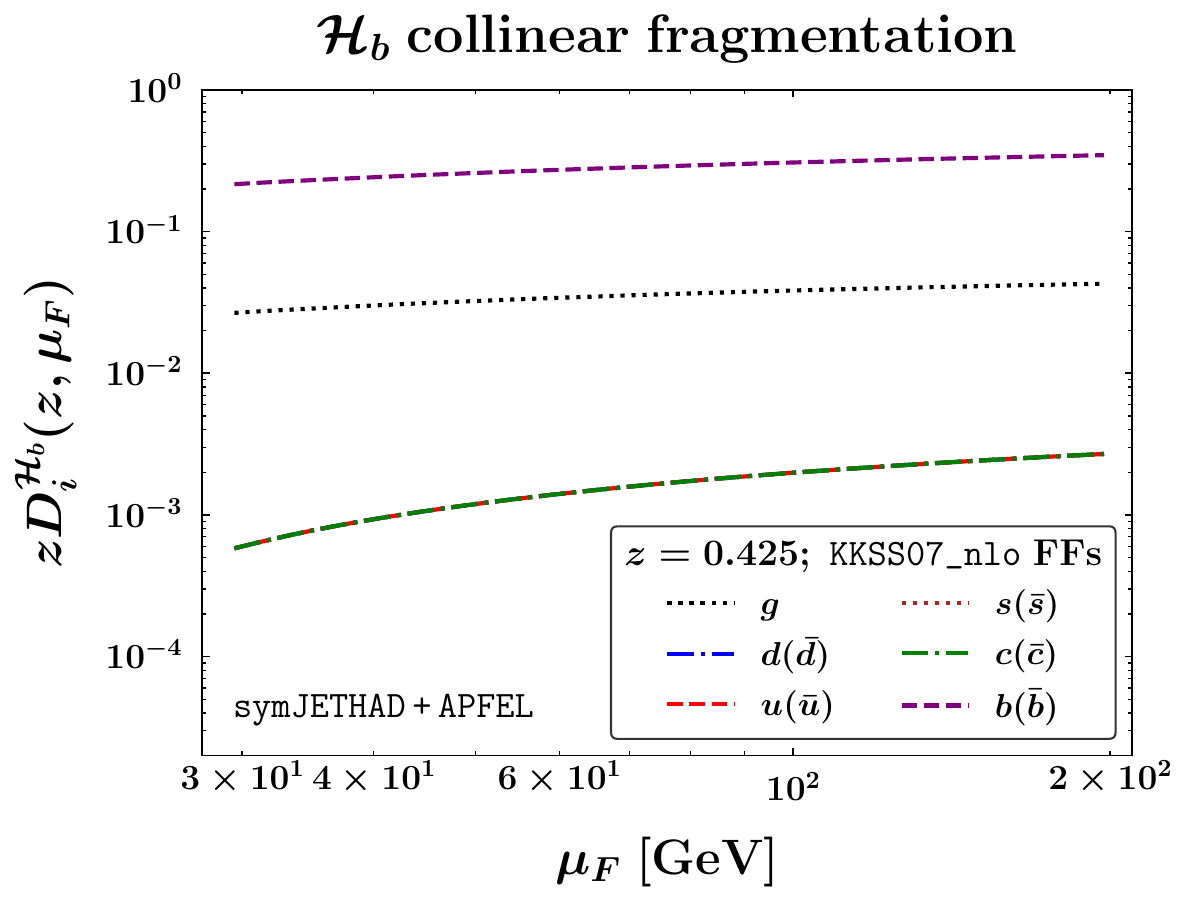}
   \includegraphics[scale=0.41,clip]{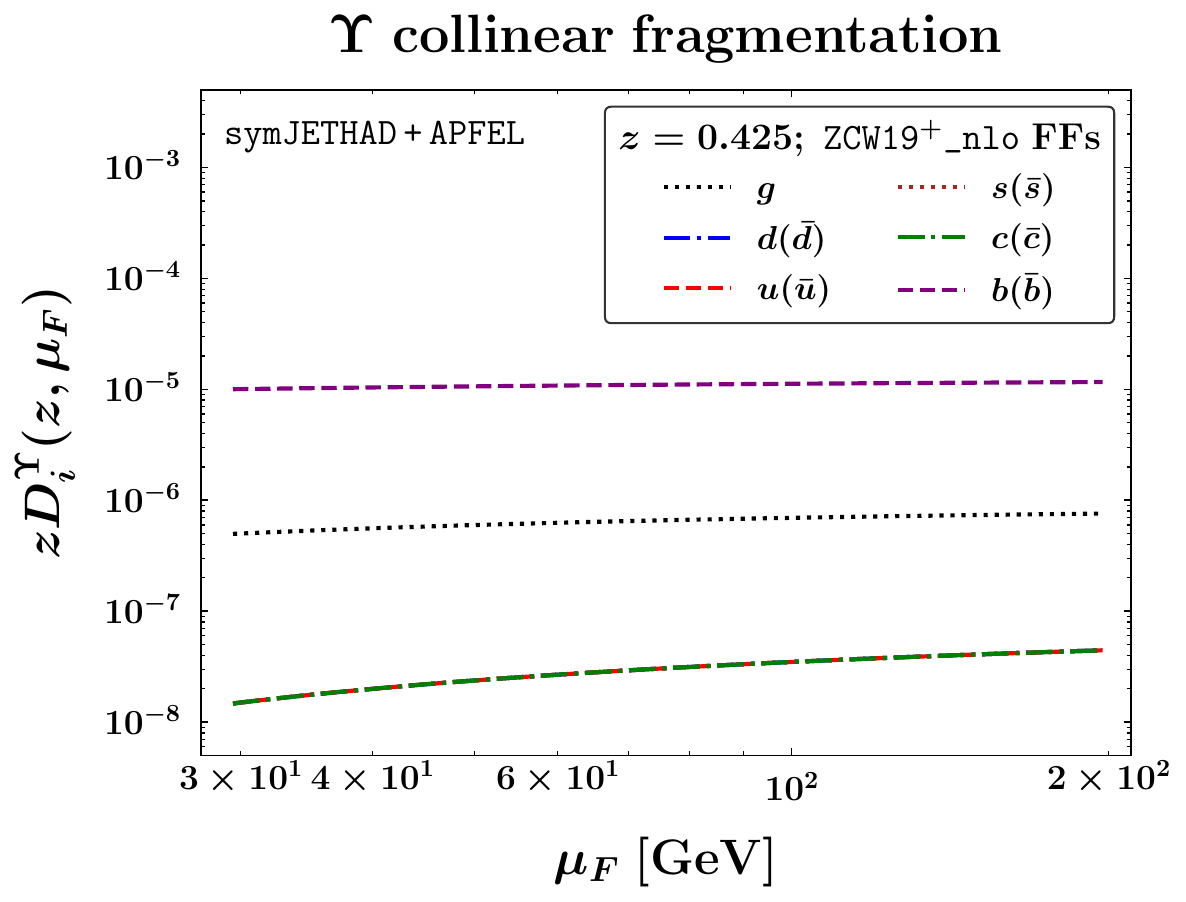}

   \vspace{0.40cm}

   \hspace{-0.25cm} \includegraphics[scale=0.41,clip]{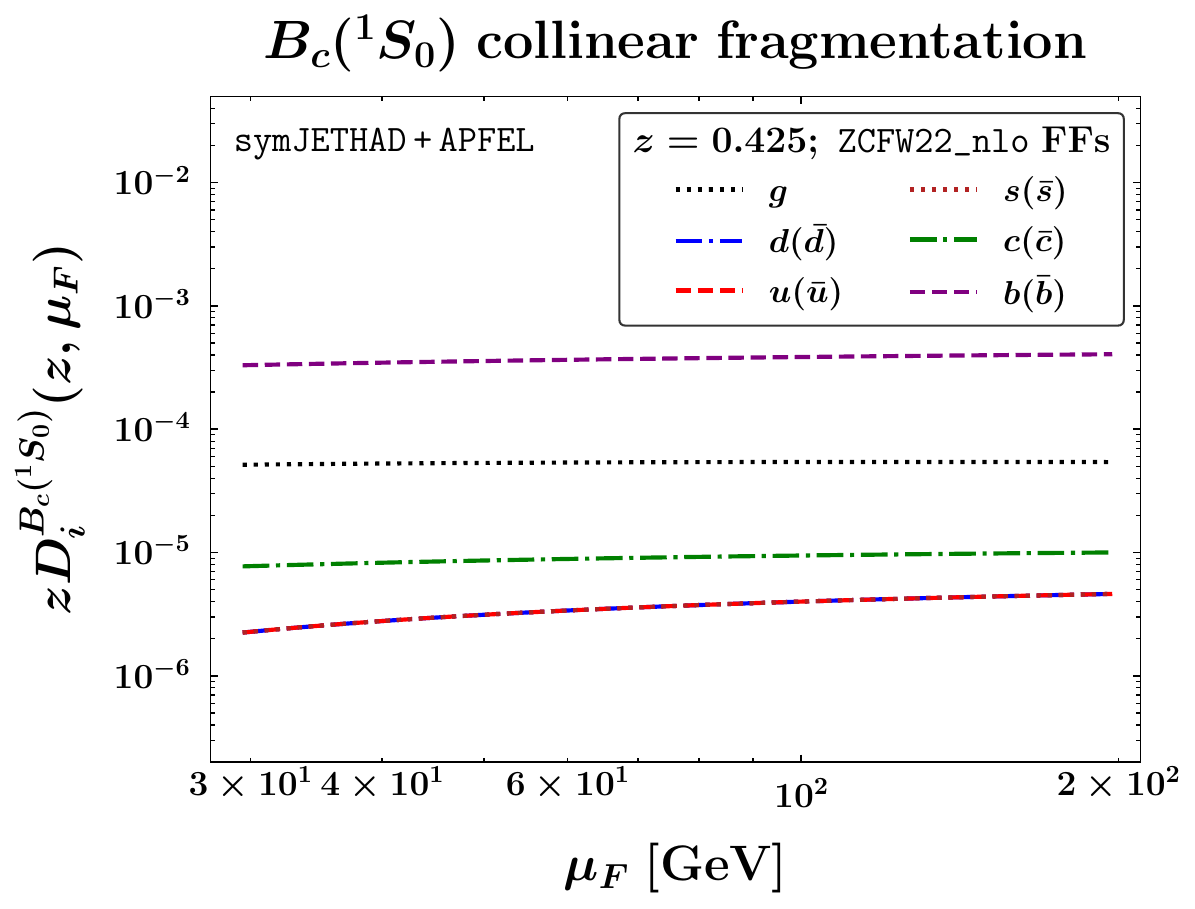}
   \includegraphics[scale=0.41,clip]{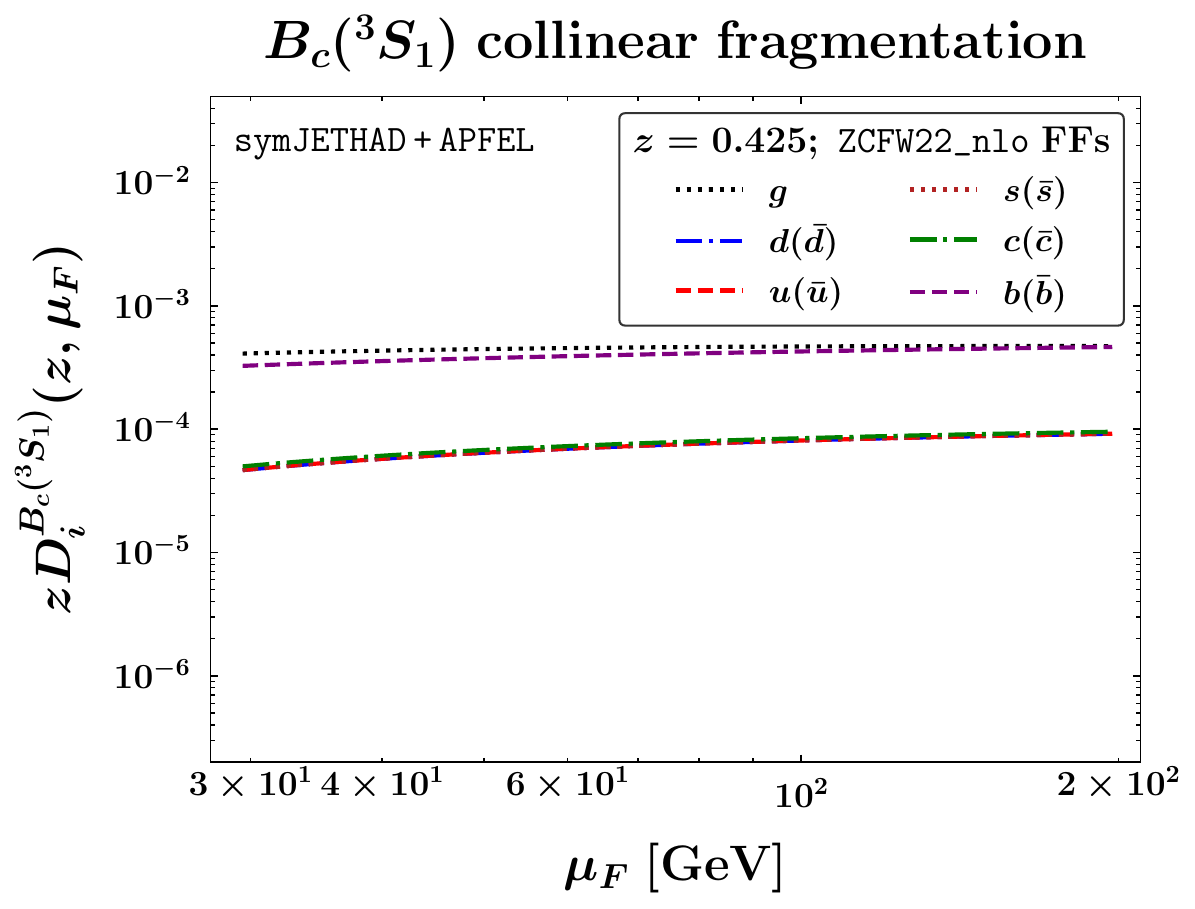}

\caption{Factorization-scale dependence of {\tt KKSS07}~\cite{Kniehl:2008zza,Kramer:2018vde}, {\tt ZCW19$^+$}~\cite{Celiberto:2022dyf,Celiberto:2023fzz}, and {\tt ZCFW22}~\cite{Celiberto:2022keu,Celiberto:2024omj} collinear FFs respectively depicting $\Hb$, $\Upsilon$, $\BCs$, and $\Bss$ particle formation at $z = 0.425 \simeq \langle z \rangle$.}
\label{fig:NLO_FFs_bHs_Jps_BCs_Bss}
\end{figure*}

Let us now focus on quarkonia, which are mesons with a lowest Fock state given by $|Q\bar{Q}\rangle$. 
The concurrent presence of a heavy quark $Q$ and its antiquark $\bar{Q}$ makes the depiction of quarkonium formation mechanisms more complex compared to that of heavy-light hadrons.
As previously mentioned, modern quarkonium theory bases on a nonrelativistic treatment of constituent heavy quarks, encapsulated in the NRQCD formalism~\cite{Caswell:1985ui,Thacker:1990bm,Bodwin:1994jh,Cho:1995vh,Cho:1995ce,Leibovich:1996pa,Bodwin:2005hm} (for a pedagogical review, we refer the reader to Refs.~\cite{Grinstein:1998xb,Kramer:2001hh,QuarkoniumWorkingGroup:2004kpm}).

NRQCD offers us a systematic approach to disengage the short-distance and long-distance dynamics in the production of quarkonia. 
By treating the heavy-quark and antiquark fields in the effective Lagrangian as nonrelativistic degrees of freedom, NRQCD allows for a consistent factorization between SDCs, which describe the perturbative production of the $(Q\bar{Q})$ system, and long-distance matrix elements (LDMEs), which capture the hadronization phase. 
LDMEs are inherently nonperturbative and must be determined from experimental data (typically from corresponding leptonic widths), estimated using potential model calculations~\cite{Eichten:1994gt}, or obtained through lattice-QCD analyses~\cite{Lepage:1992tx,Davies:1994mp}.

According to NRQCD, the physical quarkonium state reads as a linear superposition of all possible Fock states, organized by doubly expanding in the strong coupling $(\alpha_s)$ and the relative velocity ($v_{\cal Q}$) between $Q$ and $\bar Q$.
Notably, both the color-singlet dynamics~\cite{Berger:1980ni,Baier:1981uk} and the color-octet one~\cite{Bodwin:1992ye} contribute to this expansion.
The latter becomes essential to cancel divergences arising in NLO calculations of $P$-wave-quarkonium hard factors~\cite{Barbieri:1976fp,Bodwin:1992ye}.

NRQCD is based on the premise that quarkonium production begins with the \emph{short-distance} creation of a $(Q\bar{Q})$ pair through a hard-scattering process. 
This pair then hadronizes to form the final quarkonium state. 
The relative transverse separation between the quark and the antiquark is of the order of $1/\mu_E$, where $\mu_E$ represents the relevant energy scale for the process~\cite{Mangano:1995yd}. 

At large transverse momentum, where $\mu_E \sim |\vec \kappa|$, the short-distance production mechanism decreases significantly as $|\vec \kappa|$ grows. 
This decline is due to the reduced time available for the quark-antiquark pair to establish the correct color configuration, shrinking to approximately $1/|\vec \kappa|$. 
Similarly, the space in which the pair can organize is also restricted to a volume of roughly $1/|\vec \kappa|^3$, lowering the probability that the quarkonium state will form~\cite{Mangano:1995yd,Braaten:1996pv,Artoisenet:2009zwa}.

At this point, the \emph{fragmentation} mechanism becomes crucial. 
As transverse momentum increases, a single parton generated in the hard scattering gains sufficient energy to fragment into the final quarkonium state, accompanied by inclusive hadronic radiation. 
Although fragmentation usually arises at higher perturbative orders with respect to the short-distance process, it gains a significant boost due to a $(|\vec \kappa|/m_Q)^2$ factor. 
As a result, the fragmentation mechanism prevails at high energies~\cite{Braaten:1993rw,Kuhn:1981jy,Kuhn:1981jn,Cacciari:1994dr,Braaten:1994xb}. 

By extending NRQCD to the fragmentation framework, leading-order (LO) calculations for FFs of both gluon and charm quarks into a $S$-wave color-singlet charmonium were conducted in Refs.~\cite{Braaten:1993rw,Braaten:1993mp}. 
Shortly after, similar work was carried out for the $P$-wave states~\cite{Braaten:1994kd,Ma:1995ci,Yuan:1994hn}.

Notably, a nonrelativistic treatment can also be applied to charmed $B$ mesons.
The lowest Fock state of these bound states contains both the charm and the bottom quark: $|c \bar b\rangle$ for positive-charged hadrons, $|\bar c b\rangle$ for negative-charged ones.

Due to the presence of two heavy quarks, charmed $B$ mesons can be viewed as generalized quarkonium states. 
Unlike charmonia and bottomonia, however, they cannot annihilate into gluons, making them exceptionally stable with narrow decay widths~\cite{Alonso:2016oyd,Aebischer:2021eio,Aebischer:2021ilm}. 

Since top quarks are extremely short-lived and unable to hadronize, charmed $B$ mesons represent the ultimate frontier in meson spectroscopy~\cite{Ortega:2020uvc}. 
The $\BCs$ particle was first observed by the CDF Collaboration at Tevatron in 1998~\cite{CDF:1998ihx}, while the $\Bss$ resonance was detected by ATLAS in 2014~\cite{ATLAS:2014lga}.

The validity of an NRQCD-inspired approach for describing the perturbative component of the collinear fragmentation of a single parton into the observed $B_c^{(*)}$ meson, or whether it is more appropriate for modeling only the short-distance production of a charm-bottom system directly within the hard subprocess, critically depends on the transverse momenta at play.

Early phenomenological studies aimed at understanding the transition region between these two re\-gimes were primarily conducted in the context of charmonium production~\cite{Cacciari:1994dr,Roy:1994ie,Cacciari:1995yt,Cacciari:1996dg,Lansberg:2019adr}. 
These investigations revealed that the (gluon) fragmentation contribution begins to dominate at transverse momenta larger than $10 - 15$~GeV. 
A similar threshold was subsequently found for $B_c^{(*)}$ mesons~\cite{Kolodziej:1995nv}, though more recent analyses~\cite{Artoisenet:2007xi} suggest that this lower bound could be higher.

Given that a rigorous approach to the fragmentation mechanism must rely upon \emph{collinear factorization}, it is essential to establish a coherent bridge between nonrelativistic analyses and a fragmentation-correlator perspective. 
Modern advancement in heavy-flavor theory, particularly regarding quarkonium fragmentation, has been driven by recognizing NRQCD as a robust and versatile framework for modeling the initial-scale FF inputs~\cite{Kang:2011mg,Ma:2013yla,Ma:2014eja}.

This approach offers two key advantages. 
First, NRQCD permits to factorize the initial inputs into a convolution of perturbative SDCs and nonperturbative LDMEs, much like the treatment of singly heavy-flavored hadrons~\cite{Cacciari:1996wr,Cacciari:1993mq,Kniehl:2005mk}.
Additionally, NRQCD provides an efficient method for computing SDCs and offers us a clear physical interpretation of the LDMEs.
Starting from those NRQCD inputs, we can derive quarkonium VFNS FFs by switching DGLAP evolution on. 

The {\tt ZCW19$^+$}~\cite{Celiberto:2022dyf,Celiberto:2023fzz} and {\tt ZCFW22}~\cite{Celiberto:2022keu,Celiberto:2024omj} families stand as a first determination of VFNS collinear FFs for vector quarkonia [$\Jpsi$ or $\Yps(1S)$] and charmed $B$ mesons [$\BCs$ or $\Bss$] based on NLO NRQCD inputs for gluon and constituent heavy-quark fragmentation channels~\cite{Braaten:1993rw,Chang:1992bb,Braaten:1993jn,Ma:1994zt,Zheng:2019gnb,Zheng:2021sdo,Feng:2021qjm,Feng:2018ulg}.

For the sake of illustration, we show in Fig.~\ref{fig:NLO_FFs_bHs_Jps_BCs_Bss} the energy-scale dependence of {\tt ZCW19$^+$} FFs for $\Yps(1S)$ and {\tt ZCFW22} FFs for $\BCs$ and $\Bss$, and we compare them with {\tt KKSS07} determinations~\cite{Kniehl:2008zza,Kramer:2018vde} for $\Hb$ hadrons.
These latter represent an inclusive combination of singly bottomed hadrons, comprising noncharmed $B$ mesons and $\Lambda_b$ baryons.
To be concise, here we focus on a single value of the momentum fraction, specifically $z = 0.425$, which approximately corresponds to $\langle z \rangle$. This value represents the average range at which FFs are typically evaluated in semihard final states (see, \emph{e.g.}, Refs.~\cite{Celiberto:2016hae,Celiberto:2017ptm,Celiberto:2020wpk,Celiberto:2021dzy,Celiberto:2021fdp,Celiberto:2022dyf,Celiberto:2022keu,Celiberto:2022kxx,Celiberto:2024omj}).

As explained in previous studies (see, \emph{e.g.}, Refs.~\cite{Celiberto:2022grc,Celiberto:2021dzy,Celiberto:2021fdp,Celiberto:2022dyf,Celiberto:2022keu,Celiberto:2023rzw,Celiberto:2024omj}), the gluon collinear FF plays a key role in the description of semi-inclusive emissions of hadrons in high-energy proton scatterings. 
Notably, the gluon fragmentation channel controls the stability of high-energy resummed distributions under radiative corrections and missing higher-order uncertainties (MHOUs). 
Its dependence on $\mu_F$ significantly influences the behavior of the high-energy logarithmic series in our distributions.  

In the semihard regime, matter of  phenomenological applications of this work (see Sec.~\ref{sec:results}), proton PDFs are typically accessed in the range $10^{-4} \lesssim x \lesssim 10^{-2}$, where the gluon PDF dominates over the quark ones. 
Since the gluon FF combines diagonally with the gluon PDF in the LO partonic hard factors, its contribution to the cross section is amplified. This dominance persists even at NLO, where $(qg)$ and $(gq)$ nondiagonal channels also contribute~\cite{Celiberto:2021dzy,Celiberto:2021fdp}.

Those studies brought a corroborating evidence that gluon FFs exhibiting a nondecreasing, smooth behavior with $\mu_F$ act as a ``stabilizer'' for high-energy resummed cross sections sensitive to semi-inclusive emissions of hadrons.
This remarkable property, which holds for singly~\cite{Celiberto:2021dzy,Celiberto:2021fdp,Celiberto:2022zdg} as well as multiply heavy flavored~\cite{Celiberto:2022dyf,Celiberto:2022keu,Celiberto:2023rzw,Celiberto:2024mab} bound states
takes the name of \emph{natural stability} of the high-energy resummation~\cite{Celiberto:2022grc}.

Beyond the phenomenology of the high-energy resummation sector, FFs for (generalized) quarkonium states serve as valuable tools for precision studies of collinear physics and hadronization. A notable example can be found in a recent result from Ref.~\cite{Celiberto:2024omj}.
In that study, analyses of rapidity and transverse-momentum distributions of charmed $B$ mesons, described by the {\tt ZCFW22} FF determinations, confirmed the LHCb Collaboration's estimate~\cite{LHCb:2014iah,LHCb:2016qpe} that the production-rate hierarchy between $\BCs$ mesons and singly bottomed $B$ mesons does not exceed 0.1

This result provided a benchmark for the VFNS fragmentation scheme applied to these particles at large transverse momentum. 
Moreover, it reinforced the reliability of employing NRQCD initial-scale FF inputs at leading power, subsequently evolved \emph{via} DGLAP.

From explorations conducted using vector-quark\-onium {\tt ZCW19$^+$} FFs and $B_c^{(*)}$-meson {\tt ZCFW22} ones, it became evident that a consistent scheme is needed to integrate NRQCD inputs with collinear factorization, ensuring the proper definition of DGLAP evolution thresholds across all parton fragmentation channels. 

To address this requirement, the novel \emph{heavy-flavor nonrelativistic evolution} ({\HFNRevo}) method~\cite{Celiberto:2024mex,Celiberto:2024bxu} was developed, specifically designed to describe the DGLAP evolution of heavy-hadron fragmentation from nonrelativistic inputs.
{\HFNRevo} bases on three key building blocks: \emph{interpretation}, \emph{evolution}, and \emph{uncertainties}.

The first building block enables the interpretation of the short-distance mechanism, dominant at low transverse momentum, as a fixed-flavor number-scheme (FFNS, see Ref.~\cite{Alekhin:2009ni} for insights) two-parton fragmentation process, which goes beyond the leading-power approximation (see Ref.~\cite{Fleming:2012wy} for a discussion). 

This statement is reinforced by the observation that incorporating transverse-momentum dependence reveals distinct singularity behaviors, especially in the transition regions between low-$|\vec \kappa|$ shape functions~\cite{Echevarria:2019ynx} and large-$|\vec \kappa|$ FFs~\cite{Boer:2023zit}. 

Then, the DGLAP evolution of {\HFNRevo} FFs can be separated in two steps.
On one side, an \emph{expanded} and semianalytic \emph{decoupled} evolution ({\tt EDevo}) properly accounts for the evolution thresholds of all the parton channels.
On the other side, the standard \emph{all-order} evolution ({\tt AOevo}) is numerically switched on.

Finally, the third building block addresses the evaluation of MHOUs arising from scale variations associated with the evolution thresholds.  
For our exploratory study on the collinear fragmentation of doubly and fully heavy tetraquarks, we defer the investigation of the first and the second aspects to future research, focusing instead on evolution.  

Shifting our focus now to the exotic sector, recent findings suggest that NRQCD factorization can be applied to investigate the true nature of double $\Jpsi$ excitations~\cite{LHCb:2020bwg,ATLAS:2023bft,CMS:2023owd}, offering us an interpretation of these states as fully charmed tetraquarks~\cite{Zhang:2020hoh,Zhu:2020xni}.
In this framework, the formation of a $\TQc$ state begins with the short-distance emission of two charm and two anticharm quarks, occurring at a scale of approximately $1/m_c$. Asymptotic freedom then allows the fragmentation process to be described as a two-step convolution, involving a short-distance phase followed by a long-distance component.

The first calculation of the NRQCD initial-scale input for the $[g \to \TQc]$ color-singlet $S$-wave fragmentation channel was presented in Ref.~\cite{Feng:2020riv}. 
Then, in our recent work on the determination of {\tt TQ4Q1.0} FF sets, the initial-scale input for the $[c \to \TQc]$ channel was modeled by adapting the Suzuki-model-inspired calculation~\cite{Suzuki:1977km,Suzuki:1985up,Amiri:1986zv}, recently utilized to describe the fragmentation of heavy-light $\XQq$ states~\cite{Nejad:2021mmp}. Building on this approach, the first determination of VFNS FFs for such heavy-light tetraquarks, referred to as the {\tt TQHL1.0} functions, was introduced in Ref.~\cite{Celiberto:2023rzw}.

In Secs.~\ref{ssec:FFs_XQq} and~\ref{ssec:FFs_T4Q} we will describe our methodology to derive the new {\tt TQHL1.1} and {\tt TQ4Q1.1} functions, respectively.
The {\tt TQHL1.1} family contains FF sets for four distinct doubly heavy states: $\QXcu$, $\QXcs$, $\QXbu$, and $\QXbs$.
It takes, as initial-scale input, an enhanced calculation of the Suzuki-driven function obtained in Ref.~\cite{Nejad:2021mmp}.
The {\tt TQ4Q1.1} family contains FF sets for the fully heavy states, $\TQcZpp$ and $\TQbZpp$, plus their radial resonances, $\TQcTpp$ and $\TQbTpp$.
It takes, as initial-scale input, the NRQCD calculation for the $[g \to \TQc]$ channel~\cite{Feng:2020riv}, suitably adapted also to the $[g \to \TQb]$ case~\cite{Feng:2023agq}, plus the NRQCD model for the $[c \to \TQc]$ and $[b \to \TQb]$ channels, recently presented in Ref.~\cite{Bai:2024ezn}.
Both {\tt TQHL1.1} and {\tt TQ4Q1.1} FF families rely upon a threshold-consistent DGLAP evolution that incorporates fundamental aspects of the {\HFNRevo} methodology.

\subsection{Doubly heavy tetraquarks}
\label{ssec:FFs_XQq}

Our approach to modeling the initial-scale input for the constituent heavy-quark fragmentation into a color-singlet $S$-wave $\QXQq$ tetraquark bases on a calculation originally introduced in Ref.~\cite{Nejad:2021mmp}. 
This method leverages a spin-physics-inspired Suzuki framework~\cite{Suzuki:1977km,Suzuki:1985up,Amiri:1986zv} that incorporates transverse-momentum dependence. 
The collinear limit is retrieved by neglecting the relative motion of the constituent quarks within the bound state~\cite{Lepage:1980fj,Brodsky:1985cr,Amiri:1986zv}.
Calculations done in Refs.~\cite{Suzuki:1977km,Suzuki:1985up,Amiri:1986zv,Nejad:2021mmp} served as a basis for modeling the initial-scale inputs of $[c \to \QXcu]$ and $[b \to \QXbs]$ fragmentation channels encoded in our pioneering {\tt TQHL1.0} set~\cite{Celiberto:2023rzw}.
The release of $[c \to \QXcs]$ and $[b \to \QXbu]$ {\tt TQHL1.0} functions soon followed~\cite{Celiberto:2024mrq}.

\begin{figure*}[!t]
\centering
\includegraphics[width=0.475\textwidth]{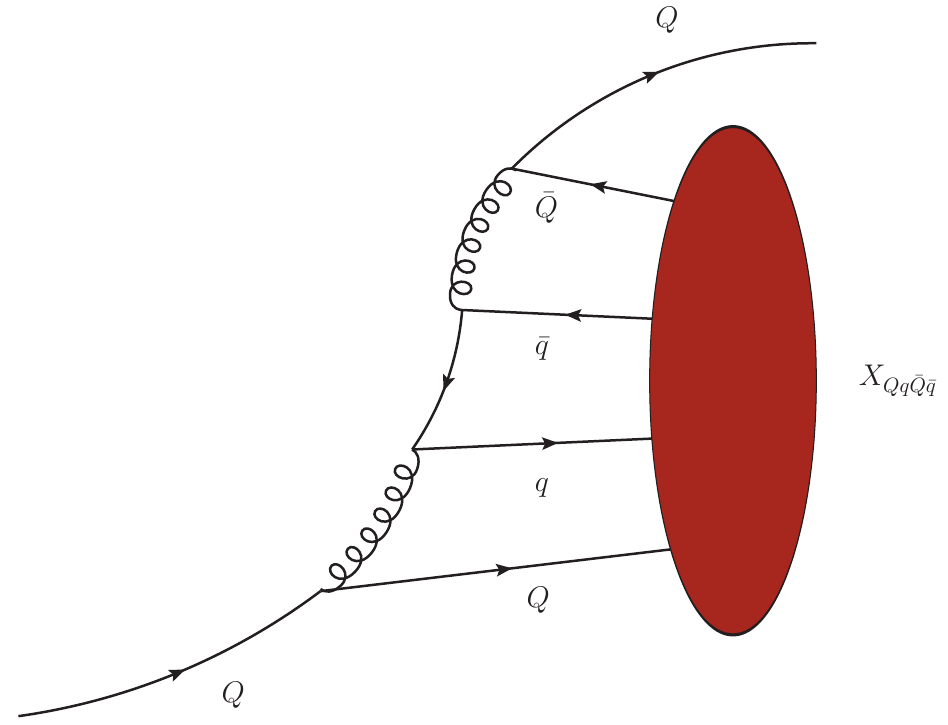}

\caption{LO representative diagram for the collinear fragmentation of a constituent heavy quark into a color-singlet $S$-wave doubly heavy tetraquark.
The firebrick blob represents the nonperturbative hadronization part of the corresponding FF.}
\label{fig:XQq_FF_diagrams}
\end{figure*}

The proposed approach parallels the factorization structure of NRQCD, where the $(Q\bar{Q})$ pair is generated perturbatively, with the subsequent hadronization described by the corresponding LDMEs. Similarly, in this framework, a $(Q\bar{Q}q\bar{q})$ system is first produced perturbatively through above-threshold splittings of the outgoing heavy (anti)quark, as illustrated in Fig.~\ref{fig:XQq_FF_diagrams}. 
The production amplitude is then convoluted with a bound-state wave function, representing the nonperturbative tetraquark hadronization-dynamics, following Suzuki's prescription.

Assuming full symmetry between $Q$ and $\bar{Q}$ fragmentation channels, the explicit form of the $[Q \to \QXQq]$ initial-scale {\tt TQHL1.1} FF is given by
\begin{equation}
\begin{split}
 \label{XQq_FF_initial-scale_Q}
 D^{\QXQq}_Q(z,\mu_{F,0}) \,&=\,
 {\cal N}_{X}^{(Q)} \,
 \frac{(1 - z)^5}{\left[\Xi^{(Q,q)}](z)\right]^2} \\
 &\,\times\,
 \sum\limits_{l=0}^3 \, z^{2(l+2)} \rho_l^{(Q,q)}(z) \left\{ \frac{\vqTTa}{m_Q^2} \right\}^{\!l}
 ,
\end{split}
\end{equation}
where
\begin{equation}
 \label{XQq_FF_initial-scale_Q_N}
 {\cal N}_{X}^{(Q)} \, = \, \left\{ 128 \, \pi^2 \, f_{\cal B} \, C_F \big[ \alpha_s(2\mu_X+m_Q) \big]^2 \right\}^2
 \,,
\end{equation}
and $\mu_X = m_Q \,+\, m_q$.
Here, $f_{\cal B} = 0.25$~GeV is the hadron decay constant~\cite{ParticleDataGroup:2020ssz}, $C_F = (N_c^2-1)/(2N_c)$ the Casimir constant related with the emission of a gluon from a quark, $m_Q = m_c = 1.5$~GeV ($m_Q = m_b = 4.9$~GeV) the charm (bottom) quark mass, $m_q = m_{u,d,s}$ the light quark mass.
Furthermore, we define
\begin{equation}
 \label{XQq_FF_initial-scale_Q_Xi}
 \Xi^{(Q,q)}(z) \,=\,
 \frac{{\cal F}_X^{(Q,q)}(z)}
 {\mu_X^2 \, m_Q^5 \, m_q}
 \;,
\end{equation}
and 
\begin{equation}
\begin{split}
 \label{XQq_FF_initial-scale_Q_F}
 {\cal F}_X^{(Q,q)}&(z) \,=\, \left\{ [ (2-z)\,m_Q + 2\,m_q ]^2 + z^2 \vqTTa \right\} \\[0.10cm]
 \,&\times\,
 \left\{ [ (2-z)\,m_Q + (1-z)\,m_q ]^2 + z^2 \vqTTa \right\}^{2} \\[0.15cm]
 \,&\times\,
 \left\{ 4 (1-z) \mu_X^2 + z^2 (m_Q^2 + \vqTTa) \right\}
 \,.
\end{split}
\end{equation}
The $\rho_l^{(Q,q)}(z)$ coefficients in Eq.~\eqref{XQq_FF_initial-scale_Q} read
\begin{equation}
\begin{split}
 \label{XQq_FF_initial-scale_Q_rho_0}
 &\rho_0^{(Q,q)}(z) \,=\,
 32\, {\cal R}_{q/Q}^6\, (2 - 3 z + z^3) \\ 
 &\,-\, 64\, {\cal R}_{q/Q}^5\, (6 - 10 z + 3 z^2 + z^3) \\ 
 &\,+\, 8\, {\cal R}_{q/Q}^4\, (-120 + 236 z - 158 z^2 \\
 &\,+\, 42 z^3 - 6 z^4 + z^5) \\ 
 &\,+\, 16\, {\cal R}_{q/Q}^3\, (-80 + 192 z - 196 z^2 \\
 &\,+\, 98 z^3 - 24 z^4 + 3 z^5) \\ 
 &\,-\, 2\, {\cal R}_{q/Q}^2\, (480 - 1424 z + 1872 z^2 \\ 
 &\,-\, 1232 z^3 + 422 z^4 - 71 z^5 + 6 z^6) \\
 &\,-\, 4\, {\cal R}_{q/Q}\, (96 - 352 z + 544 z^2 \\ 
 &\,-\, 424 z^3 + 182 z^4 - 40 z^5 + 3 z^6) \\
 &\,-\, (4 - 2 z + z^2) (4 - 8 z + 3 z^2)^2 \;,
\end{split}
\end{equation}
\begin{equation}
\begin{split}
 \label{XQq_FF_initial-scale_Q_rho_1}
 &\rho_1^{(Q,q)}(z) \,=\, 
 -8\, {\cal R}_{q/Q}^4\, (-6 + 6 z - 6 z^2 + z^3) \\ 
 &\,-\, 
 16\, {\cal R}_{q/Q}^3\, (-12 + 14 z - 12 z^2 + 3 z^3) \\
 &\,+\, 
 4\, {\cal R}_{q/Q}^2\, (72 - 112 z + 102 z^2 - 37 z^3 + 6 z^4) \\ 
 &\,+\, 
 8\, {\cal R}_{q/Q}\, (24 - 52 z + 54 z^2 - 22 z^3 + 3 z^4) \\ 
 &\,+\, 48 - 144 z + 168 z^2 - 84 z^3 + 19 z^4 \;,
\end{split}
\end{equation}
\begin{equation}
\begin{split}
 \label{XQq_FF_initial-scale_Q_rho_2}
 &\rho_2^{(Q,q)}(z) \,=\, 
 6\, {\cal R}_{q/Q}^2\, (2 - z + 2 z^2) \\ 
 &\,+\, 4\, {\cal R}_{q/Q}\, (6 - 4 z + 3 z^2) \\ 
 &\,+\, 12 - 18 z + 11 z^2 \;,
\end{split}
\end{equation}
and
\begin{equation}
\begin{split}
 \label{XQq_FF_initial-scale_Q_rho_3}
 \rho_3^{(Q,q)}(z) \,=\, 1 \;,
\end{split}
\end{equation}
where ${\cal R}_{q/Q} \equiv m_q/m_Q$\,.

Our $[Q \to \QXQq]$ initial-scale {\tt TQHL1.1} FF basically differs from the one originally introduced in Ref.~\cite{Nejad:2021mmp}, and thus from its {\tt TQHL1.0} counterpart~\cite{Celiberto:2023rzw,Celiberto:2024mrq}, in two aspects.
First, in that work the ${\cal N}_{X}^{(Q)}$ factor in Eq.~\ref{XQq_FF_initial-scale_Q_N} was not calculated, but fixed \emph{via} certain normalization conditions.
On the other hand, the selection of the $\vqTTa$ parameter in Eqs.~\eqref{XQq_FF_initial-scale_Q} and~\eqref{XQq_FF_initial-scale_Q_F} warrants closer examination. 

As mentioned earlier, the original approach proposed by Suzuki effectively integrates spin correlations and serves as a model for transverse-momentum dependent (TMD) FFs~\cite{Suzuki:1977km,Suzuki:1985up,Amiri:1986zv}. 
To achieve the collinear limit, rather than integrating over the squared modulus of the transverse momentum of the outgoing charm quark, one can substitute it with its average value, $\vqTTa$. 
This renders $\vqTTa$ a free parameter, which then must be determined through phenomenologically motivated criteria. 
As discussed in Ref.~\cite{GomshiNobary:1994eq}, increasing values of $\vqTTa$ progressively shift the peak of the FF toward the low-$z$ limit, making at the same time its bulk smaller and smaller. 

The $[Q \to \QXQq]$ FF presented in Ref.~\cite{Nejad:2021mmp} was obtained by setting $\vqTTa = 1\mbox{ GeV}^2$, which represents an upper-bound estimate for the average squared transverse momentum.
Here, we propose an improvement in choosing the value of the $\vqTTa$ parameter, which reflects a balanced and reasonable assumption, consistent with the exploratory nature of our study.
It takes inspiration from a first adjustment derived in our previous study on the collinear fragmentation of $\TQc$ states~\cite{Celiberto:2024mab}.

There, the initial-scale input of the charm FF channel was modeled by suitably adapting the original calculation in~\cite{Nejad:2021mmp} to a fully charmed state, $\TQc$.
Besides enhancing the treatment of the normalization (see Eq.~(14)), in that work we 
relied upon phenomenological indications coming from the fragmentation production of hadrons in proton collisions.
In particular, we noted that (heavy-quark) FFs of both light-flavored species~\cite{Celiberto:2016hae,Celiberto:2017ptm,Bolognino:2018oth,Celiberto:2020wpk} and heavy-flavored ones~\cite{Celiberto:2021dzy,Celiberto:2021fdp,Celiberto:2022dyf,Celiberto:2022keu} are typically probed at an average value of longitudinal fraction almost always larger than $\langle z \rangle > 0.4$.

Furthermore, we required constituent-quark FFs to be of roughly the same order of magnitude as corresponding gluon ones. 
This assumption is supported by an analogy with the simplest quarkonium case: a scalar, color-singlet $S$-wave charmonium, namely the $\eta_c$ meson. 
The production of $\eta_c$ via fragmentation can be described using NRQCD, where the SDCs have been computed up to $\mathcal{O}(\alpha_s^3)$ for gluon~\cite{Braaten:1993rw,Artoisenet:2014lpa,Zhang:2018mlo}, charm~\cite{Braaten:1993mp,Zheng:2021ylc}, and nonconstituent-quark~\cite{Zheng:2021mqr} channels. 
Notably, in the range $z > 0.4$, both the LO gluon~\cite{Braaten:1993rw} and charm~\cite{Braaten:1993mp} fragmentation are of the same order of magnitude.

From a numeric scan, we found that, for a $[c \to \TQc]$ initial-scale FF, fixing $\vqTTa \equiv \vqTTa_{\TQQ} = 70\mbox{ GeV}^2$ leads to $\langle z \rangle \gtrsim 0.4$.
In line with our requirements, that choice also makes the $[c \to \TQc]$ channels of the same order as corresponding $[g \to \TQc]$ ones taken from potential NRQCD (see Sec.~2.3 of Ref.~\cite{Celiberto:2024mab} for more details).

Coming back to doubly heavy-flavored tetraquarks, there is no currently available calculation for $[g \to \QXQq]$ initial-scale functions, unfortunately.
Therefore, to set the value of the $\vqTTa$ parameter, we can only rely upon its correlation with the peak position of the $[Q \to \QXQq]$ initial FFs.
Analogously to what we have done for the $\vqTTa_{\TQQ}$ parameter, we performed a numeric scan over the $\vqTTa \equiv \vqTTa_{\QXQq}$ range to fix its value to $\vqTTa_{\QXQq} = 4 \mbox{ GeV}^2$, so that
\begin{equation}
 \label{eq:vqTTa_XQq}
 \sqrt{\vqTTa_{\QXQq}} \simeq \frac{\sqrt{\vqTTa_{\TQQ}}}{2} \;.
\end{equation}

Beyond the heuristic nature of the previous relation, there exists a deeper rationale supporting our choice.
As it was observed in seminal studies on heavy-flavor fragmentation~\cite{Suzuki:1977km,Bjorken:1977md,Kinoshita:1985mh,Peterson:1982ak}, heavy-quark FFs peak in the large-$z$ region, while binding effects scale proportionally to the heavy-quark mass.\footnote{Because of the peak at large hadron energy fractions, heavy-quark FFs are very sensitive to soft, \emph{threshold} logarithms, which need to be resummed as well. For advancements on this topic, we refer the reader to Refs.~\cite{Mele:1990yq,Mele:1990cw,Cacciari:2001cw,Fickinger:2016rfd,Ridolfi:2019bch,Maltoni:2022bpy,Czakon:2021ohs,Czakon:2022pyz,Generet:2023vte,Aglietti:2007bp,Aglietti:2022rcm,Gaggero:2022hmv,Ghira:2023bxr,Bonino:2023icn,Cacciari:2024kaa}.}
To explain this feature, let us take as an example the fragmentation production of a singly heavy-flavored meson ${\cal D}_{Q,q}$, with lowest Fock state $|Q{\bar q}\rangle$, momentum $\kappa$, and mass $m$.

In this scenario, it is essential for the constituent heavy quark and the light antiquark to share approximately the same velocity, $v \equiv v_Q \simeq v_q$.
Consequently, their momenta are expressed as $\kappa_Q \equiv z \kappa = m_Q v$ for the heavy quark, and $\kappa_q = \Lambda_q v$ for the light antiquark, where $\Lambda_q$ stands for a hadronic mass scale of the order of $\LQCD$.
Since $m \approx m_Q$ for a heavy-light meson, we can write $m_Q v \approx \kappa = \kappa_Q + \kappa_q = z m_Q v + \Lambda_q v$. 
This leads to the relation $\langle z \rangle_Q \approx 1 - \Lambda_q/m_Q$, the `$Q$' subscript denoting the $[Q \to {\cal D}_{Q,q}]$ collinear-fragmentation channel.

As discussed in Ref.~\cite{Celiberto:2024mab}, this feature may not necessarily apply to fully heavy-flavored states, like quarkonia and $\TQQ$ particles. 
In these cases, there is no soft scale since the lowest Fock state does not involve light constituent quarks. 
Concerning $\TQQ$ tetraquarks, the interactions among the four constituent heavy quarks are expected to complicate the prediction of the FF peak position, making it unlikely to be determined solely from kinematic-based statements.

Conversely, the fragmentation of a doubly heavy tetra\-quark, $\QXQq$, represents an intriguing intermediate situation.
In this case, the simultaneous presence of a heavy-quark pair ($Q\bar{Q}$) and a light-quark pair ($q\bar{q}$) in the lowest Fock state makes it reasonable to expect the $[Q \to \QXQq]$ FFs to be peaked in a moderate to large $z$-range.

\subsubsection{Initial energy-scale inputs}
\label{sssec:FFs_XQq_Q}

For the sake of illustration, in this section we present the $z$-dependence of $[Q \to \QXQq]$ initial-scale inputs for our {\tt TQHL1.1} FFs.
To provide a first determination of uncertainties associated to our functions around their lowest energy value, we follow a procedure similar to the one adopted in our previous study on $\TQc$ states~\cite{Celiberto:2024mab}.
There, we benchmarked the $[g \to \TQc]$ initial-scale inputs of the {\tt TQ4Q1.0} functions against the original analysis in Ref.~\cite{Feng:2020riv} by performing a simplified, expanded DGLAP evolution encoding just the gluon-to-gluon timelike splitting kernel, $P_{gg}$ (for more details, see~Section 2.2 of~\cite{Celiberto:2024mab}).

\begin{figure*}[!t]
\centering

\includegraphics[scale=0.46,clip]{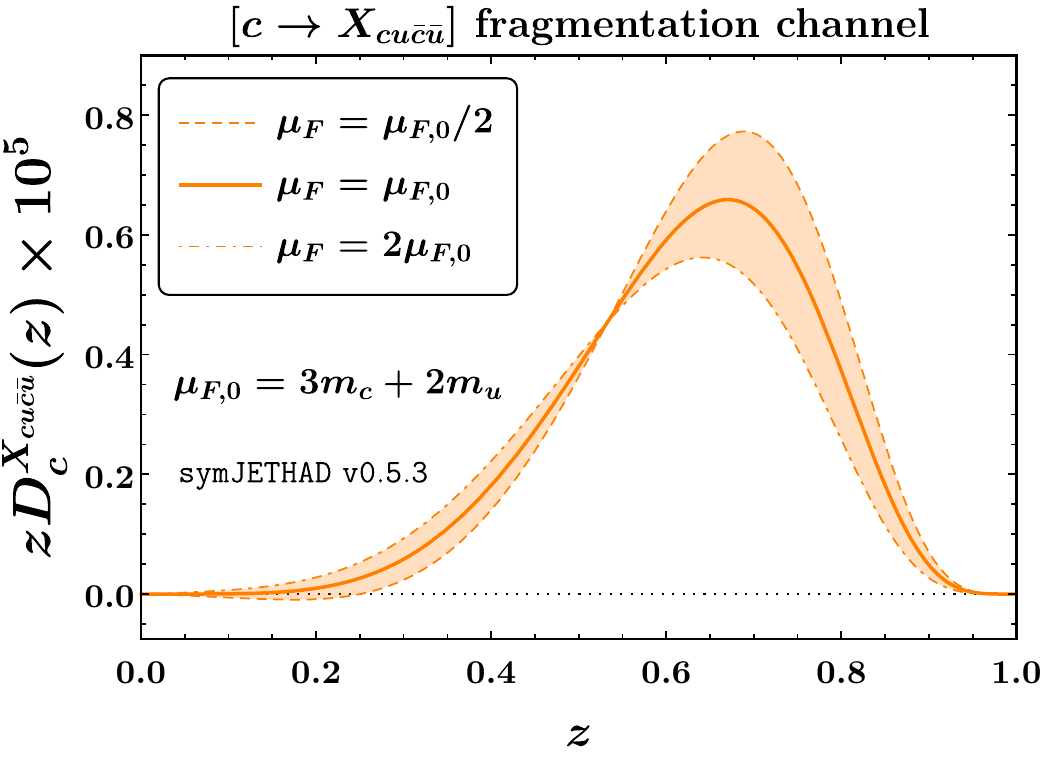}
\hspace{0.50cm}
\includegraphics[scale=0.46,clip]{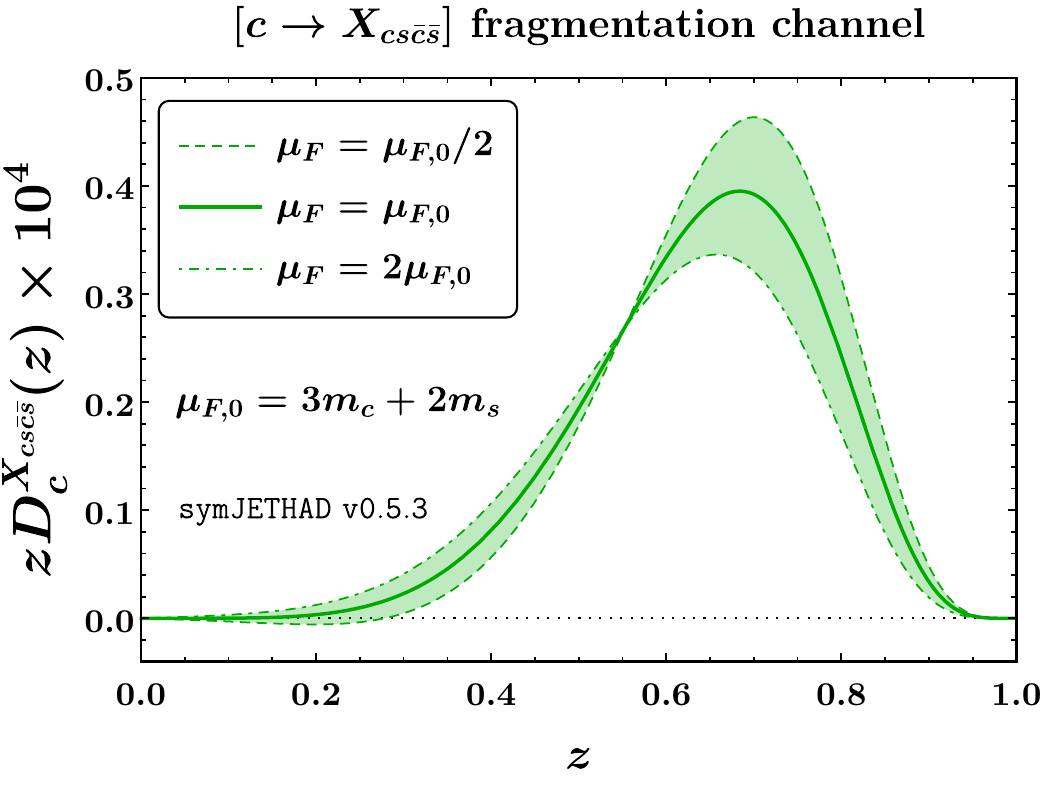}

\vspace{0.25cm}

\includegraphics[scale=0.46,clip]{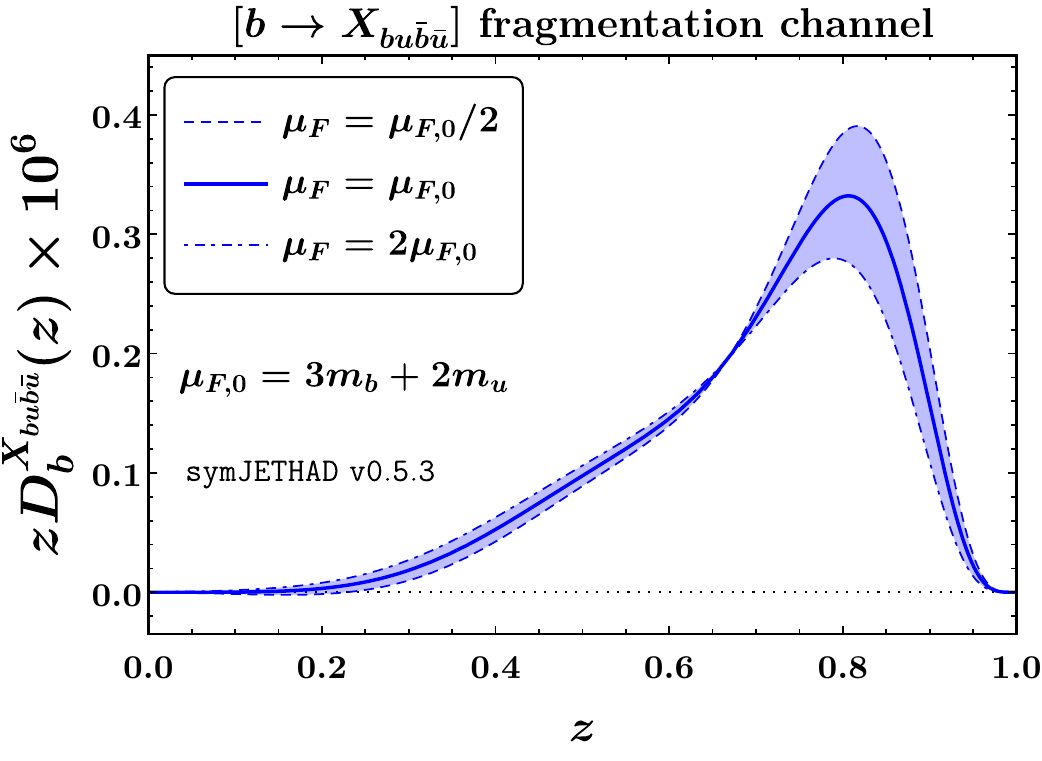}
\hspace{0.50cm}
\includegraphics[scale=0.46,clip]{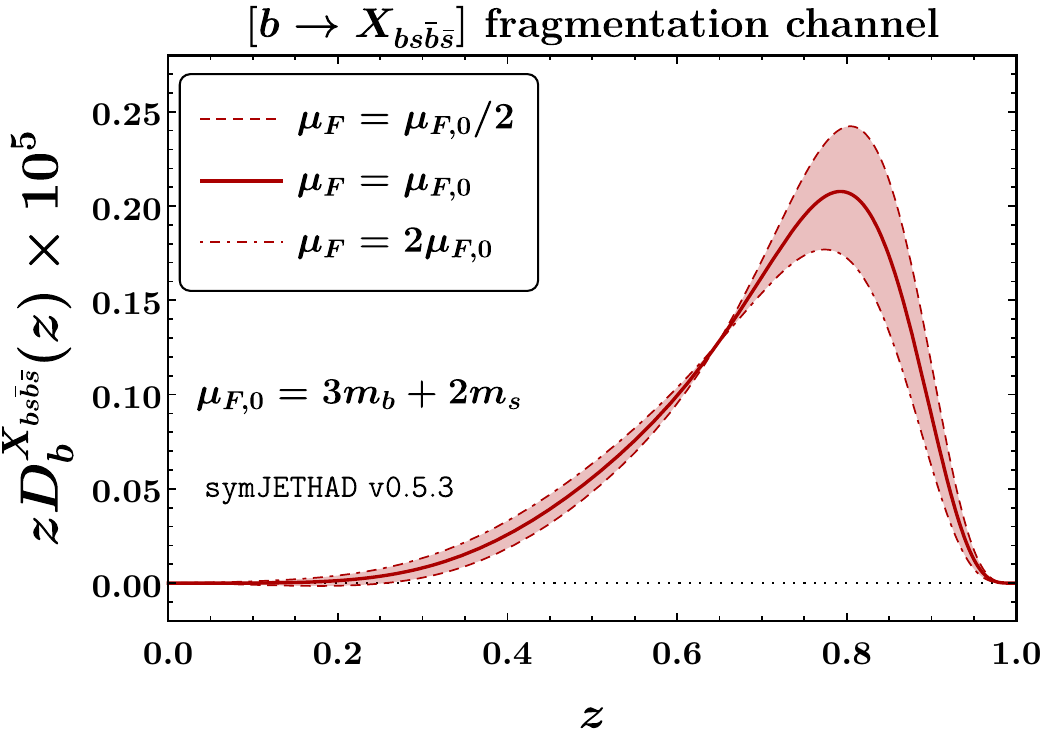}

\caption{Constituent heavy-quark to doubly charmed (upper) and bottomed (lower) tetraquark collinear fragmentation. Left and right panels are for initial-scale inputs of $[Q \to \QXQu]$ and $[Q \to \QXQs]$ channels, respectively.
For illustrative scope, an expanded DGLAP evolution is performed in the range $(3m_Q+2m_q)/2$ to $2(3m_Q+2m_q)$.}
\label{fig:XQq_FF_initial-scale_Q}
\end{figure*}

As for our $[Q \to \QXQq]$ initial-scale {\tt TQHL1.1} FFs, here we follow a similar strategy, namely we perform a simplified, expanded DGLAP evolution, where only the quark-to-quark timelike splitting kernel, $P_{qq}$, is active.
Plots of Fig.~\ref{fig:XQq_FF_initial-scale_Q} show the $z$-dependence of our constituent quark to doubly heavy tetraquark fragmentation channels.
Shaded bands are for the variation of the factorization scale in a window centered at $\mu_{F,0} = 3 m_Q + 2 m_q$, and ranging from $\mu_{F,0}/2$ to $2\mu_{F,0}$. 
As explained in Sec.~\ref{sssec:FFs_TQHL11}, the value $\mu_{F,0} = 3 m_Q + 2 m_q$ will serve as the starting scale for the {\tt TQHL1.1} constituent heavy quark fragmentation channel.

The $[Q \to \QXQq]$ initial-scale FFs, multiplied by $z$, exhibit a clear pattern, with a pronounced peak in the window $0.65 < z < 0.85$, and a vanishing behavior at both the $[z \to 0]$ and $[z \to 1]$ endpoints.
The presence of a peak in the moderate to large $z$-range is in line to our expectations, as discussed in Sec.~\ref{ssec:FFs_T4Q}.

We also note that doubly bottomed FFs (Fig.~\ref{fig:XQq_FF_initial-scale_Q}, lower plots) are almost five times larger than doubly charmed ones (Fig.~\ref{fig:XQq_FF_initial-scale_Q}, upper plots), and they are peaked at larger values of $z$.
This behavior is encoded in the dependence of the $\rho_l^{(Q,q)}(z)$ coefficient functions on the ${\cal R}_{q/Q} = m_q/m_Q$ ratio, which distinctly modulates the weight of the polynomial terms in Eqs.~\eqref{XQq_FF_initial-scale_Q_rho_0} to~\eqref{XQq_FF_initial-scale_Q_rho_3}, depending on the mass of the fragmenting heavy-quark species.

\subsubsection{The {\tt TQHL1.1} functions}
\label{sssec:FFs_TQHL11}

The final step in constructing our {\tt TQHL1.1} collinear FFs for doubly heavy tetraquarks involves performing a consistent DGLAP evolution of the $[Q \to \QXQq]$ initial-scale inputs defined in Sec.~\ref{sssec:FFs_XQq_Q} and depicted in Fig.~\ref{fig:XQq_FF_diagrams}. 
Kinematics suggests us that the minimal invariant mass for the $[Q \to (Qq\bar{q}\bar{Q}Q) + Q,\bar{Q}]$ splitting is $\mu_{F,0}(Q \to \QXQq) = 3 m_Q + 2 m_q$, which we adopt as the threshold for $Q$-quark fragmentation.
The same threshold holds for the $\bar{Q}$-antiquark, which is not shown in Fig.~\ref{fig:XQq_FF_diagrams}.

As mentioned in Sec.~\ref{ssec:FFs_quarkonia}, the DGLAP evolution within the {\HFNRevo} scheme generally consists of a first step, in which we perform an expanded and decoupled evolution ({\tt EDevo}), and of a second step, where an all-order evolution ({\tt AOevo}) is carried out.
Since, however, in our treatment only the $[Q \to \QXQq]$ channel is modeled at the initial energy scale, we can skip the {\tt EDevo} step and proceed directly with the {\tt AOevo}, which is performed numerically.\footnote{To the best of our knowledge, no models for the remaining parton fragmentation channels have been developed to date.}
In this way, starting by the $[Q \to \QXQq]$ initial-scale input of Eq.\eqref{XQq_FF_initial-scale_Q} taken at the \emph{evolution-ready} scale $Q_0 \equiv \mu_{F,0}(Q \to \QXQq)$, we build our {\tt TQHL1.1} sets via DGLAP evolution and release them in LHAPDF format.

One might contend that our methodology neglects the initial-scale contributions from light partons and nonconstituent heavy quarks, as these only emerge through evolution at scales $\mu_F > Q_0$.
On the one hand, Authors of Ref.~\cite{Nejad:2021mmp} raised arguments supporting the statement that these contributions are expected to be very small at the initial scale $Q_0$. 
A similar observation also applies to vector-quarkonium FFs, as discussed in Ref.~\cite{Celiberto:2022dyf}.

On the other hand, analyses on FFs for fully heavy states have highlighted how the presence of a nonzero initial-scale input for the gluon channel plays a role in the DGLAP evolution of the other parton channels, which can be relevant for precision studies of the fragmentation process in hadronic collisions.
We believe that, given the relevance of the $[Q \to \QXQq]$ channel over the other parton channels around $\mu_F \gtrsim Q_0$, our approach is well-defined and adequate to exploratory studies.
At the same time, further efforts are required to develop models for the remaining initial-scale inputs in the near future.

\begin{figure*}[!t]
\centering

   \hspace{-0.00cm}
   \includegraphics[scale=0.41,clip]{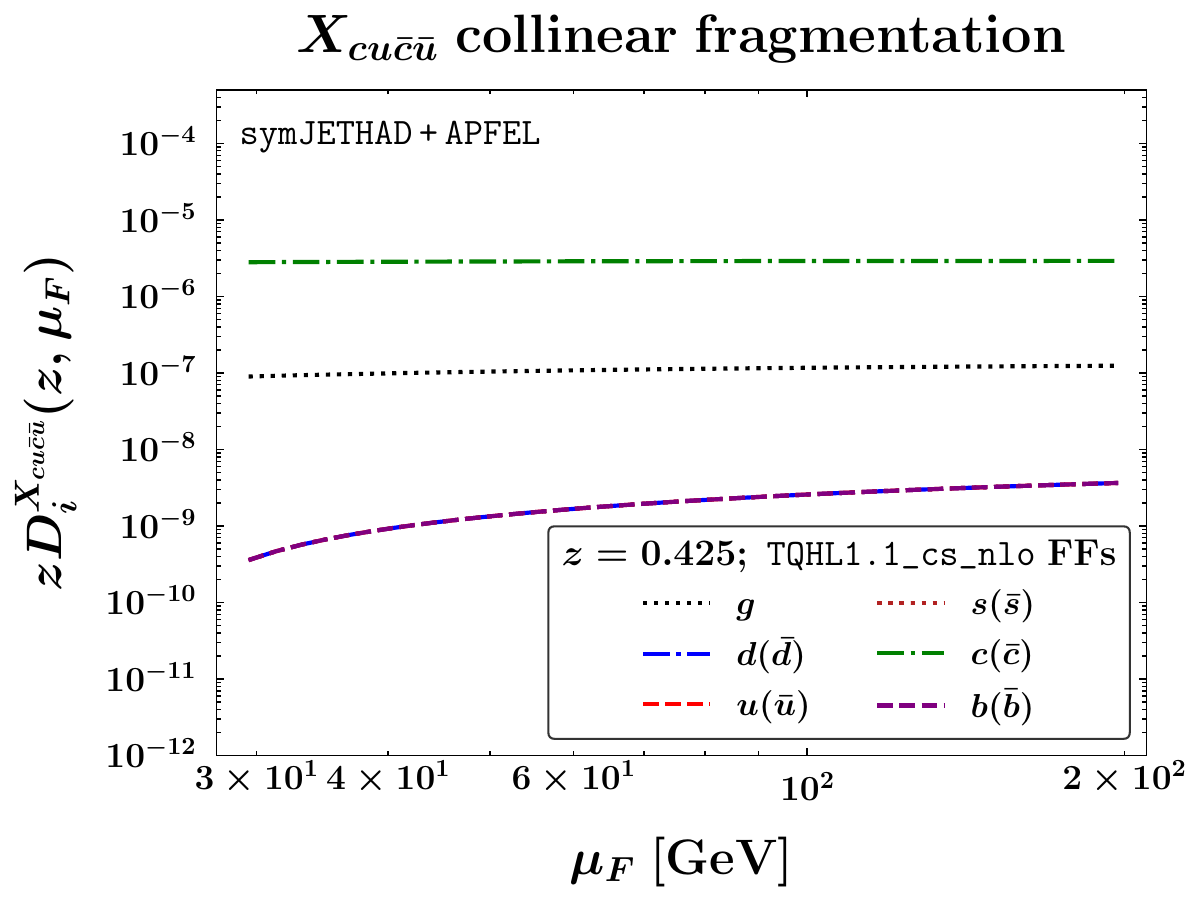}
   \includegraphics[scale=0.41,clip]{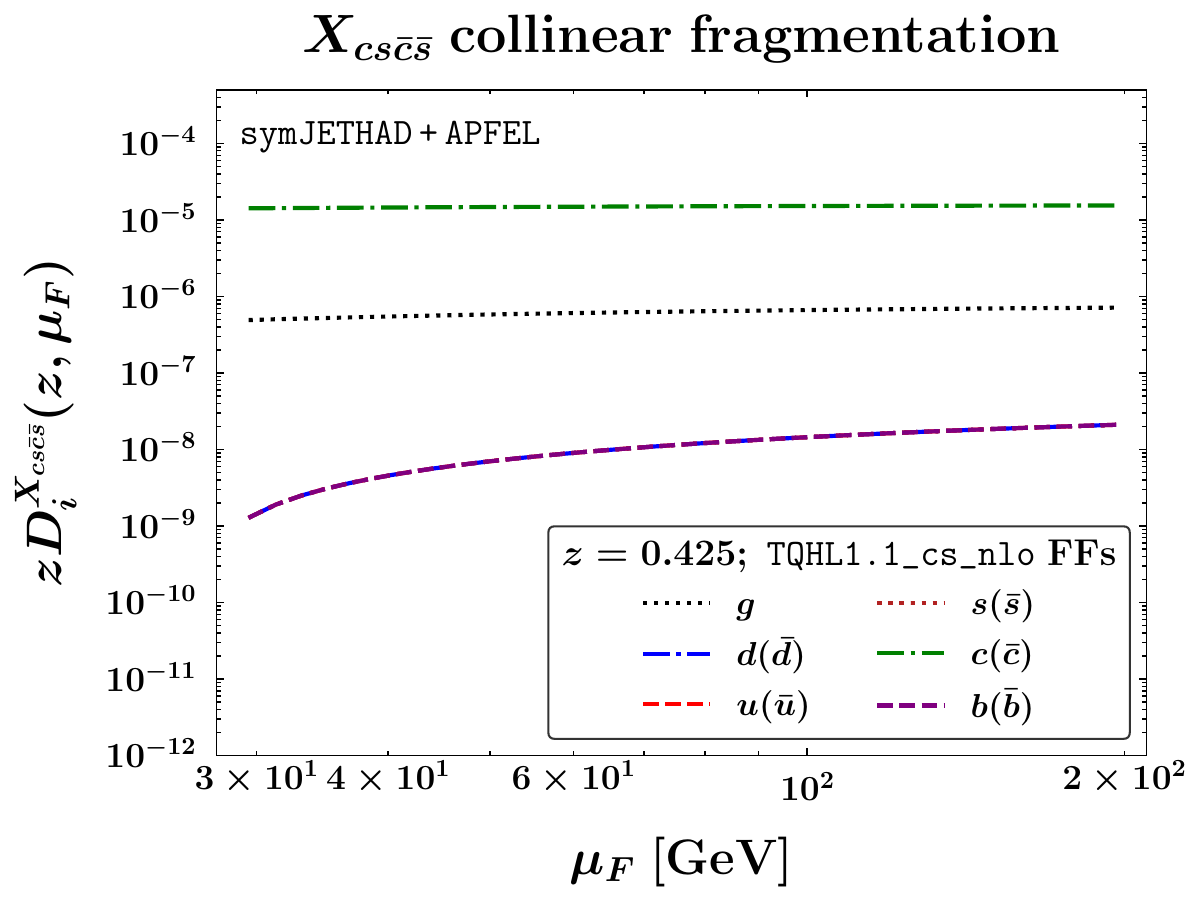}

\caption{Factorization-scale dependence of {\tt TQHL1.1} collinear FFs depicting $\QXcu$ (left) and $\QXcs$ (right) formation, at $z = 0.425 \simeq \langle z \rangle$.}
\label{fig:NLO_FFs_Xcq}
\end{figure*}

\begin{figure*}[!t]
\centering

   \hspace{-0.00cm}
   \includegraphics[scale=0.41,clip]{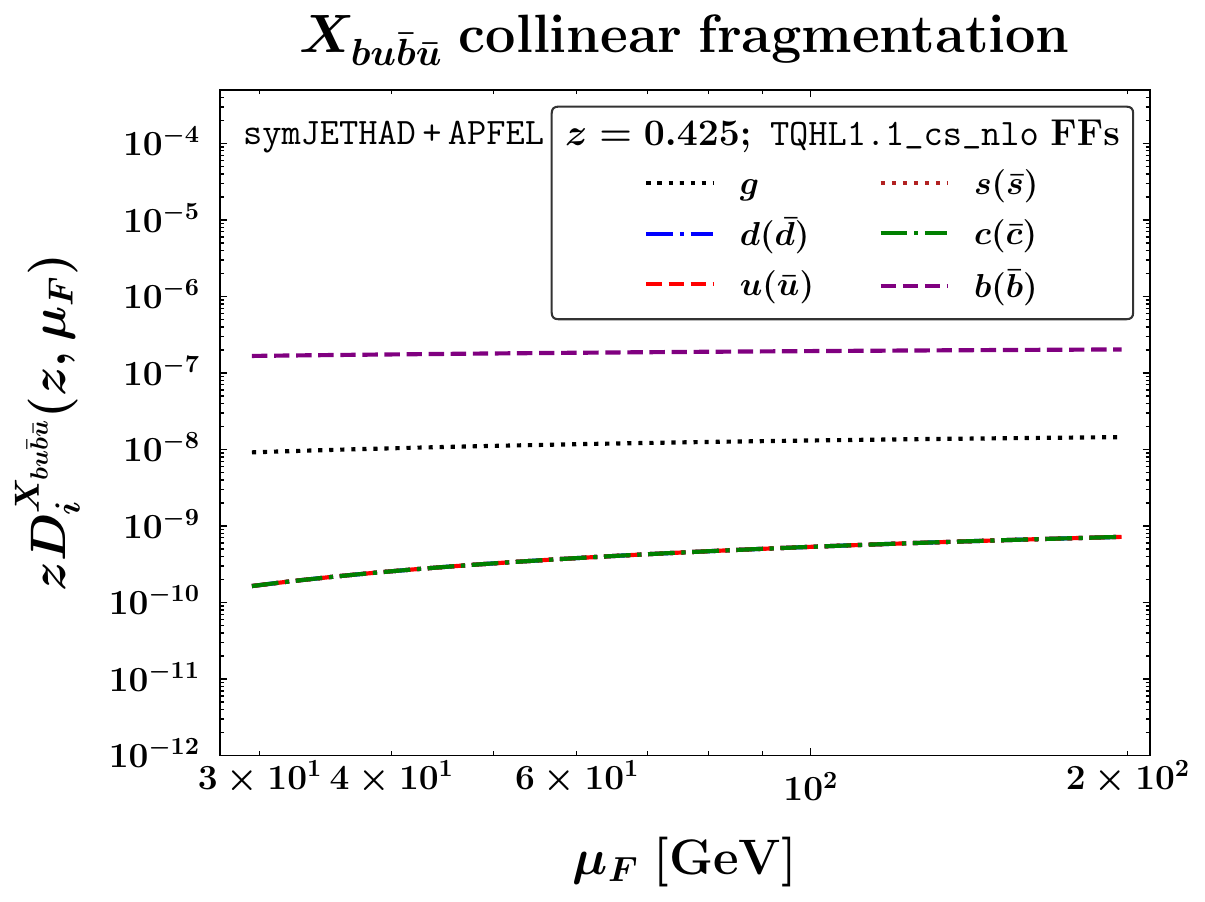}
   \includegraphics[scale=0.41,clip]{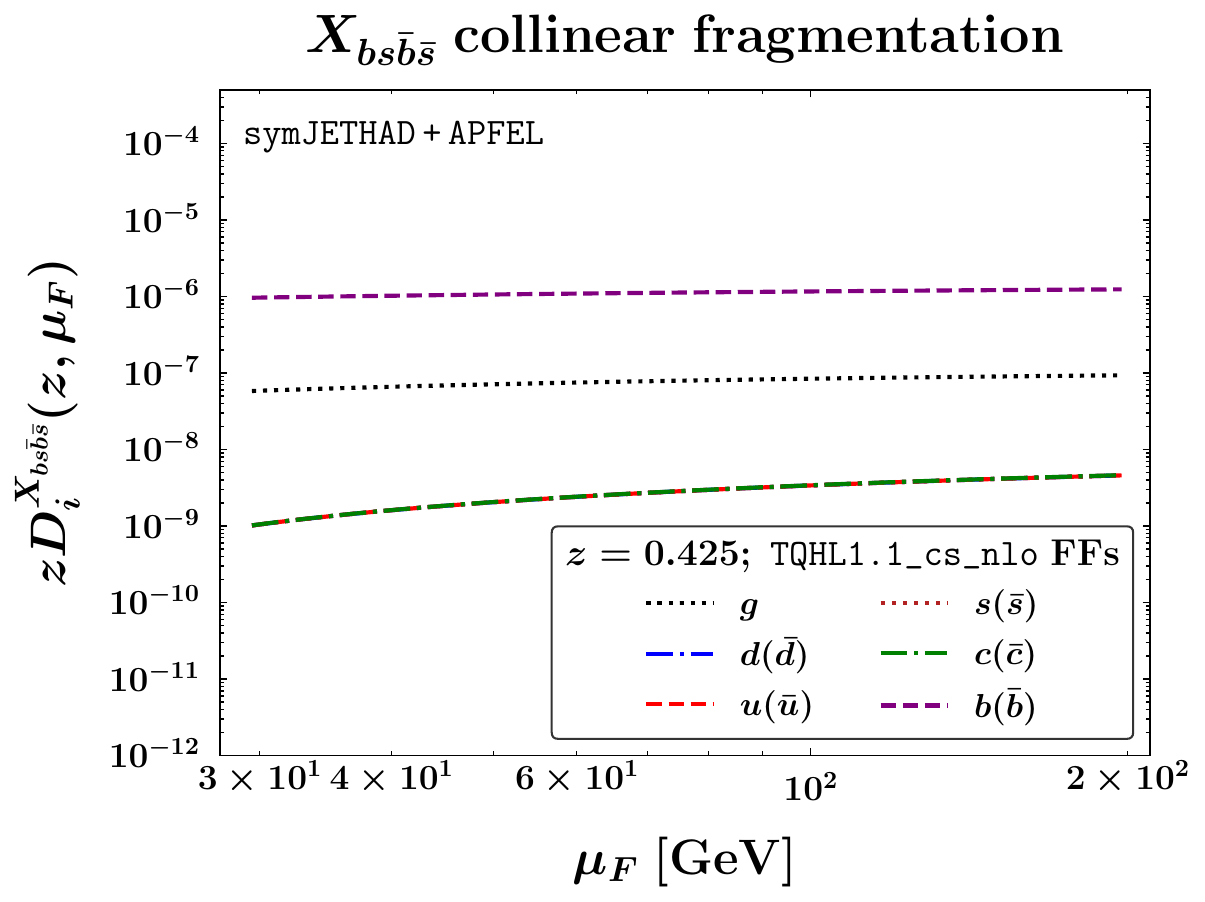}

\caption{Factorization-scale dependence of {\tt TQHL1.1} collinear FFs depicting $\QXbu$ (left) and $\QXbs$ (right) formation, at $z = 0.425 \simeq \langle z \rangle$.}
\label{fig:NLO_FFs_Xbq}
\end{figure*}

In Figs.~\ref{fig:NLO_FFs_Xcq} and~\ref{fig:NLO_FFs_Xbq} we present our {\tt TQHL1.1} FFs, as functions of $\mu_F$, for doubly charmed and doubly bottomed tetraquarks, respectively.
As done in the previous case (Fig.~\ref{fig:NLO_FFs_bHs_Jps_BCs_Bss}), we select just one value of the momentum fraction, $z = 0.425 \simeq \langle z \rangle$.
It approximately represents the average value at which FFs are typically probed in semihard final states (see, \emph{e.g.}, Refs.~\cite{Celiberto:2016hae,Celiberto:2017ptm,Celiberto:2020wpk,Celiberto:2021dzy,Celiberto:2021fdp,Celiberto:2022dyf,Celiberto:2022keu,Celiberto:2022kxx,Celiberto:2024omj}).

From the analysis of plots in Figs.~\ref{fig:NLO_FFs_Xcq} and~\ref{fig:NLO_FFs_Xbq}, the $[Q \to \QXQq]$ fragmentation channel strongly dominates over the light-parton and the nonconstituent heavy-quark channels. 
It also remains approximately one order of magnitude (or more) higher than the gluon channel across the entire energy range examined.
Moreover, in line with results for $B$ mesons, vector bottomonia, and charmed $B$ particles (Fig.~\ref{fig:NLO_FFs_bHs_Jps_BCs_Bss}), we observe that the $[g \to \QXQq]$ functions smoothly increase as $\mu_F$ grows. 
As explained in Sec.~\ref{ssec:FFs_quarkonia}, this feature is responsible for stabilizing heavy-flavor high-energy resummed distributions, matter of our phenomenological study (see Sec.~\ref{sec:results}).

\subsection{Fully heavy tetraquarks}
\label{ssec:FFs_T4Q}

\begin{figure*}[!t]
\centering
\includegraphics[width=0.475\textwidth]{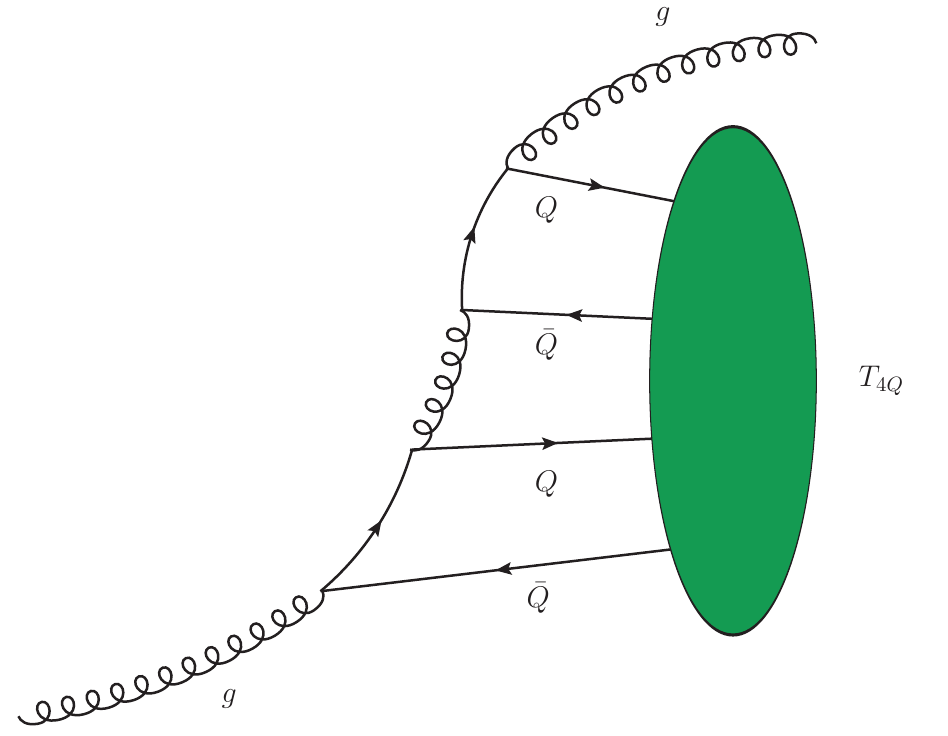}
\hspace{0.40cm}
\includegraphics[width=0.475\textwidth]{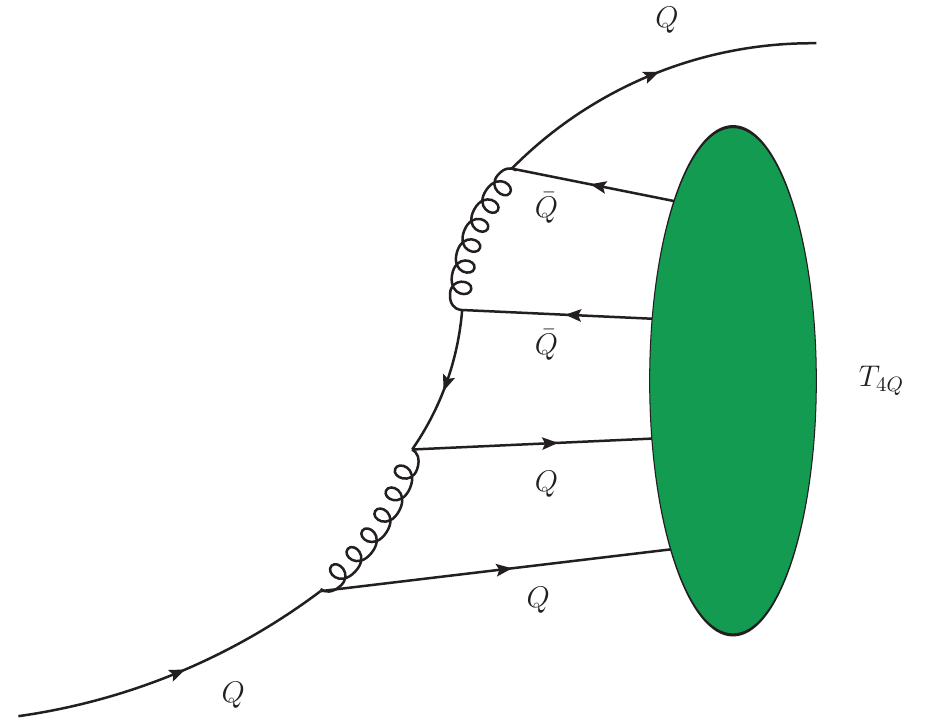}

\caption{LO representative diagrams for the collinear fragmentation of a gluon (left) or a constituent heavy quark (right) into a color-singlet $S$-wave fully heavy tetraquark.
The green blobs represent the nonperturbative hadronization part of the corresponding FFs.}
\label{fig:TQQ_FF_diagrams}
\end{figure*}

NRQCD factorization provides a robust framework for investigating the intrinsic nature of double $\Jpsi$ resonances~\cite{LHCb:2020bwg,ATLAS:2023bft,CMS:2023owd}, interpreting them as fully charmed tetraquark states~\cite{Zhang:2020hoh,Zhu:2020xni}.  
The production of a $\TQc$ state involves the short-distance emission of two charm and two anticharm quarks within a region of approximately $1/m_c$.  
Thanks to asymptotic freedom, this mechanism can be effectively described as a two-step convolution, separating the short-distance dynamics from the long-distance hadronization phase.

Working within this framework, a prime calculation for the NRQCD-based input of the $[g \to \TQc]$ color-singlet $S$-wave fragmentation channel (Fig.~\ref{fig:TQQ_FF_diagrams}, left diagram) was presented in Ref.~\cite{Feng:2020riv}.  
This result was subsequently compared with short-distance production mechanisms within LHC kinematic ranges~\cite{Feng:2023agq}.

For the sake of exploration, in Ref.~\cite{Celiberto:2024mab} the Suzuki model calculation~\cite{Suzuki:1977km,Suzuki:1985up,Amiri:1986zv,Nejad:2021mmp}, employed to describe our {\tt TQHL1.1} FFs for $\QXQq$ states (see Sec.~\ref{sssec:FFs_XQq_Q}), was suitably adapted to model the initial-scale input of the $[c \to \TQc]$ FF (Fig.~\ref{fig:TQQ_FF_diagrams}, right diagram). 
In the present work we enhance the description of the constituent heavy-quark fragmentation by relying upon a quite recent, NRQCD-based computation performed in Ref.~\cite{Bai:2024ezn}.

For a fully heavy tetraquark, $\TQQ(J^{PC})$, with total angular momentum, parity, and charge $J^{PC} = 0^{++}$ or~$2^{++}$,
considering the leading-order terms of the perturbative expansion of the nonrelativistic velocity, $v_{\cal Q}$, the initial-scale input for the collinear fragmentation of a parton $i$ into the observed exotic hadron reads
\begin{equation}
\begin{split}
 \label{TQQ_FF_initial-scale}
 D^{\TQQ(J^{PC})}_i(z,\mu_{F,0}) \, &= \,
 \frac{1}{m_Q^9}
 \sum_{[n]} 
 {\cal D}^{(J^{PC})}_i(z,[n]) \\
 &\times \, \langle {\cal O}^{\TQQ(J^{PC})}([n]) \rangle
 \;,
\end{split}
\end{equation}
where $m_Q = m_c = 1.5$~GeV ($m_Q = m_b = 4.9$~GeV) is the charm (bottom) quark mass, ${\cal D}^{(J^{PC})}_i(z,[n])$ the SDCs depicting the $[i \to (Q\bar{Q}Q\bar{Q})]$ perturbative component of the fragmentation, and $\langle {\cal O}^{\TQQ(J^{PC})}([n]) \rangle$ the \emph{color-composite} LDMEs portraying the $\TQQ(J^{PC})$ purely nonperturbative hadronization.
Furthermore, the composite quantum number $[n]$ runs over the combinations $[3,3]$, $[6,6]$, $[3,6]$, $[6,3]$.

In this context, we utilize the color diquark-anti\-diquark basis to express a color-singlet tetraquark state as either a ($\bar{3} \otimes 3$) or a ($6 \otimes \bar{6}$) configuration.  
According to Fermi--Dirac statistics, when considering the $S$-wave state, both the diquark-antidiquark systems and the overall diquark-antidiquark cluster, the ($\bar{3} \otimes 3$) configuration can exhibit spin 0, 1, or 2, while the ($6 \otimes \bar{6}$) configuration is restricted to spin 0.
Additionally, we note the symmetry relations
\begin{equation}
\begin{split}
 \label{TQQ_FF_initial-scale_symmetry}
 &{\cal D}^{(J^{PC})}_i(z,[3,6]) \, = \, {\cal D}^{(J^{PC})}_i(z,[6,3]) \,,
 \\
 &\langle {\cal O}^{\TQc(J^{PC})}([3,6]) \rangle \, = \, \langle {\cal O}^{\TQQ(J^{PC})}([6,3]) \rangle^*
 \;.
\end{split}
\end{equation}

\subsubsection{Short-distance coefficients}
\label{sssec:FFs_T4Q_SDCs}

The perturbative component of the gluon to $\TQQZpp$ fragmentation process at the initial energy scale is portrayed by the following SDCs
\begin{equation}
\begin{split}
 \label{Dg_FF_SDC_0pp_33}
 \hspace{-0.00cm}
 {\cal D}&^{(0^{++})}_g(z,[3,3]) \,=\, 
 \frac{\pi^2 \as^4(4m_Q)}{497664 \, d^{\cal D}_g(z)} \\
 & \times\, \left[186624-430272 z+511072 z^2-425814 z^3\right. \\
 & +\, 217337 z^4-61915 z^5+7466 z^6 \\
 & +\, 42(1-z)(2-z)(3-z) \\
 & \times\, (-144+634 z - 385 z^2+70 z^3 \\
 & \times\, \ln (1-z)+36(2-z)(3-z) \\
 & \times\, (144-634 z+749 z^2-364 z^3 + 74 z^4) \\
 & \times\, \ln \left(1-\frac{z}{2}\right)+12(2-z)(3-z) \\
 & \times\, \left(72-362 z+361 z^2-136 z^3+23 z^4\right) \\
 & \left.\times\, \ln \left(1-\frac{z}{3}\right)\right]
 \;,
\end{split}
\end{equation}
\\
\begin{equation}
\begin{split}
 \label{Dg_FF_SDC_0pp_66}
\hspace{-0.00cm}
 {\cal D}&^{(0^{++})}_g(z,[6,6]) \,=\,  
 \frac{\pi^2 \as^4(4m_Q)}{331776 \, d^{\cal D}_g(z)} \\
 & \times\, \left[186624-430272 z+617824 z^2-634902 z^3\right. \\
 & +\, 374489 z^4-115387 z^5+14378 z^6 \\
 & -\, 6(1-z)(2-z)(3-z) \\
 & \times\, (-144-2166 z + 1015 z^2+70 z^3) \\
 & \times\, \ln (1-z)-156(2-z)(3-z) \\
 & \times\, (144-1242 z+1693 z^2-876 z^3 + 170 z^4) \\
 & \times\, \ln \left(1-\frac{z}{2}\right)+300(2-z)(3-z) \\
 & \times\, \left(72-714 z+953 z^2-472 z^3+87 z^4\right) \\
 & \left.\times\, \ln \left(1-\frac{z}{3}\right)\right]
 \;,
\end{split}
\end{equation}
\\
\begin{equation}
\begin{split}
 \label{Dg_FF_SDC_0pp_36}
\hspace{-0.00cm}
 {\cal D}&^{(0^{++})}_g(z,[3,6]) \,=\,  
 \frac{\pi^2 \as^4(4m_Q)}{165888 \, d^{\cal D}_g(z)} \\
 & \times\, \left[186624-430272 z+490720 z^2-394422 z^3\right. \\
 & +\, 199529 z^4-57547 z^5+7082 z^6 \\
 & +\, 6(1-z)(2-z)(3-z) \\
 & \times\, (-432+3302 z - 1855 z^2+210 z^3) \\
 & \times\, \ln (1-z)-12(2-z)(3-z) \\
 & \times\, (720-2258 z+2329 z^2-1052 z^3 + 226 z^4) \\
 & \times\, \ln \left(1-\frac{z}{2}\right)+12(2-z)(3-z) \\
 & \times\, \left(936-4882 z+4989 z^2-1936 z^3+331 z^4\right) \\
 & \left.\times\, \ln \left(1-\frac{z}{3}\right)\right]
  \;,
\end{split}
\end{equation}
with $d^{\cal D}_g(z) = z (2-z)^2 (3-z)$.

As for the gluon to $\TQQTpp$ initial-scale perturbat\-ive-fragmentation component, only the $[3,3]$ term survives.
Indeed, since the NRQCD operator representing the $6 \otimes \bar{6}$ state is not compatible with the Fermi--Dirac statistics for a diquark-antidiquark system in a $2^{++}$ configuration, both the $[6,6]$ term and the $[3,6]$ interference one vanish.
We have
\begin{equation}
\begin{split}
 \label{Dg_FF_SDC_2pp_33}
\hspace{-0.00cm}
 {\cal D}&^{(2^{++})}_g(z,[3,3]) \,=\, 
 \frac{\pi^2 \as^4(4m_Q)}{622080 \, z \, d^{\cal D}_g(z)} \\
 & \times\, \left[\left(46656-490536 z \right.\right. \\
 & +\, 1162552 z^2-1156308 z^3 \\
 & \left.+\, 595421 z^4-170578 z^5+21212 z^6\right) 2z \\
 & +\, 3(1-z)(2-z)(3-z)(-20304-31788 z \\
 & +\, 73036 z^2-36574 z^3+7975 z^4) \ln (1-z) \\
 & +\, 33(2-z)(3-z) (1296+1044 z -9224 z^2 \\
 & \left.\, +9598 z^3-3943 z^4+725 z^5\right) \\
 & \left.\times\, \ln \left(1-\frac{z}{3}\right)\right]
  \;,
\end{split}
\end{equation}
whereas ${\cal D}^{(2^{++})}_g(z,[6,6]) = 0$ and ${\cal D}^{(2^{++})}_g(z,[3,6]) = 0$\,.

Analogously, the perturbative component of the constituent heavy-quark to $\TQQZpp$ fragmentation process at the initial energy scale is depicted by the following SDCs
\begin{equation}
\begin{split}
 \label{DQ_FF_SDC_0pp_33}
 \hspace{-0.00cm}
 {\cal D}&^{(0^{++})}_Q(z,[3,3]) \,=\, 
 \frac{\pi^2 \as^4(5m_Q)}{559872 \, d^{\cal D}_Q(z)} \\
 & \times\, \left[ -264 (z-4) (11 z-12) (z^2-16 z+16) \right. \\
 & \times\, (13 z^4-57 z^3-656 z^2+1424z-512) \\
 & \times\, (3 z-4)^5 \log (z^2-16 z+16)  \\
 & +\, 6 (11 z-12)(z^2-16 z+16) \\
 & \times\, (1273 z^5-16764 z^4+11840 z^3 \\
 & +\, 247808z^2-472320 z+171008) \\
 & \times\, (3 z-4)^5 \log (4-3 z) \\
 & -\, 3 (11 z-12)(z^2-16 z+16) \\
 & \times\, (129 z^5 - 7172 z^4+49504 z^3-108416z^2 \\
 & +\, 73984 z-9216) (3 z-4)^5  \\
 & \times\, \log\left[\left(4-\frac{11z}{3}\right)(4-z)\right]\\
 & +\, 16 (z-1) (657763 z^{12}-10028192z^{11} \\
 & +\, 188677968 z^{10}-2600899712 z^9 \\
 & +\, 18018056448 z^8-71685000192z^7 \\
 & +\, 179414380544 z^6-294834651136 z^5 \\
 & +\, 321642168320z^4-229388845056 z^3\\
 & +\, 102018056192 z^2-25480396800z \\
 & \left. +\, 2717908992)\right]
 \;,
\end{split}
\end{equation}
\\
\begin{equation}
\begin{split}
 \label{DQ_FF_SDC_0pp_66}
 \hspace{-0.00cm}
 {\cal D}&^{(0^{++})}_Q(z,[6,6]) \,=\, 
 \frac{\pi^2 \as^4(5m_Q)}{373248 \, d^{\cal D}_Q(z)} \\
 & \times\, \left[ -120 (z-4) (11 z-12) (z^2-16 z+16) \right. \\
 & \times\, (35 z^4-535 z^3+3472 z^2-4240z+512) \\
 & \times\, (3 z-4)^5 \log (z^2-16 z+16)  \\
 & -\, 30 (11 z-12)(z^2-16 z+16) \\
 & \times\, (3395 z^5-48020 z^4+126144 z^3 \\
 & -\, 75776^2-38656 z+62464) \\
 & \times\, (3 z-4)^5 \log (4-3 z) \\
 & +\, 75 (11 z-12)(z^2-16 z+16) \\
 & \times\, (735 z^5 - 10684 z^4+34208 z^3-44160z^2 \\
 & +\, 20224 z+9216) (3 z-4)^5  \\
 & \times\, \log\left[\left(4-\frac{11z}{3}\right)(4-z)\right]\\
 & +\, 16 (z-1) (7916587 z^{12}-263987840z^{11} \\
 & +\, 3125201872 z^{10}-16993694336 z^9 \\
 & +\, 51814689024 z^8-99638283264^7 \\
 & +\, 133459423232 z^6-140136398848 z^5 \\
 & +\, 127161204736z^4-96695746560 z^3\\
 & +\, 53372518400 z^2-17930649600z \\
 & \left. +\, 2717908992)\right]
 \;,
\end{split}
\end{equation}
\\
\begin{equation}
\begin{split}
 \label{DQ_FF_SDC_0pp_36}
 \hspace{-0.00cm}
 {\cal D}&^{(0^{++})}_Q(z,[3,6]) \,=\, 
 \frac{\pi^2 \as^4(5m_Q)}{186624 \sqrt{6} \, d^{\cal D}_Q(z)} \\
 & \times\, \left[ 24 (z-4) (11 z-12) (z^2-16 z+16) \right. \\
 & \times\, (225 z^4-3085 z^3+17456 z^2 \\
 & -\, 19760z+1536) \\
 & \times\, (3 z-4)^5 \log (z^2-16 z+16)  \\
 & -\, 6 (11 z-12)(z^2-16 z+16) \\
 & \times\, (555 z^5+52428 z^4-363328 z^3 \\
 & +\, 616448z^2-270080 z+70656) \\
 & \times\, (3 z-4)^5 \log (4-3 z) \\
 & +\, 75 (11 z-12)(z^2-16 z+16) \\
 & \times\, (1245 z^5 -84308 z^4 \\
 & +\, 601696z^3-1333120z^2 \\
 & +\, 914688 z-119808) (3 z-4)^5  \\
 & \times\, \log\left[\left(4-\frac{11z}{3}\right)(4-z)\right]\\
 & +\, 16 (z-1) (1829959z^{12}-44960912 z^{11} \\
 & +\, 285792656 z^{10}-1090093952z^9 \\
 & +\, 5123084544 z^8-24390724608 z^7 \\
 & +\, 77450817536 z^6-153897779200z^5 \\
 & +\, 194102034432 z^4-155643543552 z^3\\
 & +\, 77091307520 z^2-21705523200z \\
 & \left. +\, 2717908992)\right]
 \;,
\end{split}
\end{equation}
with $d^{\cal D}_Q(z) = (4-3 z)^6(z-4)^2 z(11 z-12)(z^2-16 z+16)$.

As mentioned before, due to the Fermi--Dirac statistics, the only surviving SDC for the $\TQQTpp$ initial-scale perturbat\-ive fragmentation is the $[3,3]$ one.
Thus we write
\begin{equation}
\begin{split}
 \label{DQ_FF_SDC_2pp_33}
 \hspace{-0.00cm}
 {\cal D}&^{(2^{++})}_Q(z,[3,3]) \,=\, 
 \frac{\pi^2 \as^4(5m_Q)}{2799360 \, z \, d^{\cal D}_Q(z)} \\
 & \times\, \left[ 672 (z-4) (11 z-12) (z^2-16 z+16) \right. \\
 & \times\, (47z^5+12186 z^4-44608 z^3 \\ 
 & +\, 40000 z^2 -7936 z+4608) \\
 & \times\, (3 z-4)^5 \log (z^2-16 z+16)  \\
 & +\, 6 (11 z-12)(z^2-16 z+16) \\
 & \times\, (107645 z^6-1088988 z^5+7805536 z^4 \\
 & -\, 20734976 z^3 +8933504z^2 \\
 & -\, 6013952 z+1695744) \\
 & \times\, (3 z-4)^5 \log (4-3 z) \\
 & -\, 33 (11 z-12)(z^2-16 z+16) \\
 & \times\, (3581 z^5 - 53216 z^4-326176 z^3+419456z^2 \\
 & -\, 6912 z+55296) (3 z-4)^6  \\
 & \times\, \log\left[\left(4-\frac{11z}{3}\right)(4-z)\right]\\
 & +\, 16 (z-1) (96449507 z^{12} - 158520388z^{11} \\
 & -\, 26228206896 z^{10} + 281743037888 z^9 \\
 & -\, 1355257362432 z^8 + 1355257362432 z^7 \\
 & -\, 6637452959744 z^6 + 7595797282816 z^5 \\
 & -\, 5643951472640 z^4 + 2662988513280 z^3\\
 & -\, 788934950912 z^2 + 161828831232 z \\
 & \left. -\, 24461180928)\right]
 \;.
\end{split}
\end{equation}

\subsubsection{Long-distance matrix elements}
\label{sssec:FFs_T4Q_LDMEs}

The color-composite LDMEs, $\langle {\cal O}^{\TQQ(J^{PC})}([n]) \rangle$, encapsulate the genuinely nonperturbative contributions to the initial-scale FFs. 
Since experimental data are currently unavailable and lattice QCD studies of tetra\-quarks are still at primeval stage, potential model calculations serve as a practical means to estimate these matrix elements.
An effective approach involves calculating the radial wave functions at the origin using potential models and then connecting them to the LDMEs through the vacuum saturation approximation~\cite{Feng:2020qee}. 

Reference~\cite{Feng:2020riv} introduced three potential-based models~\cite{Zhao:2020nwy,Lu:2020cns,liu:2020eha}.
All of them adopt a Cornell-like potential~\cite{Eichten:1974af,Eichten:1978tg} and incorporate certain spin-dependent corrections. 
The first~\cite{Zhao:2020nwy} and third models~\cite{liu:2020eha} are rooted in nonrelativistic quark dynamics, while the second~\cite{Lu:2020cns} includes relativistic effects.
The first model tends to significantly overestimate cross sections when compared to data for $\Jpsi$ production at 13~TeV~CMS~\cite{CMS:2017dju}, which are expected to exceed the $\TQc$ production rate~\cite{Feng:2020riv}.

Furthermore, numeric evaluations (not detailed in this work) reveal that FFs based on the LDMEs from the third model exhibit extreme sensitivity to even minor parameter variations, on the order of 0.1\%. 
Given these limitations, we construct the initial-scale components of our {\tt TQ4Q1.1} FFs using LDMEs derived from the second model~\cite{Lu:2020cns}. 
For a detailed comparison of these values with those from the other two models, we refer to Table~I in the published version of Ref.~\cite{Feng:2020riv}.

As for $\TQcZpp$ and $\TQcTpp$ fully charmed states, one has
\begin{equation}
\begin{split}
\label{LDMEs_T4c}
 {\cal O}^{\TQcZpp}([3,3]) &\,=\, 0.0347\mbox{ GeV}^9 \;, 
 \\[0.15cm]
 {\cal O}^{\TQcZpp}([6,6]) &\,=\, 0.0128\mbox{ GeV}^9 \;, 
 \\[0.15cm]
 {\cal O}^{\TQcZpp}([3,6]) &\,=\, 0.0211\mbox{ GeV}^9 \;, 
 \\[0.15cm]
 {\cal O}^{\TQcTpp}([3,3]) &\,=\, 0.072\mbox{ GeV}^9 \;, 
 \\[0.15cm]
 {\cal O}^{\TQcTpp}([3,6]) &\,=\, {\cal O}^{\TQcTpp}([6,6]) \,=\, 0 \;.
\end{split}
\end{equation}

Given that exact values for the LDMEs of $\TQbZpp$ and $\TQbTpp$ fully bottomed states have not been computed yet, as a reasonable \emph{Ansatz} we assume that a $\TQb$ tetraquark consists of a compact diquark-antidiquark cluster, where the binding is primarily governed by attractive color Coulomb forces. 
Under this assumption, the ratio of the four-body Schr\"odinger wave functions at the origin between the particles $\TQc$ and $\TQb$ can be estimated by means of dimensional analysis.

Following the strategy proposed in Ref.~\cite{Feng:2023agq}, we write
\begin{equation}
\label{LDMEs_T4b}
\hspace{-0.15cm}
 \frac{\langle {\cal O}^{\TQb(J^{PC})}([n]) \rangle}{\langle {\cal O}^{\TQc(J^{PC})}([n]) \rangle} \!=\! \frac{{\langle \cal O}^{\TQb}_{\rm [Coul.]} \rangle}{{\langle \cal O}^{\TQc}_{\rm [Coul.]}\rangle} \!\simeq\! \left( \frac{m_b \, \as^{(b)}}{m_c \, \as^{(c)}} \right)^{\!\!9} \!\simeq 400
 \,.
\end{equation}
Here, $\as^{(Q)}$ stands for the strong coupling, $\as(m_Q v_{\cal Q}) \sim v_{\cal Q}$, with $v_Q$ the relative velocity between the two constituent heavy quarks.
The `${\rm [Coul.]}$' label tells us that the LDME is evaluated within the Coulomb potential diquark model.

\subsubsection{Initial energy-scale inputs}
\label{sssec:FFs_T4Q_gQ}

\begin{figure*}[!t]
\centering

\includegraphics[scale=0.46,clip]{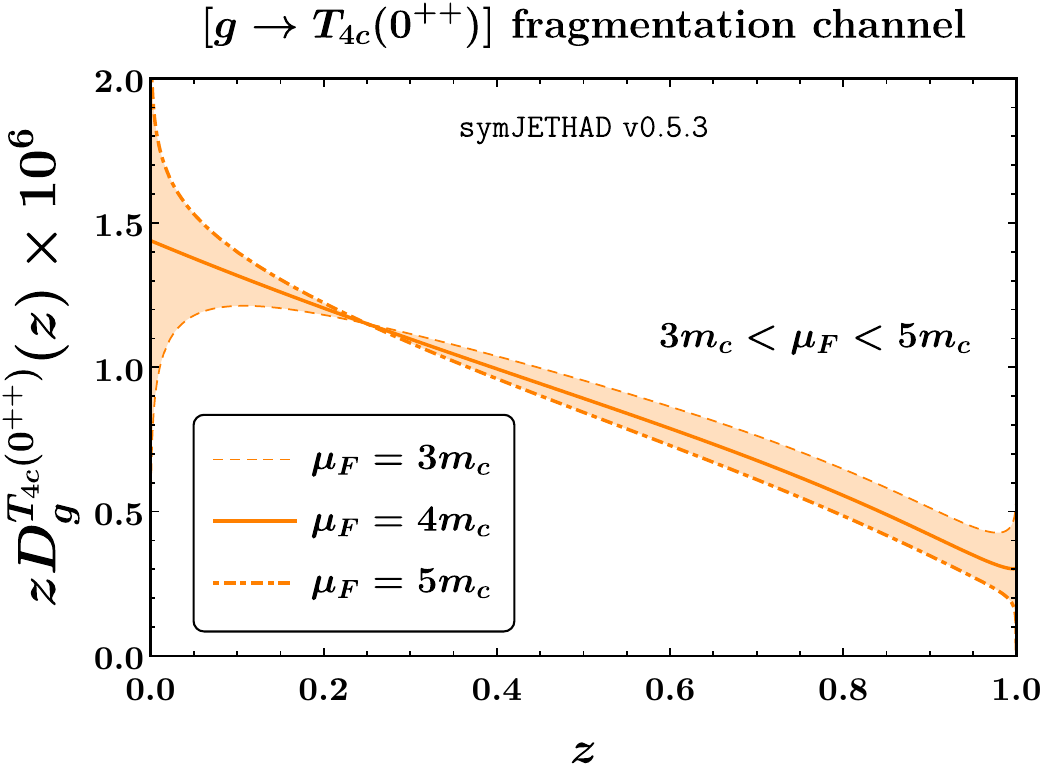}
\hspace{0.50cm}
\includegraphics[scale=0.46,clip]{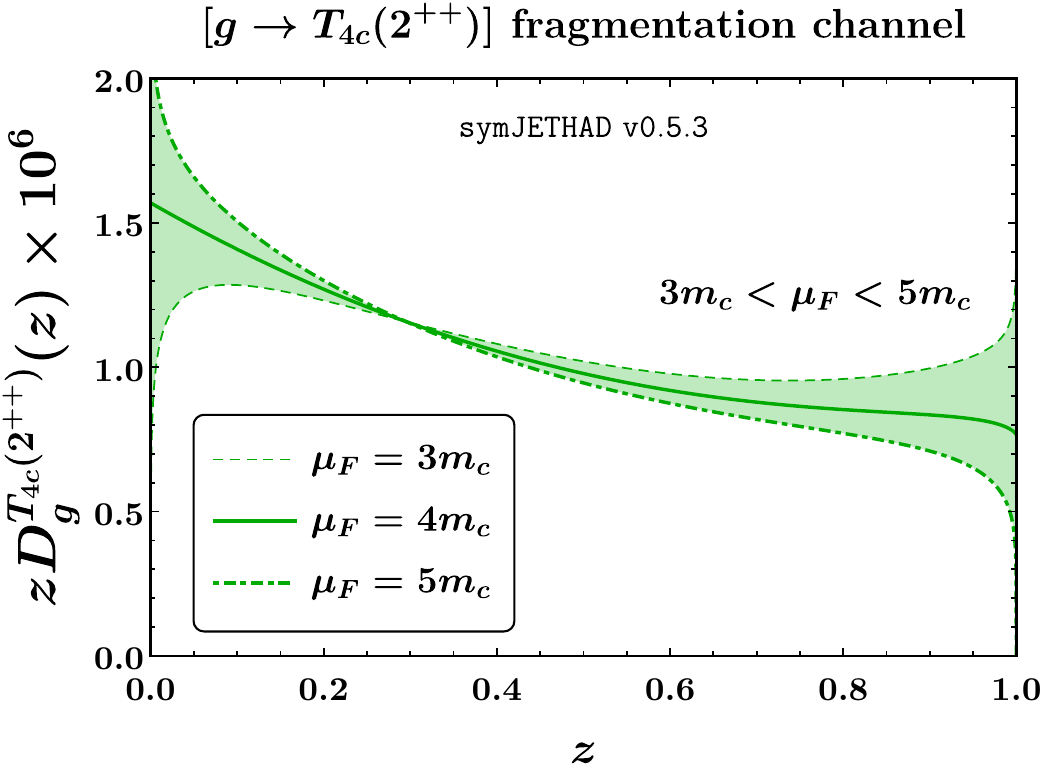}

\vspace{0.25cm}

\includegraphics[scale=0.46,clip]{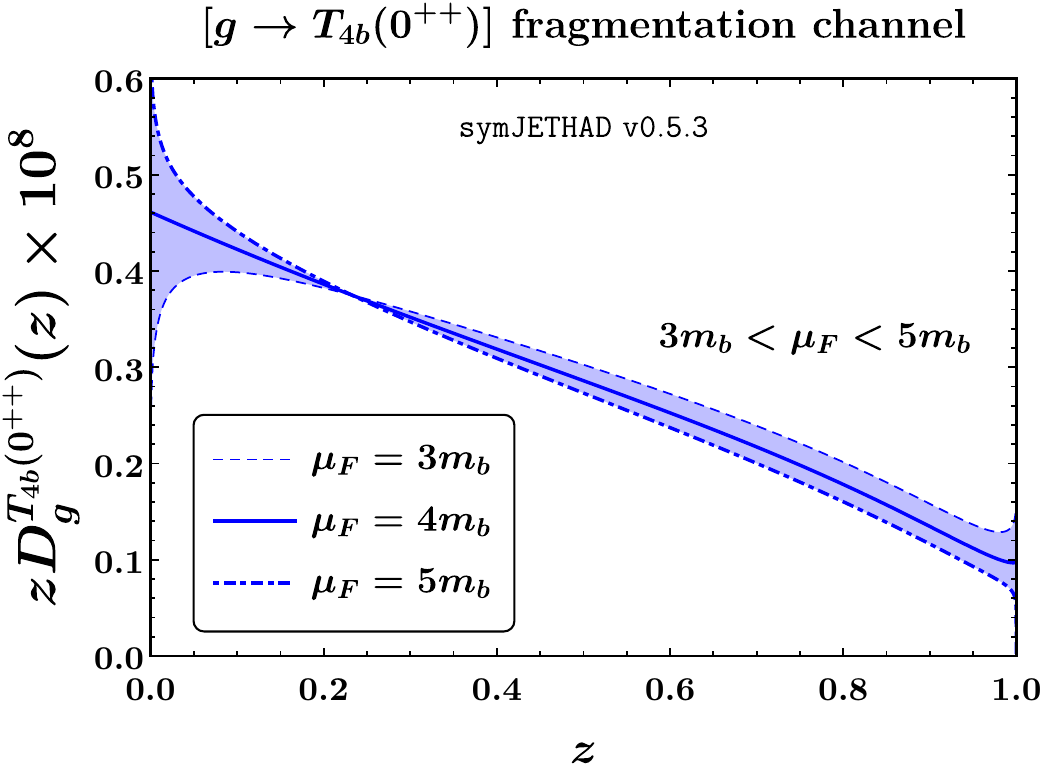}
\hspace{0.50cm}
\includegraphics[scale=0.46,clip]{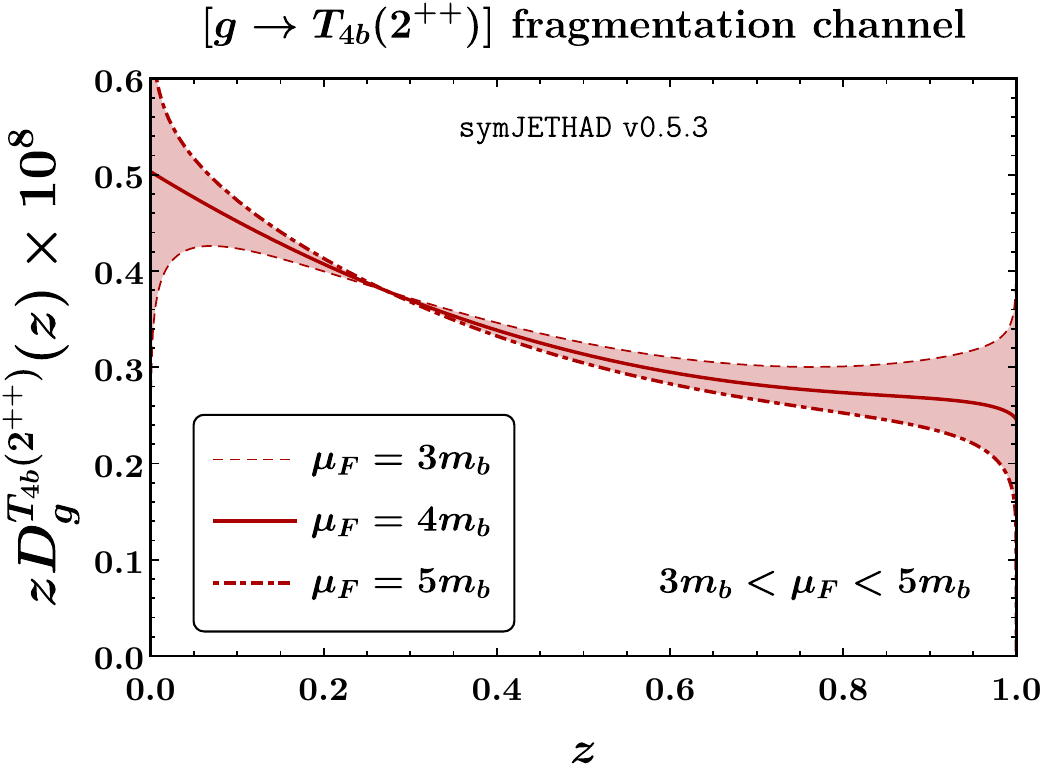}

\caption{Gluon to fully charmed (upper) and bottomed (lower) tetraquark collinear fragmentation. Left and right panels are for initial-scale inputs of $[g \to \TQQZpp]$ and $[g \to \TQQTpp]$ channels, respectively.
For illustrative scope, an expanded DGLAP evolution is performed in the range $3m_Q$ to $5m_Q$.}
\label{fig:TQQ_FF_initial-scale_gluon}
\end{figure*}

\begin{figure*}[!t]
\centering

\includegraphics[scale=0.46,clip]{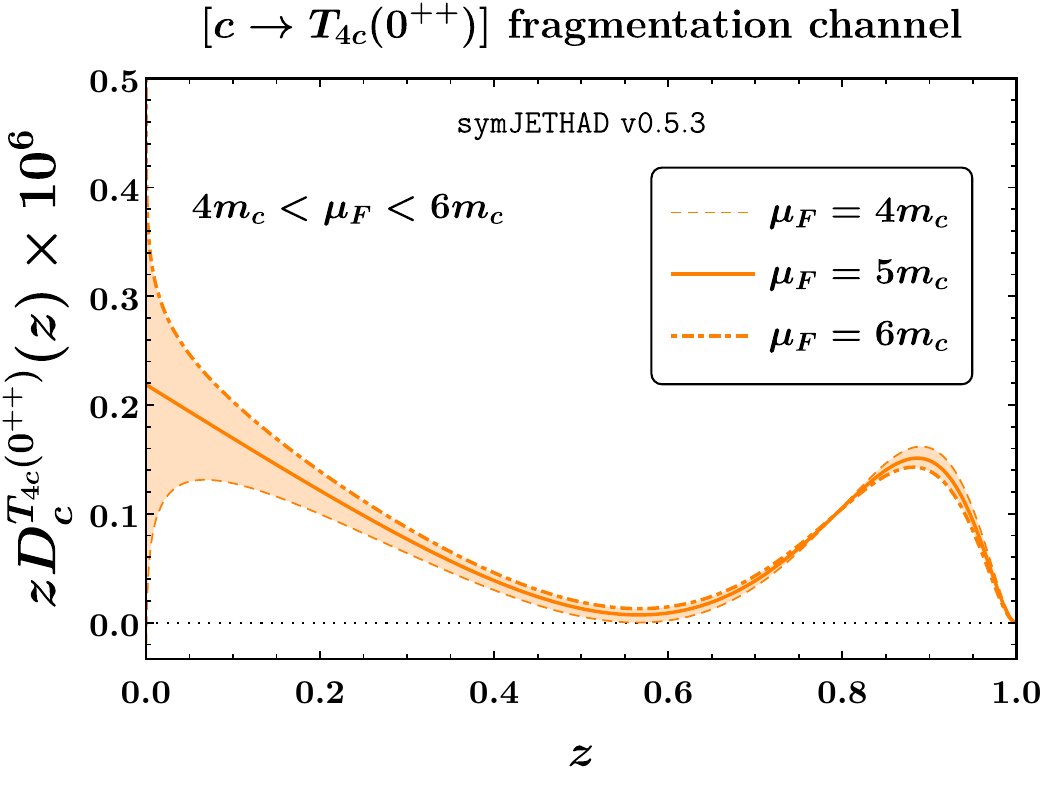}
\hspace{0.50cm}
\includegraphics[scale=0.46,clip]{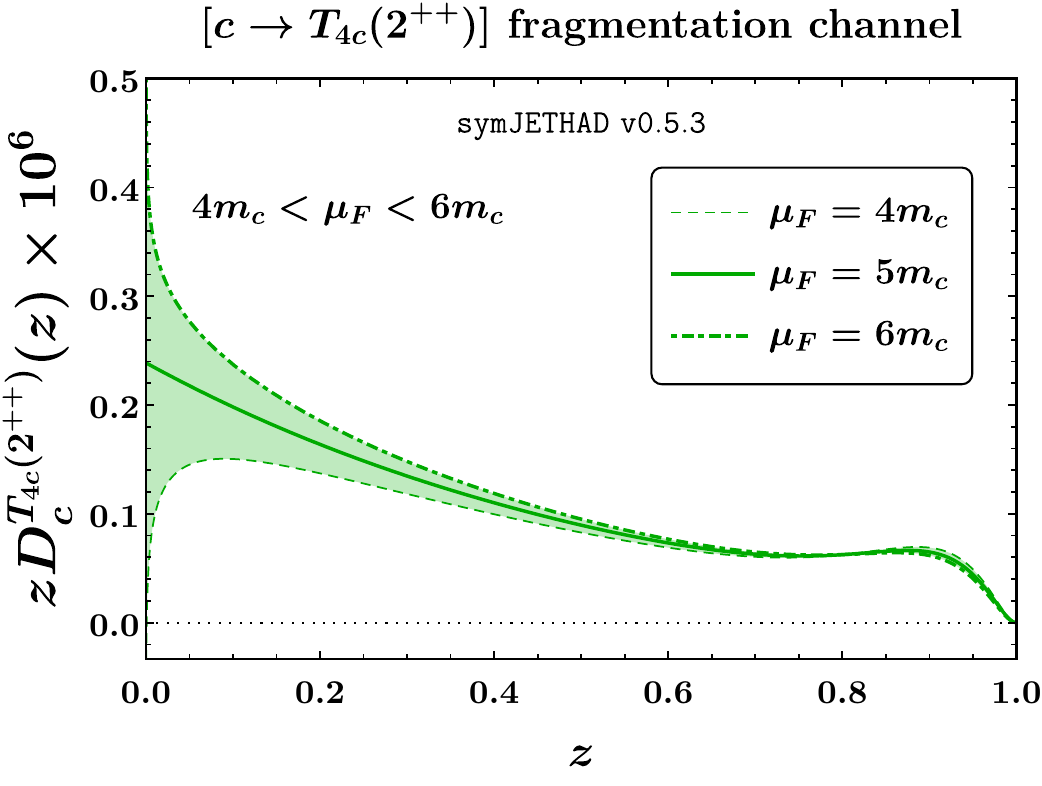}

\vspace{0.25cm}

\includegraphics[scale=0.46,clip]{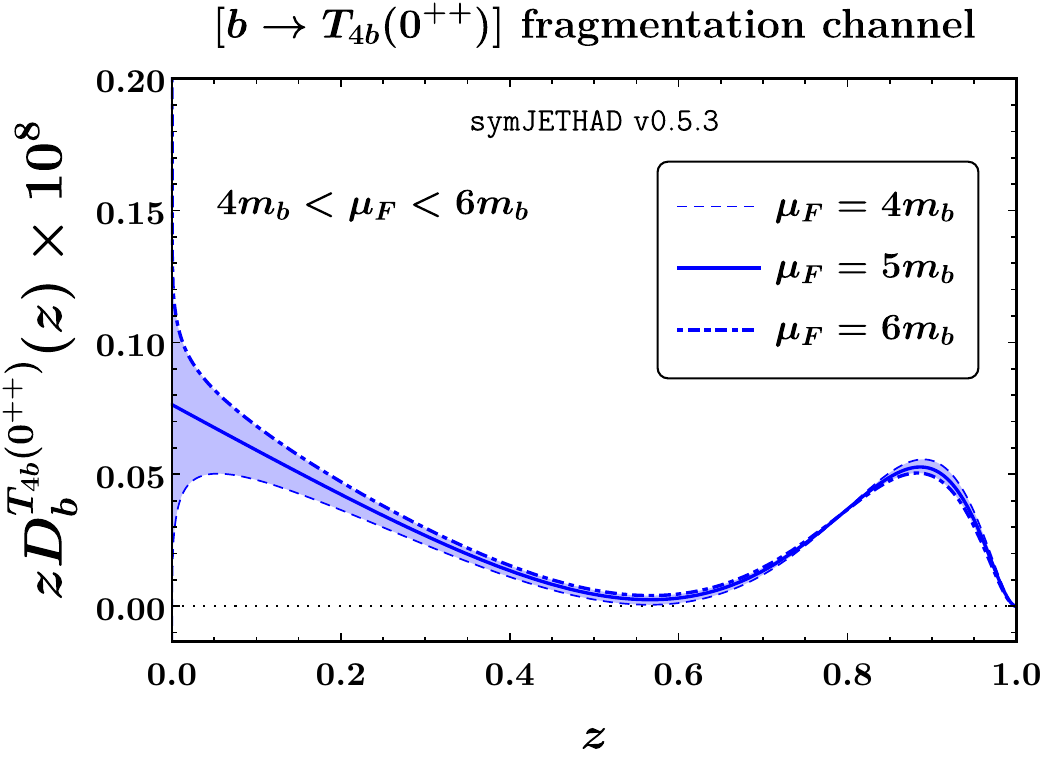}
\hspace{0.50cm}
\includegraphics[scale=0.46,clip]{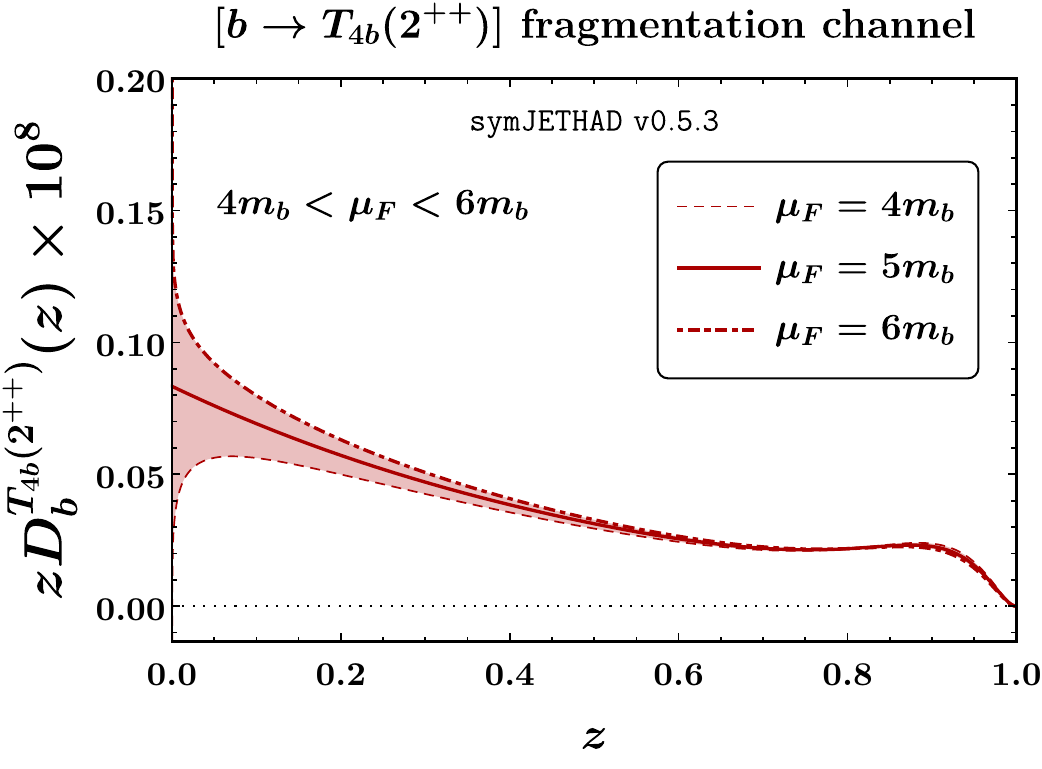}

\caption{Constituent heavy-quark to fully charmed (upper) and bottomed (lower) tetraquark collinear fragmentation. Left and right panels are for initial-scale inputs of $[Q \to \TQQZpp]$ and $[Q \to \TQQTpp]$ channels, respectively.
For illustrative scope, an expanded DGLAP evolution is performed in the range $4m_Q$ to $6m_Q$.}
\label{fig:TQQ_FF_initial-scale_Q}
\end{figure*}

As anticipated in Sec.~\ref{sssec:FFs_XQq_Q}, in Ref.~\cite{Celiberto:2024mab} the $[g \to \TQc]$ initial-scale inputs of the {\tt TQ4Q1.0} FFs were benchmarked with the original calculation of Ref.~\cite{Feng:2020riv} by performing a simplified, expanded DGLAP evolution, where only the gluon-to-gluon timelike splitting kernel, $P_{gg}$ was considered (see~Section 2.2 of~\cite{Celiberto:2024mab} for technical details).

For the sake of illustration, in Fig.~\ref{fig:TQQ_FF_initial-scale_gluon} we present the $z$-dependence of $[g \to \TQQ]$ inputs of our {\tt TQ4Q1.1} functions, multiplied by $z$.
Error bands here reflect the variation of the factorization scale in a range centered at $\mu_{F,0} = 4 m_Q$ with a width of $2 m_Q$. 
As elaborated in Sec.~\ref{sssec:FFs_TQ4Q11}, the value $\mu_{F,0} = 4 m_Q$ will serve as the starting scale for the {\tt TQ4Q1.1} gluon fragmentation channel.

Given that the gluon $[g \to \TQc]$ channels remain unchanged when passing from {\tt TQ4Q1.0} to the {\tt 1.1} update, upper plots of Fig.~\ref{fig:TQQ_FF_initial-scale_gluon} are identical, apart from the color code, to the ones of Fig.~2 of~\cite{Celiberto:2024mab}.
Novel results for the {\tt TQ4Q1.1} $[g \to \TQb]$ channels are presented in lower plots of Fig.~\ref{fig:TQQ_FF_initial-scale_gluon}.
As a consequence of the \emph{Ansatz} made in Eq.~\eqref{LDMEs_T4b}, they share the same pattern of corresponding $\TQc$ channels, but with a different magnitude.

We note that all initial-scale FFs shown in Fig.~\ref{fig:TQQ_FF_initial-scale_gluon} do not vanish as $z$ tends to one. 
This feature is somewhat surprising and might raise questions about its compatibility with collinear factorization. 
Indeed, at leading twist, the fragmentation process involves a single parton transitioning into the observed hadron. 
As such, the probability of the parton transferring its entire momentum to the hadron (as $z \to 1$) should theoretically go to zero.  

However, encountering nonzero FFs near the $z$-endpoint is not uncommon within the framework of NRQCD. 
For instance, the color-singlet $(g \to \eta_{c,b})$ FFs at LO exhibit growth with $z$, reaching a peak as $z$ reaches one~\cite{Braaten:1993rw}. 
At NLO, these same functions negatively diverge due to DGLAP evolution. 
Some authors argue that this behavior is not problematic because the collinear convolution of the divergent FF with the remainder of the cross section remains well-defined~\cite{Artoisenet:2014lpa}. 
Others interpret the endpoint singularity as indicative of perturbative instability and propose resummation techniques as a potential remedy~\cite{Zhang:2018mlo}.  

We believe that further investigation is required to establish the correct endpoint behavior of NRQCD-based FFs. 
This could involve extending or generalizing the NRQCD factorization framework as applied to fragmentation. 
While such developments are beyond the scope of our exploratory analysis, they undoubtedly warrant attention in future studies.

Let us now shift our focus to the $[Q \to \TQQ]$ initial-scale inputs for our {\tt TQ4Q1.1} FFs.
As done for the gluon case, to benchmark our functions we perform again a simplified, expanded DGLAP evolution.
This time, however, we keep both the gluon-to-gluon timelike splitting kernel, $P_{gg}$, and the quark-to-quark one, $P_{qq}$.
As explained in Sec.~\ref{sssec:FFs_TQ4Q11}, since the starting threshold for our $[Q \to \TQQ]$ fragmentation is $\mu_{F,0} = 5 m_Q$, we expect that also the $[g \to \TQQ]$ channel, whose threshold is $\mu_{F,0} = 4 m_Q$, participates into the evolution.

For the sake of illustration, in Fig.~\ref{fig:TQQ_FF_initial-scale_Q} we show the $z$-dependence of $[Q \to \TQQ]$ initial-scale inputs for our {\tt TQ4Q1.1} FFs.
In this case, the shaded bands reflect the variation of the factorization scale in a range centered at $\mu_{F,0} = 5 m_Q$ with a width of $2 m_Q$.

We observe that the $[Q \to \TQQZpp]$ initial-scale FFs, multiplied by $z$, exhibit a distinctive pattern (left plots): a decreasing trend with $z$ up to approximately $z \approx 0.6$, followed by a pronounced peak in the $z > 0.8$ region, before eventually vanishing as $z$ approaches one. 
In contrast, the $[Q \to \TQQTpp]$ FFs (right plots) display a less pronounced peak, forming a shape more akin to a plateau.
As it happens for the gluon channels, because of the \emph{Ansatz} made in Eq.~\eqref{LDMEs_T4b}, also quark to $\TQb$ initial-scale FFs share the same pattern of corresponding $\TQc$ ones, but with a different magnitude.

The $[Q \to \TQQ]$ initial-scale FFs inputs for our {\tt TQ4Q1.1} FFs, calculated by the hands of NRQCD, are quite different from {\tt TQ4Q1.0} corresponding ones, taken from Suzuki.
Comparing upper plots of Fig.~\ref{fig:TQQ_FF_initial-scale_Q} with plots of Fig.~3 of Ref.~\cite{Celiberto:2024mab}, we immediately note that they differ not only in shape but also in bulk magnitude, the latter being much smaller for the {\tt 1.1} update.

\subsubsection{The {\tt TQ4Q1.1} functions}
\label{sssec:FFs_TQ4Q11}

\begin{figure*}[!t]
\centering

   \hspace{-0.00cm}
   \includegraphics[scale=0.41,clip]{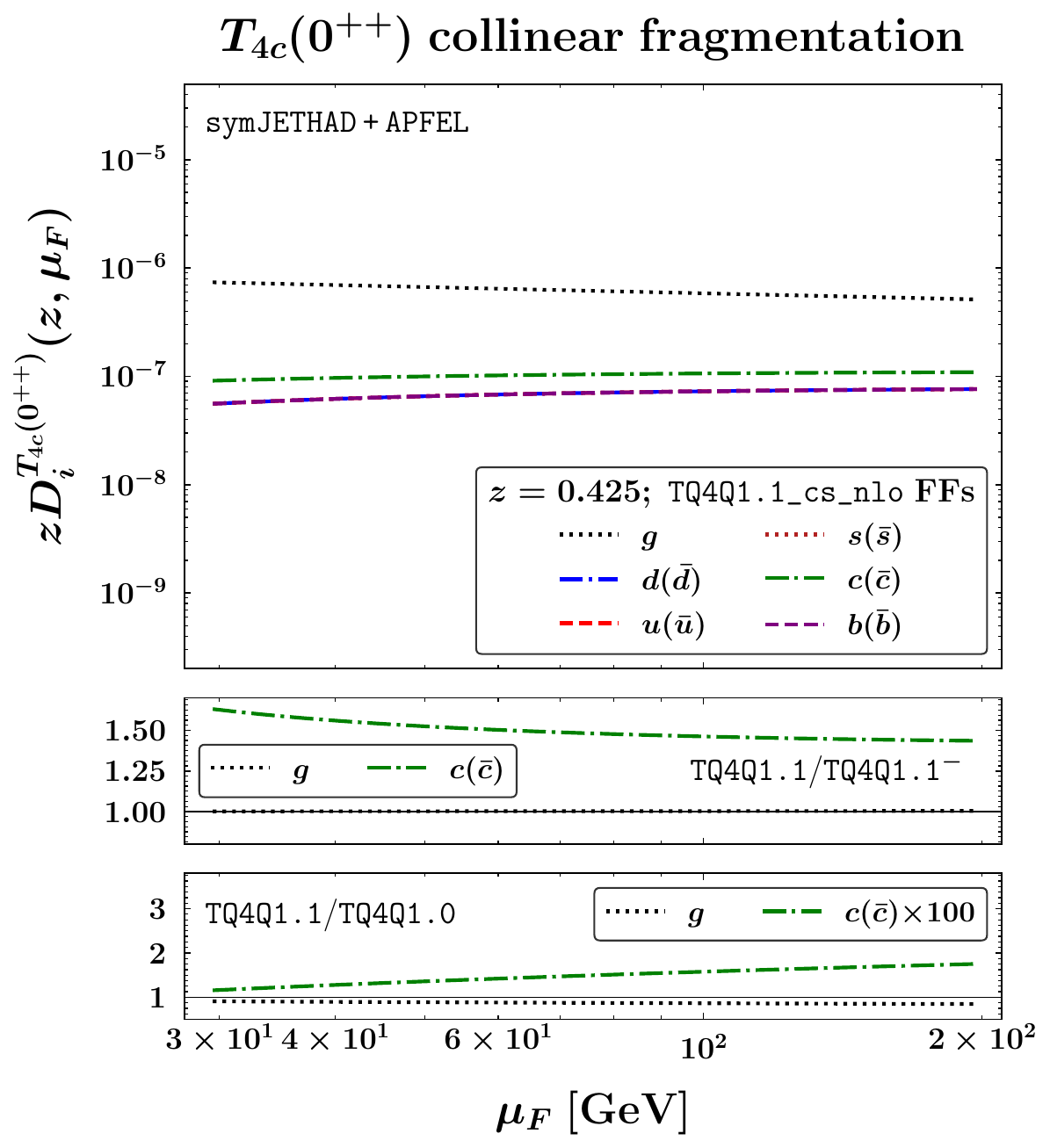}
   \hspace{-0.10cm}
   \includegraphics[scale=0.41,clip]{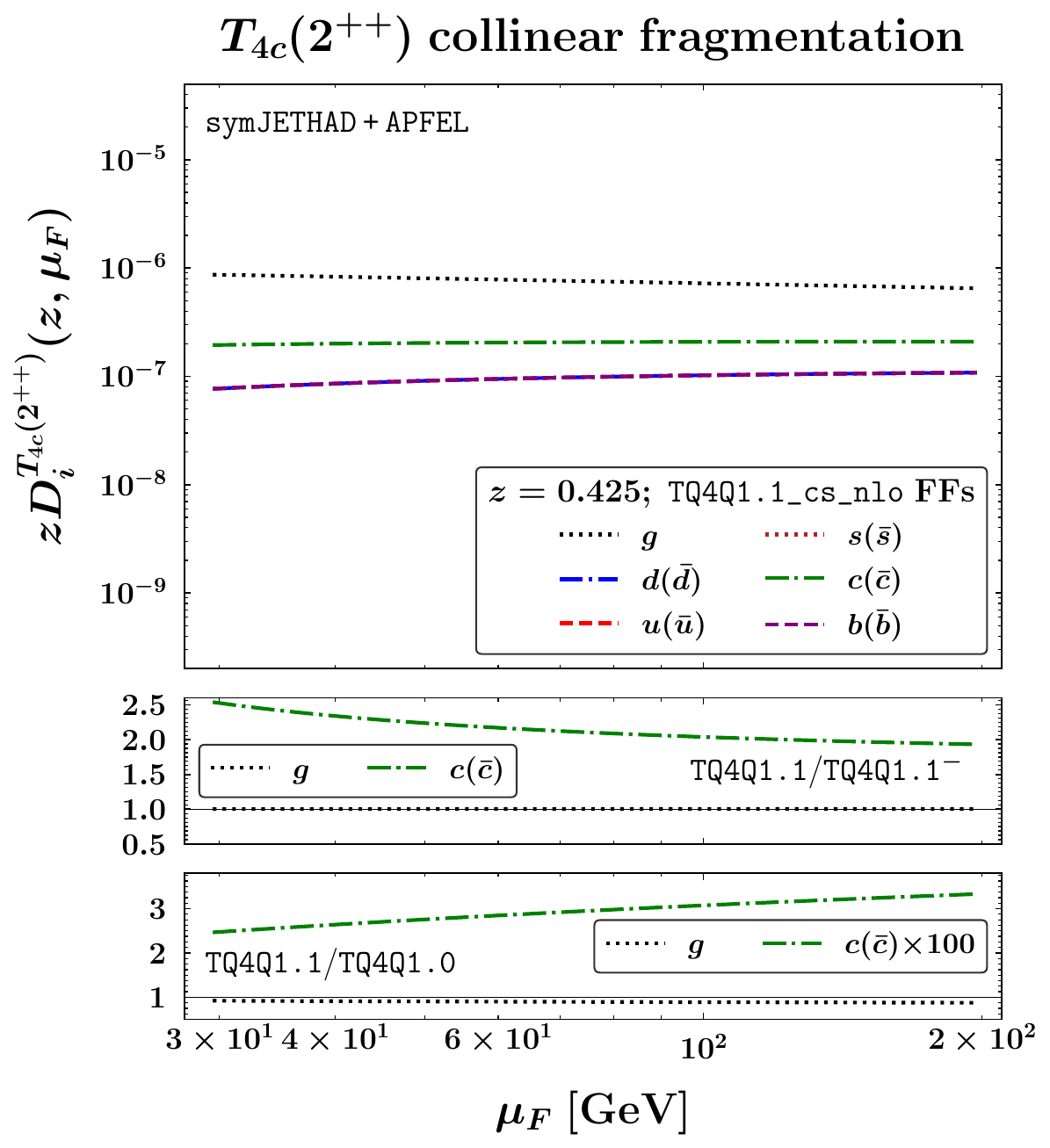}

\caption{Factorization-scale dependence of {\tt TQ4Q1.1} collinear FFs depicting $\TQcZpp$ (left) and $\TQcTpp$ (right) formation, at $z = 0.425 \simeq \langle z \rangle$.
First ancillary panels below primary plots show the ratio between {\tt TQ4Q1.1} and {\tt TQ4Q1.1}$^-$ functions.
Second ancillary panels show the ratio between {\tt TQ4Q1.1} and {\tt TQ4Q1.0} functions.
For comparison purposes, the charm ratio in the latter has been scaled down by a factor of 100.}
\label{fig:NLO_FFs_Tc0_Tc2}
\end{figure*}

\begin{figure*}[!t]
\centering

   \hspace{-0.00cm}
   \includegraphics[scale=0.41,clip]{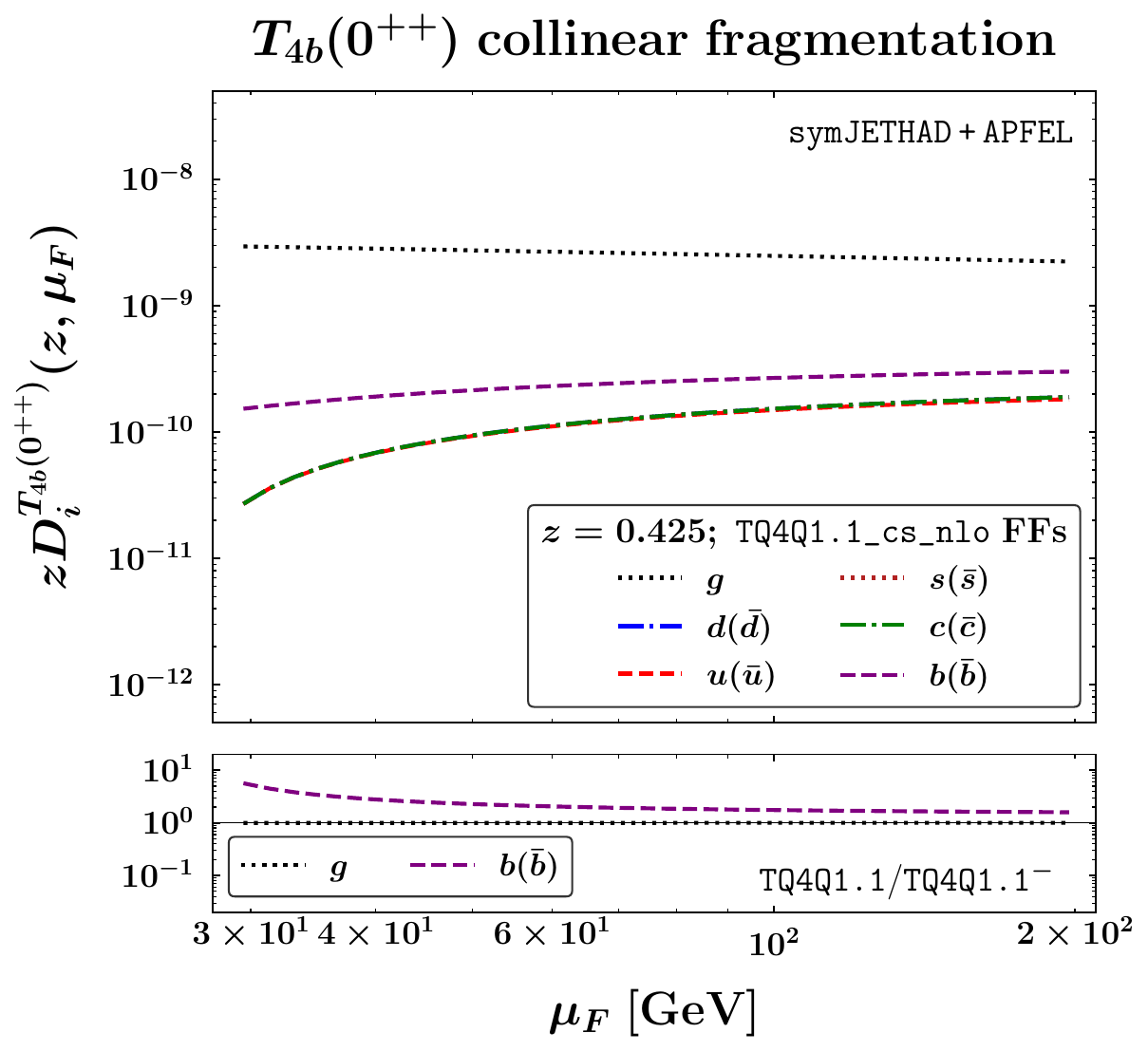}
   \hspace{-0.10cm}
   \includegraphics[scale=0.41,clip]{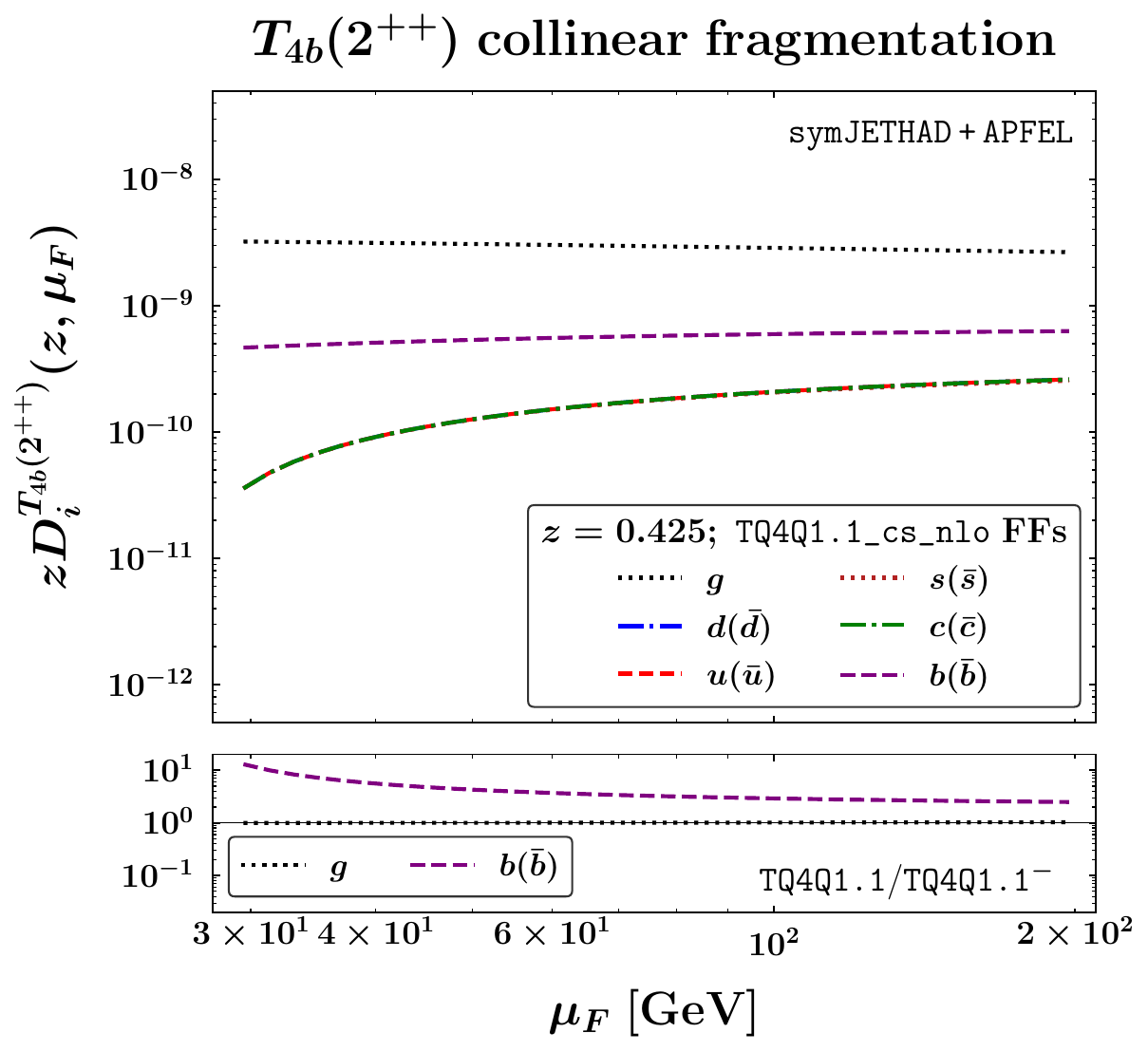}

\caption{Factorization-scale dependence of {\tt TQ4Q1.1} collinear FFs depicting $\TQbZpp$ (left) and $\TQbTpp$ (right) formation, at $z = 0.425 \simeq \langle z \rangle$.
Ancillary panels below primary plots show the ratio between {\tt TQ4Q1.1} and {\tt TQ4Q1.1}$^-$ functions.}
\label{fig:NLO_FFs_Tb0_Tb2}
\end{figure*}

The final step in constructing our {\tt TQ4Q1.1} collinear FFs for fully heavy tetraquarks involves performing a consistent DGLAP evolution of the initial-scale inputs introduced throughout this section.  
A striking distinction from light-hadron fragmentation is that, in this case, both the heavy-quark and gluon channels exhibit evolution thresholds.  
This feature arises from the $[g \to (Q\bar{Q}Q\bar{Q})]$ and $[Q,\bar{Q} \to (Q\bar{Q}Q\bar{Q}) + Q,\bar{Q}]$ perturbative splittings, which correspond to the left and right initial-scale inputs in Fig.~\ref{fig:TQQ_FF_diagrams}, respectively, and are mathematically defined by corresponding SDCs, given in Sec.~\ref{sssec:FFs_T4Q_SDCs}.  

Kinematics suggests us that the minimal invariant mass for the first splitting is $\mu_{F,0}(g \to \TQQ) = 4 m_Q$, which we adopt as the threshold for gluon fragmentation.  
Similarly, the minimal invariant mass for the second splitting is $\mu_{F,0}(Q \to \TQc) = 5 m_Q$, which we designate as the threshold for heavy-quark (or antiquark) fragmentation.

To perform a threshold-consistent DGLAP evolution, we make use of features of the novel {\HFNRevo} scheme~\cite{Celiberto:2024mex,Celiberto:2024bxu}.
As mentioned in Sec.~\ref{ssec:FFs_quarkonia}, the DGLAP evolution within {\HFNRevo} is split in two steps: first {\tt EDevo}, then {\tt AOevo}.
We start from the fragmentation channel having the lowest threshold.
It is the gluon one, with $\mu_{F,0}(g \to \TQQ) = 4 m_Q$. 
By keeping in the DGLAP equation only the $P_{gg}$ splitting kernel, we evolve the gluon FF input up the initial scale of the heavy-quark one, $\mu_{F,0}(Q \to \TQQ) = 5 m_Q$. 
Thus, we generate collinear gluons, and only gluons, between the two scales.
Because this evolution is \emph{expanded} in powers of $\alpha_s$ and \emph{decoupled} from any other quark channel, it can be done analytically by the hands of the {\symJethad} plugin.

We now proceed to the second step, where we combine the gluon FF, evolved at $Q_0 \equiv 5 m_Q$, with the corresponding heavy-quark input. 
From this point, we perform an all-order DGLAP evolution to obtain our NLO {\tt TQ4Q1.1} functions, which are then made available in {\tt LHAPDF} format.  
$Q_0$ is referred to as the \emph{evolution-ready} scale, as it represents the largest threshold among all parton species.  
It also serves as the starting point for the numerical DGLAP evolution process. 

From a first perspective, our treatment might seem incomplete because of the absence of nonconstituent light- and heavy-quark FF inputs. 
As a result, in our two-step evolution framework, nonconstituent quarks are introduced only through the evolution process, without any initial-scale inputs. However, drawing on analogies with NRQCD studies of color-singlet pseudoscalar~\cite{Braaten:1993rw,Braaten:1993mp,Artoisenet:2014lpa,Zhang:2018mlo,Zheng:2021mqr,Zheng:2021ylc} and vector~\cite{Braaten:1993rw,Braaten:1993mp,Zheng:2019dfk} quarkonia, it is reasonable to expect these channels to be significantly suppressed in comparison to gluon and charm contributions.
Nevertheless, to rely upon a more complete set of NRQCD inputs, we plan to encode the light-quark channel, recently studied~\cite{Bai:2024flh} during the preparation of this work, in a future analysis.

In Fig.~\ref{fig:NLO_FFs_Tc0_Tc2} we present the factorization-scale dependence of {\tt TQ4Q1.1} NLO FF determinations for $\TQcZpp$ and $\TQcTpp$ fully charmed tetraquarks.
Analogously, {\tt TQ4Q1.1} NLO functions for $\TQbZpp$ and $\TQbTpp$ fully bottomed tetraquarks  are shown in Fig.~\ref{fig:NLO_FFs_Tb0_Tb2}.
As done before (see Figs.~\ref{fig:NLO_FFs_bHs_Jps_BCs_Bss},~\ref{fig:NLO_FFs_Xcq}, and~\ref{fig:NLO_FFs_Xbq}), we set $z = 0.425 \simeq \langle z \rangle$.
This roughly corresponds to the mean value at which FFs are generally sounded in semihard final states (see, \emph{e.g.}, Refs.~\cite{Celiberto:2016hae,Celiberto:2017ptm,Celiberto:2020wpk,Celiberto:2021dzy,Celiberto:2021fdp,Celiberto:2022dyf,Celiberto:2022keu,Celiberto:2022kxx,Celiberto:2024omj}).

As a general trend, the constituent heavy-quark FF is always five to 20 times larger than the other ones in the scanned window of $\mu_F$.
Although the contribution of $[g \to \TQQ]$ fragmentation is significantly smaller than that of $[Q \to \TQQ]$ one, the former is still crucial for accurately describing production rates at hadron colliders. 
This is primarily due to the much larger gluon PDF compared to the quark PDFs, which makes the $[gg \to gg]$ partonic channel far more important than the $[gg \to c\bar{c}]$ channel.

Given that our {\tt TQ4Q1.1} NLO functions include the charm FF at the initial scale, we can assess its influence on the resulting DGLAP-evolved sets. 
To this end, we follow the same strategy adopted for the {\tt TQ4Q1.0} case (see Sec.~2.4 of Ref.~\cite{Celiberto:2024mab}). 
For testing purposes, we derive supplementary sets NLO FFs, named {\tt TQ4Q1.1$^-$}. 
These sets have been obtained by using the same methodology as the main {\tt TQ4Q1.1} functions, but with the $[Q \to \TQQ]$ input excluded. 
Consequently, only the gluon FF is present at the initial scale, while the charm channel, like the other quark species, is generated entirely through evolution.

The first ancillary panels beneath the primary plots in Figs.~\ref{fig:NLO_FFs_Tc0_Tc2} and~\ref{fig:NLO_FFs_Tb0_Tb2} display the ratio of {\tt 1.1} to {\tt 1.1$^-$} FFs for both gluon and constituent heavy-quark FFs. 
We note that the gluon FF remains almost unchanged in all cases. 
On the other hand, {\tt 1.1} $[Q \to \TQQ]$ channels are roughly 1.5 to 10 times higher than corresponding {\tt 1.1$^-$} counterparts. 

This significant difference could critically affect the accurate modeling of $\TQQ$ production rates at lepton and lepton-hadron colliders, where $[\gamma^{(*)}\gamma^{()*} \to c\bar{c}]$ and $[\gamma^{(*)}g \to c\bar{c}]$ subprocesses play an essential role. 
These findings underscore that, by incorporating both gluon and charm contributions at the initial scale, the {\tt TQ4Q1.1} FFs offer a versatile tool for describing a wide array of processes across hadron, lepton, and lepton-hadron collider environments.

For the sake of comparison, the second ancillary panels below plots of Fig.~\ref{fig:NLO_FFs_Tc0_Tc2} show the ratio between {\tt TQ4Q1.1} gluon and charm FFs and their previous versions, embodied in corresponding {\tt TQ4Q1.0} sets~\cite{Celiberto:2024mab}.
Notably, in all cases the {\tt 1.1} NRQCD-updated gluon channel loses not more than 10\% of magnitude with respect to the {\tt 1.0} case.
Conversely, the {\tt 1.1} charm channel is smaller than the corresponding {\tt 1.0} one.
For $\TQcZpp$ states it roughly goes from 0.2\% to 2\% as $\mu_F$ increases, while for $\TQcTpp$ ones it grows from 2.5\% to 3.5\% with $\mu_F$.
This is not surprising, since we already noted that NRQCD and Suzuki models lead to distinct patterns, both in the shape as in the bulk magnitude (see the discussion at the end of Sec.~\ref{sssec:FFs_T4Q_gQ}).

The {\tt TQ4Q1.1} gluon FFs in Figs.~\ref{fig:NLO_FFs_Tc0_Tc2} and~\ref{fig:NLO_FFs_Tb0_Tb2} show a very soft decline with increasing $\mu_F$. 
A similar behavior is partially mirrored in the gluon-to-$B_c^{(*)}$ fragmentation channel (Fig.~\ref{fig:NLO_FFs_bHs_Jps_BCs_Bss}) as well as in gluon-to-$\QXQq$ ones (Figs.~\ref{fig:NLO_FFs_Xcq} and.~\ref{fig:NLO_FFs_Xbq}). 
As discussed in Sec.~\ref{ssec:FFs_quarkonia}, a smooth $\mu_F$-dependence of the gluon FF acts as a ``natural stabilizer'' for high-energy resummed distributions sensitive to semi-inclusive emissions of heavy-flavored hadrons (see Sec.~\ref{sec:results} for phenomenological applications). 

\section{Hybrid factorization at NLL/NLO$^+$}
\label{sec:hybrid_factorization}

The first part of this Section (\ref{ssec:HE_resummation}) contains a short overview of recent phenomenological studies of the QCD semihard sector at hadron colliders. 
The second part (\ref{ssec:NLL_cross_section}) provides us with the core ingredients of the $\NLLp$ hybrid factorization well adapted to the description of bottomoniumlike states.

\subsection{High-energy resummation at a glance}
\label{ssec:HE_resummation}

Final states sensitive to heavy-flavored hadron production are essential for probing high-energy QCD, where large energy logarithms significantly impact the perturbative expansion in the strong coupling, and need to be resummed to all orders. 
The Balitsky–Fadin–Kuraev–Lipatov (BFKL) formalism~\cite{Fadin:1975cb,Kuraev:1977fs,Balitsky:1978ic} resums these logarithms both at the leading level (LL), including terms proportional to $[\alpha_s \ln (s)]^n$, and the next-to-leading level (NLL), handling contributions proportional to $\alpha_s [\alpha_s^n \ln (s)]^n$.

High-energy resummed production rates for hadron-initiated reactions read as a transverse-momentum convolution of a universal Green's function, known at NLO~\cite{Fadin:1998py,Ciafaloni:1998gs}, and two process-dependent, singly off-shell emission functions, also named forward impact factors.
These emission functions embody collinear PDFs and FFs. 
This collinear convolution, nested inside the aforementioned BFKL one, makes our formalism a \emph{hybrid} collinear and high-energy factorization~\cite{Colferai:2010wu,Celiberto:2020wpk,Celiberto:2020tmb,Bolognino:2021mrc,Celiberto:2022rfj,Celiberto:2022dyf} (see also~\cite{vanHameren:2022mtk,Giachino:2023loc,Bonvini:2018ixe,Silvetti:2022hyc,Silvetti:2023suu,Rinaudo:2024hdb} for similar approaches to single-particle detections).

BFKL studies within a full or partial $\NLLp$ accuracy where done \emph{via} the following processes: the Mueller–Navelet~\cite{Mueller:1986ey} dijet production~\cite{Colferai:2010wu,Ducloue:2013hia,Ducloue:2013bva,Caporale:2014gpa,Celiberto:2015yba,Celiberto:2015mpa,Celiberto:2016ygs,Celiberto:2022gji,Caporale:2018qnm,deLeon:2021ecb,Baldenegro:2024ndr,Egorov:2023duz}, light dihadron~\cite{Celiberto:2016hae,Celiberto:2017ptm,Celiberto:2020rxb,Celiberto:2022rfj}, and hadron-jet~\cite{Bolognino:2018oth,Bolognino:2019cac,Bolognino:2019yqj,Celiberto:2020wpk,Celiberto:2020rxb,Celiberto:2022kxx} emissions, Drell–Yan \cite{Celiberto:2018muu,Golec-Biernat:2018kem}, Higgs \cite{Celiberto:2020tmb,Celiberto:2023rtu,Celiberto:2023uuk,Celiberto:2023eba,Celiberto:2023nym,Celiberto:2023rqp,Celiberto:2022zdg,Nefedov:2019mrg}, and heavy-flavor tags~\cite{Celiberto:2017nyx,Boussarie:2017oae,Bolognino:2019ouc,Bolognino:2019yls,Celiberto:2021dzy,Celiberto:2021fdp,Celiberto:2022dyf,Celiberto:2023fzz,Celiberto:2022grc,Bolognino:2022paj,Celiberto:2022keu,Celiberto:2022kza,Celiberto:2024omj}.

Additionally, forward emissions of single particles are direct probes of the gluon content of the proton at low-$x$ through the unintegrated gluon distribution (UGD), which relies upon the BFKL evolution kernel. Phenomenological studies of the UGD have been performed through exclusive light vector-meson electroproduction at HERA~\cite{Anikin:2011sa,Bolognino:2018rhb,Bolognino:2018mlw,Bolognino:2019bko,Bolognino:2019pba,Celiberto:2019slj,Luszczak:2022fkf} and the EIC~\cite{Bolognino:2021niq,Bolognino:2021gjm,Bolognino:2022uty,Bolognino:2022ndh}, as well as through vector-quarkonium photoemission~\cite{Garcia:2019tne,Hentschinski:2020yfm,Peredo:2023oym}. These analyses enhance our understanding of the UGD and its role in low-$x$ physics.

Accessing the gluon content \emph{via} the UGD has been crucial in enhancing the collinear factorization framework, particularly in determining resummed low-$x$ PDFs~\cite{Ball:2017otu,Abdolmaleki:2018jln,Bonvini:2019wxf}. It has also been used to improve spin-dependent TMD PDFs at low~$x$~\cite{Bacchetta:2020vty,Bacchetta:2024fci,Celiberto:2021zww,Bacchetta:2021oht,Bacchetta:2021lvw,Bacchetta:2021twk,Bacchetta:2022esb,Bacchetta:2022crh,Bacchetta:2022nyv,Celiberto:2022omz,Bacchetta:2023zir,Bacchetta:2024uxb}. 
Refs.~\cite{Hentschinski:2021lsh,Mukherjee:2023snp} explore the interplay between BFKL and TMD dynamics, while Refs.~\cite{Boroun:2023goy,Boroun:2023ldq} link color-dipole production rates to the UGD.

A key advancement came recently out from high-energy emissions of singly heavy-flavored bound states, like $\Lambda_c$ baryons~\cite{Celiberto:2021dzy} or $b$-hadrons~\cite{Celiberto:2021fdp}, which allowed us to mitigate the well-known issues affecting the BFKL description of semihard final states at natural scales.
These issues are particularly pronounced when lighter particles are involved~\cite{Ducloue:2013bva,Caporale:2014gpa,Bolognino:2018oth,Celiberto:2020wpk}. 
In such cases, large negative NLL corrections, together with unresummed \emph{threshold} logarithms, hinder the convergence of the resummed series. 
This issue becomes manifest when examining the impact of MHOU uncertainties through variations in factorization and renormalization scales.

Conversely, as anticipated in Sec.~\ref{ssec:FFs_quarkonia}, a \emph{natural stabilization} trend~\cite{Celiberto:2022grc} has been observed in reactions involving the (semi)inclusive production of heavy hadrons at high transverse masses. 
In these cases, the primary production mechanism is VFNS collinear fragmentation. 
This stability was further tested through doubly heavy-flavored mesons using a collinearly enhanced nonrelativistic fragmentation approach. 
As part of this research, new VFNS, DGLAP-evolving FFs were built on NRQCD inputs \cite{Braaten:1993mp,Zheng:2019dfk,Braaten:1993rw,Chang:1992bb,Braaten:1993jn,Ma:1994zt,Zheng:2019gnb,Zheng:2021sdo,Feng:2021qjm,Feng:2018ulg}, first for vector quarkonia~\cite{Celiberto:2022dyf,Celiberto:2023fzz}, and later for charmed $B$ mesons~\cite{Celiberto:2022keu,Celiberto:2024omj}.

Thus, the natural stability of the high-energy resummation acted as a phenomenological bridge between the high-energy QCD regime and the exotic-matter domain.
In Ref.~\cite{Celiberto:2023rzw} (see Ref.~\cite{Celiberto:2024mrq} for a review), novel VFNS FF determinations, named {\tt TQHL1.0} functions, were built to address the formation mechanism of neutrally charged, doubly heavy tetraquarks: $\QXQq$.
The analysis was subsequently extended to fully heavy tetraquarks, $\TQQ$, with the determination of the corresponding {\tt TQ4Q1.0} FFs.

\subsection{Resummed cross section}
\label{ssec:NLL_cross_section}

\begin{figure*}[!t]
\centering
\includegraphics[width=0.475\textwidth]{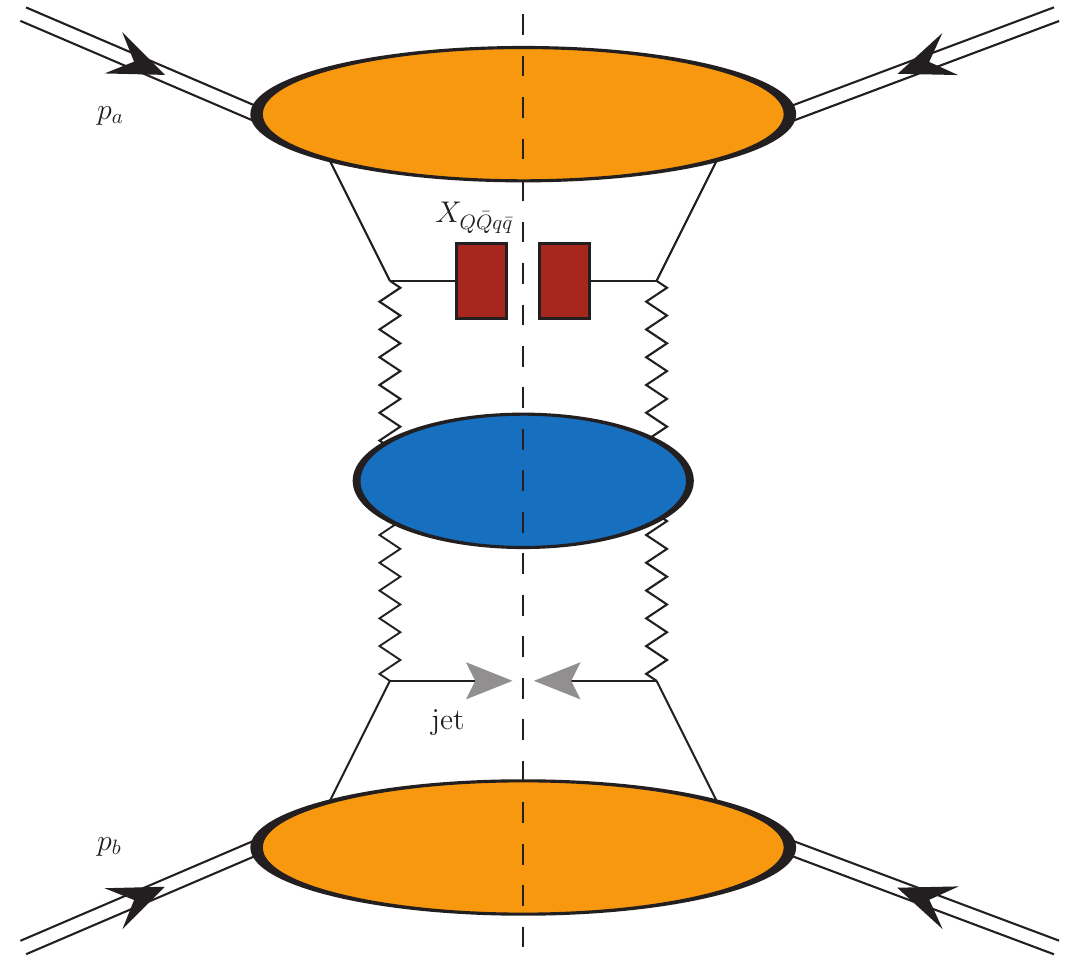}
\hspace{0.55cm}
\includegraphics[width=0.475\textwidth]{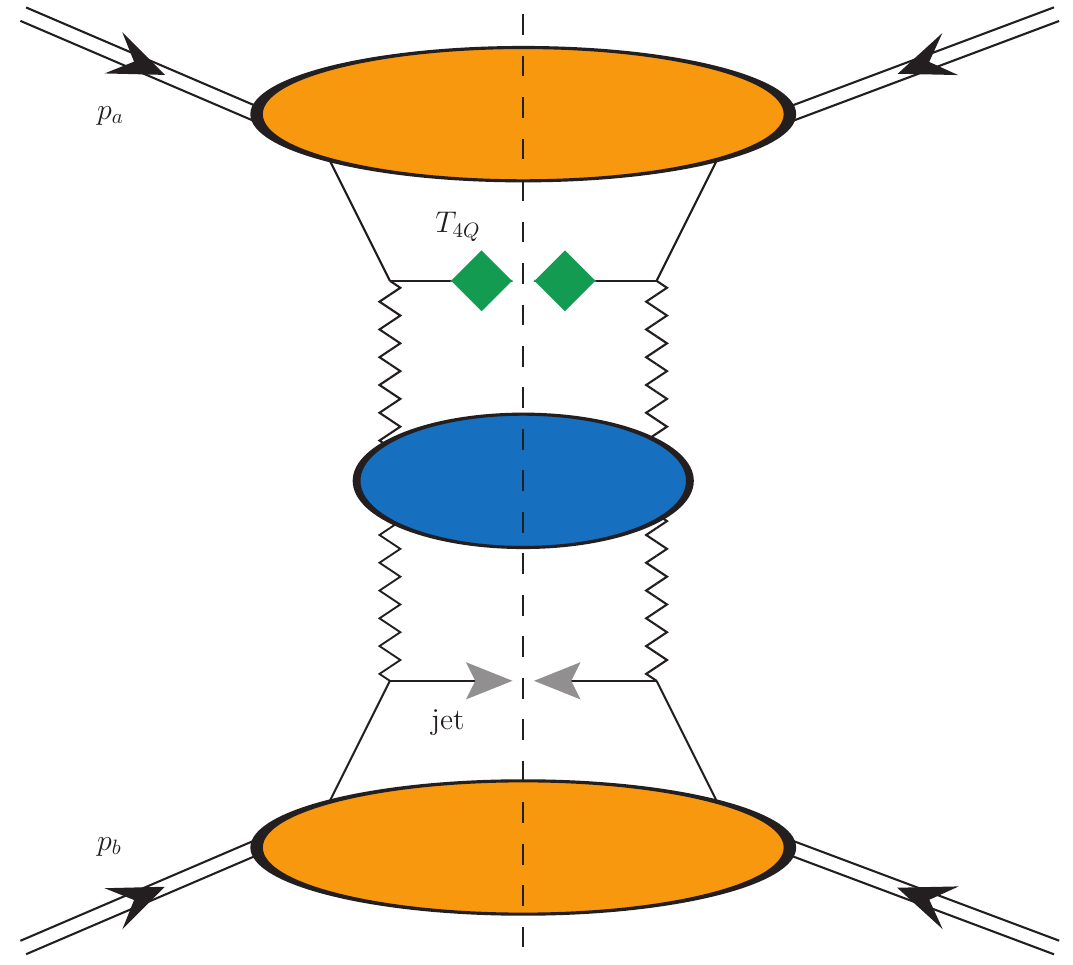}

\caption{Hybrid factorization for the $\TQQ$ plus jet (left) and $\QXQq$ plus jet (right) semi-inclusive hadroproduction.
Firebrick squares (green rhombi) represent $\QXQq$ ($\TQQ$) tetraquark
collinear FFs. Gray arrows depict light-flavored jets. 
Orange ovals stand for proton collinear PDFs.
The BFKL Green’s function (blue blob) is connected to the two off-shell emission functions through Reggeon lines.}
\label{fig:process}
\end{figure*}

As an application to LHC/FCC phenomenology, we focus on the following semi-inclusive hadroproduction (see Fig.~\ref{fig:process})
\begin{equation}
\label{eq:process}
 {\rm p}(p_a) + {\rm p}(p_b) \,\to\, {\B}(\kappa_1, y_1) + {\cal X} + {\rm jet}(\kappa_2 , y_2) \;,
\end{equation}
where a bottomoniumlike state, ${\B}$, is detected with four-momentum $\kappa_1$, rapidity $y_1$, and azimuthal angle $\varphi_1$.
Furthermore, a light jet is tagged with four-momentum $\kappa_2$, rapidity $y_2$, and azimuthal angle $\varphi_2$.
The two objects feature high transverse momenta, say $|\vec \kappa_{1,2}| \gg \Lambda_{\rm QCD}$, and large rapidity separation, $\DY \equiv y_1 - y_2$. 
An undetected gluon system, ${\cal X}$, is inclusively produced.
We decompose the final-state transverse momenta on the Sudakov-vector basis generated by parent protons' momenta, $p_{a,b}$, with $p_{a,b}^2 = 0$ and $({p_a} \cdot {p_b}) = s/2$, thus having
\begin{equation}
\label{Sudakov}
 \kappa_{1,2} = x_{1,2} \, p_{a,b} - \frac{\kappa_{1,2 \perp}^2}{x_{1,2} s} \, p_{b,a} + \kappa_{{1,2 \perp}} \;,
\end{equation}
with $\kappa_{1,2 \perp}^2 \equiv - \vec \kappa_{1,2}^{\,2}$.
In the center-of-mass frame the following relations between final-state longitudinal-momentum fractions, rapidities, and transverse momenta hold
\begin{equation}
\label{xyp}
 x_{1,2} = \frac{|\vec \kappa_{1,2}|}{\sqrt{s}} e^{\pm y_{1,2}} \;, \qquad \drv y_{1,2} = \pm \frac{d x_{1,2}}{x_{1,2}} \;,
\end{equation}
and thus
\begin{equation}
\label{Delta_Y}
 \qquad \DY \equiv y_1 - y_2 = \ln\frac{x_1 x_2 s}{|\vec \kappa_1| |\vec \kappa_2|} \;.
\end{equation}

In a pure collinear factorization vision, the LO differential cross section for our reactions would be cast as a one-dimensional convolution between the on-shell hard factor, proton PDFs, and $\B$ FFs
\begin{eqnarray}
\label{sigma_collinear}
&&\hspace{-0.25cm}
\frac{\drv\sigma^{\rm LO}_{\rm [coll.]}}{\drv x_1\drv x_2\drv ^2\vec \kappa_1\drv ^2\vec \kappa_2}
= \hspace{-0.25cm} \sum_{i,j=q,{\bar q},g}\int_0^1 \hspace{-0.20cm} \drv x_a \!\! \int_0^1 \hspace{-0.20cm} \drv x_b\ f_i\left(x_a\right) f_j\left(x_b\right) 
\nonumber \\
&&\quad\times \, 
\int_{x_1}^1 \hspace{-0.15cm} \frac{\drv \zeta}{\zeta} \, D^{\B}_i\left(\frac{x_1}{\zeta}\right) 
\frac{\drv {\hat\sigma}_{i,j}\left(\hat s\right)}
{\drv x_1\drv x_2\drv ^2\vec \kappa_1\drv ^2\vec \kappa_2}\;,
\end{eqnarray}
where the $i,j$ indices run over all partons except for the $t$ quark, which does not hadronize.
For brevity, the explicit dependence on the factorization scale, $\mu_F$, is not shown in Eq.~\eqref{sigma_collinear}.
The $f_{i,j}\left(x_{a,b}, \mu_F \right)$ functions are the proton PDFs, whereas the $D^{\B}_i\left(x_1/\zeta, \mu_F \right)$ functions stand for the $\B$ FFs.
Then, $x_{a,b}$ denote the parent partons' longitudinal fractions, while $\zeta$ is the momentum fraction of the outgoing parton fragmenting into the $\B$ particle. 
Finally, $\drv\hat\sigma_{i,j}\left(\hat s\right)$ depicts the partonic hard factor, with $\hat s \equiv x_a x_b s$ the partonic center-of-mass energy squared.

On the other side, the differential cross section within our hybrid collinear and high-energy factorization builds on a transverse-momentum convolution between the BFKL Green's function and the two forward-production, singly off-shell emission functions.
Collinear elements, namely PDFs and FFs, are embodied in the latter ones.
We suitably recast the cross section as a Fourier series of the azimuthal-angle coefficients, ${\cal C}_{, \ge 0}$,
\begin{equation}
 \label{dsigma_Fourier}
 \hspace{-0.22cm}
 \frac{(2\pi)^2 \, \drv \sigma}{\drv y_1 \drv y_2 \drv |\vec \kappa_1| \drv |\vec \kappa_2| \drv \varphi_1 \drv \varphi_2} \!=\!
 \left[ {\cal C}_0 + 2 \! \sum_{m=1}^\infty \! \cos (m \varphi)\,
 {\cal C}_m \right]\, ,
\end{equation}
with $\varphi \equiv \varphi_1 - \varphi_2 - \pi$.

Working in the $\MSb$ renormalization scheme and using BFKL, we get (see Ref.~\cite{Caporale:2012ih} for technical details)
\begin{equation}
\label{Cm_NLLp_MSb}
 \CmNLLp \!\!\!\!=\!\! \frac{x_1 x_2}{|\vec \kappa_1| |\vec \kappa_2|} 
 \int_{-\infty}^{+\infty} \!\!\! \drv \nu \, e^{{\DY} \bar \alpha_s(\mu_R)
 \chi^{\rm NLO}(m,\nu)} 
 \end{equation}
\[ 
 \times \, \alpha_s^2(\mu_R) \biggl\{ \E_\B^{\rm NLO}(m,\nu,|\vec \kappa_1|, x_1)[\E_J^{\rm NLO}(m,\nu,|\vec \kappa_2|,x_2)]^* 
\]
\[
 +
 \left.
 \alpha_s^2(\mu_R) \DY \frac{\beta_0}{4 \pi} \,
 \chi(m,\nu)
 \left[\ln\left(|\vec \kappa_1| |\vec \kappa_2|\right) + \frac{i}{2} \, \frac{\drv}{\drv \nu} \ln\frac{\E_\B}{\E_J^*}\right]
 \right\}
 .
\]
with $\bar \alpha_s(\mu_R) \equiv \alpha_s(\mu_R) N_c/\pi$ the QCD running coupling, $N_c = 3$ the number of colors, and $\beta_0 = 11N_c/3 - 2 n_f/3$ the QCD-$\beta$-function first coefficient.
We select a two-loop running-coupling setup with $\alpha_s\left(m_Z\right)=0.118$ and a dynamic number of flavors, $n_f$.
The high-energy kernel at the exponent of Eq.~\eqref{Cm_NLLp_MSb} encodes the resummation of energy logarithms at NLL
\begin{eqnarray}
 \label{chi}
 \chi^{\rm NLO}(m,\nu) = \chi(m,\nu) + \bar\alpha_s \hat \chi(m,\nu) \;,
\end{eqnarray}
where $\chi(m,\nu)$ are the eigenvalues of the LO kernel
\begin{eqnarray}
 \label{kernel_LO}
 \chi\left(m,\nu\right) = -2\gamma_{\rm E} - 2 \, {\rm Re} \left\{ \psi\left(\frac{1+m}{2} + i \nu \right) \right\} \, ,
\end{eqnarray}
with $\gamma_{\rm E}$ being the Euler-Mascheroni constant and $\psi(z) \equiv \Gamma^\prime
(z)/\Gamma(z)$ the gamma-function logarithmic derivative. 
The $\hat\chi(m,\nu)$ function in Eq.~\eqref{chi} stands for the NLO correction to the high-energy kernel
\begin{eqnarray}
\label{chi_NLO}
&\hat \chi&\left(m,\nu\right) = \bar\chi(m,\nu)+\frac{\beta_0}{8 N_c}\chi(m,\nu)
\\ \nonumber &\times& 
\left\{-\chi(m,\nu)+10/3+2\ln\left[\mu_R^2/\left(|\vec \kappa_1| |\vec \kappa_2|\right)\right]\right\} \;,
\end{eqnarray}
the characteristic $\bar\chi(m,\nu)$ function being computed in Ref.~\cite{Kotikov:2000pm}.

The two expressions
\begin{eqnarray}
\label{EFs}
\E_{\B,J}^{\rm NLO}(m,\nu,|\vec \kappa_{1,2}|,x_{1,2}) =
\E_{\B,J} +
\alpha_s(\mu_R) \, \hat \E_{\B,J}
\end{eqnarray}
respectively represent the bottomed-tetraquark and the light-jet NLO emission functions, obtained in Mellin space after taking their projections onto the eigenfunctions of the LO kernel.
As for the $\B$-particle emission function, we make use of the NLO computation performed in Ref.~\cite{Ivanov:2012iv}. 
While designed for studying the production of light-flavored hadrons, it also fits our VFNS approach for heavy-flavored tetraquarks, provided that the transverse momenta are well above heavy-quark thresholds for the DGLAP evolution.
At LO, one has
\begin{equation}
\begin{split}
 \label{LOBEF}
 \E_\B(m,\nu,|\vec \kappa_1|,x_1) = \upsilon_c \, |\vec \kappa_1|^{2i\nu-1}\int_x^1 \frac{\drv \xi}{\xi} \; \hat{x}^{1-2i\nu} 
 \\
 \times \, \Big[\tau_c f_g(\xi)D_g^\B\left(\hat{x}\right)
 +\sum_{i=q,\bar q}f_i(\xi)D_i^\B\left(\hat{x}\right)\Big] \;,
\end{split}
\end{equation}
with $\hat{x} = x/\xi$, $\upsilon_c = 2 \sqrt{C_F/C_A}$, and $\tau_c = C_A/C_F$, where $C_F = (N_c^2-1)/(2N_c)$ and $C_A = N_c$ are the Casimir constants related with gluon emission from quark and gluon, respectively.
The full NLO formula for $\E_\B^{\rm NLO}$ can be found in Ref.~\cite{Ivanov:2012iv}.
The light-jet LO emission function reads
\begin{equation}
 \label{LOJEF}
 c_J(n,\nu,|\vec \kappa_2|,x) = \upsilon_c
 |\vec \kappa_2|^{2i\nu-1}\,\hspace{-0.05cm}\Big[\tau_c f_g(x)
 +\hspace{-0.15cm}\sum_{j=q,\bar q}\hspace{-0.10cm}f_j(x)\Big] \;.
\end{equation}
Its NLO correction is got by combining Eq.~(36) of Ref.~\cite{Caporale:2012ih} with Eqs.~(4.19)-(4.20) of Ref.~\cite{Colferai:2015zfa}.
It bases upon results presented in Refs.~\cite{Ivanov:2012iv,Ivanov:2012ms}, suited to numerical analyses, which rely on a ``small-cone'' jet selection function~\cite{Furman:1981kf,Aversa:1988vb} with a cone-type algorithm~\cite{Colferai:2015zfa}.

Equations~(\ref{Cm_NLLp_MSb}) and~(\ref{LOJEF}) elegantly illustrate the realization of our hybrid collinear and high-energy factorization scheme. 
Within this framework, the cross section is factorized in the BFKL formalism, where the gluon Green's function and emission functions play central roles. 
The gluon Green's function accounts for the resummation of large logarithms in the high-energy limit, while the emission functions encode the PDFs and FFs, \emph{de facto} combining collinear factorization with high-energy dynamics.

The notation with a `$+$' superscript in the $\CmNLLp$ label indicates that the expression for the azimuthal coefficients in Eq.~(\ref{Cm_NLLp_MSb}) includes contributions beyond the NLL accuracy. 
These additional contributions arise from two sources: the exponentiated NLO corrections to the high-energy kernel and the cross product of the NLO corrections to the impact factors. 
This results in a more precise representation of the azimuthal coefficients, capturing subtle effects that are essential for accurate predictions in processes where both collinear and high-energy logarithms play significant roles.

Then, if one discards all the NLO contributions in Eq.~\eqref{Cm_NLLp_MSb}, the pure LL limit of our angular coefficients is obtained.
We have
\begin{equation}
\begin{split}
 \label{Cm_LL_MSb}
 &\CmLL = \frac{x_1 x_2}{|\vec \kappa_1| |\vec \kappa_2|} 
 \int_{-\infty}^{+\infty} \drv \nu \, e^{{\DY} \bar \alpha_s(\mu_R)\chi(m,\nu)}
 \\[0.18cm] &\hspace{0.15cm}
 \times \, \alpha_s^2(\mu_R) \, \E_\B(m,\nu,|\vec \kappa_1|, x_1)[\E_J(m,\nu,|\vec \kappa_2|,x_2)]^* \;.
\end{split}
\end{equation}

To properly compare high-energy resummed predictions with those from a pure collinear, DGLAP-inspired setup, it is essential to evaluate observables using both our hybrid factorization approach and pure fixed-order computations. 
However, given the current limitations, no numerical code is available for calculating fixed-order distributions at NLO in the context of inclusive semihard hadron-plus-jet production. 
To bridge this gap and assess the impact of high-energy resummation on top of DGLAP predictions, we employ an alternative approach.

Our methodology, originally tailored for studies of Mueller-Navelet~\cite{Celiberto:2015yba,Celiberto:2015mpa} and hadron-jet~\cite{Celiberto:2020wpk} angularities, prescribes the truncation of the high-energy series at NLO accuracy. 
By doing so, we can mimic the high-energy signal that would emerge from a pure NLO calculation.
Specifically, we achieve this by expanding the azimuthal coefficients only up to the order ${\cal O}(\alpha_s^3)$, effectively yielding a high-energy fixed-order ($\HENLOp$) expression. This $\HENLOp$ approximation serves as a practical tool for our phenomenological program, allowing us to systematically compare the effects of the BFKL resummation with the high-energy limit of fixed-order predictions.

Our $\HENLOp$ angular coefficients in the $\MSb$ renormalization scheme read
\begin{align}
\label{Cm_HENLOp_MSb}
 &\CmHENLOp = 
 \frac{e^{\DY}}{s} 
 \int_{-\infty}^{+\infty} \drv \nu \, 
 \alpha_s^2(\mu_R)
 \nonumber \\[0.75em]
 &\hspace{0.50cm}\times \,
 \left[ 1 + \bar \alpha_s(\mu_R) \DY \chi(m,\nu) \right]
 \\[0.75em] \nonumber
 &\hspace{0.50cm}\times \,
 \E_\B^{\rm NLO}(m,\nu,|\vec \kappa_1|, x_1)[\E_J^{\rm NLO}(m,\nu,|\vec \kappa_2|,x_2)]^* \;.
\end{align}

In our study, the renormalization scale ($\mu_R$) and factorization one ($\mu_F$) are set to \emph{natural} energies, which are determined by the kinematics of the final state. 
Specifically, we take $\mu_R = \mu_F \equiv \mu_N$, where $\mu_N$ is defined as
\begin{eqnarray}
\label{mu_N}
 \mu_N = m_{\B \perp} + |\vec{\kappa}_2| \;.
\end{eqnarray}
Here, $m_{\B \perp} = \sqrt{m_\B^2 + |\vec{\kappa}_1|^2}$ denotes the transverse mass of the produced bottomed tetraquark, with its mass set to the sum of the masses of the four constituent quarks. 
The transverse mass of the light-flavor\-ed jet coincides with its transverse momentum, $|\vec{\kappa}_2|$.

Although the emission of two particles naturally introduces two distinct energy scales, in Eq.~\eqref{mu_N} we have adopted a simplified approach by combining these into a single natural reference scale, chosen as the sum of the transverse masses of the two particles.
This selection is consistent with the strategy used in other precision QCD codes and calculations, such as in Refs.~\cite{Alioli:2010xd,Campbell:2012am,Hamilton:2012rf}. 
It facilitates the comparison of our results with predictions from different approaches, while maintaining consistency with typical conventions in QCD phenomenology.

To explore the effect of MHOUs, we will vary both $\mu_R$ and $\mu_F$ in a range from $\mu_N/2$ to $2\mu_N$, controlled by the $C_\mu$ parameter. 
This variation allows us to assess the sensitivity of our results to changes in the energy scales, providing a reliable estimation of theoretical uncertainties in our predictions.

\section{Phenomenology}
\label{sec:results}

All the predictions presented in this section were computed using the \textsc{Python}+\textsc{Fortran} {\Jethad} multimodular code~\cite{Celiberto:2020wpk,Celiberto:2022rfj,Celiberto:2023fzz,Celiberto:2024mrq,Celiberto:2024swu}. 
Proton PDFs are described \emph{via} {\tt NNPDF4.0} NLO set~\cite{NNPDF:2021uiq,NNPDF:2021njg} from {\tt LHAPDF v6.5.4} \cite{Buckley:2014ana}. 
The impact of MHOUs on our observables was evaluated by varying factorization and renormalization scales, $\mu_F$ and $\mu_R$, around the \emph{natural} scale determined by kinematics, adjusting them by a factor between 1/2 and 2 through the $C_\mu$ parameter. 
Uncertainty bands in the plots reflect the combined effects of MHOUs and errors from multidimensional numerical integrations, which were kept constantly below 1\% by the {\Jethad} integration routines.

\subsection{Rapidity-interval rates}
\label{ssec:DY}

\begin{figure*}[!t]
\centering

   \hspace{0.00cm}
   \includegraphics[scale=0.395,clip]{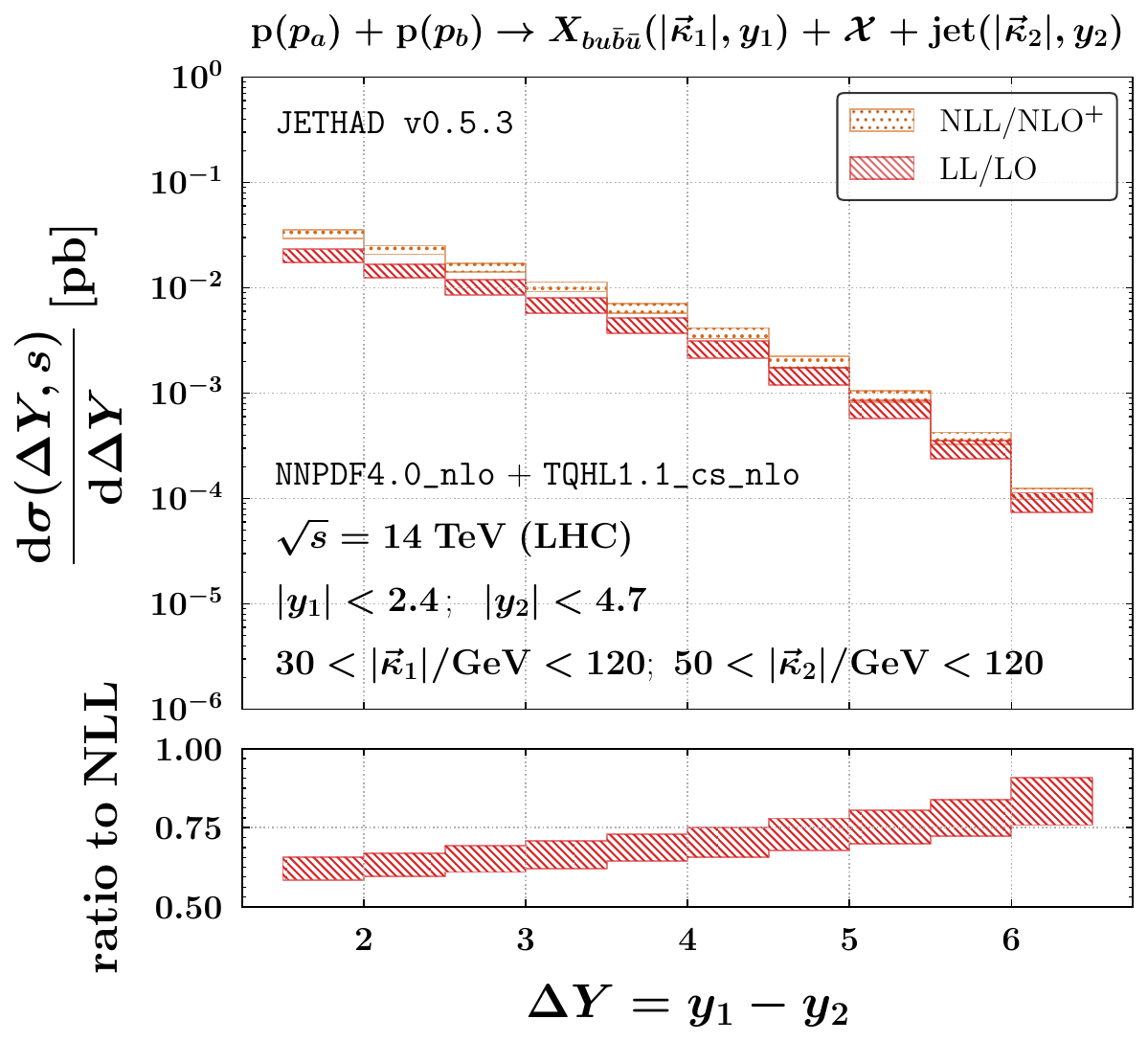}
   \hspace{-0.00cm}
   \includegraphics[scale=0.395,clip]{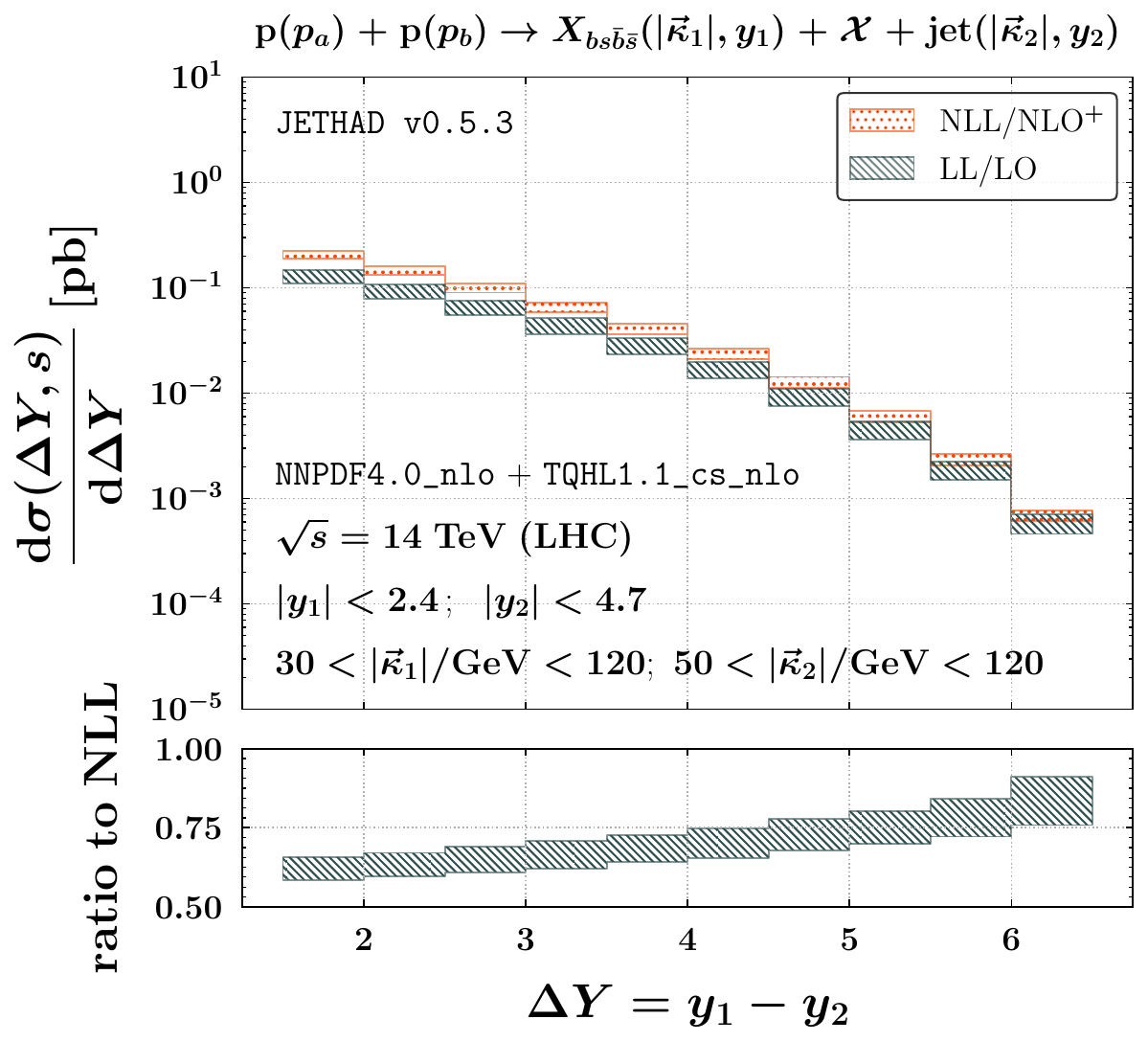}

   \vspace{0.50cm}

   \includegraphics[scale=0.395,clip]{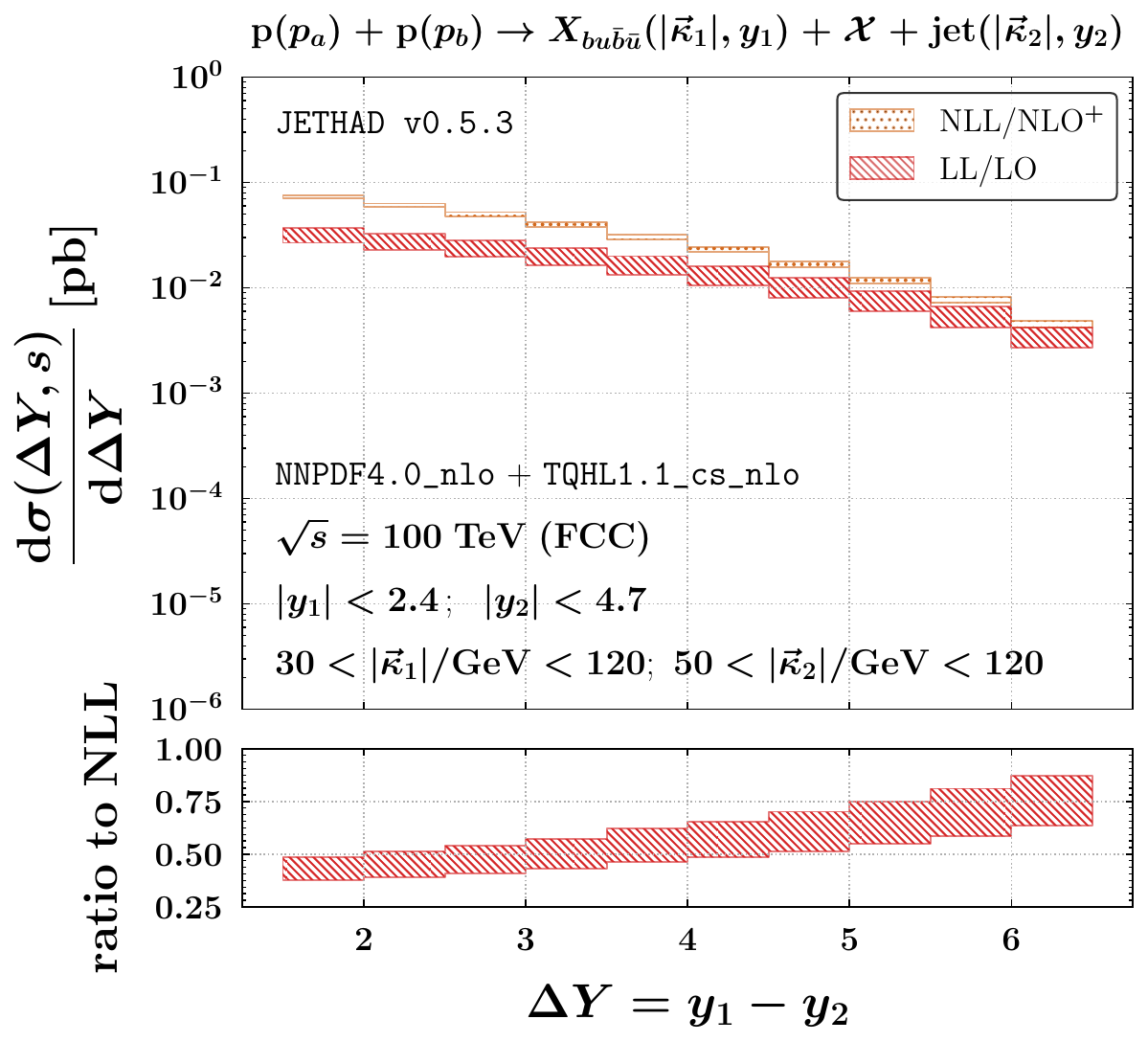}
   \hspace{-0.00cm}
   \includegraphics[scale=0.395,clip]{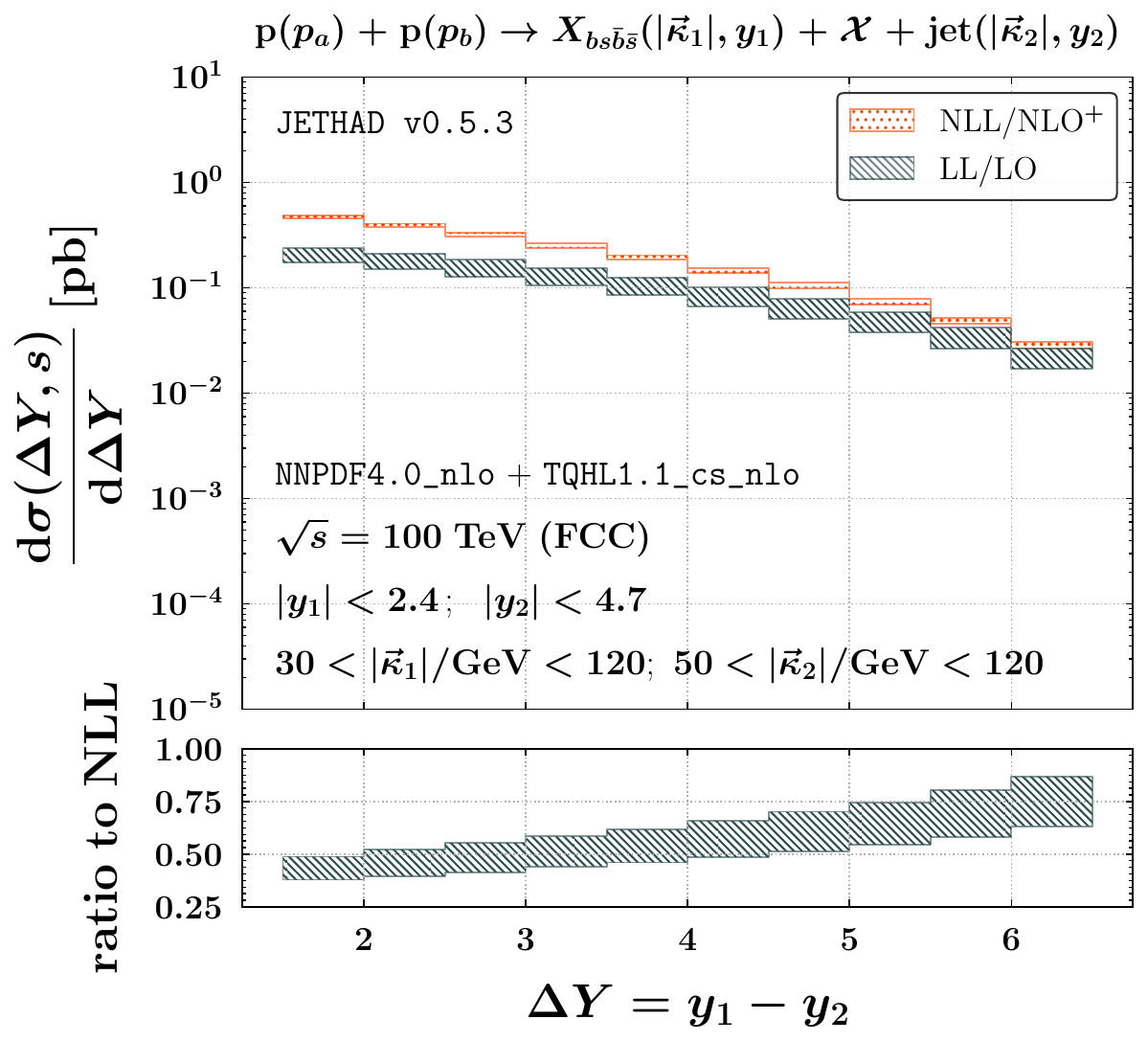}

\caption{Rapidity-interval rates for $\QXbu$ (left) and $\QXbs$ (right) plus jet hadroproduction at $\sqrt{s} = 14$ TeV (LHC, upper) or $100$ TeV (nominal FCC, lower).
Ancillary panels below primary plots exhibit the ratio between $\LL$ and $\NLLp$ predictions.
Uncertainty bands capture the total effect of MHOUs and phase-space multidimensional numeric integration.}
\label{fig:I_Xbq}
\end{figure*}

\begin{figure*}[!t]
\centering

   \hspace{0.00cm}
   \includegraphics[scale=0.395,clip]{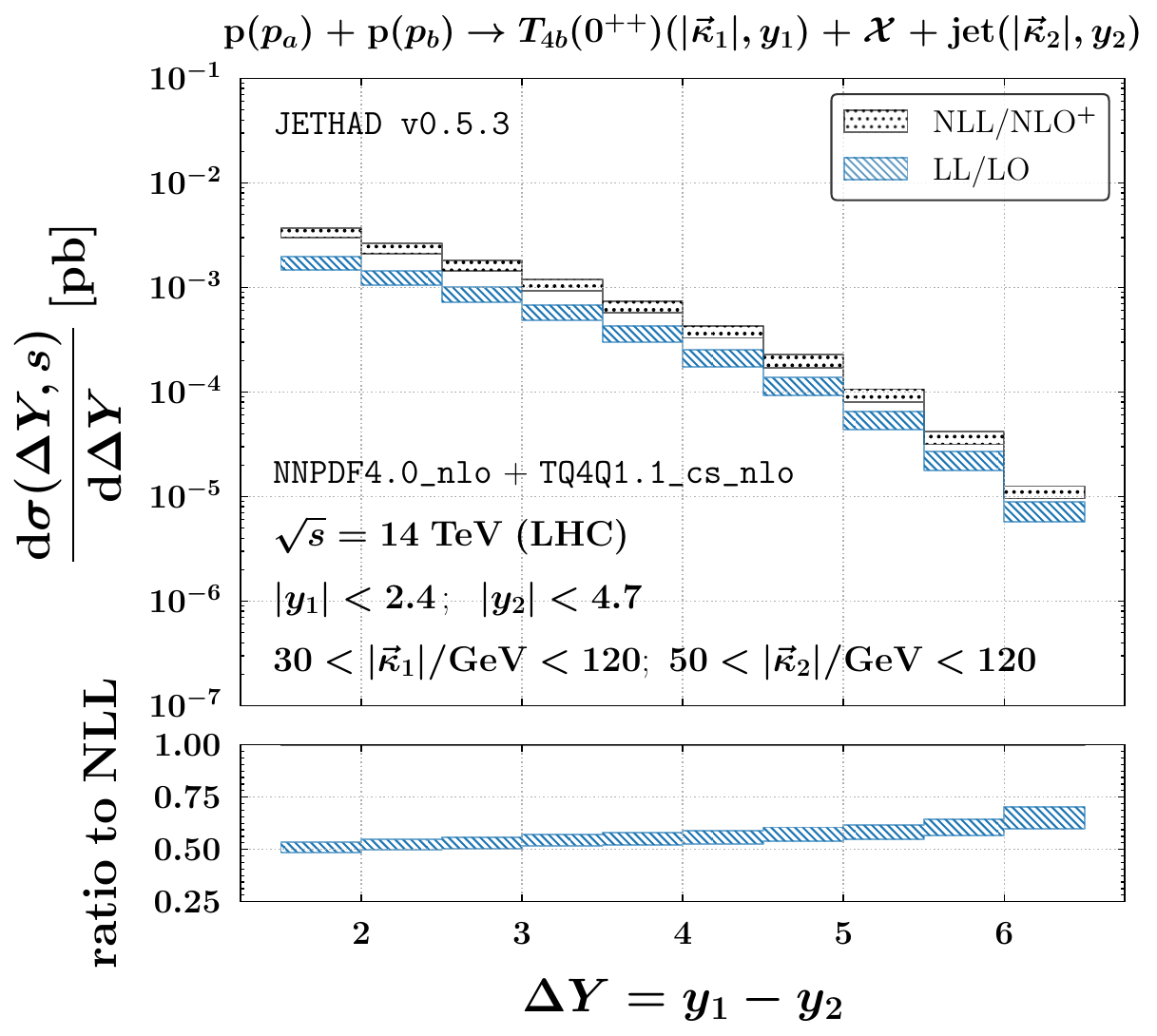}
   \hspace{-0.00cm}
   \includegraphics[scale=0.395,clip]{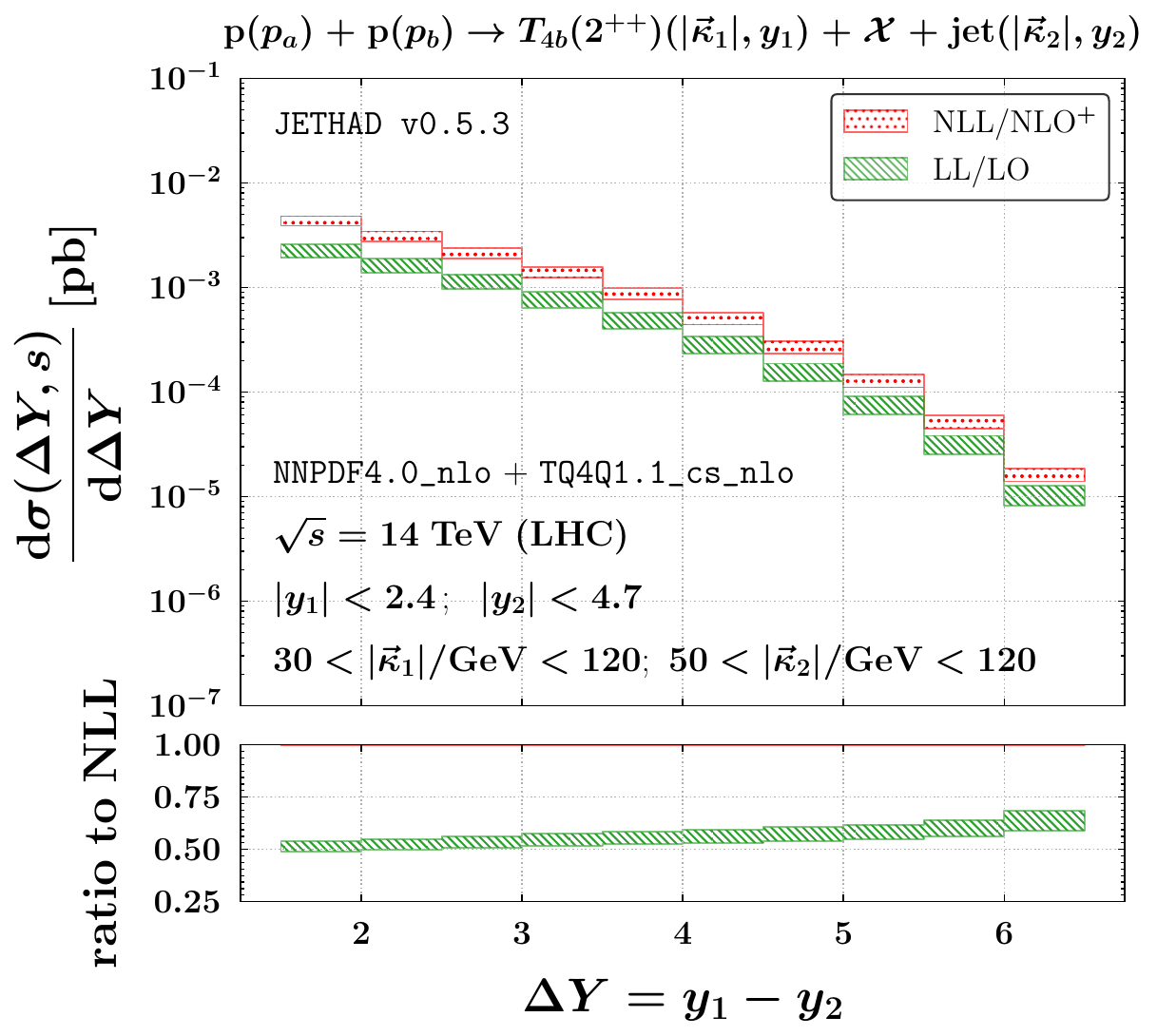}

   \vspace{0.50cm}

   \includegraphics[scale=0.395,clip]{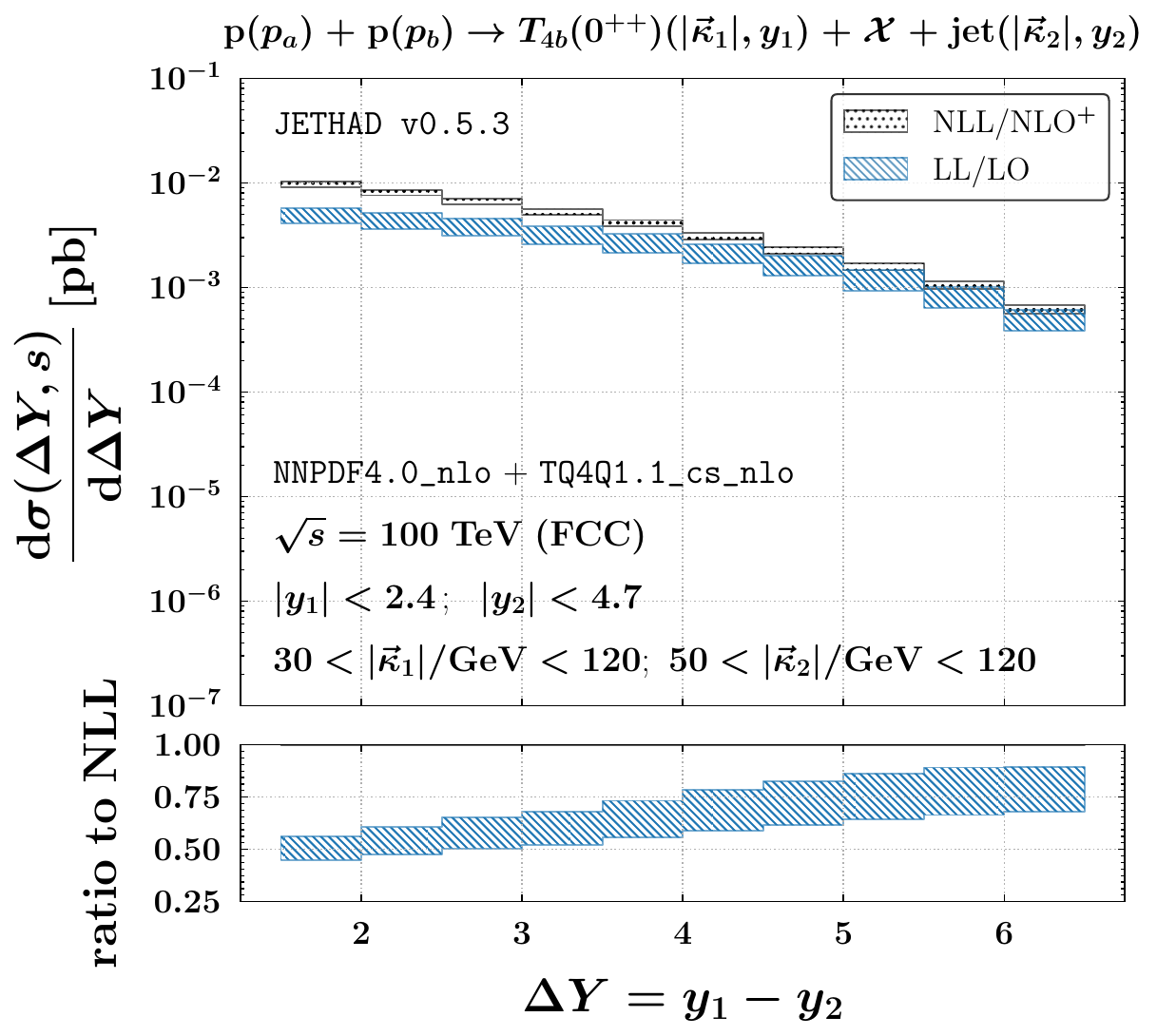}
   \hspace{-0.00cm}
   \includegraphics[scale=0.395,clip]{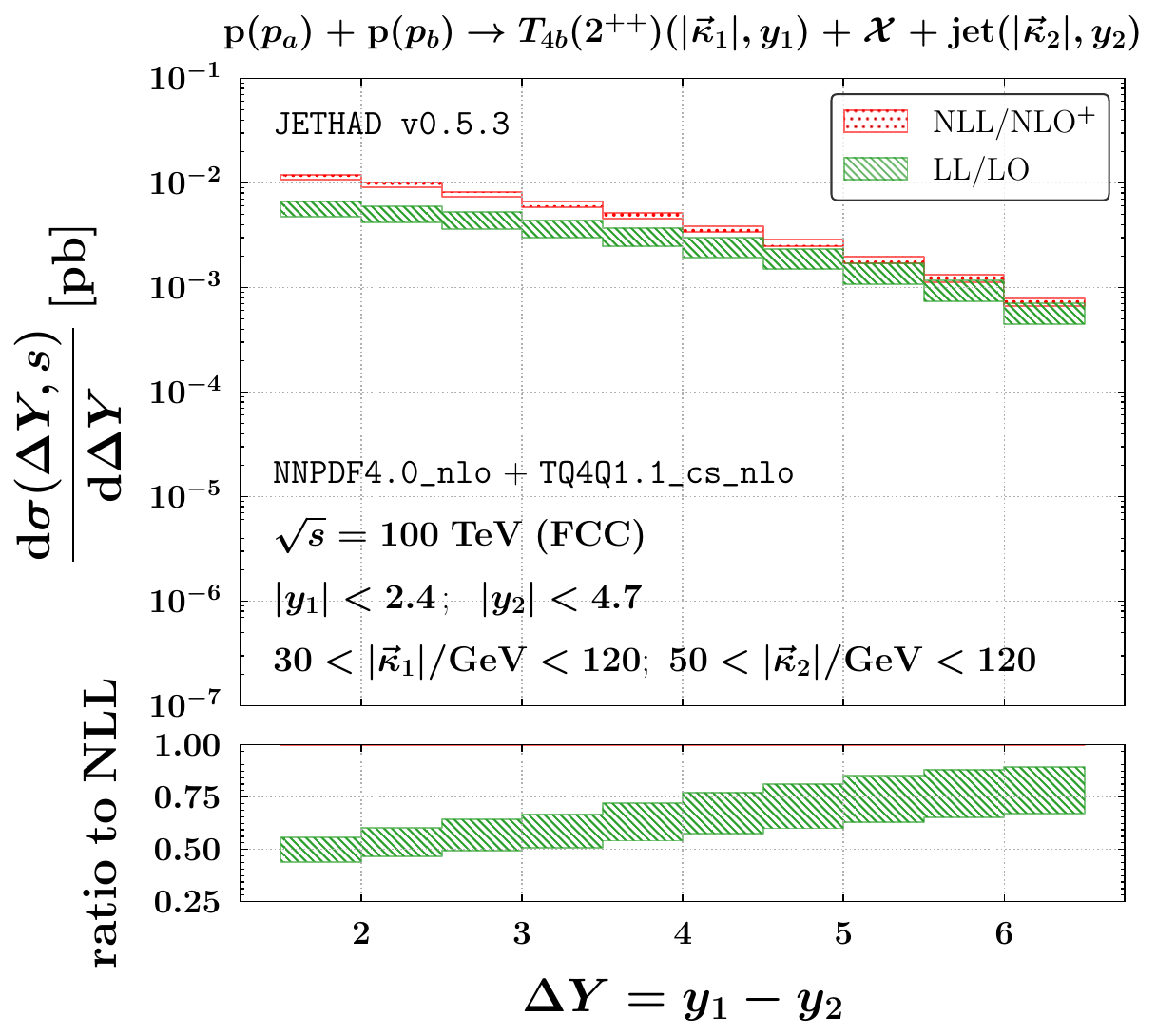}

\caption{Rapidity-interval rates for $\TQbZpp$ (left) and $\TQbTpp$ (right) plus jet hadroproduction at $\sqrt{s} = 14$ TeV (LHC, upper) or $100$ TeV (nominal FCC, lower).
Ancillary panels below primary plots exhibit the ratio between $\LL$ and $\NLLp$ predictions.
Uncertainty bands capture the total effect of MHOUs and phase-space multidimensional numeric integration.}
\label{fig:I_T4b}
\end{figure*}

The first observable we analyze is the rapidity-interval rate, which corresponds to the cross section differential in the rapidity distance, $\DY = y_1 - y_2$, between the tetraquark and the jet.
One has
\begin{equation}
\begin{split}
\label{DY_distribution}
 \hspace{-0.12cm}
 \frac{\drv \sigma(\DY, s)}{\drv \DY} &=
 \int_{y_1^{\rm min}}^{y_1^{\rm max}} \!\!\!\!\! \drv y_1
 \int_{y_2^{\rm min}}^{y_2^{\rm max}} \!\!\!\!\! \drv y_2
 \, \,
 \delta (\DY - (y_1 - y_2))
 \\
 &\times \,
 \int_{|\vec \kappa_1|^{\rm min}}^{|\vec \kappa_1|^{\rm max}} 
 \!\!\!\!\! \drv |\vec \kappa_1|
 \int_{|\vec \kappa_2|^{\rm min}}^{|\vec \kappa_2|^{\rm max}} 
 \!\!\!\!\! \drv |\vec \kappa_2|
 \, \,
 {\cal C}_{0}^{\rm [resum]}
 \;,
\end{split}
\end{equation}
with ${\cal C}_{0}$ is the first azimuthal-angle coefficient defined in Sec.~\ref{ssec:NLL_cross_section}.
Here, the `${\rm [resum]}$' superscript inclusively refers to: $\NLLp$, $\LL$, or $\HENLOp$.
Transverse momenta of the $\B$ hadron range from $30$ to $120$~GeV, and the jet ones stay from $50$ to $120$~GeV. 
They are compatible with current and future analyses of jets and hadrons at the LHC~\cite{Khachatryan:2016udy,CMS:2020ldm,Khachatryan:2020mpd}.

Adopting \emph{asymmetric} windows for the observed transverse momenta helps to magnify the onset the high-energy signal on top of the fixed-order background~\cite{Celiberto:2015yba,Celiberto:2015mpa,Celiberto:2020wpk}. 
It also quenches large Sudakov logarithms generated by nearly back-to-back events, which otherwise should be resummed~\cite{Mueller:2013wwa,Marzani:2015oyb,Mueller:2015ael,Xiao:2018esv,Hatta:2020bgy,Hatta:2021jcd}.
Finally, it controls radiative-correction instabilities~\cite{Andersen:2001kta,Fontannaz:2001nq} and dampens violations of the energy-momentum conservation~\cite{Ducloue:2014koa}.
Our selection of rapidity intervals adheres to the established criteria in current LHC studies. 
Hadron detections, limited to the barrel calorimeter like in the CMS experiment~\cite{Chatrchyan:2012xg}, are confined to the rapidity interval $|y_1| < 2.4$. 
As for jets, which can also be traced in the end cap regions~\cite{Khachatryan:2016udy}, we consider a wider rapidity range, $|y_2| < 4.7$.

Our results for rapidity-interval rates sensitive to the semi-inclusive emission of doubly bottomed tetra\-quarks and fully bottomed ones are shown in Figs.~\ref{fig:I_Xbq} and~\ref{fig:I_T4b}, respectively. 
Left (right) panels of Fig.~\ref{fig:I_Xbq} are for $\QXbu$ ($\QXbs$) states, whereas left (right) panels of Fig.~\ref{fig:I_T4b} are for $\TQbZpp$ [$\TQbTpp$] ones.
Upper and lower plots of both figures exhibit $\DY$-rates at $14$~TeV~LHC or $100$~TeV~FCC.
To facilitate direct comparisons with future experimental data, we propose using uniform $\DY$ bins with a fixed length of 0.5. 

The ancillary panels beneath the main plots display the ratio between pure $\LL$ predictions and $\NLLp$ resummed results.
Statistics is promising: for $\QXbq$ states it ranges from around one~pb to around $10^{-4}$~pb, while for $\TQb$ particles it stays between $10^{-2}$~pb to around $10^{-5}$~pb. 
As a general pattern, our distributions increase by approximately one order of magnitude as the center-of-mass energy, $\sqrt{s}$, runs from $14$~TeV~LHC or $100$~TeV~FCC.

We observe a consistent trend across all $\DY$-distri\-but\-ions: they fall off with $\DY$. 
This behavior is the result of two opposing features. 
First, the partonic hard factor tends to rise with energy, and thus with $\DY$, as predicted by the high-energy resummation. 
However, this increase is substantially dampened by the collinear convolution with PDFs and FFs in the emission functions [refer to Eqs.~\eqref{LOBEF} and~\eqref{LOJEF}].

The analysis of our plots reveals two key outcomes. 
First, there is robust stability concerning MHOUs, with uncertainty bands consistently remaining below $1.5$ in relative size. 
Second, under higher-order logarithmic corrections, the $\NLLp$ bands are uniformly narrower than the $\LL$ ones, gradually converging and even partially overlapping in the large-$\DY$ region. 
This observation aligns with findings from studies on doubly~\cite{Celiberto:2023rzw} and fully charmed~\cite{Celiberto:2024mab} tetraquarks and corroborates the statement that single-parton-fragmentation production of bottomoniumlike states provides us with a reliable channel whereby probing high-energy QCD dynamics.

\subsection{Transverse-momentum distributions}
\label{ssec:pT}

\begin{figure*}[!t]
\centering

   \hspace{0.00cm}
   \includegraphics[scale=0.395,clip]{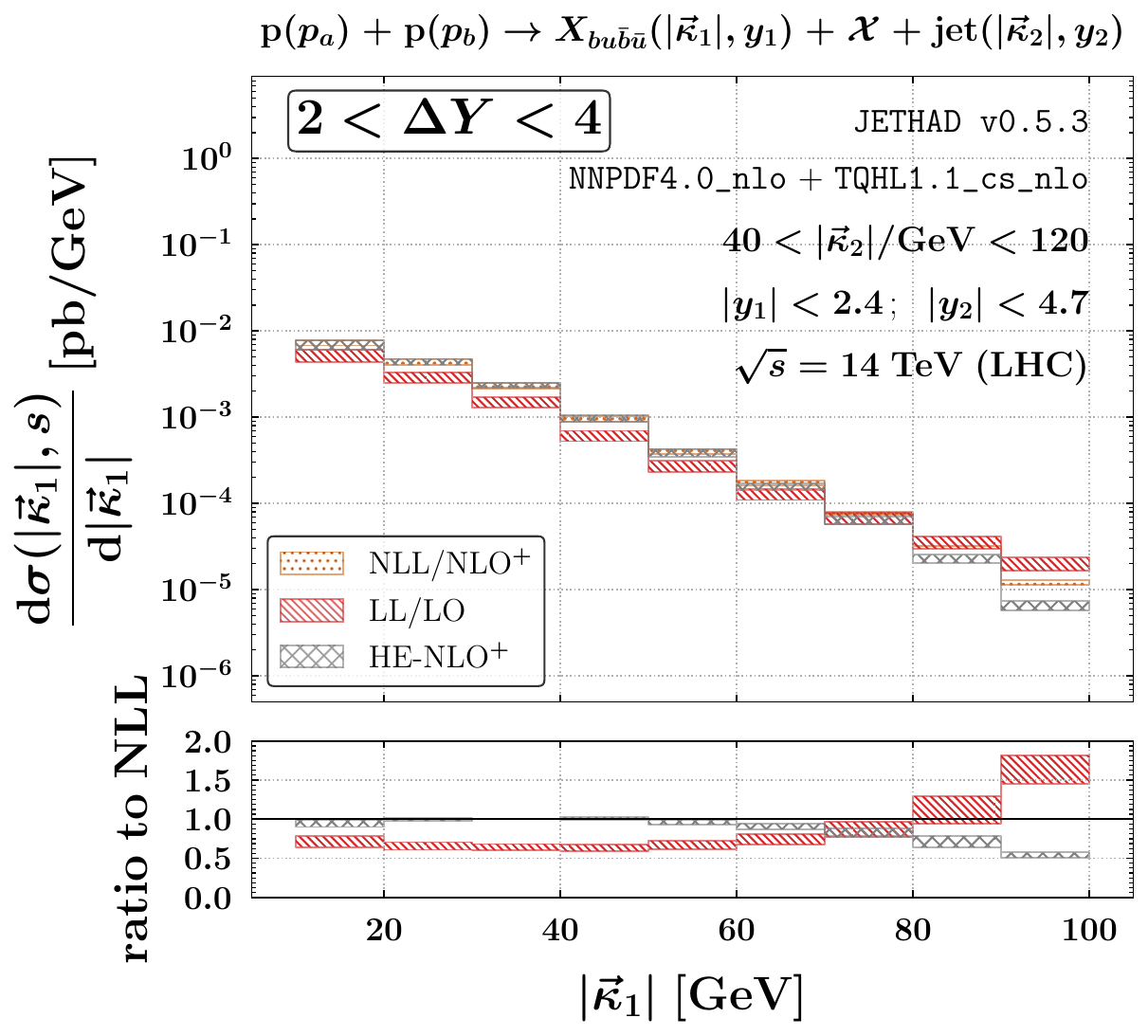}
   \hspace{-0.00cm}
   \includegraphics[scale=0.395,clip]{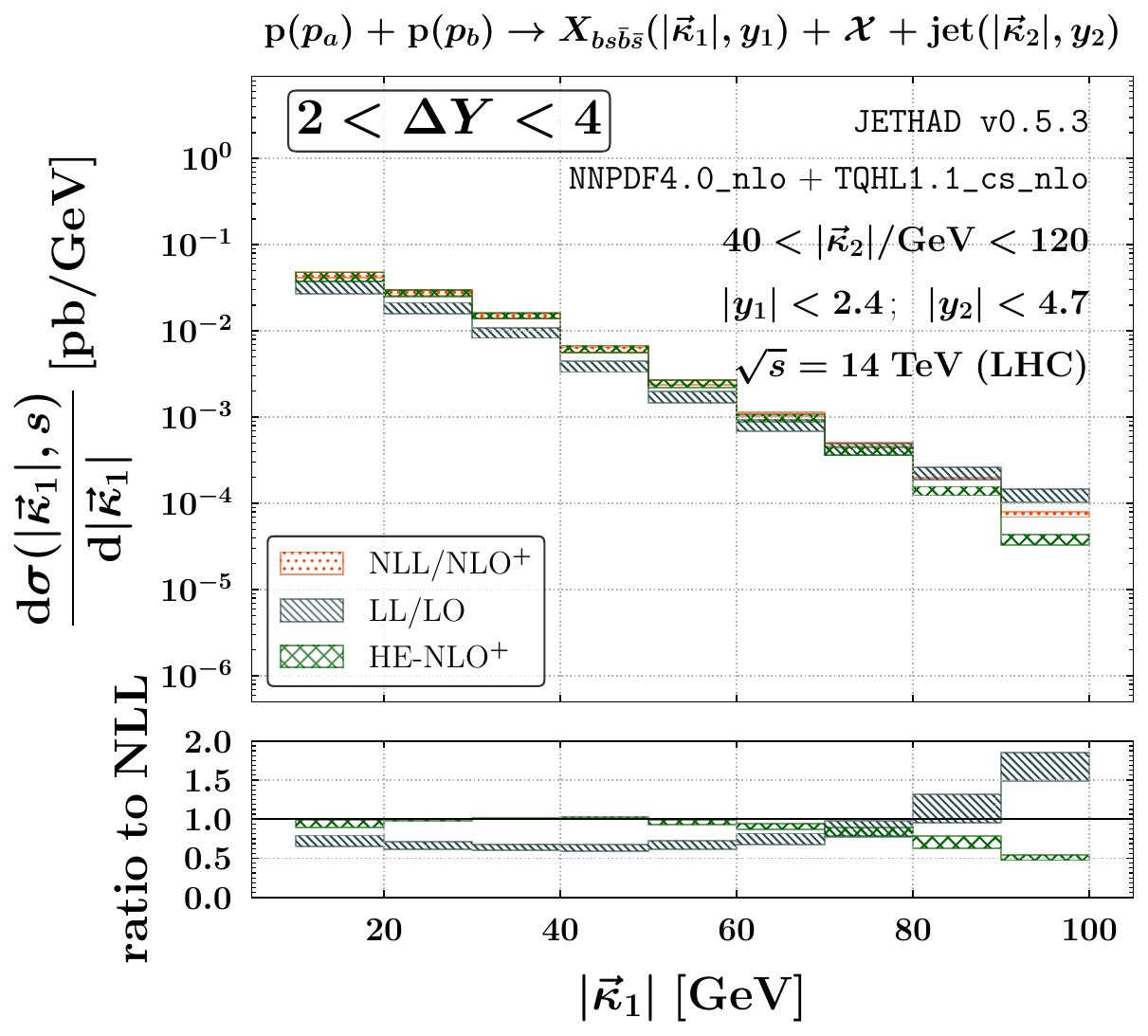}

   \vspace{0.50cm}

   \hspace{-0.00cm}
   \includegraphics[scale=0.395,clip]{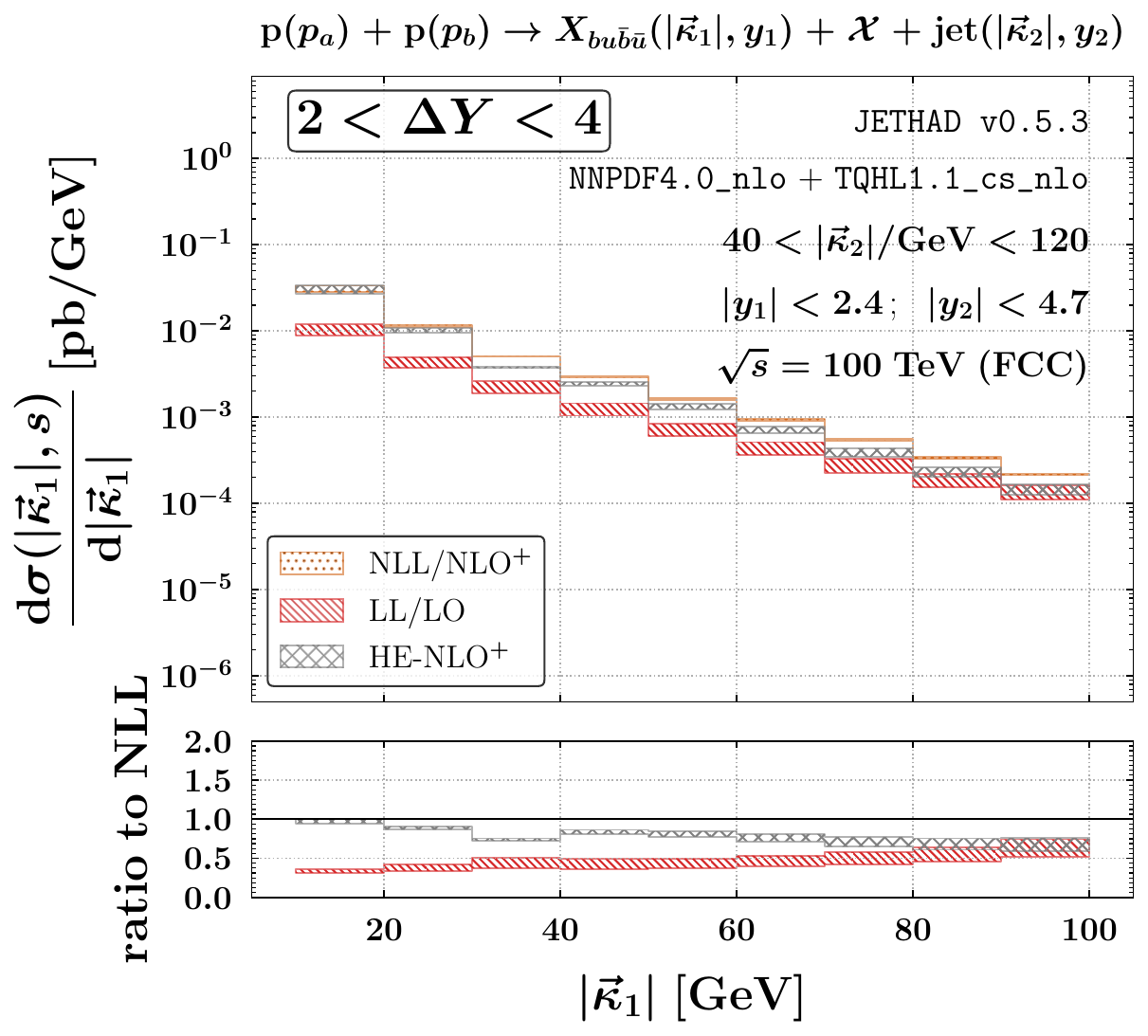}
   \includegraphics[scale=0.395,clip]{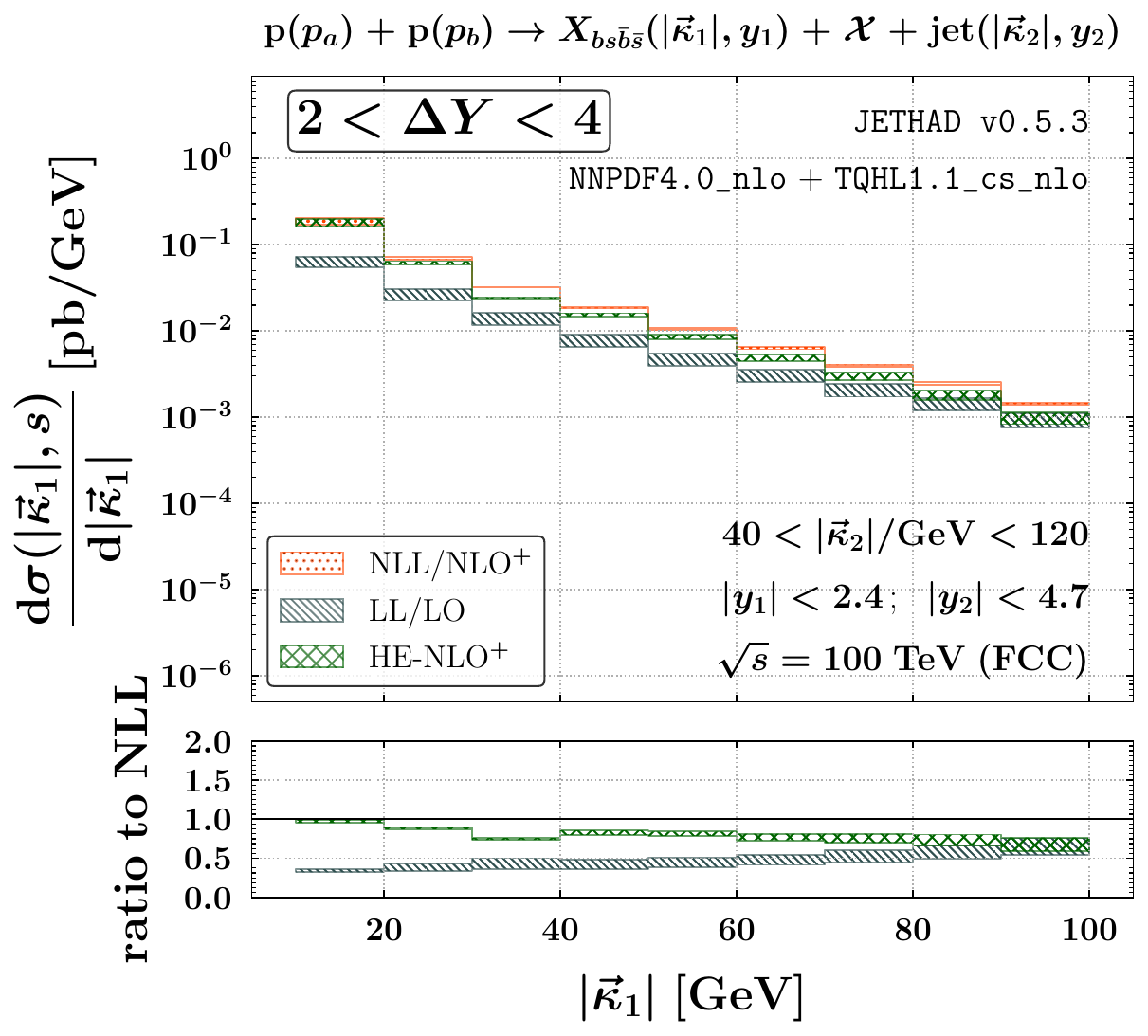}

\caption{Transverse-momentum distributions for $\QXbu$ (left) and $\QXbs$ (right) plus jet hadroproduction at $\sqrt{s} = 14$ TeV (LHC, upper) or $100$ TeV (nominal FCC, lower), and for $2 < \DY < 4$.
Ancillary panels below primary plots exhibit the ratio between $\LL$ or $\HENLOp$ and $\NLLp$ predictions.
Uncertainty bands capture the total effect of MHOUs and phase-space multidimensional numeric integration.}
\label{fig:I-k1b-S_Xbq}
\end{figure*}

\begin{figure*}[!t]
\centering

   \hspace{0.00cm}
   \includegraphics[scale=0.395,clip]{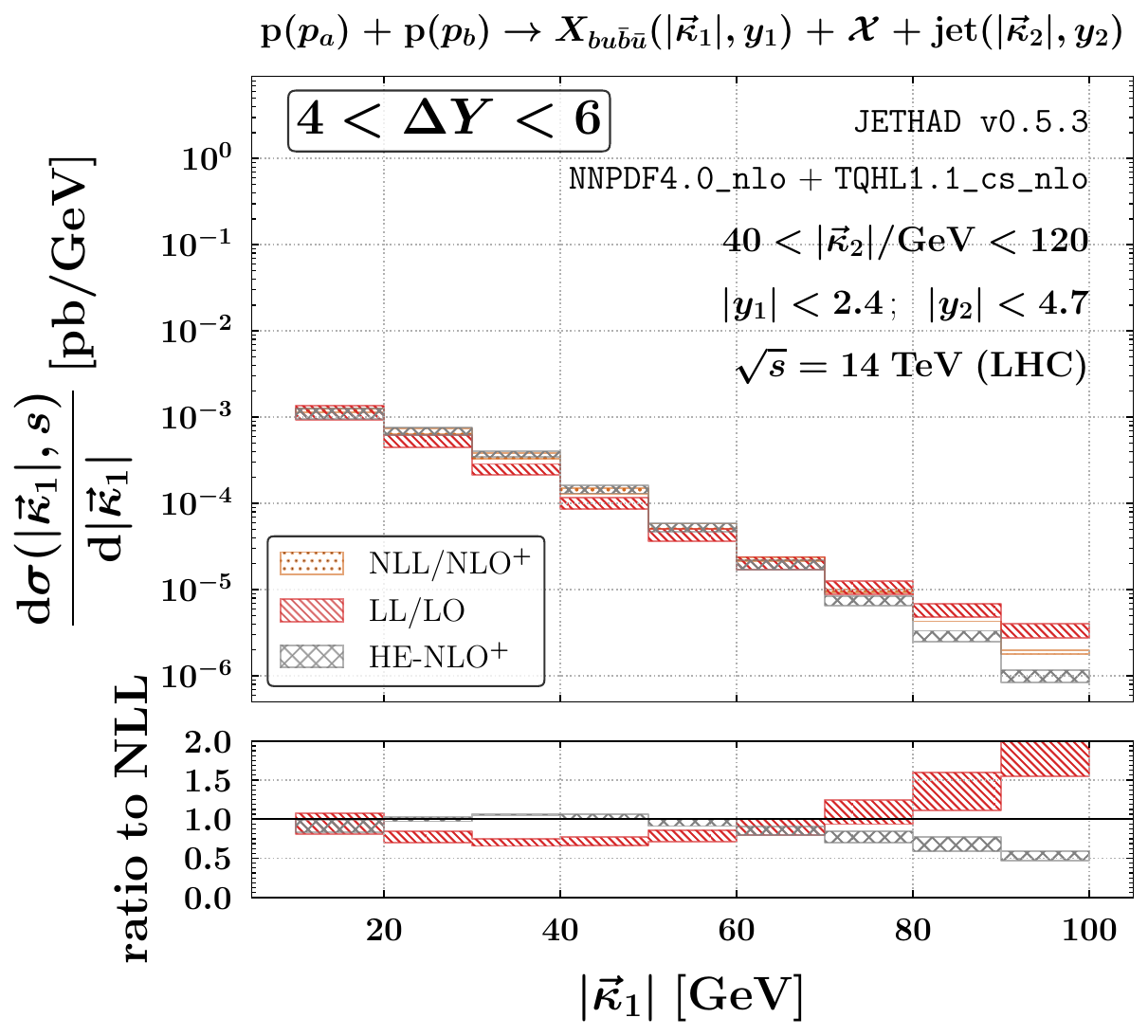}
   \hspace{-0.00cm}
   \includegraphics[scale=0.395,clip]{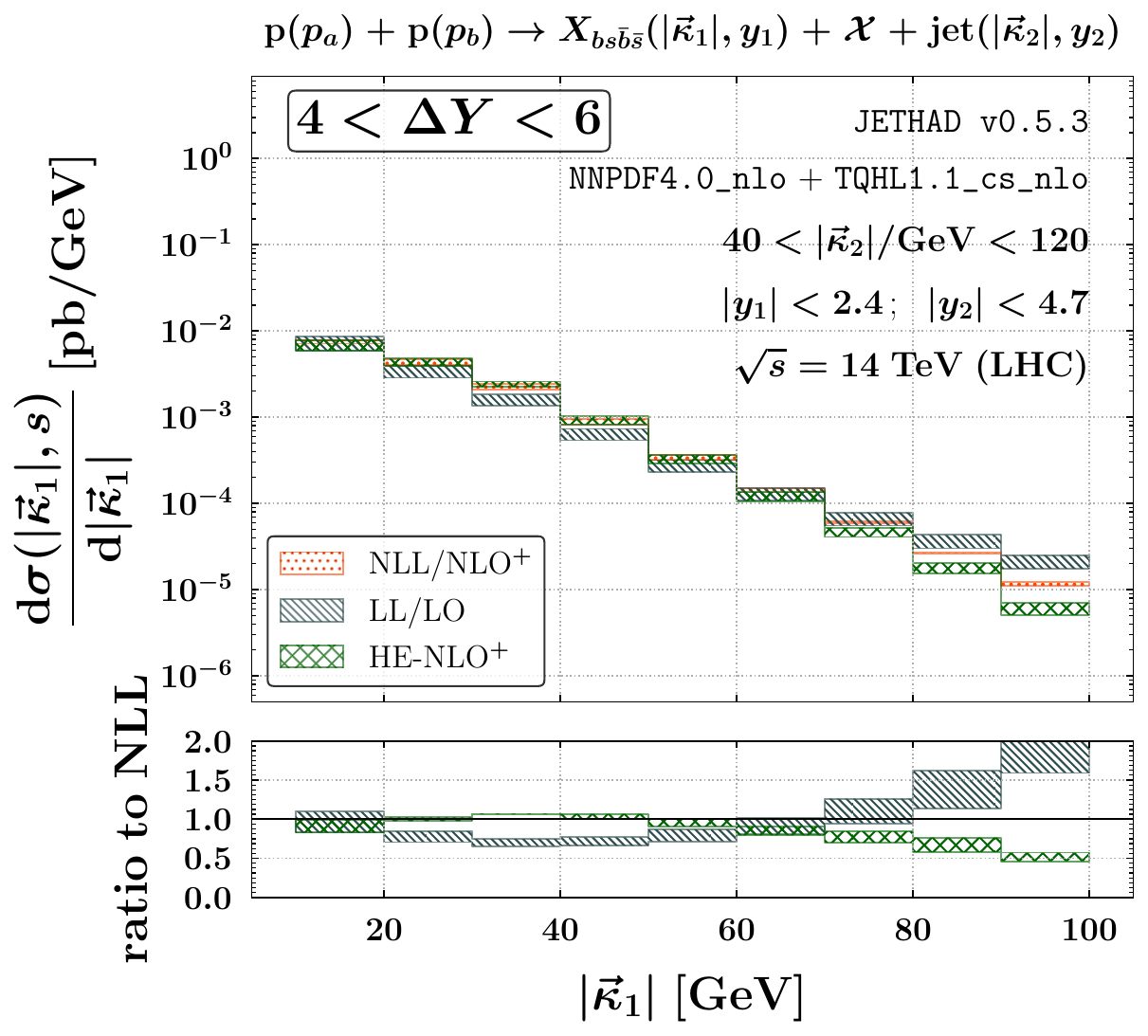}

   \vspace{0.50cm}

   \hspace{-0.00cm}
   \includegraphics[scale=0.395,clip]{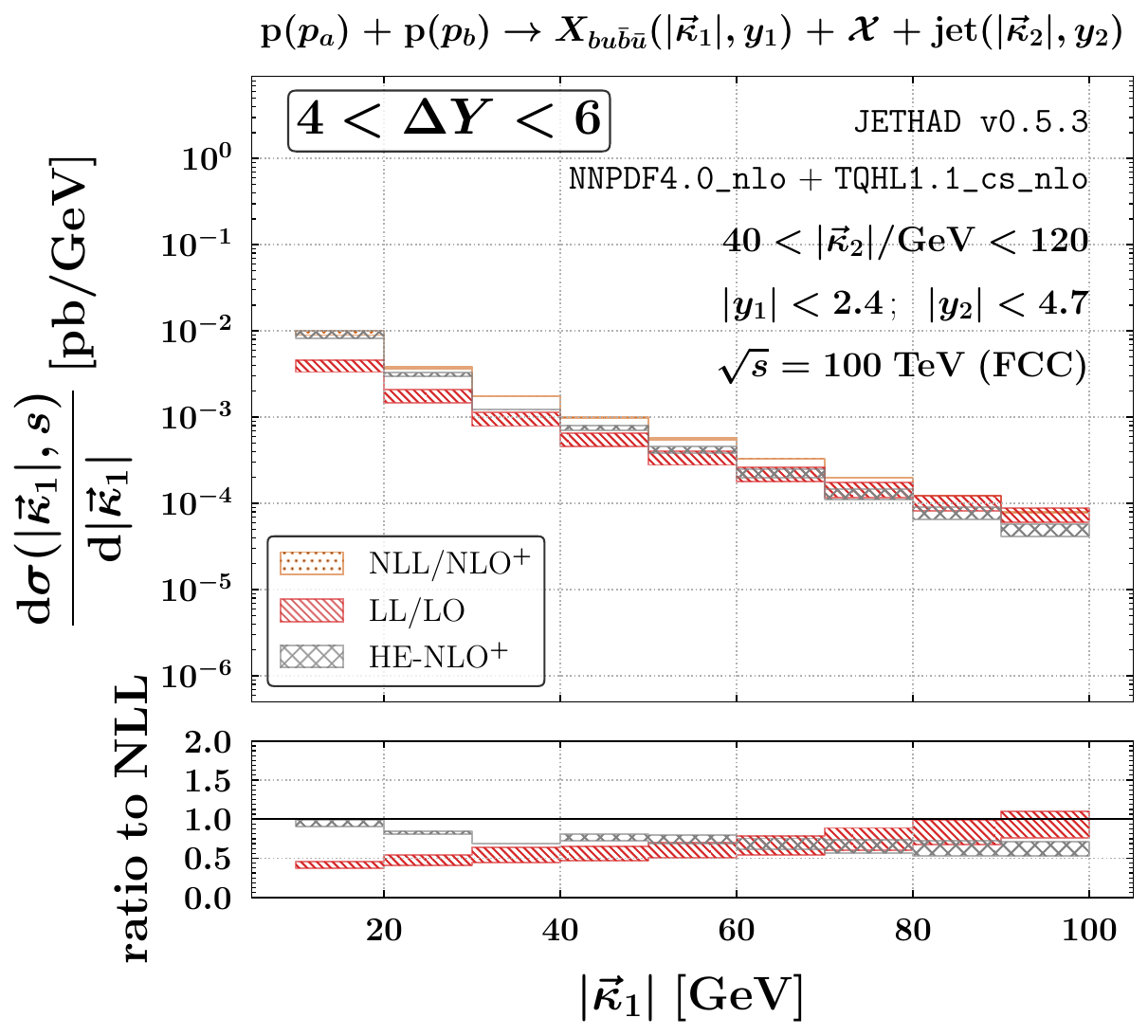}
   \includegraphics[scale=0.395,clip]{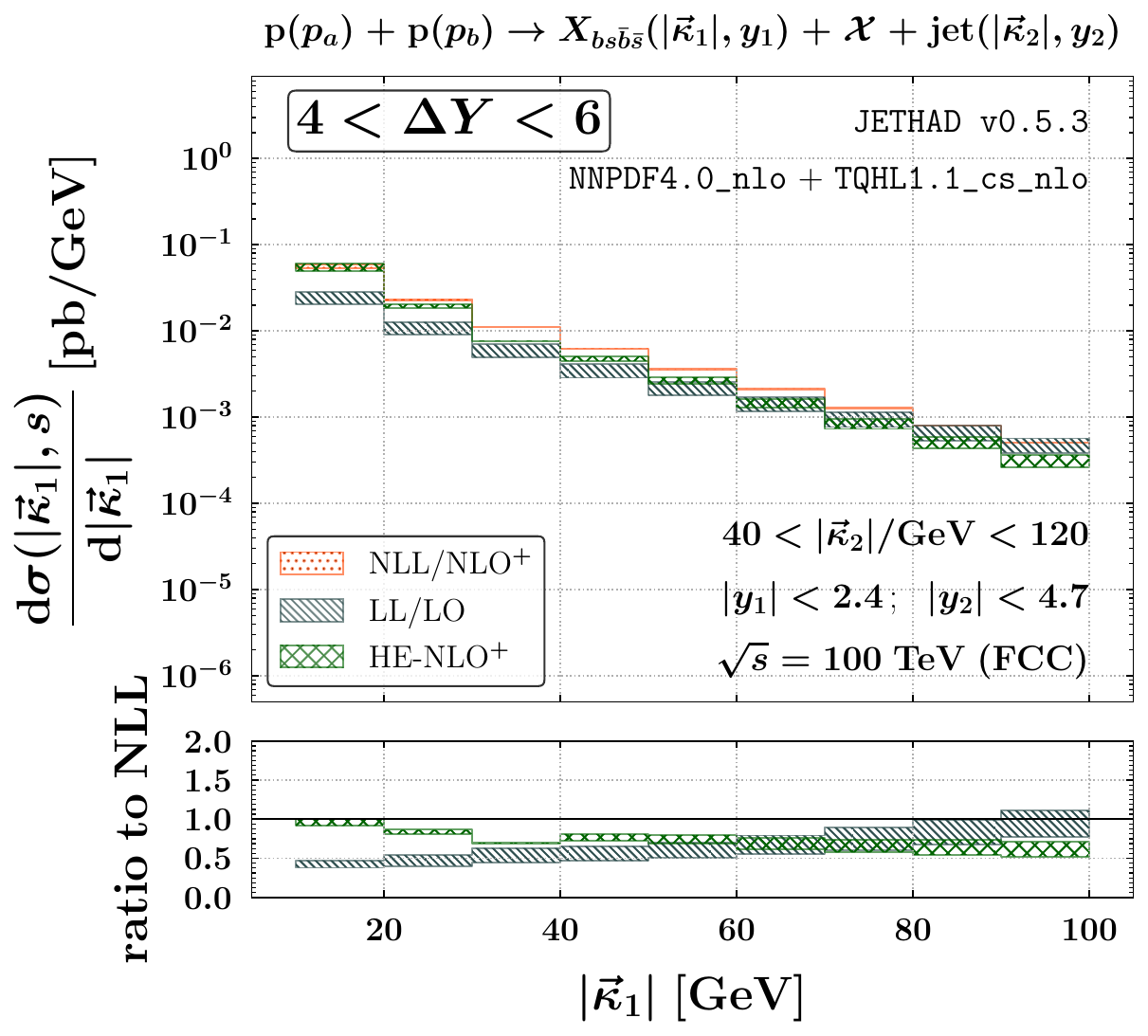}

\caption{Transverse-momentum distributions for $\QXbu$ (left) and $\QXbs$ (right) plus jet hadroproduction at $\sqrt{s} = 14$ TeV (LHC, upper) or $100$ TeV (nominal FCC, lower), and for $4 < \DY < 6$.
Ancillary panels below primary plots exhibit the ratio between $\LL$ or $\HENLOp$ or $\HENLOp$ and $\NLLp$ predictions.
Uncertainty bands capture the total effect of MHOUs and phase-space multidimensional numeric integration.}
\label{fig:I-k1b-M_Xbq}
\end{figure*}

\begin{figure*}[!t]
\centering

   \hspace{0.00cm}
   \includegraphics[scale=0.395,clip]{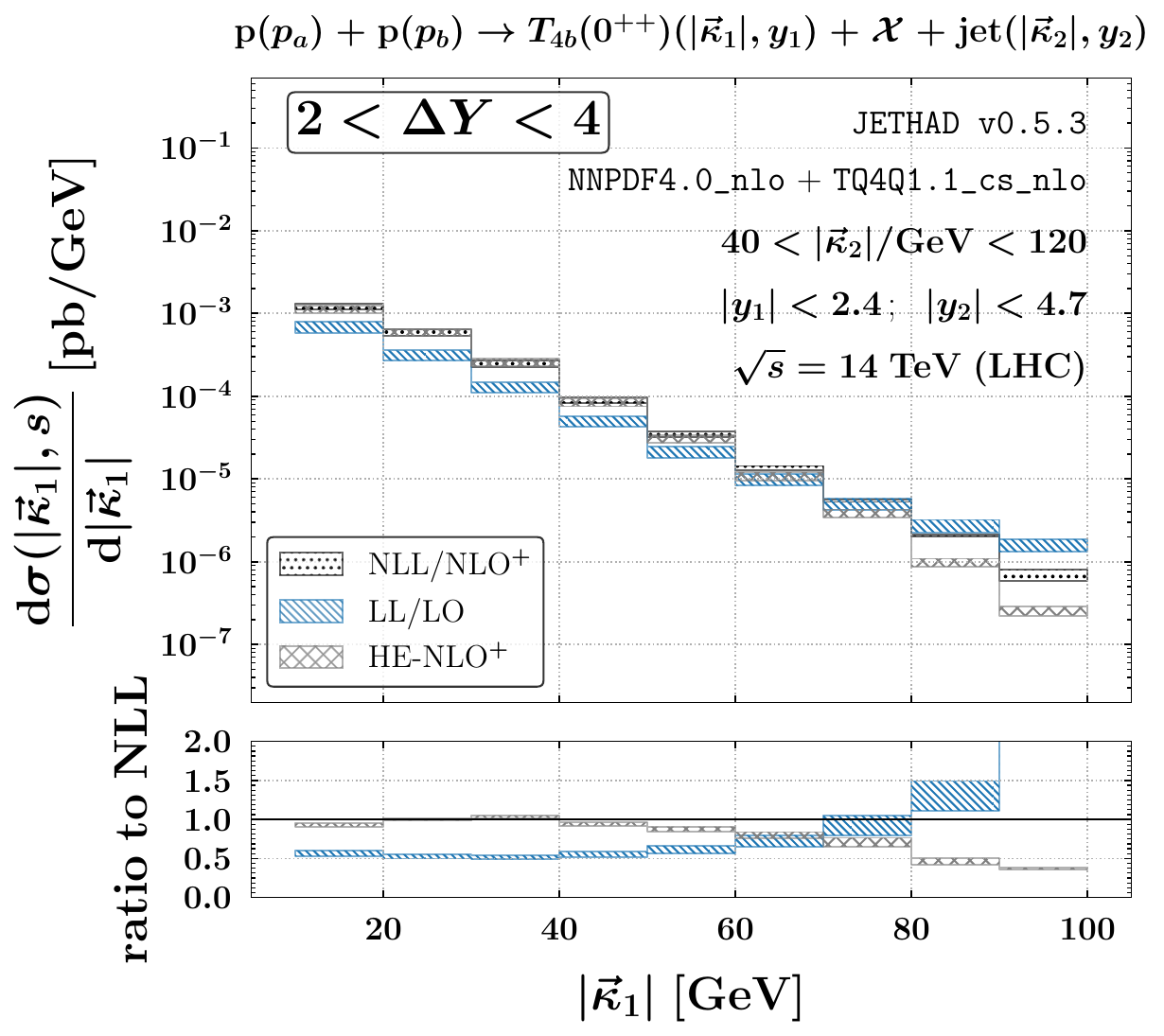}
   \hspace{-0.00cm}
   \includegraphics[scale=0.395,clip]{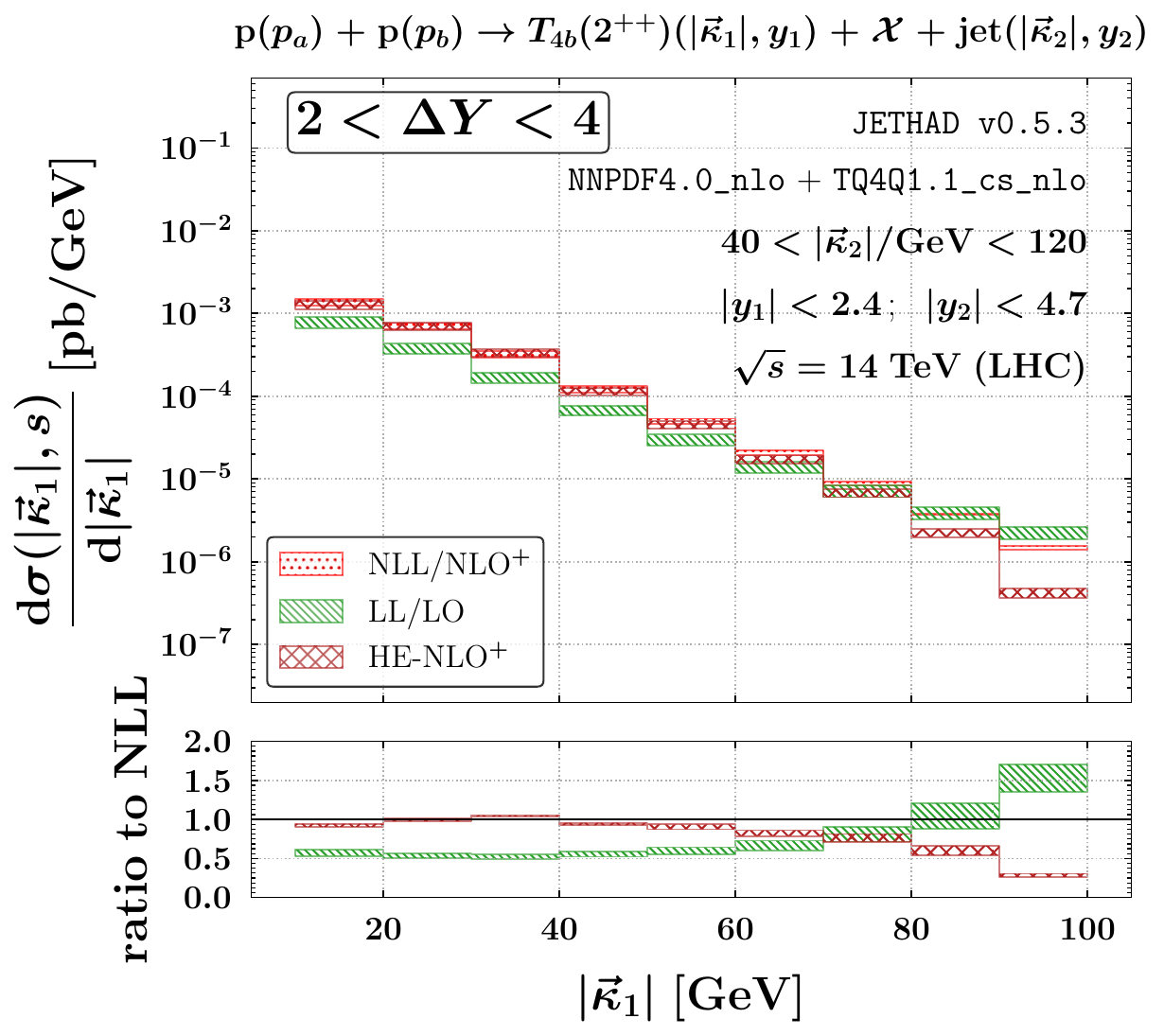}

   \vspace{0.50cm}

   \hspace{-0.00cm}
   \includegraphics[scale=0.395,clip]{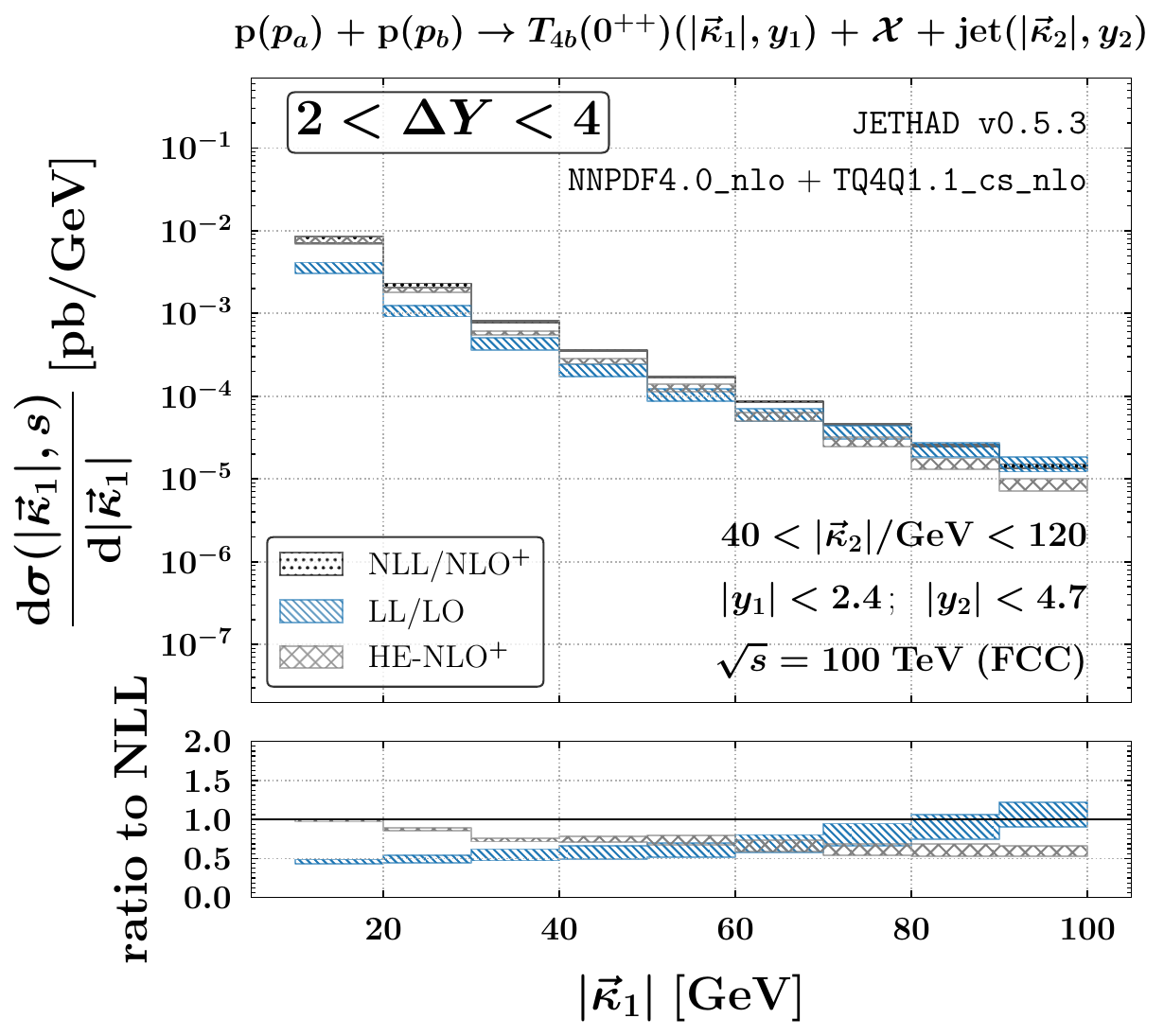}
   \includegraphics[scale=0.395,clip]{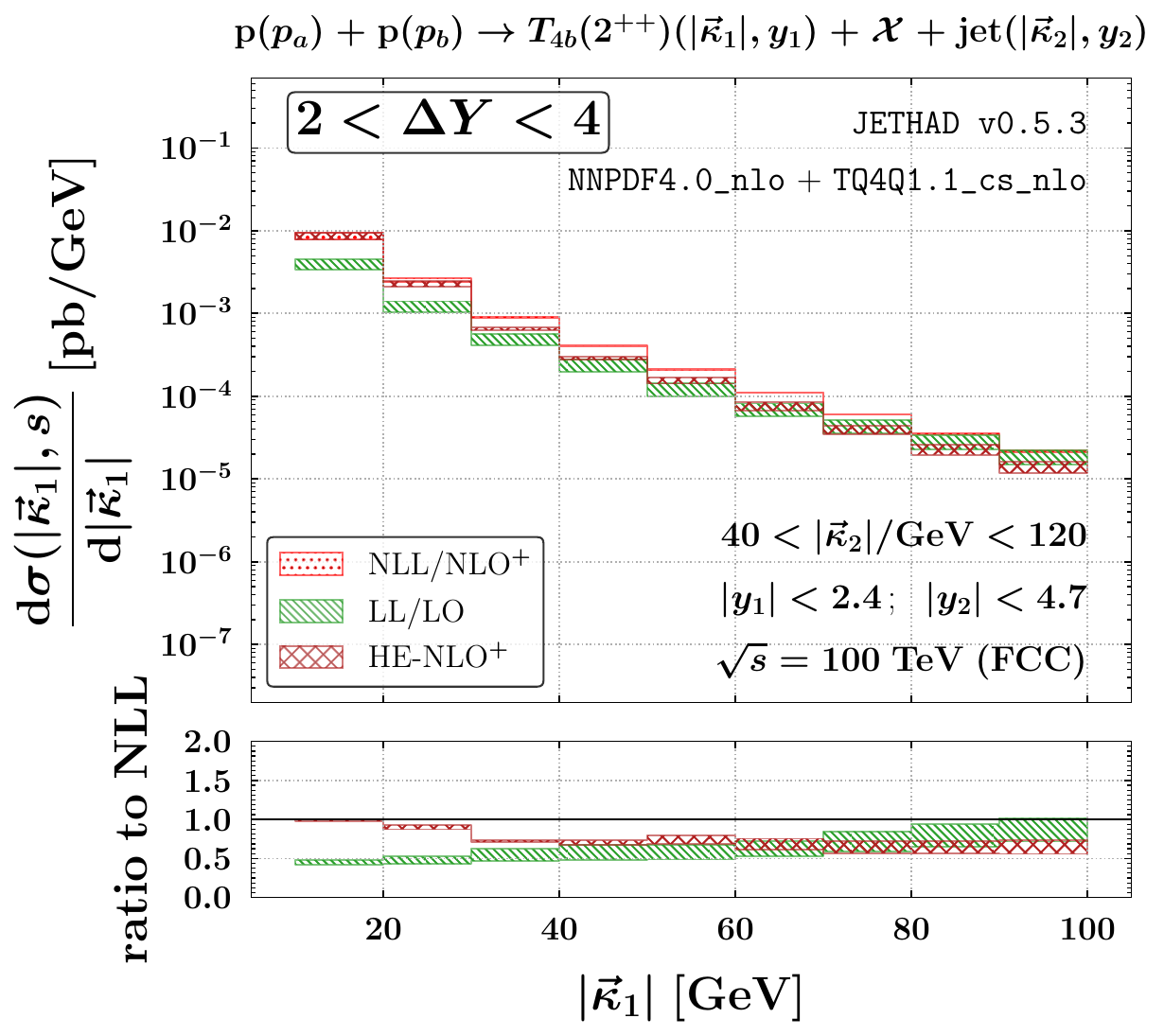}

\caption{Transverse-momentum distributions for $\TQbZpp$ (left) and $\TQbTpp$ (right) plus jet hadroproduction at $\sqrt{s} = 14$ TeV (LHC, upper) or $100$ TeV (nominal FCC, lower), and for $2 < \DY < 4$.
Ancillary panels below primary plots exhibit the ratio between $\LL$ or $\HENLOp$ and $\NLLp$ predictions.
Uncertainty bands capture the total effect of MHOUs and phase-space multidimensional numeric integration.}
\label{fig:I-k1b-S_T4b}
\end{figure*}

\begin{figure*}[!t]
\centering

   \hspace{0.00cm}
   \includegraphics[scale=0.395,clip]{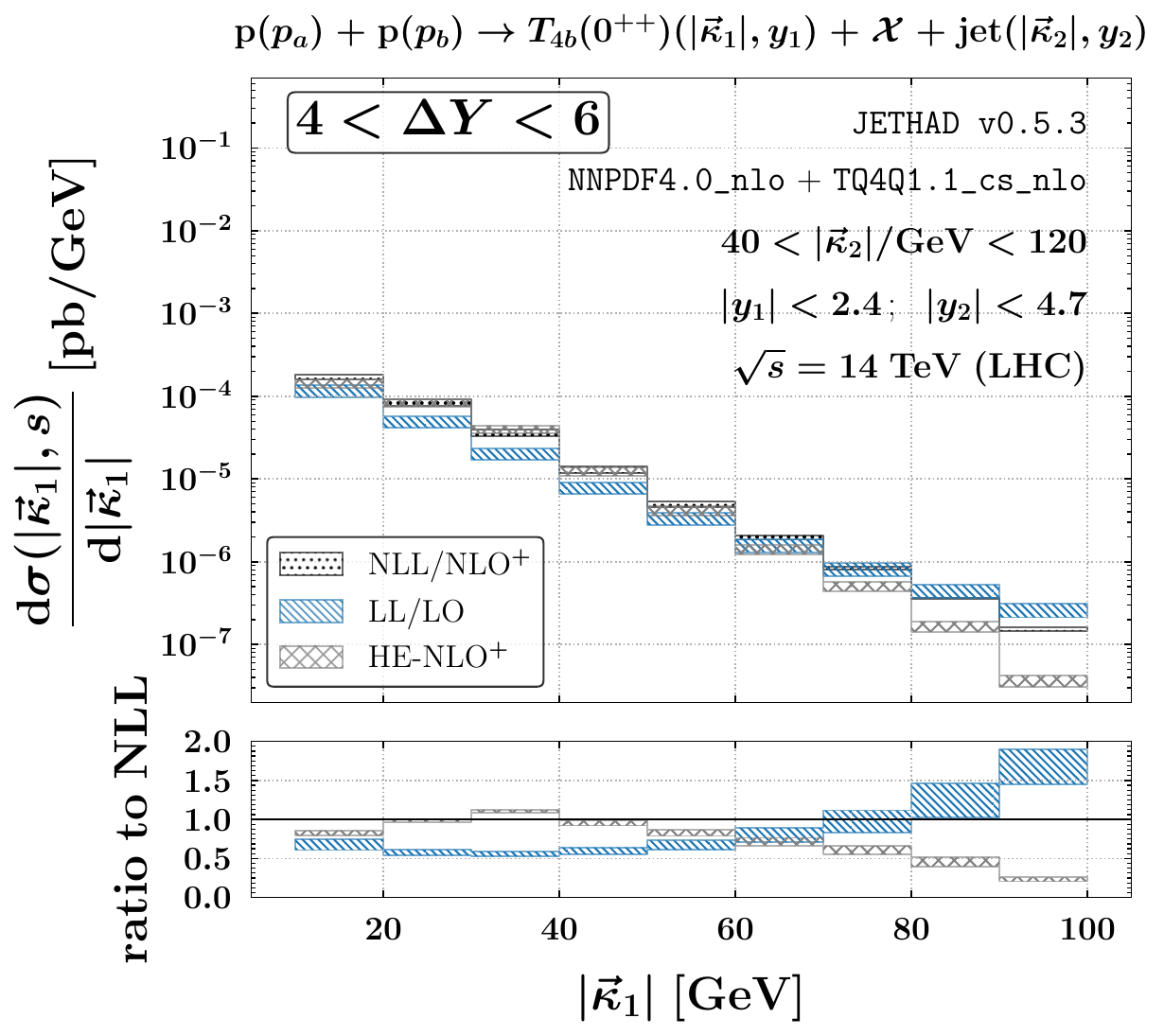}
   \hspace{-0.00cm}
   \includegraphics[scale=0.395,clip]{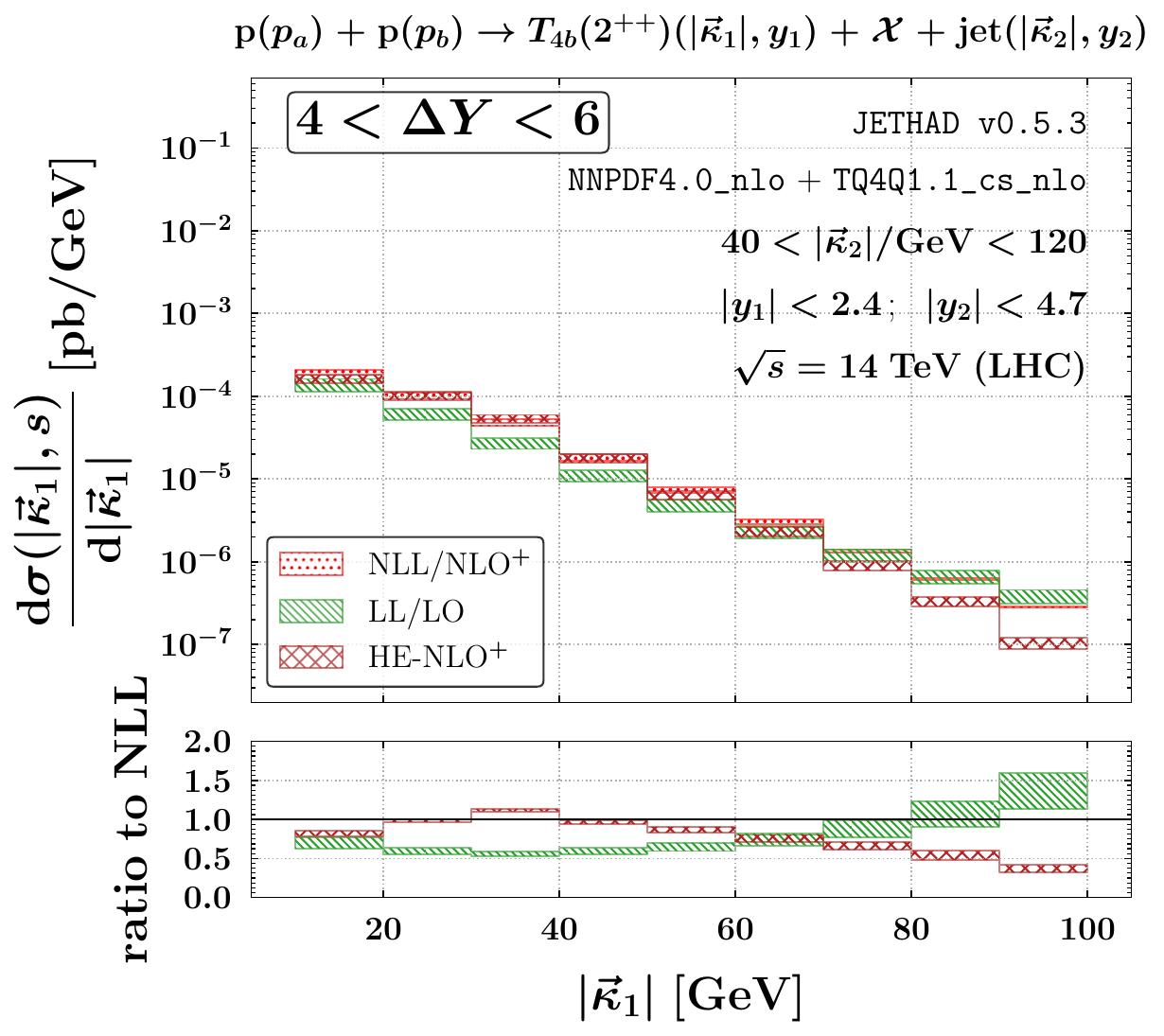}

   \vspace{0.50cm}

   \hspace{-0.00cm}
   \includegraphics[scale=0.395,clip]{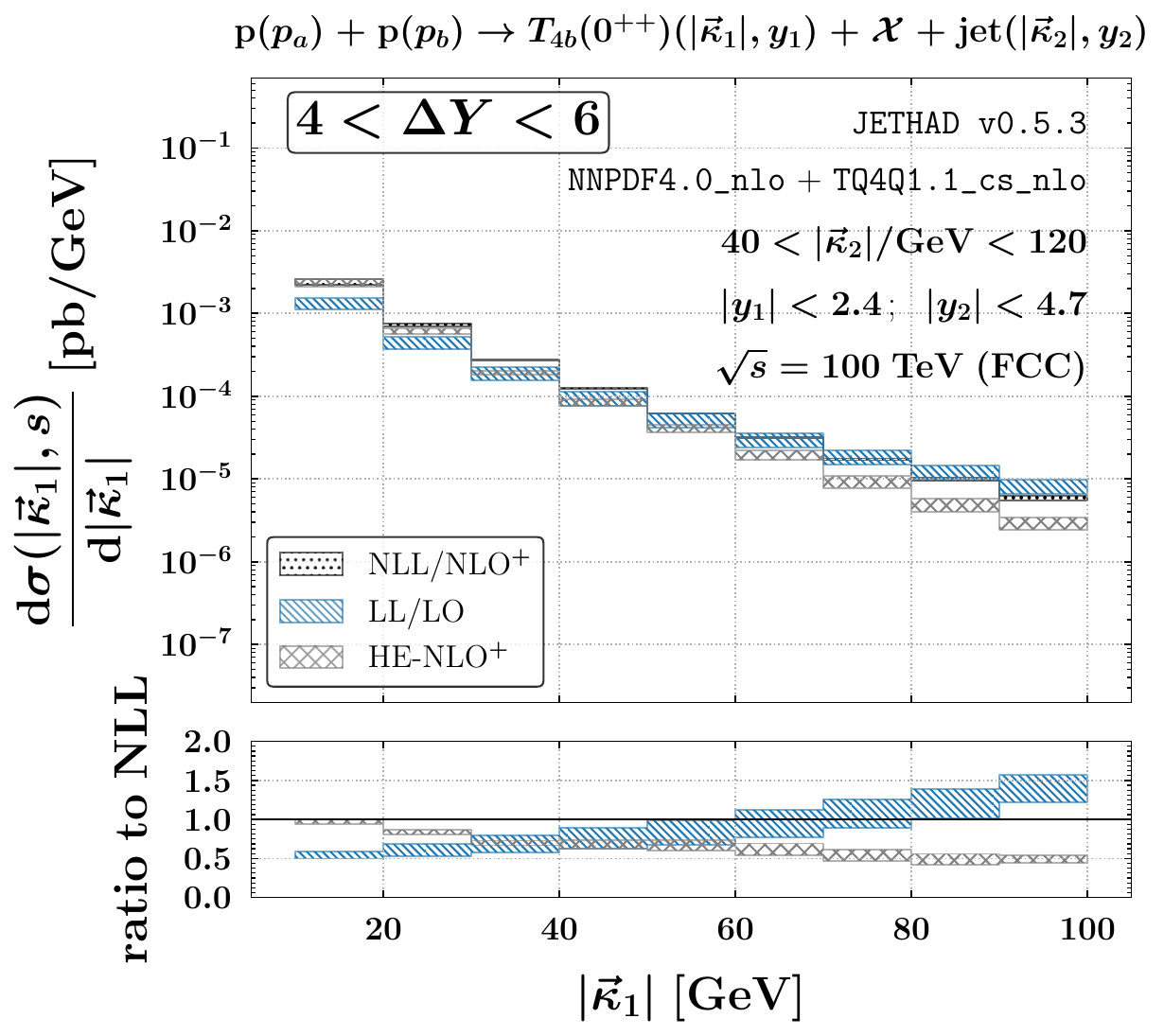}
   \includegraphics[scale=0.395,clip]{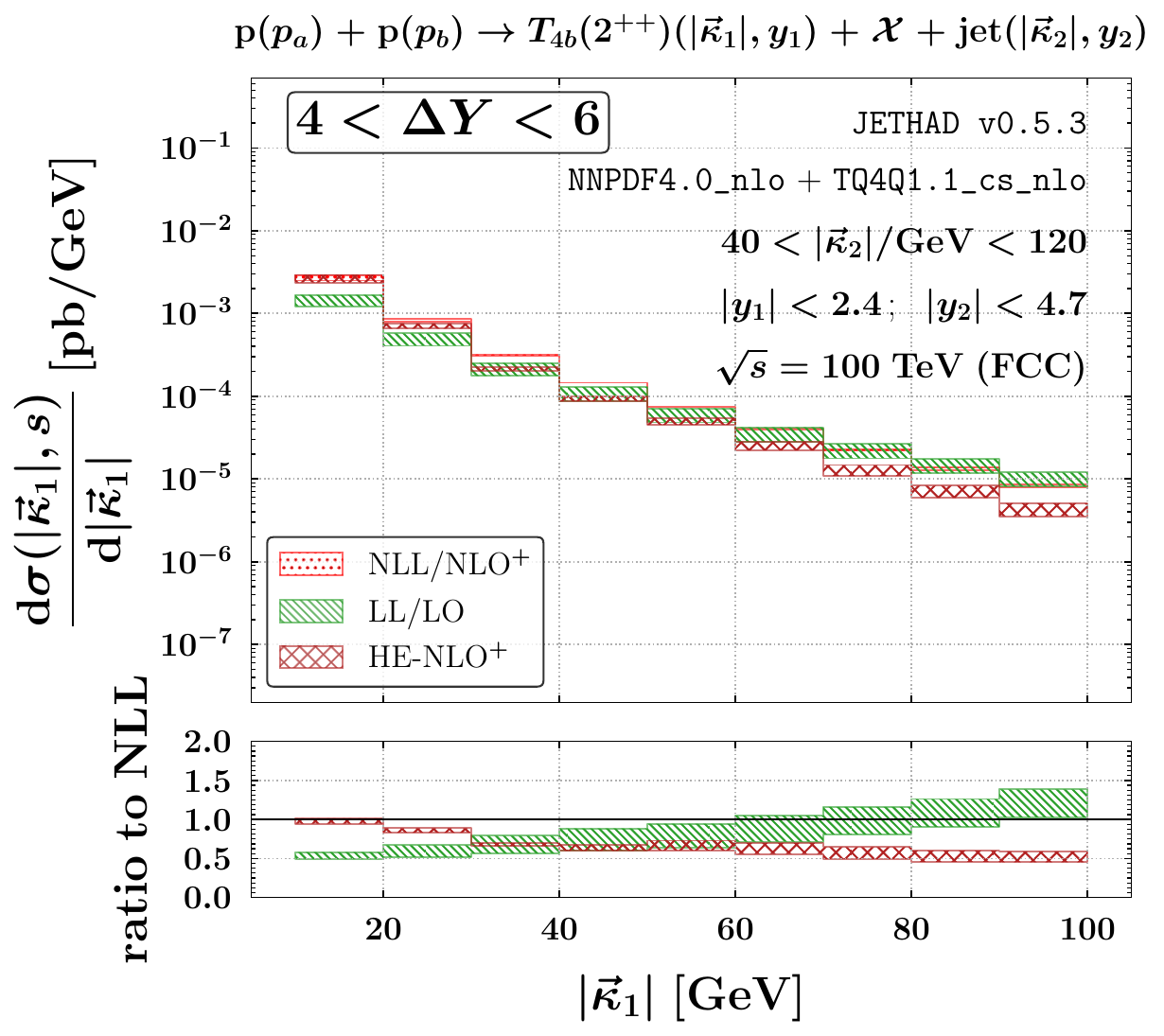}

\caption{Transverse-momentum distributions for $\TQbZpp$ (left) and $\TQbTpp$ (right) plus jet hadroproduction at $\sqrt{s} = 14$ TeV (LHC, upper) or $100$ TeV (nominal FCC, lower), and for $4 < \DY < 6$.
Ancillary panels below primary plots exhibit the ratio between $\LL$ or $\HENLOp$ and $\NLLp$ predictions.
Uncertainty bands capture the total effect of MHOUs and phase-space multidimensional numeric integration.}
\label{fig:I-k1b-M_T4b}
\end{figure*}

While rapidity-differential observables are crucial for revealing high-energy dynamics at hadron colliders, to investigate kinematic regions where other resummation mechanisms are also relevant, we must consider transverse momentum-dependent observables.
High transverse momenta or significant imbalances between them enhance DGLAP-type logarithms and soft, \emph{threshold} logarithms~\cite{Sterman:1986aj,Catani:1989ne,Catani:1996yz,Bonciani:2003nt,deFlorian:2005fzc,Ahrens:2009cxz,deFlorian:2012yg,Forte:2021wxe,Mukherjee:2006uu,Bolzoni:2006ky,Becher:2006nr,Becher:2007ty,Bonvini:2010tp,Ahmed:2014era,Muselli:2017bad,Banerjee:2018vvb,Duhr:2022cob,Shi:2021hwx,Wang:2022zdu,Bonvini:2023mfj}, which differ in nature from high-energy logarithms and also require resummation.

Simultaneously resumming threshold and energy logarithms is challenging. 
A pioneering double resummation was achieved in inclusive Higgs hadroproduction \emph{via} gluon fusion~\cite{Bonvini:2018ixe}, where the possibility of decoupling the two dynamics in Mellin space allowed us for the separate resummation of small-$x$ (high-energy) and large-$x$ (threshold) logarithms, controlling the small-$N$ and large-$N$ tails, respectively~\cite{Ball:2013bra,Bonvini:2014joa}. 
However, in our scenario involving the semi-inclusive emission of two particles, the rapidity-differential nature of observables complicates this approach, and a suitable framework for performing a double resummation is not yet available.

On the other hand, low transverse momenta lead to the rise of large Sudakov-type logarithms, which our hybrid factorization fails to capture. 
Additionally, the effects linked to the so-called \emph{diffusion pattern} become significant~\cite{Bartels:1993du,Caporale:2013bva,Ross:2016zwl}. 
The most effective approach to resumming these logarithms to all orders is the transverse-momentum resummation formalism (see Refs.~\cite{Catani:2000vq,Bozzi:2005wk,Bozzi:2008bb,Catani:2010pd,Catani:2011kr,Catani:2013tia,Catani:2015vma,Duhr:2022yyp} and references therein).

Final states sensitive to the semi-inclusive tagging of two identified objects in hadronic scatterings, like photon~\cite{Cieri:2015rqa,Alioli:2020qrd,Becher:2020ugp,Neumann:2021zkb}, Higgs~\cite{Ferrera:2016prr}, and $W^\pm$ boson~\cite{Ju:2021lah} pairs, or plus jet~\cite{Monni:2019yyr,Buonocore:2021akg} and $Z$ plus photon~\cite{Wiesemann:2020gbm} events, have been identified as promising channels for accessing the core dynamics of the transverse-momentum resummation. 
Third-order resummed differential rates for Drell-Yan and Higgs final states were recently explored (see, \emph{e.g.}, Refs.~\cite{Ebert:2020dfc,Re:2021con,Chen:2022cgv,Neumann:2022lft} and~\cite{Bizon:2017rah,Billis:2021ecs,Re:2021con,Caola:2022ayt}). 
A joint resummation of transverse-momentum logarithms arising from two-particle detections was presented in Ref.~\cite{Monni:2019yyr}, where the transverse momenta of the Higgs boson and the leading jet were analyzed up to next-to-NLL accuracy with {\RadISH}~\cite{Bizon:2017rah}. 
Similar distributions for fully leptonic $W^+W^-$ production at the LHC were analyzed in Ref.~\cite{Kallweit:2020gva}.

An additional layer of complexity is introduced when observables sensitive to heavy flavor(s) are considered. 
Specifically, as the transverse momentum of a heavy hadron decreases, its transverse mass approaches and eventually crosses the heavy-quark mass thresholds that govern the DGLAP evolution. 
Under these conditions, the applicability of a pure VFNS approach becomes questionable, as it may not accurately account for the dynamics near or below these mass thresholds.

With the aim of unveiling possible common ground between the $\NLLp$ hybrid formalism and other resummations, singly as well as doubly differential transverse-momentum rates were recently investigated in the context of inclusive semihard jet~\cite{Bolognino:2021mrc}, Higgs \cite{Celiberto:2020tmb}, $b$-hadrons~\cite{Celiberto:2021fdp}, cascade baryons~\cite{Celiberto:2022kxx}, and charm\-ed-tetraquark~\cite{Celiberto:2023rzw,Celiberto:2024mab} productions.
For the sake of brevity, in this section we focus on the high-energy behavior of cross sections differential in the transverse momentum of the tetraquark, $\vec \kappa_1$, and integrated over the 40~GeV~$< |\vec \kappa_2| <$~120~GeV jet transverse-momentum range as well as over two possible bins of $\DY$.
We write
\begin{equation}
\begin{split}
\label{pT1_distribution}
 &\frac{\drv \sigma(|\vec \kappa_1|, s)}{\drv |\vec \kappa_1|} = 
 \int_{\DY^{\rm min}}^{\DY^{\rm max}} \drv \DY
 \int_{y_1^{\rm min}}^{y_1^{\rm max}} \drv y_1
 \int_{y_2^{\rm min}}^{y_2^{\rm max}}  \drv y_2
 \\
 &\hspace{0.55cm} \times \,
 \delta (\DY - (y_1 - y_2))
 \int_{|\vec \kappa_2|^{\rm min}}^{|\vec \kappa_2|^{\rm max}} 
 \!\!\drv |\vec \kappa_2|
 \, \,
 {\cal C}_{0}^{\rm [resum]}
 \;.
\end{split}
\end{equation}

Figures~\ref{fig:I-k1b-S_Xbq} and~\ref{fig:I-k1b-M_Xbq} contain results for the $\QXbu$ (left panels) or $\QXbs$ (right panels) plus light-jet $\vec \kappa_1$-rates at $2 < \DY < 4$ and $4 < \DY < 6$, respectively.
Analogously, plots of Fig.~\ref{fig:I-k1b-S_T4b} and~\ref{fig:I-k1b-M_T4b} show predictions for the $\TQbZpp$ (left panels) or $\TQbTpp$ (right panels) plus light-jet $\vec \kappa_1$-rates at $2 < \DY < 4$ and $4 < \DY < 6$, respectively.
Upper (lower) plots of these figures refer to results taken at 14~TeV~LHC (100~TeV~FCC).
We uniformly adopt transverse-momentum bins with a length of 10~GeV.
Ancillary panels right below primary plots highlight the ratio between $\LL$ or $\HENLOp$ predictions and $\NLLp$ ones.

The general trend shared by all our distributions is a steep decline as $|\vec \kappa_1|$ increases. 
The results exhibit strong stability against MHOUs, with uncertainty bands showing a maximum width of 20\%.
We observe that $\HENLOp$ to $\NLLp$ ratios generally remain below one, diminishing as $\vec \kappa_1$ grows. 
This turn-down is less manifest at nominal FCC energies compared to typical LHC ones. 
In contrast, the $\LL$ to $\NLLp$ ratio shows an almost opposite pattern: it starts below one in the low-$|\vec \kappa_1|$ region but steadily grows as $|\vec \kappa_1|$ rises, eventually reaching a maximum value between 1.5 and two.
Explaining these trends is challenging, as they result from a combination of several interacting effects.

On the one side, previous studies on semihard processes have revealed that the behavior of the NLL-resummed signal relative to its NLO high-energy background in singly differential transverse-momentum rates tends to vary depending on the process under consideration. 
For instance, the $\HENLOp$ to $\NLLp$ ratio for the cascade-baryon plus jet channel consistently exceeds one, as shown in Fig.~7 of Ref.~\cite{Celiberto:2022kxx}. 
However, preliminary analyses of Higgs plus jet distributions, performed within a partially NLL-to-NLO matched accuracy, have exhibited a more complex pattern~\cite{Celiberto:2023dkr}.
Therefore, the observation that $\HENLOp$ to $\NLLp$ ratios are less than one in the context of $\B$ plus jet tags appears to be a distinctive characteristic of this particular process, and it is shared also by the same differential observables sensitive to the production of fully charmed tetraquarks~\cite{Celiberto:2024mab}. 
This suggests that the dynamics governing these emissions are different from those observed in other semihard reactions, highlighting the unique interplay of high-energy and NLL resummation effects in these channels.

Over there, the behavior of the $\LL$/$\NLLp$ ratio is influenced by an intricate combination of distinct factors, particularly the nature of NLO corrections associated with different emission functions. 
For jet emissions, it is well established that NLO corrections to the jet function are generally negative~\cite{Bartels:2001ge,Ivanov:2012ms,Colferai:2015zfa}. 
In contrast, the hadron function presents a different trend: the NLO corrections coming from the perturbative $C_{gg}$ coefficient function are positive, while those from other coefficient functions are negative~\cite{Ivanov:2012iv}. 
This dichotomy suggests that, depending on the transverse-momentum phase space, these corrections might partially counterbalance each other, leading to varying effects on the $\LL$ over $\NLLp$ ratio.

As an example, in the case of cascade-baryon plus jet emissions, the $\LL$ over $\NLLp$ ratio is observed to be larger than one~\cite{Celiberto:2022kxx}. 
However, this behavior is less pronounced in other processes, such as doubly charmed tetraquark plus jet production~\cite{Celiberto:2023rzw}. 
This variability underscores how the specific dynamics of each process affects the relative importance of leading logarithms versus next-to-leading ones, resulting in different patterns in the ratio of $\LL$ to $\NLLp$ predictions across various reactions.

The solid stability observed under both MHOUs and NLL corrections in the predictions displayed in Figs.~\ref{fig:I-k1b-S_Xbq} to~\ref{fig:I-k1b-M_T4b} makes the transverse-momentum spectrum for tetraquark-plus-jet production one of the most promising observables for probing the fundamental dynamics of high-energy QCD. This stability, which is intrinsically linked to the fragmentation of $\TQc$ within a Variable Flavor Number Scheme (VFNS), is not only prominent at typical LHC energies but also remains robust at the higher, nominal energies of the Future Circular Collider (FCC).

One of the most remarkable features of results presented in this section is the emerging stabilizing pattern of our transverse-momentum rates under NLL corrections and MHOU analyses. 
This pattern suggests that high transverse-momentum regions could provide a particularly advantageous opportunity to clearly discriminate high-energy-resummation effects from the fixed-order signal. 
Here, $\kappa_1$-distributions come as powerful tools for experimentally testing and validating the predictions of high-energy dynamics and, more generally, for advancing our understanding of QCD in the semihard sector.

\section{Summary and Outlook}
\label{sec:conclusions}

In this work, we addressed the semi-inclusive hadro\-production of doubly bottomed tetraquarks, $X_{b\bar{b}u\bar{u}}$ and $X_{b\bar{b}u\bar{s}}$, as well as fully bottomed ones, $\TQbZpp$ and its $\TQbTpp$ radial resonance, to which we collectively referred as ``bottomoniumlike'' states.
We relied upon a VFNS-fragmentation approach at leading power, where a single parton perturbatively splits into the corresponding Fock state, which then hadronizes into the color-neutral, observed tetraquark.
To this extent, we built two new families of DGLAP-evolving, hadron-structure oriented collinear FFs, which we na\-med {\tt TQHL1.1} and {\tt TQ4Q1.1} functions.

These sets extend and supersede the corresponding {\tt 1.0} versions derived in recent studies~\cite{Celiberto:2023rzw,Celiberto:2024mab}.
The {\tt TQHL1.1} family describes the fragmentation of $\QXbq$ tetraquarks and was built on the basis of an improved version of the Suzuki model for the heavy-quark channel.
The {\tt TQ4Q1.1} family depicts the fragmentation of $\TQb$ states and contains, as building blocks, initial-scale inputs for gluon and heavy-quark channels, both of them calculated by the hands of potential NRQCD.
A threshold-consistent DGLAP evolution of these inputs was done by taking advantage of basic features of the novel {\HFNRevo} methodology~\cite{Celiberto:2024mex,Celiberto:2024bxu}.

As a phenomenological application, we made use of the $\NLLp$ hybrid collinear and high-energy factorization as implemented in the {\bsymJethad} code~\cite{Celiberto:2020wpk,Celiberto:2022rfj,Celiberto:2023fzz,Celiberto:2024mrq,Celiberto:2024swu}.
Working within this scheme, we provided predictions for rapidity-interval and transverse-momentum rates sensitive to the associated emission of a $\B$ tetra\-quark and a jet at center-of-mass energies running from 14~TeV~LHC to 100~TeV~FCC.

The use of the VFNS collinear fragmentation to depict the production of $\B$ states at large transverse momentum stabilized our high-energy resummation, protecting it from potential instabilities due to NLL corrections and nonresummed threshold logarithms. 
The emerging \emph{natural stability} ensured the validity and convergence of our formalism across a wide range of center-of-mass energies, from LHC to FCC.
To achieve precision, our $\NLLp$ hybrid approach must evolve into a \emph{multilateral} formalism that incorporates other resummation techniques. 
Initial steps will include establishing connections with soft-gluon~\cite{Hatta:2020bgy,Hatta:2021jcd,Caucal:2022ulg,Taels:2022tza} and jet-radius resummations~\cite{Dasgupta:2014yra,Dasgupta:2016bnd,Banfi:2012jm,Banfi:2015pju,Liu:2017pbb}. 
Additionally, exploring potential synergies with research on jet angularities~\cite{Luisoni:2015xha,Caletti:2021oor,Reichelt:2021svh} is a compelling future direction.

A key advancement in the study of exotic states with bottom flavor will be the exploration of single inclusive $\B$-hadron detections in forward-rapidity regions \emph{via} the $\NLLp$ factorization. 
These channels offer us a direct probe of the small-$x$ UGD in the proton, which is currently known only at a very qualitative level and strongly depends on models.
Here, our {\tt TQHL1.1} and {\tt TQ4Q1.1} publicly released determinations will guide us toward exploratory studies on the production of heavy tetraquark states across a wide range of reactions, spanning from semi-inclusive tags at hadron machines to emissions at lepton-hadron colliders, such as the forthcoming EIC~\cite{AbdulKhalek:2021gbh,Khalek:2022bzd,Hentschinski:2022xnd,Amoroso:2022eow,Abir:2023fpo,Allaire:2023fgp}.

Progress in understanding the hadron structure will be driven by advancements in our knowledge of the fundamental dynamics behind exotic-matter formation. 
These advancements will be fueled by data from the FCC~\cite{FCC:2018byv,FCC:2018evy,FCC:2018vvp,FCC:2018bvk} and other upcoming colliders~\cite{Chapon:2020heu,Anchordoqui:2021ghd,Feng:2022inv,AlexanderAryshev:2022pkx,Arbuzov:2020cqg,Accettura:2023ked,InternationalMuonCollider:2024jyv,MuCoL:2024oxj,Black:2022cth,Accardi:2023chb}.
To support this program, we aim to improve our description of tetraquark fragmentation by determining quantitative uncertainties, potentially linked to a rigorous examination of MHOU effects~\cite{Kassabov:2022orn,Harland-Lang:2018bxd,Ball:2021icz,McGowan:2022nag,NNPDF:2024dpb,Pasquini:2023aaf}. 
Additionally, we plan to incorporate color-octet contributions in future studies.

The recently found evidence~\cite{Ball:2022qks} of intrinsic charm quarks~\cite{Brodsky:1980pb,Brodsky:2015fna} in the proton (see Refs.~\cite{Jimenez-Delgado:2014zga,Ball:2016neh,Hou:2017khm,Guzzi:2022rca} for other studies), further supported by recent findings on its valence density~\cite{NNPDF:2023tyk}, paves the way toward exploring the potential existence of an intrinsic bottom component. 
Advancing our understanding of ordinary as well as exotic bottom physics is crucial in this context, just as it has been for the exotic charm sector~\cite{Vogt:2024fky}. 
We believe that research focused on the formation mechanism of bottom-flavored tetraquarks \emph{via} the VFNS collinear fragmentation will be essential in pursuing these new directions.

\section*{Data availability, Acknowledgments, Funding}
\label{sec:data_availability}
\addcontentsline{toc}{section}{\nameref{sec:data_availability}}

The {\tt TQHL1.1}~\cite{Celiberto:2024_TQHL11} and {\tt TQ4Q1.1}~\cite{Celiberto:2024_TQ4Q11} collinear FFs are released in {\tt LHAPDF} format
\begin{itemize}

    \item[$\bullet$] \; NLO, \,$\QXQq$\,: \vskip0.15cm

    \begin{itemize}

        \item[$\diamond$] \,{\tt TQHL11\_cs\_Xcu\_nlo}\,; \vskip0.15cm

        \item[$\diamond$] \,{\tt TQHL11\_cs\_Xcs\_nlo}\,; \vskip0.15cm

        \item[$\diamond$] \,{\tt TQHL11\_cs\_Xbu\_nlo}\,; \vskip0.15cm

        \item[$\diamond$] \,{\tt TQHL11\_cs\_Xbs\_nlo}\,; \vskip0.15cm

    \end{itemize} \vskip0.25cm

    \item[$\bullet$] \; NLO, \,$\TQQZpp$\,: \vskip0.15cm

    \begin{itemize}

        \item[$\diamond$] \,{\tt TQ4Q11\_cs\_T4c-0pp\_nlo}\,; \vskip0.15cm

        \item[$\diamond$] \,{\tt TQ4Q11mQ\_cs\_T4c-0pp\_nlo}\,; \vskip0.15cm

        \item[$\diamond$] \,{\tt TQ4Q11\_cs\_T4b-0pp\_nlo}\,; \vskip0.15cm

        \item[$\diamond$] \,{\tt TQ4Q11mQ\_cs\_T4b-0pp\_nlo}\,; \vskip0.15cm

    \end{itemize} \vskip0.25cm

    \item[$\bullet$] \; NLO, \,$\TQQTpp$\,: \vskip0.15cm

    \begin{itemize}

        \item[$\diamond$] \,{\tt TQ4Q11\_cs\_T4c-2pp\_nlo}\,; \vskip0.15cm

        \item[$\diamond$] \,{\tt TQ4Q11mQ\_cs\_T4c-2pp\_nlo}\,; \vskip0.15cm

        \item[$\diamond$] \,{\tt TQ4Q11\_cs\_T4b-2pp\_nlo}\,; \vskip0.15cm

        \item[$\diamond$] \,{\tt TQ4Q11mQ\_cs\_T4b-2pp\_nlo}\,. \vskip0.15cm

    \end{itemize}

\end{itemize}

The ``{\tt cs}'' sublabel specifies that the given initial-scale inputs are calculated within the color-singlet configuration, whereas the ``{\tt mQ}'' one refers to {\tt TQ4Q1.1$^-$} FFs evolved from the gluon input only.
These sets can be publicly accessed from the following url: \url{https://github.com/FGCeliberto/Collinear_FFs/}.


We would like to express our gratitude to Alessandro Papa 
for a critical reading of the manuscript, for useful suggestions, and for encouragement.
We thank Alessandro Pilloni for a discussion on the physics of exotic states.

F.G.C. acknowledges support from the Atracci\'on de Talento Grant n. 2022-T1/TIC-24176 of the Comunidad Aut\'onoma de Madrid, Spain.
G.G. acknowledges support from the INFN/QFT@\-COL\-LI\-DERS project, Italy.

Feynman diagrams in this work were realized with {\tt JaxoDraw 2.0}~\cite{Binosi:2008ig}.

\bibliographystyle{apsrev}
\bibliography{references}

\begin{thebibliography}{433}
\expandafter\ifx\csname natexlab\endcsname\relax\def\natexlab#1{#1}\fi
\expandafter\ifx\csname bibnamefont\endcsname\relax
  \def\bibnamefont#1{#1}\fi
\expandafter\ifx\csname bibfnamefont\endcsname\relax
  \def\bibfnamefont#1{#1}\fi
\expandafter\ifx\csname citenamefont\endcsname\relax
  \def\citenamefont#1{#1}\fi
\expandafter\ifx\csname url\endcsname\relax
  \def\url#1{\texttt{#1}}\fi
\expandafter\ifx\csname urlprefix\endcsname\relax\def\urlprefix{URL }\fi
\providecommand{\bibinfo}[2]{#2}
\providecommand{\eprint}[2][]{\url{#2}}

\bibitem[{\citenamefont{Abdul~Khalek et~al.}(2022{\natexlab{a}})}]{AbdulKhalek:2021gbh}
\bibinfo{author}{\bibfnamefont{R.}~\bibnamefont{Abdul~Khalek}} \bibnamefont{et~al.}, \bibinfo{journal}{Nucl. Phys. A} \textbf{\bibinfo{volume}{1026}}, \bibinfo{pages}{122447} (\bibinfo{year}{2022}{\natexlab{a}}), \eprint{2103.05419}.

\bibitem[{\citenamefont{Abdul~Khalek et~al.}(2022{\natexlab{b}})}]{Khalek:2022bzd}
\bibinfo{author}{\bibfnamefont{R.}~\bibnamefont{Abdul~Khalek}} \bibnamefont{et~al.}, in \emph{\bibinfo{booktitle}{{2022 Snowmass Summer Study}}} (\bibinfo{year}{2022}{\natexlab{b}}), \eprint{2203.13199}.

\bibitem[{\citenamefont{Hentschinski et~al.}(2023)}]{Hentschinski:2022xnd}
\bibinfo{author}{\bibfnamefont{M.}~\bibnamefont{Hentschinski}} \bibnamefont{et~al.}, \bibinfo{journal}{Acta Phys. Polon. B} \textbf{\bibinfo{volume}{54}}, \bibinfo{pages}{2} (\bibinfo{year}{2023}), \eprint{2203.08129}.

\bibitem[{\citenamefont{Amoroso et~al.}(2022)}]{Amoroso:2022eow}
\bibinfo{author}{\bibfnamefont{S.}~\bibnamefont{Amoroso}} \bibnamefont{et~al.}, \bibinfo{journal}{Acta Phys. Polon. B} \textbf{\bibinfo{volume}{53}}, \bibinfo{pages}{A1} (\bibinfo{year}{2022}), \eprint{2203.13923}.

\bibitem[{\citenamefont{Abir et~al.}(2023)}]{Abir:2023fpo}
\bibinfo{author}{\bibfnamefont{R.}~\bibnamefont{Abir}} \bibnamefont{et~al.} (\bibinfo{year}{2023}), \eprint{2305.14572}.

\bibitem[{\citenamefont{Allaire et~al.}(2024)}]{Allaire:2023fgp}
\bibinfo{author}{\bibfnamefont{C.}~\bibnamefont{Allaire}} \bibnamefont{et~al.}, \bibinfo{journal}{Comput. Softw. Big Sci.} \textbf{\bibinfo{volume}{8}}, \bibinfo{pages}{5} (\bibinfo{year}{2024}), \eprint{2307.08593}.

\bibitem[{\citenamefont{Abada et~al.}(2019{\natexlab{a}})}]{FCC:2018byv}
\bibinfo{author}{\bibfnamefont{A.}~\bibnamefont{Abada}} \bibnamefont{et~al.} (\bibinfo{collaboration}{FCC}), \bibinfo{journal}{Eur. Phys. J. C} \textbf{\bibinfo{volume}{79}}, \bibinfo{pages}{474} (\bibinfo{year}{2019}{\natexlab{a}}).

\bibitem[{\citenamefont{Abada et~al.}(2019{\natexlab{b}})}]{FCC:2018evy}
\bibinfo{author}{\bibfnamefont{A.}~\bibnamefont{Abada}} \bibnamefont{et~al.} (\bibinfo{collaboration}{FCC}), \bibinfo{journal}{Eur. Phys. J. ST} \textbf{\bibinfo{volume}{228}}, \bibinfo{pages}{261} (\bibinfo{year}{2019}{\natexlab{b}}).

\bibitem[{\citenamefont{Abada et~al.}(2019{\natexlab{c}})}]{FCC:2018vvp}
\bibinfo{author}{\bibfnamefont{A.}~\bibnamefont{Abada}} \bibnamefont{et~al.} (\bibinfo{collaboration}{FCC}), \bibinfo{journal}{Eur. Phys. J. ST} \textbf{\bibinfo{volume}{228}}, \bibinfo{pages}{755} (\bibinfo{year}{2019}{\natexlab{c}}).

\bibitem[{\citenamefont{Abada et~al.}(2019{\natexlab{d}})}]{FCC:2018bvk}
\bibinfo{author}{\bibfnamefont{A.}~\bibnamefont{Abada}} \bibnamefont{et~al.} (\bibinfo{collaboration}{FCC}), \bibinfo{journal}{Eur. Phys. J. ST} \textbf{\bibinfo{volume}{228}}, \bibinfo{pages}{1109} (\bibinfo{year}{2019}{\natexlab{d}}).

\bibitem[{\citenamefont{Kou and Pene}(2005)}]{Kou:2005gt}
\bibinfo{author}{\bibfnamefont{E.}~\bibnamefont{Kou}} \bibnamefont{and} \bibinfo{author}{\bibfnamefont{O.}~\bibnamefont{Pene}}, \bibinfo{journal}{Phys. Lett. B} \textbf{\bibinfo{volume}{631}}, \bibinfo{pages}{164} (\bibinfo{year}{2005}), \eprint{hep-ph/0507119}.

\bibitem[{\citenamefont{Braaten}(2013)}]{Braaten:2013boa}
\bibinfo{author}{\bibfnamefont{E.}~\bibnamefont{Braaten}}, \bibinfo{journal}{Phys. Rev. Lett.} \textbf{\bibinfo{volume}{111}}, \bibinfo{pages}{162003} (\bibinfo{year}{2013}), \eprint{1305.6905}.

\bibitem[{\citenamefont{Berwein et~al.}(2015)\citenamefont{Berwein, Brambilla, Tarr\'us~Castell\`a, and Vairo}}]{Berwein:2015vca}
\bibinfo{author}{\bibfnamefont{M.}~\bibnamefont{Berwein}}, \bibinfo{author}{\bibfnamefont{N.}~\bibnamefont{Brambilla}}, \bibinfo{author}{\bibfnamefont{J.}~\bibnamefont{Tarr\'us~Castell\`a}}, \bibnamefont{and} \bibinfo{author}{\bibfnamefont{A.}~\bibnamefont{Vairo}}, \bibinfo{journal}{Phys. Rev. D} \textbf{\bibinfo{volume}{92}}, \bibinfo{pages}{114019} (\bibinfo{year}{2015}), \eprint{1510.04299}.

\bibitem[{\citenamefont{Minkowski and Ochs}(1999)}]{Minkowski:1998mf}
\bibinfo{author}{\bibfnamefont{P.}~\bibnamefont{Minkowski}} \bibnamefont{and} \bibinfo{author}{\bibfnamefont{W.}~\bibnamefont{Ochs}}, \bibinfo{journal}{Eur. Phys. J. C} \textbf{\bibinfo{volume}{9}}, \bibinfo{pages}{283} (\bibinfo{year}{1999}), \eprint{hep-ph/9811518}.

\bibitem[{\citenamefont{Mathieu et~al.}(2009)\citenamefont{Mathieu, Kochelev, and Vento}}]{Mathieu:2008me}
\bibinfo{author}{\bibfnamefont{V.}~\bibnamefont{Mathieu}}, \bibinfo{author}{\bibfnamefont{N.}~\bibnamefont{Kochelev}}, \bibnamefont{and} \bibinfo{author}{\bibfnamefont{V.}~\bibnamefont{Vento}}, \bibinfo{journal}{Int. J. Mod. Phys. E} \textbf{\bibinfo{volume}{18}}, \bibinfo{pages}{1} (\bibinfo{year}{2009}), \eprint{0810.4453}.

\bibitem[{\citenamefont{Abazov et~al.}(2021)}]{D0:2020tig}
\bibinfo{author}{\bibfnamefont{V.~M.} \bibnamefont{Abazov}} \bibnamefont{et~al.} (\bibinfo{collaboration}{D0, TOTEM}), \bibinfo{journal}{Phys. Rev. Lett.} \textbf{\bibinfo{volume}{127}}, \bibinfo{pages}{062003} (\bibinfo{year}{2021}), \eprint{2012.03981}.

\bibitem[{\citenamefont{Cs\"org\H{o} et~al.}(2021)\citenamefont{Cs\"org\H{o}, Novak, Pasechnik, Ster, and Szanyi}}]{Csorgo:2019ewn}
\bibinfo{author}{\bibfnamefont{T.}~\bibnamefont{Cs\"org\H{o}}}, \bibinfo{author}{\bibfnamefont{T.}~\bibnamefont{Novak}}, \bibinfo{author}{\bibfnamefont{R.}~\bibnamefont{Pasechnik}}, \bibinfo{author}{\bibfnamefont{A.}~\bibnamefont{Ster}}, \bibnamefont{and} \bibinfo{author}{\bibfnamefont{I.}~\bibnamefont{Szanyi}}, \bibinfo{journal}{Eur. Phys. J. C} \textbf{\bibinfo{volume}{81}}, \bibinfo{pages}{180} (\bibinfo{year}{2021}), \eprint{1912.11968}.

\bibitem[{\citenamefont{Gell-Mann}(1964)}]{Gell-Mann:1964ewy}
\bibinfo{author}{\bibfnamefont{M.}~\bibnamefont{Gell-Mann}}, \bibinfo{journal}{Phys. Lett.} \textbf{\bibinfo{volume}{8}}, \bibinfo{pages}{214} (\bibinfo{year}{1964}).

\bibitem[{\citenamefont{Jaffe}(1977)}]{Jaffe:1976ig}
\bibinfo{author}{\bibfnamefont{R.~L.} \bibnamefont{Jaffe}}, \bibinfo{journal}{Phys. Rev. D} \textbf{\bibinfo{volume}{15}}, \bibinfo{pages}{267} (\bibinfo{year}{1977}).

\bibitem[{\citenamefont{Ader et~al.}(1982)\citenamefont{Ader, Richard, and Taxil}}]{Ader:1981db}
\bibinfo{author}{\bibfnamefont{J.~P.} \bibnamefont{Ader}}, \bibinfo{author}{\bibfnamefont{J.~M.} \bibnamefont{Richard}}, \bibnamefont{and} \bibinfo{author}{\bibfnamefont{P.}~\bibnamefont{Taxil}}, \bibinfo{journal}{Phys. Rev. D} \textbf{\bibinfo{volume}{25}}, \bibinfo{pages}{2370} (\bibinfo{year}{1982}).

\bibitem[{\citenamefont{Choi et~al.}(2003)}]{Belle:2003nnu}
\bibinfo{author}{\bibfnamefont{S.~K.} \bibnamefont{Choi}} \bibnamefont{et~al.} (\bibinfo{collaboration}{Belle}), \bibinfo{journal}{Phys. Rev. Lett.} \textbf{\bibinfo{volume}{91}}, \bibinfo{pages}{262001} (\bibinfo{year}{2003}), \eprint{hep-ex/0309032}.

\bibitem[{\citenamefont{Acosta et~al.}(2004)}]{CDF:2003cab}
\bibinfo{author}{\bibfnamefont{D.}~\bibnamefont{Acosta}} \bibnamefont{et~al.} (\bibinfo{collaboration}{CDF}), \bibinfo{journal}{Phys. Rev. Lett.} \textbf{\bibinfo{volume}{93}}, \bibinfo{pages}{072001} (\bibinfo{year}{2004}), \eprint{hep-ex/0312021}.

\bibitem[{\citenamefont{Aaij et~al.}(2013)}]{LHCb:2013kgk}
\bibinfo{author}{\bibfnamefont{R.}~\bibnamefont{Aaij}} \bibnamefont{et~al.} (\bibinfo{collaboration}{LHCb}), \bibinfo{journal}{Phys. Rev. Lett.} \textbf{\bibinfo{volume}{110}}, \bibinfo{pages}{222001} (\bibinfo{year}{2013}), \eprint{1302.6269}.

\bibitem[{\citenamefont{Sirunyan et~al.}(2022)}]{CMS:2021znk}
\bibinfo{author}{\bibfnamefont{A.~M.} \bibnamefont{Sirunyan}} \bibnamefont{et~al.} (\bibinfo{collaboration}{CMS}), \bibinfo{journal}{Phys. Rev. Lett.} \textbf{\bibinfo{volume}{128}}, \bibinfo{pages}{032001} (\bibinfo{year}{2022}), \eprint{2102.13048}.

\bibitem[{\citenamefont{Swanson}(2006)}]{Swanson:2006st}
\bibinfo{author}{\bibfnamefont{E.~S.} \bibnamefont{Swanson}}, \bibinfo{journal}{Phys. Rept.} \textbf{\bibinfo{volume}{429}}, \bibinfo{pages}{243} (\bibinfo{year}{2006}), \eprint{hep-ph/0601110}.

\bibitem[{\citenamefont{Augustin et~al.}(1974)}]{SLAC-SP-017:1974ind}
\bibinfo{author}{\bibfnamefont{J.~E.} \bibnamefont{Augustin}} \bibnamefont{et~al.} (\bibinfo{collaboration}{SLAC-SP-017}), \bibinfo{journal}{Phys. Rev. Lett.} \textbf{\bibinfo{volume}{33}}, \bibinfo{pages}{1406} (\bibinfo{year}{1974}).

\bibitem[{\citenamefont{Aubert et~al.}(1974)}]{E598:1974sol}
\bibinfo{author}{\bibfnamefont{J.~J.} \bibnamefont{Aubert}} \bibnamefont{et~al.} (\bibinfo{collaboration}{E598}), \bibinfo{journal}{Phys. Rev. Lett.} \textbf{\bibinfo{volume}{33}}, \bibinfo{pages}{1404} (\bibinfo{year}{1974}).

\bibitem[{\citenamefont{Bacci et~al.}(1974)}]{Bacci:1974za}
\bibinfo{author}{\bibfnamefont{C.}~\bibnamefont{Bacci}} \bibnamefont{et~al.}, \bibinfo{journal}{Phys. Rev. Lett.} \textbf{\bibinfo{volume}{33}}, \bibinfo{pages}{1408} (\bibinfo{year}{1974}), \bibinfo{note}{[Erratum: Phys.Rev.Lett. 33, 1649 (1974)]}.

\bibitem[{\citenamefont{Chen et~al.}(2016)\citenamefont{Chen, Chen, Liu, and Zhu}}]{Chen:2016qju}
\bibinfo{author}{\bibfnamefont{H.-X.} \bibnamefont{Chen}}, \bibinfo{author}{\bibfnamefont{W.}~\bibnamefont{Chen}}, \bibinfo{author}{\bibfnamefont{X.}~\bibnamefont{Liu}}, \bibnamefont{and} \bibinfo{author}{\bibfnamefont{S.-L.} \bibnamefont{Zhu}}, \bibinfo{journal}{Phys. Rept.} \textbf{\bibinfo{volume}{639}}, \bibinfo{pages}{1} (\bibinfo{year}{2016}), \eprint{1601.02092}.

\bibitem[{\citenamefont{Liu et~al.}(2019)\citenamefont{Liu, Chen, Chen, Liu, and Zhu}}]{Liu:2019zoy}
\bibinfo{author}{\bibfnamefont{Y.-R.} \bibnamefont{Liu}}, \bibinfo{author}{\bibfnamefont{H.-X.} \bibnamefont{Chen}}, \bibinfo{author}{\bibfnamefont{W.}~\bibnamefont{Chen}}, \bibinfo{author}{\bibfnamefont{X.}~\bibnamefont{Liu}}, \bibnamefont{and} \bibinfo{author}{\bibfnamefont{S.-L.} \bibnamefont{Zhu}}, \bibinfo{journal}{Prog. Part. Nucl. Phys.} \textbf{\bibinfo{volume}{107}}, \bibinfo{pages}{237} (\bibinfo{year}{2019}), \eprint{1903.11976}.

\bibitem[{\citenamefont{Aaij et~al.}(2020{\natexlab{a}})}]{LHCb:2020bls}
\bibinfo{author}{\bibfnamefont{R.}~\bibnamefont{Aaij}} \bibnamefont{et~al.} (\bibinfo{collaboration}{LHCb}), \bibinfo{journal}{Phys. Rev. Lett.} \textbf{\bibinfo{volume}{125}}, \bibinfo{pages}{242001} (\bibinfo{year}{2020}{\natexlab{a}}), \eprint{2009.00025}.

\bibitem[{\citenamefont{Maiani et~al.}(2005)\citenamefont{Maiani, Piccinini, Polosa, and Riquer}}]{Maiani:2004vq}
\bibinfo{author}{\bibfnamefont{L.}~\bibnamefont{Maiani}}, \bibinfo{author}{\bibfnamefont{F.}~\bibnamefont{Piccinini}}, \bibinfo{author}{\bibfnamefont{A.~D.} \bibnamefont{Polosa}}, \bibnamefont{and} \bibinfo{author}{\bibfnamefont{V.}~\bibnamefont{Riquer}}, \bibinfo{journal}{Phys. Rev. D} \textbf{\bibinfo{volume}{71}}, \bibinfo{pages}{014028} (\bibinfo{year}{2005}), \eprint{hep-ph/0412098}.

\bibitem[{\citenamefont{'t~Hooft et~al.}(2008)\citenamefont{'t~Hooft, Isidori, Maiani, Polosa, and Riquer}}]{tHooft:2008rus}
\bibinfo{author}{\bibfnamefont{G.}~\bibnamefont{'t~Hooft}}, \bibinfo{author}{\bibfnamefont{G.}~\bibnamefont{Isidori}}, \bibinfo{author}{\bibfnamefont{L.}~\bibnamefont{Maiani}}, \bibinfo{author}{\bibfnamefont{A.~D.} \bibnamefont{Polosa}}, \bibnamefont{and} \bibinfo{author}{\bibfnamefont{V.}~\bibnamefont{Riquer}}, \bibinfo{journal}{Phys. Lett. B} \textbf{\bibinfo{volume}{662}}, \bibinfo{pages}{424} (\bibinfo{year}{2008}), \eprint{0801.2288}.

\bibitem[{\citenamefont{Maiani et~al.}(2013)\citenamefont{Maiani, Riquer, Faccini, Piccinini, Pilloni, and Polosa}}]{Maiani:2013nmn}
\bibinfo{author}{\bibfnamefont{L.}~\bibnamefont{Maiani}}, \bibinfo{author}{\bibfnamefont{V.}~\bibnamefont{Riquer}}, \bibinfo{author}{\bibfnamefont{R.}~\bibnamefont{Faccini}}, \bibinfo{author}{\bibfnamefont{F.}~\bibnamefont{Piccinini}}, \bibinfo{author}{\bibfnamefont{A.}~\bibnamefont{Pilloni}}, \bibnamefont{and} \bibinfo{author}{\bibfnamefont{A.~D.} \bibnamefont{Polosa}}, \bibinfo{journal}{Phys. Rev. D} \textbf{\bibinfo{volume}{87}}, \bibinfo{pages}{111102} (\bibinfo{year}{2013}), \eprint{1303.6857}.

\bibitem[{\citenamefont{Maiani et~al.}(2014)\citenamefont{Maiani, Piccinini, Polosa, and Riquer}}]{Maiani:2014aja}
\bibinfo{author}{\bibfnamefont{L.}~\bibnamefont{Maiani}}, \bibinfo{author}{\bibfnamefont{F.}~\bibnamefont{Piccinini}}, \bibinfo{author}{\bibfnamefont{A.~D.} \bibnamefont{Polosa}}, \bibnamefont{and} \bibinfo{author}{\bibfnamefont{V.}~\bibnamefont{Riquer}}, \bibinfo{journal}{Phys. Rev. D} \textbf{\bibinfo{volume}{89}}, \bibinfo{pages}{114010} (\bibinfo{year}{2014}), \eprint{1405.1551}.

\bibitem[{\citenamefont{Maiani et~al.}(2018)\citenamefont{Maiani, Polosa, and Riquer}}]{Maiani:2017kyi}
\bibinfo{author}{\bibfnamefont{L.}~\bibnamefont{Maiani}}, \bibinfo{author}{\bibfnamefont{A.~D.} \bibnamefont{Polosa}}, \bibnamefont{and} \bibinfo{author}{\bibfnamefont{V.}~\bibnamefont{Riquer}}, \bibinfo{journal}{Phys. Lett. B} \textbf{\bibinfo{volume}{778}}, \bibinfo{pages}{247} (\bibinfo{year}{2018}), \eprint{1712.05296}.

\bibitem[{\citenamefont{Mutuk}(2021)}]{Mutuk:2021hmi}
\bibinfo{author}{\bibfnamefont{H.}~\bibnamefont{Mutuk}}, \bibinfo{journal}{Eur. Phys. J. C} \textbf{\bibinfo{volume}{81}}, \bibinfo{pages}{367} (\bibinfo{year}{2021}), \eprint{2104.11823}.

\bibitem[{\citenamefont{Wang and Huang}(2014{\natexlab{a}})}]{Wang:2013vex}
\bibinfo{author}{\bibfnamefont{Z.-G.} \bibnamefont{Wang}} \bibnamefont{and} \bibinfo{author}{\bibfnamefont{T.}~\bibnamefont{Huang}}, \bibinfo{journal}{Phys. Rev. D} \textbf{\bibinfo{volume}{89}}, \bibinfo{pages}{054019} (\bibinfo{year}{2014}{\natexlab{a}}), \eprint{1310.2422}.

\bibitem[{\citenamefont{Wang}(2014{\natexlab{a}})}]{Wang:2013exa}
\bibinfo{author}{\bibfnamefont{Z.-G.} \bibnamefont{Wang}}, \bibinfo{journal}{Eur. Phys. J. C} \textbf{\bibinfo{volume}{74}}, \bibinfo{pages}{2874} (\bibinfo{year}{2014}{\natexlab{a}}), \eprint{1311.1046}.

\bibitem[{\citenamefont{Grinstein et~al.}(2024)\citenamefont{Grinstein, Maiani, and Polosa}}]{Grinstein:2024rcu}
\bibinfo{author}{\bibfnamefont{B.}~\bibnamefont{Grinstein}}, \bibinfo{author}{\bibfnamefont{L.}~\bibnamefont{Maiani}}, \bibnamefont{and} \bibinfo{author}{\bibfnamefont{A.~D.} \bibnamefont{Polosa}}, \bibinfo{journal}{Phys. Rev. D} \textbf{\bibinfo{volume}{109}}, \bibinfo{pages}{074009} (\bibinfo{year}{2024}), \eprint{2401.11623}.

\bibitem[{\citenamefont{Tornqvist}(1994)}]{Tornqvist:1993ng}
\bibinfo{author}{\bibfnamefont{N.~A.} \bibnamefont{Tornqvist}}, \bibinfo{journal}{Z. Phys. C} \textbf{\bibinfo{volume}{61}}, \bibinfo{pages}{525} (\bibinfo{year}{1994}), \eprint{hep-ph/9310247}.

\bibitem[{\citenamefont{Braaten and Kusunoki}(2004)}]{Braaten:2003he}
\bibinfo{author}{\bibfnamefont{E.}~\bibnamefont{Braaten}} \bibnamefont{and} \bibinfo{author}{\bibfnamefont{M.}~\bibnamefont{Kusunoki}}, \bibinfo{journal}{Phys. Rev. D} \textbf{\bibinfo{volume}{69}}, \bibinfo{pages}{074005} (\bibinfo{year}{2004}), \eprint{hep-ph/0311147}.

\bibitem[{\citenamefont{Guo et~al.}(2013)\citenamefont{Guo, Hidalgo-Duque, Nieves, and Valderrama}}]{Guo:2013sya}
\bibinfo{author}{\bibfnamefont{F.-K.} \bibnamefont{Guo}}, \bibinfo{author}{\bibfnamefont{C.}~\bibnamefont{Hidalgo-Duque}}, \bibinfo{author}{\bibfnamefont{J.}~\bibnamefont{Nieves}}, \bibnamefont{and} \bibinfo{author}{\bibfnamefont{M.~P.} \bibnamefont{Valderrama}}, \bibinfo{journal}{Phys. Rev. D} \textbf{\bibinfo{volume}{88}}, \bibinfo{pages}{054007} (\bibinfo{year}{2013}), \eprint{1303.6608}.

\bibitem[{\citenamefont{Mutuk}(2022)}]{Mutuk:2022ckn}
\bibinfo{author}{\bibfnamefont{H.}~\bibnamefont{Mutuk}}, \bibinfo{journal}{Eur. Phys. J. C} \textbf{\bibinfo{volume}{82}}, \bibinfo{pages}{1142} (\bibinfo{year}{2022}), \eprint{2211.14836}.

\bibitem[{\citenamefont{Wang and Huang}(2014{\natexlab{b}})}]{Wang:2013daa}
\bibinfo{author}{\bibfnamefont{Z.-G.} \bibnamefont{Wang}} \bibnamefont{and} \bibinfo{author}{\bibfnamefont{T.}~\bibnamefont{Huang}}, \bibinfo{journal}{Eur. Phys. J. C} \textbf{\bibinfo{volume}{74}}, \bibinfo{pages}{2891} (\bibinfo{year}{2014}{\natexlab{b}}), \eprint{1312.7489}.

\bibitem[{\citenamefont{Wang}(2014{\natexlab{b}})}]{Wang:2014gwa}
\bibinfo{author}{\bibfnamefont{Z.-G.} \bibnamefont{Wang}}, \bibinfo{journal}{Eur. Phys. J. C} \textbf{\bibinfo{volume}{74}}, \bibinfo{pages}{2963} (\bibinfo{year}{2014}{\natexlab{b}}), \eprint{1403.0810}.

\bibitem[{\citenamefont{Esposito et~al.}(2023)\citenamefont{Esposito, Germani, Glioti, Polosa, Rattazzi, and Tarquini}}]{Esposito:2023mxw}
\bibinfo{author}{\bibfnamefont{A.}~\bibnamefont{Esposito}}, \bibinfo{author}{\bibfnamefont{D.}~\bibnamefont{Germani}}, \bibinfo{author}{\bibfnamefont{A.}~\bibnamefont{Glioti}}, \bibinfo{author}{\bibfnamefont{A.~D.} \bibnamefont{Polosa}}, \bibinfo{author}{\bibfnamefont{R.}~\bibnamefont{Rattazzi}}, \bibnamefont{and} \bibinfo{author}{\bibfnamefont{M.}~\bibnamefont{Tarquini}}, \bibinfo{journal}{Phys. Lett. B} \textbf{\bibinfo{volume}{847}}, \bibinfo{pages}{138285} (\bibinfo{year}{2023}), \eprint{2307.11400}.

\bibitem[{\citenamefont{Dubynskiy and Voloshin}(2008)}]{Dubynskiy:2008mq}
\bibinfo{author}{\bibfnamefont{S.}~\bibnamefont{Dubynskiy}} \bibnamefont{and} \bibinfo{author}{\bibfnamefont{M.~B.} \bibnamefont{Voloshin}}, \bibinfo{journal}{Phys. Lett. B} \textbf{\bibinfo{volume}{666}}, \bibinfo{pages}{344} (\bibinfo{year}{2008}), \eprint{0803.2224}.

\bibitem[{\citenamefont{Voloshin}(2013)}]{Voloshin:2013dpa}
\bibinfo{author}{\bibfnamefont{M.~B.} \bibnamefont{Voloshin}}, \bibinfo{journal}{Phys. Rev. D} \textbf{\bibinfo{volume}{87}}, \bibinfo{pages}{091501} (\bibinfo{year}{2013}), \eprint{1304.0380}.

\bibitem[{\citenamefont{Guo et~al.}(2018)\citenamefont{Guo, Hanhart, Mei\ss{}ner, Wang, Zhao, and Zou}}]{Guo:2017jvc}
\bibinfo{author}{\bibfnamefont{F.-K.} \bibnamefont{Guo}}, \bibinfo{author}{\bibfnamefont{C.}~\bibnamefont{Hanhart}}, \bibinfo{author}{\bibfnamefont{U.-G.} \bibnamefont{Mei\ss{}ner}}, \bibinfo{author}{\bibfnamefont{Q.}~\bibnamefont{Wang}}, \bibinfo{author}{\bibfnamefont{Q.}~\bibnamefont{Zhao}}, \bibnamefont{and} \bibinfo{author}{\bibfnamefont{B.-S.} \bibnamefont{Zou}}, \bibinfo{journal}{Rev. Mod. Phys.} \textbf{\bibinfo{volume}{90}}, \bibinfo{pages}{015004} (\bibinfo{year}{2018}), \bibinfo{note}{[Erratum: Rev.Mod.Phys. 94, 029901 (2022)]}, \eprint{1705.00141}.

\bibitem[{\citenamefont{Ferretti et~al.}(2019)\citenamefont{Ferretti, Santopinto, Naeem~Anwar, and Bedolla}}]{Ferretti:2018ojb}
\bibinfo{author}{\bibfnamefont{J.}~\bibnamefont{Ferretti}}, \bibinfo{author}{\bibfnamefont{E.}~\bibnamefont{Santopinto}}, \bibinfo{author}{\bibfnamefont{M.}~\bibnamefont{Naeem~Anwar}}, \bibnamefont{and} \bibinfo{author}{\bibfnamefont{M.~A.} \bibnamefont{Bedolla}}, \bibinfo{journal}{Phys. Lett. B} \textbf{\bibinfo{volume}{789}}, \bibinfo{pages}{562} (\bibinfo{year}{2019}), \eprint{1807.01207}.

\bibitem[{\citenamefont{Ferretti and Santopinto}(2019)}]{Ferretti:2018tco}
\bibinfo{author}{\bibfnamefont{J.}~\bibnamefont{Ferretti}} \bibnamefont{and} \bibinfo{author}{\bibfnamefont{E.}~\bibnamefont{Santopinto}}, \bibinfo{journal}{Phys. Lett. B} \textbf{\bibinfo{volume}{789}}, \bibinfo{pages}{550} (\bibinfo{year}{2019}), \eprint{1806.02489}.

\bibitem[{\citenamefont{Ferretti and Santopinto}(2020)}]{Ferretti:2020ewe}
\bibinfo{author}{\bibfnamefont{J.}~\bibnamefont{Ferretti}} \bibnamefont{and} \bibinfo{author}{\bibfnamefont{E.}~\bibnamefont{Santopinto}}, \bibinfo{journal}{JHEP} \textbf{\bibinfo{volume}{04}}, \bibinfo{pages}{119} (\bibinfo{year}{2020}), \eprint{2001.01067}.

\bibitem[{\citenamefont{Esposito et~al.}(2021)\citenamefont{Esposito, Ferreiro, Pilloni, Polosa, and Salgado}}]{Esposito:2020ywk}
\bibinfo{author}{\bibfnamefont{A.}~\bibnamefont{Esposito}}, \bibinfo{author}{\bibfnamefont{E.~G.} \bibnamefont{Ferreiro}}, \bibinfo{author}{\bibfnamefont{A.}~\bibnamefont{Pilloni}}, \bibinfo{author}{\bibfnamefont{A.~D.} \bibnamefont{Polosa}}, \bibnamefont{and} \bibinfo{author}{\bibfnamefont{C.~A.} \bibnamefont{Salgado}}, \bibinfo{journal}{Eur. Phys. J. C} \textbf{\bibinfo{volume}{81}}, \bibinfo{pages}{669} (\bibinfo{year}{2021}), \eprint{2006.15044}.

\bibitem[{\citenamefont{Armesto et~al.}(2024)\citenamefont{Armesto, Ferreiro, Escobedo, and L\'opez-Pardo}}]{Armesto:2024zad}
\bibinfo{author}{\bibfnamefont{N.}~\bibnamefont{Armesto}}, \bibinfo{author}{\bibfnamefont{E.~G.} \bibnamefont{Ferreiro}}, \bibinfo{author}{\bibfnamefont{M.~A.} \bibnamefont{Escobedo}}, \bibnamefont{and} \bibinfo{author}{\bibfnamefont{V.}~\bibnamefont{L\'opez-Pardo}}, \bibinfo{journal}{Phys. Lett. B} \textbf{\bibinfo{volume}{854}}, \bibinfo{pages}{138760} (\bibinfo{year}{2024}), \eprint{2401.10125}.

\bibitem[{\citenamefont{Aaij et~al.}(2022{\natexlab{a}})}]{LHCb:2021vvq}
\bibinfo{author}{\bibfnamefont{R.}~\bibnamefont{Aaij}} \bibnamefont{et~al.} (\bibinfo{collaboration}{LHCb}), \bibinfo{journal}{Nature Phys.} \textbf{\bibinfo{volume}{18}}, \bibinfo{pages}{751} (\bibinfo{year}{2022}{\natexlab{a}}), \eprint{2109.01038}.

\bibitem[{\citenamefont{Aaij et~al.}(2022{\natexlab{b}})}]{LHCb:2021auc}
\bibinfo{author}{\bibfnamefont{R.}~\bibnamefont{Aaij}} \bibnamefont{et~al.} (\bibinfo{collaboration}{LHCb}), \bibinfo{journal}{Nature Commun.} \textbf{\bibinfo{volume}{13}}, \bibinfo{pages}{3351} (\bibinfo{year}{2022}{\natexlab{b}}), \eprint{2109.01056}.

\bibitem[{\citenamefont{Fleming et~al.}(2021)\citenamefont{Fleming, Hodges, and Mehen}}]{Fleming:2021wmk}
\bibinfo{author}{\bibfnamefont{S.}~\bibnamefont{Fleming}}, \bibinfo{author}{\bibfnamefont{R.}~\bibnamefont{Hodges}}, \bibnamefont{and} \bibinfo{author}{\bibfnamefont{T.}~\bibnamefont{Mehen}}, \bibinfo{journal}{Phys. Rev. D} \textbf{\bibinfo{volume}{104}}, \bibinfo{pages}{116010} (\bibinfo{year}{2021}), \eprint{2109.02188}.

\bibitem[{\citenamefont{Dai et~al.}(2023)\citenamefont{Dai, Fleming, Hodges, and Mehen}}]{Dai:2023mxm}
\bibinfo{author}{\bibfnamefont{L.}~\bibnamefont{Dai}}, \bibinfo{author}{\bibfnamefont{S.}~\bibnamefont{Fleming}}, \bibinfo{author}{\bibfnamefont{R.}~\bibnamefont{Hodges}}, \bibnamefont{and} \bibinfo{author}{\bibfnamefont{T.}~\bibnamefont{Mehen}}, \bibinfo{journal}{Phys. Rev. D} \textbf{\bibinfo{volume}{107}}, \bibinfo{pages}{076001} (\bibinfo{year}{2023}), \eprint{2301.11950}.

\bibitem[{\citenamefont{Hodges}(2024)}]{Hodges:2024awq}
\bibinfo{author}{\bibfnamefont{R.}~\bibnamefont{Hodges}}, \bibinfo{type}{Phd thesis} (\bibinfo{year}{2024}), \eprint{2404.18907}.

\bibitem[{\citenamefont{Fleming et~al.}(2007)\citenamefont{Fleming, Kusunoki, Mehen, and van Kolck}}]{Fleming:2007rp}
\bibinfo{author}{\bibfnamefont{S.}~\bibnamefont{Fleming}}, \bibinfo{author}{\bibfnamefont{M.}~\bibnamefont{Kusunoki}}, \bibinfo{author}{\bibfnamefont{T.}~\bibnamefont{Mehen}}, \bibnamefont{and} \bibinfo{author}{\bibfnamefont{U.}~\bibnamefont{van Kolck}}, \bibinfo{journal}{Phys. Rev. D} \textbf{\bibinfo{volume}{76}}, \bibinfo{pages}{034006} (\bibinfo{year}{2007}), \eprint{hep-ph/0703168}.

\bibitem[{\citenamefont{Fleming and Mehen}(2008)}]{Fleming:2008yn}
\bibinfo{author}{\bibfnamefont{S.}~\bibnamefont{Fleming}} \bibnamefont{and} \bibinfo{author}{\bibfnamefont{T.}~\bibnamefont{Mehen}}, \bibinfo{journal}{Phys. Rev. D} \textbf{\bibinfo{volume}{78}}, \bibinfo{pages}{094019} (\bibinfo{year}{2008}), \eprint{0807.2674}.

\bibitem[{\citenamefont{Braaten et~al.}(2010)\citenamefont{Braaten, Hammer, and Mehen}}]{Braaten:2010mg}
\bibinfo{author}{\bibfnamefont{E.}~\bibnamefont{Braaten}}, \bibinfo{author}{\bibfnamefont{H.-W.} \bibnamefont{Hammer}}, \bibnamefont{and} \bibinfo{author}{\bibfnamefont{T.}~\bibnamefont{Mehen}}, \bibinfo{journal}{Phys. Rev. D} \textbf{\bibinfo{volume}{82}}, \bibinfo{pages}{034018} (\bibinfo{year}{2010}), \eprint{1005.1688}.

\bibitem[{\citenamefont{Fleming and Mehen}(2012)}]{Fleming:2011xa}
\bibinfo{author}{\bibfnamefont{S.}~\bibnamefont{Fleming}} \bibnamefont{and} \bibinfo{author}{\bibfnamefont{T.}~\bibnamefont{Mehen}}, \bibinfo{journal}{Phys. Rev. D} \textbf{\bibinfo{volume}{85}}, \bibinfo{pages}{014016} (\bibinfo{year}{2012}), \eprint{1110.0265}.

\bibitem[{\citenamefont{Mehen}(2015)}]{Mehen:2015efa}
\bibinfo{author}{\bibfnamefont{T.}~\bibnamefont{Mehen}}, \bibinfo{journal}{Phys. Rev. D} \textbf{\bibinfo{volume}{92}}, \bibinfo{pages}{034019} (\bibinfo{year}{2015}), \eprint{1503.02719}.

\bibitem[{\citenamefont{Braaten et~al.}(2020)\citenamefont{Braaten, He, Ingles, and Jiang}}]{Braaten:2020iye}
\bibinfo{author}{\bibfnamefont{E.}~\bibnamefont{Braaten}}, \bibinfo{author}{\bibfnamefont{L.-P.} \bibnamefont{He}}, \bibinfo{author}{\bibfnamefont{K.}~\bibnamefont{Ingles}}, \bibnamefont{and} \bibinfo{author}{\bibfnamefont{J.}~\bibnamefont{Jiang}}, \bibinfo{journal}{Phys. Rev. D} \textbf{\bibinfo{volume}{101}}, \bibinfo{pages}{096020} (\bibinfo{year}{2020}), \eprint{2004.12841}.

\bibitem[{\citenamefont{Aaij et~al.}(2020{\natexlab{b}})}]{LHCb:2020bwg}
\bibinfo{author}{\bibfnamefont{R.}~\bibnamefont{Aaij}} \bibnamefont{et~al.} (\bibinfo{collaboration}{LHCb}), \bibinfo{journal}{Sci. Bull.} \textbf{\bibinfo{volume}{65}}, \bibinfo{pages}{1983} (\bibinfo{year}{2020}{\natexlab{b}}), \eprint{2006.16957}.

\bibitem[{\citenamefont{Chen et~al.}(2023)\citenamefont{Chen, Chen, Liu, Liu, and Zhu}}]{Chen:2022asf}
\bibinfo{author}{\bibfnamefont{H.-X.} \bibnamefont{Chen}}, \bibinfo{author}{\bibfnamefont{W.}~\bibnamefont{Chen}}, \bibinfo{author}{\bibfnamefont{X.}~\bibnamefont{Liu}}, \bibinfo{author}{\bibfnamefont{Y.-R.} \bibnamefont{Liu}}, \bibnamefont{and} \bibinfo{author}{\bibfnamefont{S.-L.} \bibnamefont{Zhu}}, \bibinfo{journal}{Rept. Prog. Phys.} \textbf{\bibinfo{volume}{86}}, \bibinfo{pages}{026201} (\bibinfo{year}{2023}), \eprint{2204.02649}.

\bibitem[{\citenamefont{Pineda}(2012)}]{Pineda:2011dg}
\bibinfo{author}{\bibfnamefont{A.}~\bibnamefont{Pineda}}, \bibinfo{journal}{Prog. Part. Nucl. Phys.} \textbf{\bibinfo{volume}{67}}, \bibinfo{pages}{735} (\bibinfo{year}{2012}), \eprint{1111.0165}.

\bibitem[{\citenamefont{Maiani et~al.}(2019)\citenamefont{Maiani, Polosa, and Riquer}}]{Maiani:2019cwl}
\bibinfo{author}{\bibfnamefont{L.}~\bibnamefont{Maiani}}, \bibinfo{author}{\bibfnamefont{A.~D.} \bibnamefont{Polosa}}, \bibnamefont{and} \bibinfo{author}{\bibfnamefont{V.}~\bibnamefont{Riquer}}, \bibinfo{journal}{Phys. Rev. D} \textbf{\bibinfo{volume}{100}}, \bibinfo{pages}{014002} (\bibinfo{year}{2019}), \eprint{1903.10253}.

\bibitem[{\citenamefont{Maciu\l{}a et~al.}(2021)\citenamefont{Maciu\l{}a, Sch\"afer, and Szczurek}}]{Maciula:2020wri}
\bibinfo{author}{\bibfnamefont{R.}~\bibnamefont{Maciu\l{}a}}, \bibinfo{author}{\bibfnamefont{W.}~\bibnamefont{Sch\"afer}}, \bibnamefont{and} \bibinfo{author}{\bibfnamefont{A.}~\bibnamefont{Szczurek}}, \bibinfo{journal}{Phys. Lett. B} \textbf{\bibinfo{volume}{812}}, \bibinfo{pages}{136010} (\bibinfo{year}{2021}), \eprint{2009.02100}.

\bibitem[{\citenamefont{Berezhnoy et~al.}(2011)\citenamefont{Berezhnoy, Likhoded, Luchinsky, and Novoselov}}]{Berezhnoy:2011xy}
\bibinfo{author}{\bibfnamefont{A.~V.} \bibnamefont{Berezhnoy}}, \bibinfo{author}{\bibfnamefont{A.~K.} \bibnamefont{Likhoded}}, \bibinfo{author}{\bibfnamefont{A.~V.} \bibnamefont{Luchinsky}}, \bibnamefont{and} \bibinfo{author}{\bibfnamefont{A.~A.} \bibnamefont{Novoselov}}, \bibinfo{journal}{Phys. Rev. D} \textbf{\bibinfo{volume}{84}}, \bibinfo{pages}{094023} (\bibinfo{year}{2011}), \eprint{1101.5881}.

\bibitem[{\citenamefont{Karliner et~al.}(2017)\citenamefont{Karliner, Nussinov, and Rosner}}]{Karliner:2016zzc}
\bibinfo{author}{\bibfnamefont{M.}~\bibnamefont{Karliner}}, \bibinfo{author}{\bibfnamefont{S.}~\bibnamefont{Nussinov}}, \bibnamefont{and} \bibinfo{author}{\bibfnamefont{J.~L.} \bibnamefont{Rosner}}, \bibinfo{journal}{Phys. Rev. D} \textbf{\bibinfo{volume}{95}}, \bibinfo{pages}{034011} (\bibinfo{year}{2017}), \eprint{1611.00348}.

\bibitem[{\citenamefont{Becchi et~al.}(2020)\citenamefont{Becchi, Giachino, Maiani, and Santopinto}}]{Becchi:2020mjz}
\bibinfo{author}{\bibfnamefont{C.}~\bibnamefont{Becchi}}, \bibinfo{author}{\bibfnamefont{A.}~\bibnamefont{Giachino}}, \bibinfo{author}{\bibfnamefont{L.}~\bibnamefont{Maiani}}, \bibnamefont{and} \bibinfo{author}{\bibfnamefont{E.}~\bibnamefont{Santopinto}}, \bibinfo{journal}{Phys. Lett. B} \textbf{\bibinfo{volume}{806}}, \bibinfo{pages}{135495} (\bibinfo{year}{2020}), \eprint{2002.11077}.

\bibitem[{\citenamefont{Carvalho et~al.}(2016)\citenamefont{Carvalho, Cazaroto, Gon\c{c}alves, and Navarra}}]{Carvalho:2015nqf}
\bibinfo{author}{\bibfnamefont{F.}~\bibnamefont{Carvalho}}, \bibinfo{author}{\bibfnamefont{E.~R.} \bibnamefont{Cazaroto}}, \bibinfo{author}{\bibfnamefont{V.~P.} \bibnamefont{Gon\c{c}alves}}, \bibnamefont{and} \bibinfo{author}{\bibfnamefont{F.~S.} \bibnamefont{Navarra}}, \bibinfo{journal}{Phys. Rev. D} \textbf{\bibinfo{volume}{93}}, \bibinfo{pages}{034004} (\bibinfo{year}{2016}), \eprint{1511.05209}.

\bibitem[{\citenamefont{Abreu et~al.}(2024)\citenamefont{Abreu, Carvalho, Oliveira, and Gon\c{c}alves}}]{Abreu:2023wwg}
\bibinfo{author}{\bibfnamefont{L.~M.} \bibnamefont{Abreu}}, \bibinfo{author}{\bibfnamefont{F.}~\bibnamefont{Carvalho}}, \bibinfo{author}{\bibfnamefont{J.~V.~C.} \bibnamefont{Oliveira}}, \bibnamefont{and} \bibinfo{author}{\bibfnamefont{V.~P.} \bibnamefont{Gon\c{c}alves}}, \bibinfo{journal}{Eur. Phys. J. C} \textbf{\bibinfo{volume}{84}}, \bibinfo{pages}{470} (\bibinfo{year}{2024}), \eprint{2306.12731}.

\bibitem[{\citenamefont{Cisek et~al.}(2022)\citenamefont{Cisek, Sch\"afer, and Szczurek}}]{Cisek:2022uqx}
\bibinfo{author}{\bibfnamefont{A.}~\bibnamefont{Cisek}}, \bibinfo{author}{\bibfnamefont{W.}~\bibnamefont{Sch\"afer}}, \bibnamefont{and} \bibinfo{author}{\bibfnamefont{A.}~\bibnamefont{Szczurek}}, \bibinfo{journal}{Eur. Phys. J. C} \textbf{\bibinfo{volume}{82}}, \bibinfo{pages}{1062} (\bibinfo{year}{2022}), \eprint{2203.07827}.

\bibitem[{\citenamefont{Feng et~al.}(2021)\citenamefont{Feng, Huang, Jia, Sang, and Zhang}}]{Feng:2020qee}
\bibinfo{author}{\bibfnamefont{F.}~\bibnamefont{Feng}}, \bibinfo{author}{\bibfnamefont{Y.}~\bibnamefont{Huang}}, \bibinfo{author}{\bibfnamefont{Y.}~\bibnamefont{Jia}}, \bibinfo{author}{\bibfnamefont{W.-L.} \bibnamefont{Sang}}, \bibnamefont{and} \bibinfo{author}{\bibfnamefont{J.-Y.} \bibnamefont{Zhang}}, \bibinfo{journal}{Phys. Lett. B} \textbf{\bibinfo{volume}{818}}, \bibinfo{pages}{136368} (\bibinfo{year}{2021}), \eprint{2011.03039}.

\bibitem[{\citenamefont{Feng et~al.}(2024)\citenamefont{Feng, Huang, Jia, Sang, Yang, and Zhang}}]{Feng:2023ghc}
\bibinfo{author}{\bibfnamefont{F.}~\bibnamefont{Feng}}, \bibinfo{author}{\bibfnamefont{Y.}~\bibnamefont{Huang}}, \bibinfo{author}{\bibfnamefont{Y.}~\bibnamefont{Jia}}, \bibinfo{author}{\bibfnamefont{W.-L.} \bibnamefont{Sang}}, \bibinfo{author}{\bibfnamefont{D.-S.} \bibnamefont{Yang}}, \bibnamefont{and} \bibinfo{author}{\bibfnamefont{J.-Y.} \bibnamefont{Zhang}}, \bibinfo{journal}{Phys. Rev. D} \textbf{\bibinfo{volume}{110}}, \bibinfo{pages}{054007} (\bibinfo{year}{2024}), \eprint{2311.08292}.

\bibitem[{\citenamefont{Bondar et~al.}(2012)}]{Belle:2011aa}
\bibinfo{author}{\bibfnamefont{A.}~\bibnamefont{Bondar}} \bibnamefont{et~al.} (\bibinfo{collaboration}{Belle}), \bibinfo{journal}{Phys. Rev. Lett.} \textbf{\bibinfo{volume}{108}}, \bibinfo{pages}{122001} (\bibinfo{year}{2012}), \eprint{1110.2251}.

\bibitem[{\citenamefont{Bland et~al.}(2019)}]{ANDY:2019bfn}
\bibinfo{author}{\bibfnamefont{L.~C.} \bibnamefont{Bland}} \bibnamefont{et~al.} (\bibinfo{collaboration}{ANDY}) (\bibinfo{year}{2019}), \eprint{1909.03124}.

\bibitem[{\citenamefont{Vogt and Angerami}(2021)}]{Vogt:2021lei}
\bibinfo{author}{\bibfnamefont{R.}~\bibnamefont{Vogt}} \bibnamefont{and} \bibinfo{author}{\bibfnamefont{A.}~\bibnamefont{Angerami}}, \bibinfo{journal}{Phys. Rev. D} \textbf{\bibinfo{volume}{104}}, \bibinfo{pages}{094025} (\bibinfo{year}{2021}), \eprint{2109.00706}.

\bibitem[{\citenamefont{Francis et~al.}(2019)\citenamefont{Francis, Hudspith, Lewis, and Maltman}}]{Francis:2018jyb}
\bibinfo{author}{\bibfnamefont{A.}~\bibnamefont{Francis}}, \bibinfo{author}{\bibfnamefont{R.~J.} \bibnamefont{Hudspith}}, \bibinfo{author}{\bibfnamefont{R.}~\bibnamefont{Lewis}}, \bibnamefont{and} \bibinfo{author}{\bibfnamefont{K.}~\bibnamefont{Maltman}}, \emph{\bibinfo{title}{{Evidence for charm-bottom tetraquarks and the mass dependence of heavy-light tetraquark states from lattice QCD}}} (\bibinfo{year}{2019}), \eprint{1810.10550}.

\bibitem[{\citenamefont{Padmanath et~al.}(2024)\citenamefont{Padmanath, Radhakrishnan, and Mathur}}]{Padmanath:2023rdu}
\bibinfo{author}{\bibfnamefont{M.}~\bibnamefont{Padmanath}}, \bibinfo{author}{\bibfnamefont{A.}~\bibnamefont{Radhakrishnan}}, \bibnamefont{and} \bibinfo{author}{\bibfnamefont{N.}~\bibnamefont{Mathur}}, \bibinfo{journal}{Phys. Rev. Lett.} \textbf{\bibinfo{volume}{132}}, \bibinfo{pages}{201902} (\bibinfo{year}{2024}), \eprint{2307.14128}.

\bibitem[{\citenamefont{Bicudo et~al.}(2015)\citenamefont{Bicudo, Cichy, Peters, Wagenbach, and Wagner}}]{Bicudo:2015vta}
\bibinfo{author}{\bibfnamefont{P.}~\bibnamefont{Bicudo}}, \bibinfo{author}{\bibfnamefont{K.}~\bibnamefont{Cichy}}, \bibinfo{author}{\bibfnamefont{A.}~\bibnamefont{Peters}}, \bibinfo{author}{\bibfnamefont{B.}~\bibnamefont{Wagenbach}}, \bibnamefont{and} \bibinfo{author}{\bibfnamefont{M.}~\bibnamefont{Wagner}}, \bibinfo{journal}{Phys. Rev. D} \textbf{\bibinfo{volume}{92}}, \bibinfo{pages}{014507} (\bibinfo{year}{2015}), \eprint{1505.00613}.

\bibitem[{\citenamefont{Leskovec et~al.}(2019)\citenamefont{Leskovec, Meinel, Pflaumer, and Wagner}}]{Leskovec:2019ioa}
\bibinfo{author}{\bibfnamefont{L.}~\bibnamefont{Leskovec}}, \bibinfo{author}{\bibfnamefont{S.}~\bibnamefont{Meinel}}, \bibinfo{author}{\bibfnamefont{M.}~\bibnamefont{Pflaumer}}, \bibnamefont{and} \bibinfo{author}{\bibfnamefont{M.}~\bibnamefont{Wagner}}, \bibinfo{journal}{Phys. Rev. D} \textbf{\bibinfo{volume}{100}}, \bibinfo{pages}{014503} (\bibinfo{year}{2019}), \eprint{1904.04197}.

\bibitem[{\citenamefont{Alexandrou et~al.}(2024)\citenamefont{Alexandrou, Finkenrath, Leontiou, Meinel, Pflaumer, and Wagner}}]{Alexandrou:2024iwi}
\bibinfo{author}{\bibfnamefont{C.}~\bibnamefont{Alexandrou}}, \bibinfo{author}{\bibfnamefont{J.}~\bibnamefont{Finkenrath}}, \bibinfo{author}{\bibfnamefont{T.}~\bibnamefont{Leontiou}}, \bibinfo{author}{\bibfnamefont{S.}~\bibnamefont{Meinel}}, \bibinfo{author}{\bibfnamefont{M.}~\bibnamefont{Pflaumer}}, \bibnamefont{and} \bibinfo{author}{\bibfnamefont{M.}~\bibnamefont{Wagner}}, \bibinfo{journal}{Phys. Rev. D} \textbf{\bibinfo{volume}{110}}, \bibinfo{pages}{054510} (\bibinfo{year}{2024}), \eprint{2404.03588}.

\bibitem[{\citenamefont{Chatrchyan et~al.}(2013)}]{CMS:2013fpt}
\bibinfo{author}{\bibfnamefont{S.}~\bibnamefont{Chatrchyan}} \bibnamefont{et~al.} (\bibinfo{collaboration}{CMS}), \bibinfo{journal}{JHEP} \textbf{\bibinfo{volume}{04}}, \bibinfo{pages}{154} (\bibinfo{year}{2013}), \eprint{1302.3968}.

\bibitem[{\citenamefont{Aaboud et~al.}(2017)}]{ATLAS:2016kwu}
\bibinfo{author}{\bibfnamefont{M.}~\bibnamefont{Aaboud}} \bibnamefont{et~al.} (\bibinfo{collaboration}{ATLAS}), \bibinfo{journal}{JHEP} \textbf{\bibinfo{volume}{01}}, \bibinfo{pages}{117} (\bibinfo{year}{2017}), \eprint{1610.09303}.

\bibitem[{\citenamefont{Aaij et~al.}(2022{\natexlab{c}})}]{LHCb:2021ten}
\bibinfo{author}{\bibfnamefont{R.}~\bibnamefont{Aaij}} \bibnamefont{et~al.} (\bibinfo{collaboration}{LHCb}), \bibinfo{journal}{JHEP} \textbf{\bibinfo{volume}{01}}, \bibinfo{pages}{131} (\bibinfo{year}{2022}{\natexlab{c}}), \eprint{2109.07360}.

\bibitem[{\citenamefont{Suzuki}(1977)}]{Suzuki:1977km}
\bibinfo{author}{\bibfnamefont{M.}~\bibnamefont{Suzuki}}, \bibinfo{journal}{Phys. Lett. B} \textbf{\bibinfo{volume}{71}}, \bibinfo{pages}{139} (\bibinfo{year}{1977}).

\bibitem[{\citenamefont{Suzuki}(1986)}]{Suzuki:1985up}
\bibinfo{author}{\bibfnamefont{M.}~\bibnamefont{Suzuki}}, \bibinfo{journal}{Phys. Rev. D} \textbf{\bibinfo{volume}{33}}, \bibinfo{pages}{676} (\bibinfo{year}{1986}).

\bibitem[{\citenamefont{Amiri and Ji}(1987)}]{Amiri:1986zv}
\bibinfo{author}{\bibfnamefont{F.}~\bibnamefont{Amiri}} \bibnamefont{and} \bibinfo{author}{\bibfnamefont{C.-R.} \bibnamefont{Ji}}, \bibinfo{journal}{Phys. Lett. B} \textbf{\bibinfo{volume}{195}}, \bibinfo{pages}{593} (\bibinfo{year}{1987}).

\bibitem[{\citenamefont{{\relax Moosavi}~Nejad and Amiri}(2022)}]{Nejad:2021mmp}
\bibinfo{author}{\bibfnamefont{S.~M.} \bibnamefont{{\relax Moosavi}~Nejad}} \bibnamefont{and} \bibinfo{author}{\bibfnamefont{N.}~\bibnamefont{Amiri}}, \bibinfo{journal}{Phys. Rev. D} \textbf{\bibinfo{volume}{105}}, \bibinfo{pages}{034001} (\bibinfo{year}{2022}), \eprint{2110.15251}.

\bibitem[{\citenamefont{Celiberto and Papa}(2024)}]{Celiberto:2023rzw}
\bibinfo{author}{\bibfnamefont{F.~G.} \bibnamefont{Celiberto}} \bibnamefont{and} \bibinfo{author}{\bibfnamefont{A.}~\bibnamefont{Papa}}, \bibinfo{journal}{Phys. Lett. B} \textbf{\bibinfo{volume}{848}}, \bibinfo{pages}{138406} (\bibinfo{year}{2024}), \eprint{2308.00809}.

\bibitem[{\citenamefont{Celiberto}(2024{\natexlab{a}})}]{Celiberto:2024mrq}
\bibinfo{author}{\bibfnamefont{F.~G.} \bibnamefont{Celiberto}}, \bibinfo{journal}{Symmetry} \textbf{\bibinfo{volume}{16}}, \bibinfo{pages}{550} (\bibinfo{year}{2024}{\natexlab{a}}), \eprint{2403.15639}.

\bibitem[{\citenamefont{Caswell and Lepage}(1986)}]{Caswell:1985ui}
\bibinfo{author}{\bibfnamefont{W.~E.} \bibnamefont{Caswell}} \bibnamefont{and} \bibinfo{author}{\bibfnamefont{G.~P.} \bibnamefont{Lepage}}, \bibinfo{journal}{Phys. Lett. B} \textbf{\bibinfo{volume}{167}}, \bibinfo{pages}{437} (\bibinfo{year}{1986}).

\bibitem[{\citenamefont{Thacker and Lepage}(1991)}]{Thacker:1990bm}
\bibinfo{author}{\bibfnamefont{B.~A.} \bibnamefont{Thacker}} \bibnamefont{and} \bibinfo{author}{\bibfnamefont{G.~P.} \bibnamefont{Lepage}}, \bibinfo{journal}{Phys. Rev. D} \textbf{\bibinfo{volume}{43}}, \bibinfo{pages}{196} (\bibinfo{year}{1991}).

\bibitem[{\citenamefont{Bodwin et~al.}(1995)\citenamefont{Bodwin, Braaten, and Lepage}}]{Bodwin:1994jh}
\bibinfo{author}{\bibfnamefont{G.~T.} \bibnamefont{Bodwin}}, \bibinfo{author}{\bibfnamefont{E.}~\bibnamefont{Braaten}}, \bibnamefont{and} \bibinfo{author}{\bibfnamefont{G.~P.} \bibnamefont{Lepage}}, \bibinfo{journal}{Phys. Rev. D} \textbf{\bibinfo{volume}{51}}, \bibinfo{pages}{1125} (\bibinfo{year}{1995}), \bibinfo{note}{[Erratum: Phys.Rev.D 55, 5853 (1997)]}, \eprint{hep-ph/9407339}.

\bibitem[{\citenamefont{Cho and Leibovich}(1996{\natexlab{a}})}]{Cho:1995vh}
\bibinfo{author}{\bibfnamefont{P.~L.} \bibnamefont{Cho}} \bibnamefont{and} \bibinfo{author}{\bibfnamefont{A.~K.} \bibnamefont{Leibovich}}, \bibinfo{journal}{Phys. Rev. D} \textbf{\bibinfo{volume}{53}}, \bibinfo{pages}{150} (\bibinfo{year}{1996}{\natexlab{a}}), \eprint{hep-ph/9505329}.

\bibitem[{\citenamefont{Cho and Leibovich}(1996{\natexlab{b}})}]{Cho:1995ce}
\bibinfo{author}{\bibfnamefont{P.~L.} \bibnamefont{Cho}} \bibnamefont{and} \bibinfo{author}{\bibfnamefont{A.~K.} \bibnamefont{Leibovich}}, \bibinfo{journal}{Phys. Rev. D} \textbf{\bibinfo{volume}{53}}, \bibinfo{pages}{6203} (\bibinfo{year}{1996}{\natexlab{b}}), \eprint{hep-ph/9511315}.

\bibitem[{\citenamefont{Leibovich}(1997)}]{Leibovich:1996pa}
\bibinfo{author}{\bibfnamefont{A.~K.} \bibnamefont{Leibovich}}, \bibinfo{journal}{Phys. Rev. D} \textbf{\bibinfo{volume}{56}}, \bibinfo{pages}{4412} (\bibinfo{year}{1997}), \eprint{hep-ph/9610381}.

\bibitem[{\citenamefont{Bodwin et~al.}(2005)\citenamefont{Bodwin, Braaten, and Lee}}]{Bodwin:2005hm}
\bibinfo{author}{\bibfnamefont{G.~T.} \bibnamefont{Bodwin}}, \bibinfo{author}{\bibfnamefont{E.}~\bibnamefont{Braaten}}, \bibnamefont{and} \bibinfo{author}{\bibfnamefont{J.}~\bibnamefont{Lee}}, \bibinfo{journal}{Phys. Rev. D} \textbf{\bibinfo{volume}{72}}, \bibinfo{pages}{014004} (\bibinfo{year}{2005}), \eprint{hep-ph/0504014}.

\bibitem[{\citenamefont{Feng et~al.}(2022{\natexlab{a}})\citenamefont{Feng, Huang, Jia, Sang, Xiong, and Zhang}}]{Feng:2020riv}
\bibinfo{author}{\bibfnamefont{F.}~\bibnamefont{Feng}}, \bibinfo{author}{\bibfnamefont{Y.}~\bibnamefont{Huang}}, \bibinfo{author}{\bibfnamefont{Y.}~\bibnamefont{Jia}}, \bibinfo{author}{\bibfnamefont{W.-L.} \bibnamefont{Sang}}, \bibinfo{author}{\bibfnamefont{X.}~\bibnamefont{Xiong}}, \bibnamefont{and} \bibinfo{author}{\bibfnamefont{J.-Y.} \bibnamefont{Zhang}}, \bibinfo{journal}{Phys. Rev. D} \textbf{\bibinfo{volume}{106}}, \bibinfo{pages}{114029} (\bibinfo{year}{2022}{\natexlab{a}}), \eprint{2009.08450}.

\bibitem[{\citenamefont{Bai et~al.}(2024{\natexlab{a}})\citenamefont{Bai, Feng, Gan, Huang, Sang, and Zhang}}]{Bai:2024ezn}
\bibinfo{author}{\bibfnamefont{X.-W.} \bibnamefont{Bai}}, \bibinfo{author}{\bibfnamefont{F.}~\bibnamefont{Feng}}, \bibinfo{author}{\bibfnamefont{C.-M.} \bibnamefont{Gan}}, \bibinfo{author}{\bibfnamefont{Y.}~\bibnamefont{Huang}}, \bibinfo{author}{\bibfnamefont{W.-L.} \bibnamefont{Sang}}, \bibnamefont{and} \bibinfo{author}{\bibfnamefont{H.-F.} \bibnamefont{Zhang}}, \bibinfo{journal}{JHEP} \textbf{\bibinfo{volume}{09}}, \bibinfo{pages}{002} (\bibinfo{year}{2024}{\natexlab{a}}), \eprint{2404.13889}.

\bibitem[{\citenamefont{Celiberto et~al.}(2024{\natexlab{a}})\citenamefont{Celiberto, Gatto, and Papa}}]{Celiberto:2024mab}
\bibinfo{author}{\bibfnamefont{F.~G.} \bibnamefont{Celiberto}}, \bibinfo{author}{\bibfnamefont{G.}~\bibnamefont{Gatto}}, \bibnamefont{and} \bibinfo{author}{\bibfnamefont{A.}~\bibnamefont{Papa}}, \bibinfo{journal}{Eur. Phys. J. C} \textbf{\bibinfo{volume}{84}}, \bibinfo{pages}{1071} (\bibinfo{year}{2024}{\natexlab{a}}), \eprint{2405.14773}.

\bibitem[{\citenamefont{Gribov and Lipatov}(1972{\natexlab{a}})}]{Gribov:1972ri}
\bibinfo{author}{\bibfnamefont{V.}~\bibnamefont{Gribov}} \bibnamefont{and} \bibinfo{author}{\bibfnamefont{L.}~\bibnamefont{Lipatov}}, \bibinfo{journal}{Sov. J. Nucl. Phys.} \textbf{\bibinfo{volume}{15}}, \bibinfo{pages}{438} (\bibinfo{year}{1972}{\natexlab{a}}).

\bibitem[{\citenamefont{Gribov and Lipatov}(1972{\natexlab{b}})}]{Gribov:1972rt}
\bibinfo{author}{\bibfnamefont{V.~N.} \bibnamefont{Gribov}} \bibnamefont{and} \bibinfo{author}{\bibfnamefont{L.~N.} \bibnamefont{Lipatov}}, \bibinfo{journal}{Sov. J. Nucl. Phys.} \textbf{\bibinfo{volume}{15}}, \bibinfo{pages}{675} (\bibinfo{year}{1972}{\natexlab{b}}).

\bibitem[{\citenamefont{Lipatov}(1974)}]{Lipatov:1974qm}
\bibinfo{author}{\bibfnamefont{L.~N.} \bibnamefont{Lipatov}}, \bibinfo{journal}{Yad. Fiz.} \textbf{\bibinfo{volume}{20}}, \bibinfo{pages}{181} (\bibinfo{year}{1974}).

\bibitem[{\citenamefont{Altarelli and Parisi}(1977)}]{Altarelli:1977zs}
\bibinfo{author}{\bibfnamefont{G.}~\bibnamefont{Altarelli}} \bibnamefont{and} \bibinfo{author}{\bibfnamefont{G.}~\bibnamefont{Parisi}}, \bibinfo{journal}{Nucl. Phys. B} \textbf{\bibinfo{volume}{126}}, \bibinfo{pages}{298} (\bibinfo{year}{1977}).

\bibitem[{\citenamefont{Dokshitzer}(1977)}]{Dokshitzer:1977sg}
\bibinfo{author}{\bibfnamefont{Y.~L.} \bibnamefont{Dokshitzer}}, \bibinfo{journal}{Sov. Phys. JETP} \textbf{\bibinfo{volume}{46}}, \bibinfo{pages}{641} (\bibinfo{year}{1977}).

\bibitem[{\citenamefont{Mele and Nason}(1991)}]{Mele:1990cw}
\bibinfo{author}{\bibfnamefont{B.}~\bibnamefont{Mele}} \bibnamefont{and} \bibinfo{author}{\bibfnamefont{P.}~\bibnamefont{Nason}}, \bibinfo{journal}{Nucl. Phys. B} \textbf{\bibinfo{volume}{361}}, \bibinfo{pages}{626} (\bibinfo{year}{1991}), \bibinfo{note}{[Erratum: Nucl.Phys.B 921, 841--842 (2017)]}.

\bibitem[{\citenamefont{Cacciari and Greco}(1994{\natexlab{a}})}]{Cacciari:1993mq}
\bibinfo{author}{\bibfnamefont{M.}~\bibnamefont{Cacciari}} \bibnamefont{and} \bibinfo{author}{\bibfnamefont{M.}~\bibnamefont{Greco}}, \bibinfo{journal}{Nucl. Phys. B} \textbf{\bibinfo{volume}{421}}, \bibinfo{pages}{530} (\bibinfo{year}{1994}{\natexlab{a}}), \eprint{hep-ph/9311260}.

\bibitem[{\citenamefont{Buza et~al.}(1998)\citenamefont{Buza, Matiounine, Smith, and van Neerven}}]{Buza:1996wv}
\bibinfo{author}{\bibfnamefont{M.}~\bibnamefont{Buza}}, \bibinfo{author}{\bibfnamefont{Y.}~\bibnamefont{Matiounine}}, \bibinfo{author}{\bibfnamefont{J.}~\bibnamefont{Smith}}, \bibnamefont{and} \bibinfo{author}{\bibfnamefont{W.~L.} \bibnamefont{van Neerven}}, \bibinfo{journal}{Eur. Phys. J. C} \textbf{\bibinfo{volume}{1}}, \bibinfo{pages}{301} (\bibinfo{year}{1998}), \eprint{hep-ph/9612398}.

\bibitem[{\citenamefont{Celiberto}(2024{\natexlab{b}})}]{Celiberto:2024mex}
\bibinfo{author}{\bibfnamefont{F.~G.} \bibnamefont{Celiberto}}, in \emph{\bibinfo{booktitle}{{58th Rencontres de Moriond on QCD and High Energy Interactions}}} (\bibinfo{year}{2024}{\natexlab{b}}), \eprint{2405.08221}.

\bibitem[{\citenamefont{Celiberto}(2025)}]{Celiberto:2024bxu}
\bibinfo{author}{\bibfnamefont{F.~G.} \bibnamefont{Celiberto}}, \bibinfo{journal}{PoS} \textbf{\bibinfo{volume}{DIS2024}}, \bibinfo{pages}{168} (\bibinfo{year}{2025}), \eprint{2406.10779}.

\bibitem[{\citenamefont{Celiberto}(2021{\natexlab{a}})}]{Celiberto:2020wpk}
\bibinfo{author}{\bibfnamefont{F.~G.} \bibnamefont{Celiberto}}, \bibinfo{journal}{Eur. Phys. J. C} \textbf{\bibinfo{volume}{81}}, \bibinfo{pages}{691} (\bibinfo{year}{2021}{\natexlab{a}}), \eprint{2008.07378}.

\bibitem[{\citenamefont{Celiberto}(2022{\natexlab{a}})}]{Celiberto:2022rfj}
\bibinfo{author}{\bibfnamefont{F.~G.} \bibnamefont{Celiberto}}, \bibinfo{journal}{Phys. Rev. D} \textbf{\bibinfo{volume}{105}}, \bibinfo{pages}{114008} (\bibinfo{year}{2022}{\natexlab{a}}), \eprint{2204.06497}.

\bibitem[{\citenamefont{Celiberto}(2023{\natexlab{a}})}]{Celiberto:2023fzz}
\bibinfo{author}{\bibfnamefont{F.~G.} \bibnamefont{Celiberto}}, \bibinfo{journal}{Universe} \textbf{\bibinfo{volume}{9}}, \bibinfo{pages}{324} (\bibinfo{year}{2023}{\natexlab{a}}), \eprint{2305.14295}.

\bibitem[{\citenamefont{Celiberto}(2024{\natexlab{c}})}]{Celiberto:2024swu}
\bibinfo{author}{\bibfnamefont{F.~G.} \bibnamefont{Celiberto}}, \bibinfo{journal}{Particles} \textbf{\bibinfo{volume}{7}}, \bibinfo{pages}{502} (\bibinfo{year}{2024}{\natexlab{c}}), \eprint{2405.09526}.

\bibitem[{\citenamefont{Celiberto and Fucilla}(2022{\natexlab{a}})}]{Celiberto:2022dyf}
\bibinfo{author}{\bibfnamefont{F.~G.} \bibnamefont{Celiberto}} \bibnamefont{and} \bibinfo{author}{\bibfnamefont{M.}~\bibnamefont{Fucilla}}, \bibinfo{journal}{Eur. Phys. J. C} \textbf{\bibinfo{volume}{82}}, \bibinfo{pages}{929} (\bibinfo{year}{2022}{\natexlab{a}}), \eprint{2202.12227}.

\bibitem[{\citenamefont{Celiberto}(2022{\natexlab{b}})}]{Celiberto:2022keu}
\bibinfo{author}{\bibfnamefont{F.~G.} \bibnamefont{Celiberto}}, \bibinfo{journal}{Phys. Lett. B} \textbf{\bibinfo{volume}{835}}, \bibinfo{pages}{137554} (\bibinfo{year}{2022}{\natexlab{b}}), \eprint{2206.09413}.

\bibitem[{\citenamefont{Celiberto}(2024{\natexlab{d}})}]{Celiberto:2024omj}
\bibinfo{author}{\bibfnamefont{F.~G.} \bibnamefont{Celiberto}}, \bibinfo{journal}{Eur. Phys. J. C} \textbf{\bibinfo{volume}{84}}, \bibinfo{pages}{384} (\bibinfo{year}{2024}{\natexlab{d}}), \eprint{2401.01410}.

\bibitem[{\citenamefont{Cacciari et~al.}(1997)\citenamefont{Cacciari, Greco, Rolli, and Tanzini}}]{Cacciari:1996wr}
\bibinfo{author}{\bibfnamefont{M.}~\bibnamefont{Cacciari}}, \bibinfo{author}{\bibfnamefont{M.}~\bibnamefont{Greco}}, \bibinfo{author}{\bibfnamefont{S.}~\bibnamefont{Rolli}}, \bibnamefont{and} \bibinfo{author}{\bibfnamefont{A.}~\bibnamefont{Tanzini}}, \bibinfo{journal}{Phys. Rev. D} \textbf{\bibinfo{volume}{55}}, \bibinfo{pages}{2736} (\bibinfo{year}{1997}), \eprint{hep-ph/9608213}.

\bibitem[{\citenamefont{Kniehl et~al.}(2005)\citenamefont{Kniehl, Kramer, Schienbein, and Spiesberger}}]{Kniehl:2005mk}
\bibinfo{author}{\bibfnamefont{B.~A.} \bibnamefont{Kniehl}}, \bibinfo{author}{\bibfnamefont{G.}~\bibnamefont{Kramer}}, \bibinfo{author}{\bibfnamefont{I.}~\bibnamefont{Schienbein}}, \bibnamefont{and} \bibinfo{author}{\bibfnamefont{H.}~\bibnamefont{Spiesberger}}, \bibinfo{journal}{Eur. Phys. J. C} \textbf{\bibinfo{volume}{41}}, \bibinfo{pages}{199} (\bibinfo{year}{2005}), \eprint{hep-ph/0502194}.

\bibitem[{\citenamefont{Helenius and Paukkunen}(2018)}]{Helenius:2018uul}
\bibinfo{author}{\bibfnamefont{I.}~\bibnamefont{Helenius}} \bibnamefont{and} \bibinfo{author}{\bibfnamefont{H.}~\bibnamefont{Paukkunen}}, \bibinfo{journal}{JHEP} \textbf{\bibinfo{volume}{05}}, \bibinfo{pages}{196} (\bibinfo{year}{2018}), \eprint{1804.03557}.

\bibitem[{\citenamefont{Helenius and Paukkunen}(2023)}]{Helenius:2023wkn}
\bibinfo{author}{\bibfnamefont{I.}~\bibnamefont{Helenius}} \bibnamefont{and} \bibinfo{author}{\bibfnamefont{H.}~\bibnamefont{Paukkunen}}, \bibinfo{journal}{JHEP} \textbf{\bibinfo{volume}{07}}, \bibinfo{pages}{054} (\bibinfo{year}{2023}), \eprint{2303.17864}.

\bibitem[{\citenamefont{Mele and Nason}(1990)}]{Mele:1990yq}
\bibinfo{author}{\bibfnamefont{B.}~\bibnamefont{Mele}} \bibnamefont{and} \bibinfo{author}{\bibfnamefont{P.}~\bibnamefont{Nason}}, \bibinfo{journal}{Phys. Lett. B} \textbf{\bibinfo{volume}{245}}, \bibinfo{pages}{635} (\bibinfo{year}{1990}).

\bibitem[{\citenamefont{Rijken and van Neerven}(1996)}]{Rijken:1996vr}
\bibinfo{author}{\bibfnamefont{P.~J.} \bibnamefont{Rijken}} \bibnamefont{and} \bibinfo{author}{\bibfnamefont{W.~L.} \bibnamefont{van Neerven}}, \bibinfo{journal}{Phys. Lett. B} \textbf{\bibinfo{volume}{386}}, \bibinfo{pages}{422} (\bibinfo{year}{1996}), \eprint{hep-ph/9604436}.

\bibitem[{\citenamefont{Mitov and Moch}(2006)}]{Mitov:2006wy}
\bibinfo{author}{\bibfnamefont{A.}~\bibnamefont{Mitov}} \bibnamefont{and} \bibinfo{author}{\bibfnamefont{S.-O.} \bibnamefont{Moch}}, \bibinfo{journal}{Nucl. Phys. B} \textbf{\bibinfo{volume}{751}}, \bibinfo{pages}{18} (\bibinfo{year}{2006}), \eprint{hep-ph/0604160}.

\bibitem[{\citenamefont{Blumlein and Ravindran}(2006)}]{Blumlein:2006rr}
\bibinfo{author}{\bibfnamefont{J.}~\bibnamefont{Blumlein}} \bibnamefont{and} \bibinfo{author}{\bibfnamefont{V.}~\bibnamefont{Ravindran}}, \bibinfo{journal}{Nucl. Phys. B} \textbf{\bibinfo{volume}{749}}, \bibinfo{pages}{1} (\bibinfo{year}{2006}), \eprint{hep-ph/0604019}.

\bibitem[{\citenamefont{Melnikov and Mitov}(2004)}]{Melnikov:2004bm}
\bibinfo{author}{\bibfnamefont{K.}~\bibnamefont{Melnikov}} \bibnamefont{and} \bibinfo{author}{\bibfnamefont{A.}~\bibnamefont{Mitov}}, \bibinfo{journal}{Phys. Rev. D} \textbf{\bibinfo{volume}{70}}, \bibinfo{pages}{034027} (\bibinfo{year}{2004}), \eprint{hep-ph/0404143}.

\bibitem[{\citenamefont{Mitov}(2005)}]{Mitov:2004du}
\bibinfo{author}{\bibfnamefont{A.}~\bibnamefont{Mitov}}, \bibinfo{journal}{Phys. Rev. D} \textbf{\bibinfo{volume}{71}}, \bibinfo{pages}{054021} (\bibinfo{year}{2005}), \eprint{hep-ph/0410205}.

\bibitem[{\citenamefont{Biello and Bonino}(2024)}]{Biello:2024zti}
\bibinfo{author}{\bibfnamefont{C.}~\bibnamefont{Biello}} \bibnamefont{and} \bibinfo{author}{\bibfnamefont{L.}~\bibnamefont{Bonino}}, \bibinfo{journal}{Eur. Phys. J. C} \textbf{\bibinfo{volume}{84}}, \bibinfo{pages}{1192} (\bibinfo{year}{2024}), \eprint{2407.07623}.

\bibitem[{\citenamefont{Kartvelishvili et~al.}(1978)\citenamefont{Kartvelishvili, Likhoded, and Petrov}}]{Kartvelishvili:1977pi}
\bibinfo{author}{\bibfnamefont{V.}~\bibnamefont{Kartvelishvili}}, \bibinfo{author}{\bibfnamefont{A.}~\bibnamefont{Likhoded}}, \bibnamefont{and} \bibinfo{author}{\bibfnamefont{V.}~\bibnamefont{Petrov}}, \bibinfo{journal}{Phys. Lett. B} \textbf{\bibinfo{volume}{78}}, \bibinfo{pages}{615} (\bibinfo{year}{1978}).

\bibitem[{\citenamefont{Bowler}(1981)}]{Bowler:1981sb}
\bibinfo{author}{\bibfnamefont{M.~G.} \bibnamefont{Bowler}}, \bibinfo{journal}{Z. Phys. C} \textbf{\bibinfo{volume}{11}}, \bibinfo{pages}{169} (\bibinfo{year}{1981}).

\bibitem[{\citenamefont{Peterson et~al.}(1983)\citenamefont{Peterson, Schlatter, Schmitt, and Zerwas}}]{Peterson:1982ak}
\bibinfo{author}{\bibfnamefont{C.}~\bibnamefont{Peterson}}, \bibinfo{author}{\bibfnamefont{D.}~\bibnamefont{Schlatter}}, \bibinfo{author}{\bibfnamefont{I.}~\bibnamefont{Schmitt}}, \bibnamefont{and} \bibinfo{author}{\bibfnamefont{P.~M.} \bibnamefont{Zerwas}}, \bibinfo{journal}{Phys. Rev. D} \textbf{\bibinfo{volume}{27}}, \bibinfo{pages}{105} (\bibinfo{year}{1983}).

\bibitem[{\citenamefont{Andersson et~al.}(1983)\citenamefont{Andersson, Gustafson, and Soderberg}}]{Andersson:1983jt}
\bibinfo{author}{\bibfnamefont{B.}~\bibnamefont{Andersson}}, \bibinfo{author}{\bibfnamefont{G.}~\bibnamefont{Gustafson}}, \bibnamefont{and} \bibinfo{author}{\bibfnamefont{B.}~\bibnamefont{Soderberg}}, \bibinfo{journal}{Z. Phys. C} \textbf{\bibinfo{volume}{20}}, \bibinfo{pages}{317} (\bibinfo{year}{1983}).

\bibitem[{\citenamefont{Collins and Spiller}(1985)}]{Collins:1984ms}
\bibinfo{author}{\bibfnamefont{P.~D.~B.} \bibnamefont{Collins}} \bibnamefont{and} \bibinfo{author}{\bibfnamefont{T.~P.} \bibnamefont{Spiller}}, \bibinfo{journal}{J. Phys. G} \textbf{\bibinfo{volume}{11}}, \bibinfo{pages}{1289} (\bibinfo{year}{1985}).

\bibitem[{\citenamefont{Colangelo and Nason}(1992)}]{Colangelo:1992kh}
\bibinfo{author}{\bibfnamefont{G.}~\bibnamefont{Colangelo}} \bibnamefont{and} \bibinfo{author}{\bibfnamefont{P.}~\bibnamefont{Nason}}, \bibinfo{journal}{Phys. Lett. B} \textbf{\bibinfo{volume}{285}}, \bibinfo{pages}{167} (\bibinfo{year}{1992}).

\bibitem[{\citenamefont{Georgi}(1990)}]{Georgi:1990um}
\bibinfo{author}{\bibfnamefont{H.}~\bibnamefont{Georgi}}, \bibinfo{journal}{Phys. Lett. B} \textbf{\bibinfo{volume}{240}}, \bibinfo{pages}{447} (\bibinfo{year}{1990}).

\bibitem[{\citenamefont{Eichten and Hill}(1990)}]{Eichten:1989zv}
\bibinfo{author}{\bibfnamefont{E.}~\bibnamefont{Eichten}} \bibnamefont{and} \bibinfo{author}{\bibfnamefont{B.~R.} \bibnamefont{Hill}}, \bibinfo{journal}{Phys. Lett. B} \textbf{\bibinfo{volume}{234}}, \bibinfo{pages}{511} (\bibinfo{year}{1990}).

\bibitem[{\citenamefont{Grinstein}(1992)}]{Grinstein:1992ss}
\bibinfo{author}{\bibfnamefont{B.}~\bibnamefont{Grinstein}}, \bibinfo{journal}{Ann. Rev. Nucl. Part. Sci.} \textbf{\bibinfo{volume}{42}}, \bibinfo{pages}{101} (\bibinfo{year}{1992}).

\bibitem[{\citenamefont{Neubert}(1994)}]{Neubert:1993mb}
\bibinfo{author}{\bibfnamefont{M.}~\bibnamefont{Neubert}}, \bibinfo{journal}{Phys. Rept.} \textbf{\bibinfo{volume}{245}}, \bibinfo{pages}{259} (\bibinfo{year}{1994}), \eprint{hep-ph/9306320}.

\bibitem[{\citenamefont{Jaffe and Randall}(1994)}]{Jaffe:1993ie}
\bibinfo{author}{\bibfnamefont{R.~L.} \bibnamefont{Jaffe}} \bibnamefont{and} \bibinfo{author}{\bibfnamefont{L.}~\bibnamefont{Randall}}, \bibinfo{journal}{Nucl. Phys. B} \textbf{\bibinfo{volume}{412}}, \bibinfo{pages}{79} (\bibinfo{year}{1994}), \eprint{hep-ph/9306201}.

\bibitem[{\citenamefont{Kniehl et~al.}(2008)\citenamefont{Kniehl, Kramer, Schienbein, and Spiesberger}}]{Kniehl:2008zza}
\bibinfo{author}{\bibfnamefont{B.~A.} \bibnamefont{Kniehl}}, \bibinfo{author}{\bibfnamefont{G.}~\bibnamefont{Kramer}}, \bibinfo{author}{\bibfnamefont{I.}~\bibnamefont{Schienbein}}, \bibnamefont{and} \bibinfo{author}{\bibfnamefont{H.}~\bibnamefont{Spiesberger}}, \bibinfo{journal}{Phys. Rev. D} \textbf{\bibinfo{volume}{77}}, \bibinfo{pages}{014011} (\bibinfo{year}{2008}), \eprint{0705.4392}.

\bibitem[{\citenamefont{Kramer and Spiesberger}(2018)}]{Kramer:2018vde}
\bibinfo{author}{\bibfnamefont{G.}~\bibnamefont{Kramer}} \bibnamefont{and} \bibinfo{author}{\bibfnamefont{H.}~\bibnamefont{Spiesberger}}, \bibinfo{journal}{Phys. Rev. D} \textbf{\bibinfo{volume}{98}}, \bibinfo{pages}{114010} (\bibinfo{year}{2018}), \eprint{1809.04297}.

\bibitem[{\citenamefont{Grinstein}(2000)}]{Grinstein:1998xb}
\bibinfo{author}{\bibfnamefont{B.}~\bibnamefont{Grinstein}}, \bibinfo{journal}{Int. J. Mod. Phys. A} \textbf{\bibinfo{volume}{15}}, \bibinfo{pages}{461} (\bibinfo{year}{2000}), \eprint{hep-ph/9811264}.

\bibitem[{\citenamefont{Kr\"amer}(2001)}]{Kramer:2001hh}
\bibinfo{author}{\bibfnamefont{M.}~\bibnamefont{Kr\"amer}}, \bibinfo{journal}{Prog. Part. Nucl. Phys.} \textbf{\bibinfo{volume}{47}}, \bibinfo{pages}{141} (\bibinfo{year}{2001}), \eprint{hep-ph/0106120}.

\bibitem[{\citenamefont{Brambilla et~al.}(2004)}]{QuarkoniumWorkingGroup:2004kpm}
\bibinfo{author}{\bibfnamefont{N.}~\bibnamefont{Brambilla}} \bibnamefont{et~al.} (\bibinfo{collaboration}{Quarkonium Working Group}) (\bibinfo{year}{2004}), \eprint{hep-ph/0412158}.

\bibitem[{\citenamefont{Eichten and Quigg}(1994)}]{Eichten:1994gt}
\bibinfo{author}{\bibfnamefont{E.~J.} \bibnamefont{Eichten}} \bibnamefont{and} \bibinfo{author}{\bibfnamefont{C.}~\bibnamefont{Quigg}}, \bibinfo{journal}{Phys. Rev. D} \textbf{\bibinfo{volume}{49}}, \bibinfo{pages}{5845} (\bibinfo{year}{1994}), \eprint{hep-ph/9402210}.

\bibitem[{\citenamefont{Lepage et~al.}(1992)\citenamefont{Lepage, Magnea, Nakhleh, Magnea, and Hornbostel}}]{Lepage:1992tx}
\bibinfo{author}{\bibfnamefont{G.~P.} \bibnamefont{Lepage}}, \bibinfo{author}{\bibfnamefont{L.}~\bibnamefont{Magnea}}, \bibinfo{author}{\bibfnamefont{C.}~\bibnamefont{Nakhleh}}, \bibinfo{author}{\bibfnamefont{U.}~\bibnamefont{Magnea}}, \bibnamefont{and} \bibinfo{author}{\bibfnamefont{K.}~\bibnamefont{Hornbostel}}, \bibinfo{journal}{Phys. Rev. D} \textbf{\bibinfo{volume}{46}}, \bibinfo{pages}{4052} (\bibinfo{year}{1992}), \eprint{hep-lat/9205007}.

\bibitem[{\citenamefont{Davies et~al.}(1994)\citenamefont{Davies, Hornbostel, Langnau, Lepage, Lidsey, Shigemitsu, and Sloan}}]{Davies:1994mp}
\bibinfo{author}{\bibfnamefont{C.~T.~H.} \bibnamefont{Davies}}, \bibinfo{author}{\bibfnamefont{K.}~\bibnamefont{Hornbostel}}, \bibinfo{author}{\bibfnamefont{A.}~\bibnamefont{Langnau}}, \bibinfo{author}{\bibfnamefont{G.~P.} \bibnamefont{Lepage}}, \bibinfo{author}{\bibfnamefont{A.}~\bibnamefont{Lidsey}}, \bibinfo{author}{\bibfnamefont{J.}~\bibnamefont{Shigemitsu}}, \bibnamefont{and} \bibinfo{author}{\bibfnamefont{J.~H.} \bibnamefont{Sloan}}, \bibinfo{journal}{Phys. Rev. D} \textbf{\bibinfo{volume}{50}}, \bibinfo{pages}{6963} (\bibinfo{year}{1994}), \eprint{hep-lat/9406017}.

\bibitem[{\citenamefont{Berger and Jones}(1981)}]{Berger:1980ni}
\bibinfo{author}{\bibfnamefont{E.~L.} \bibnamefont{Berger}} \bibnamefont{and} \bibinfo{author}{\bibfnamefont{D.~L.} \bibnamefont{Jones}}, \bibinfo{journal}{Phys. Rev. D} \textbf{\bibinfo{volume}{23}}, \bibinfo{pages}{1521} (\bibinfo{year}{1981}).

\bibitem[{\citenamefont{Baier and Ruckl}(1981)}]{Baier:1981uk}
\bibinfo{author}{\bibfnamefont{R.}~\bibnamefont{Baier}} \bibnamefont{and} \bibinfo{author}{\bibfnamefont{R.}~\bibnamefont{Ruckl}}, \bibinfo{journal}{Phys. Lett. B} \textbf{\bibinfo{volume}{102}}, \bibinfo{pages}{364} (\bibinfo{year}{1981}).

\bibitem[{\citenamefont{Bodwin et~al.}(1992)\citenamefont{Bodwin, Braaten, and Lepage}}]{Bodwin:1992ye}
\bibinfo{author}{\bibfnamefont{G.~T.} \bibnamefont{Bodwin}}, \bibinfo{author}{\bibfnamefont{E.}~\bibnamefont{Braaten}}, \bibnamefont{and} \bibinfo{author}{\bibfnamefont{G.~P.} \bibnamefont{Lepage}}, \bibinfo{journal}{Phys. Rev. D} \textbf{\bibinfo{volume}{46}}, \bibinfo{pages}{R1914} (\bibinfo{year}{1992}), \eprint{hep-lat/9205006}.

\bibitem[{\citenamefont{Barbieri et~al.}(1976)\citenamefont{Barbieri, Gatto, and Remiddi}}]{Barbieri:1976fp}
\bibinfo{author}{\bibfnamefont{R.}~\bibnamefont{Barbieri}}, \bibinfo{author}{\bibfnamefont{R.}~\bibnamefont{Gatto}}, \bibnamefont{and} \bibinfo{author}{\bibfnamefont{E.}~\bibnamefont{Remiddi}}, \bibinfo{journal}{Phys. Lett. B} \textbf{\bibinfo{volume}{61}}, \bibinfo{pages}{465} (\bibinfo{year}{1976}).

\bibitem[{\citenamefont{Mangano}(1996)}]{Mangano:1995yd}
\bibinfo{author}{\bibfnamefont{M.~L.} \bibnamefont{Mangano}}, \bibinfo{journal}{AIP Conf. Proc.} \textbf{\bibinfo{volume}{357}}, \bibinfo{pages}{120} (\bibinfo{year}{1996}), \eprint{hep-ph/9507353}.

\bibitem[{\citenamefont{Braaten et~al.}(1996)\citenamefont{Braaten, Fleming, and Yuan}}]{Braaten:1996pv}
\bibinfo{author}{\bibfnamefont{E.}~\bibnamefont{Braaten}}, \bibinfo{author}{\bibfnamefont{S.}~\bibnamefont{Fleming}}, \bibnamefont{and} \bibinfo{author}{\bibfnamefont{T.~C.} \bibnamefont{Yuan}}, \bibinfo{journal}{Ann. Rev. Nucl. Part. Sci.} \textbf{\bibinfo{volume}{46}}, \bibinfo{pages}{197} (\bibinfo{year}{1996}), \eprint{hep-ph/9602374}.

\bibitem[{\citenamefont{Artoisenet}(2009)}]{Artoisenet:2009zwa}
\bibinfo{author}{\bibfnamefont{P.}~\bibnamefont{Artoisenet}}, Ph.D. thesis, \bibinfo{school}{Louvain U., CP3} (\bibinfo{year}{2009}).

\bibitem[{\citenamefont{Braaten and Yuan}(1993)}]{Braaten:1993rw}
\bibinfo{author}{\bibfnamefont{E.}~\bibnamefont{Braaten}} \bibnamefont{and} \bibinfo{author}{\bibfnamefont{T.~C.} \bibnamefont{Yuan}}, \bibinfo{journal}{Phys. Rev. Lett.} \textbf{\bibinfo{volume}{71}}, \bibinfo{pages}{1673} (\bibinfo{year}{1993}), \eprint{hep-ph/9303205}.

\bibitem[{\citenamefont{Kuhn and Schneider}(1981{\natexlab{a}})}]{Kuhn:1981jy}
\bibinfo{author}{\bibfnamefont{J.~H.} \bibnamefont{Kuhn}} \bibnamefont{and} \bibinfo{author}{\bibfnamefont{H.}~\bibnamefont{Schneider}}, \bibinfo{journal}{Z. Phys. C} \textbf{\bibinfo{volume}{11}}, \bibinfo{pages}{263} (\bibinfo{year}{1981}{\natexlab{a}}).

\bibitem[{\citenamefont{Kuhn and Schneider}(1981{\natexlab{b}})}]{Kuhn:1981jn}
\bibinfo{author}{\bibfnamefont{J.~H.} \bibnamefont{Kuhn}} \bibnamefont{and} \bibinfo{author}{\bibfnamefont{H.}~\bibnamefont{Schneider}}, \bibinfo{journal}{Phys. Rev. D} \textbf{\bibinfo{volume}{24}}, \bibinfo{pages}{2996} (\bibinfo{year}{1981}{\natexlab{b}}).

\bibitem[{\citenamefont{Cacciari and Greco}(1994{\natexlab{b}})}]{Cacciari:1994dr}
\bibinfo{author}{\bibfnamefont{M.}~\bibnamefont{Cacciari}} \bibnamefont{and} \bibinfo{author}{\bibfnamefont{M.}~\bibnamefont{Greco}}, \bibinfo{journal}{Phys. Rev. Lett.} \textbf{\bibinfo{volume}{73}}, \bibinfo{pages}{1586} (\bibinfo{year}{1994}{\natexlab{b}}), \eprint{hep-ph/9405241}.

\bibitem[{\citenamefont{Braaten et~al.}(1994)\citenamefont{Braaten, Doncheski, Fleming, and Mangano}}]{Braaten:1994xb}
\bibinfo{author}{\bibfnamefont{E.}~\bibnamefont{Braaten}}, \bibinfo{author}{\bibfnamefont{M.~A.} \bibnamefont{Doncheski}}, \bibinfo{author}{\bibfnamefont{S.}~\bibnamefont{Fleming}}, \bibnamefont{and} \bibinfo{author}{\bibfnamefont{M.~L.} \bibnamefont{Mangano}}, \bibinfo{journal}{Phys. Lett. B} \textbf{\bibinfo{volume}{333}}, \bibinfo{pages}{548} (\bibinfo{year}{1994}), \eprint{hep-ph/9405407}.

\bibitem[{\citenamefont{Braaten et~al.}(1993{\natexlab{a}})\citenamefont{Braaten, Cheung, and Yuan}}]{Braaten:1993mp}
\bibinfo{author}{\bibfnamefont{E.}~\bibnamefont{Braaten}}, \bibinfo{author}{\bibfnamefont{K.-m.} \bibnamefont{Cheung}}, \bibnamefont{and} \bibinfo{author}{\bibfnamefont{T.~C.} \bibnamefont{Yuan}}, \bibinfo{journal}{Phys. Rev. D} \textbf{\bibinfo{volume}{48}}, \bibinfo{pages}{4230} (\bibinfo{year}{1993}{\natexlab{a}}), \eprint{hep-ph/9302307}.

\bibitem[{\citenamefont{Braaten and Yuan}(1994)}]{Braaten:1994kd}
\bibinfo{author}{\bibfnamefont{E.}~\bibnamefont{Braaten}} \bibnamefont{and} \bibinfo{author}{\bibfnamefont{T.~C.} \bibnamefont{Yuan}}, \bibinfo{journal}{Phys. Rev. D} \textbf{\bibinfo{volume}{50}}, \bibinfo{pages}{3176} (\bibinfo{year}{1994}), \eprint{hep-ph/9403401}.

\bibitem[{\citenamefont{Ma}(1995)}]{Ma:1995ci}
\bibinfo{author}{\bibfnamefont{J.~P.} \bibnamefont{Ma}}, \bibinfo{journal}{Nucl. Phys. B} \textbf{\bibinfo{volume}{447}}, \bibinfo{pages}{405} (\bibinfo{year}{1995}), \eprint{hep-ph/9503346}.

\bibitem[{\citenamefont{Yuan}(1994)}]{Yuan:1994hn}
\bibinfo{author}{\bibfnamefont{T.~C.} \bibnamefont{Yuan}}, \bibinfo{journal}{Phys. Rev. D} \textbf{\bibinfo{volume}{50}}, \bibinfo{pages}{5664} (\bibinfo{year}{1994}), \eprint{hep-ph/9405348}.

\bibitem[{\citenamefont{Alonso et~al.}(2017)\citenamefont{Alonso, Grinstein, and Martin~Camalich}}]{Alonso:2016oyd}
\bibinfo{author}{\bibfnamefont{R.}~\bibnamefont{Alonso}}, \bibinfo{author}{\bibfnamefont{B.}~\bibnamefont{Grinstein}}, \bibnamefont{and} \bibinfo{author}{\bibfnamefont{J.}~\bibnamefont{Martin~Camalich}}, \bibinfo{journal}{Phys. Rev. Lett.} \textbf{\bibinfo{volume}{118}}, \bibinfo{pages}{081802} (\bibinfo{year}{2017}), \eprint{1611.06676}.

\bibitem[{\citenamefont{Aebischer and Grinstein}(2022)}]{Aebischer:2021eio}
\bibinfo{author}{\bibfnamefont{J.}~\bibnamefont{Aebischer}} \bibnamefont{and} \bibinfo{author}{\bibfnamefont{B.}~\bibnamefont{Grinstein}}, \bibinfo{journal}{Phys. Lett. B} \textbf{\bibinfo{volume}{834}}, \bibinfo{pages}{137435} (\bibinfo{year}{2022}), \eprint{2108.10285}.

\bibitem[{\citenamefont{Aebischer and Grinstein}(2021)}]{Aebischer:2021ilm}
\bibinfo{author}{\bibfnamefont{J.}~\bibnamefont{Aebischer}} \bibnamefont{and} \bibinfo{author}{\bibfnamefont{B.}~\bibnamefont{Grinstein}}, \bibinfo{journal}{JHEP} \textbf{\bibinfo{volume}{07}}, \bibinfo{pages}{130} (\bibinfo{year}{2021}), \eprint{2105.02988}.

\bibitem[{\citenamefont{Ortega et~al.}(2020)\citenamefont{Ortega, Segovia, Entem, and Fernandez}}]{Ortega:2020uvc}
\bibinfo{author}{\bibfnamefont{P.~G.} \bibnamefont{Ortega}}, \bibinfo{author}{\bibfnamefont{J.}~\bibnamefont{Segovia}}, \bibinfo{author}{\bibfnamefont{D.~R.} \bibnamefont{Entem}}, \bibnamefont{and} \bibinfo{author}{\bibfnamefont{F.}~\bibnamefont{Fernandez}}, \bibinfo{journal}{Eur. Phys. J. C} \textbf{\bibinfo{volume}{80}}, \bibinfo{pages}{223} (\bibinfo{year}{2020}), \eprint{2001.08093}.

\bibitem[{\citenamefont{Abe et~al.}(1998)}]{CDF:1998ihx}
\bibinfo{author}{\bibfnamefont{F.}~\bibnamefont{Abe}} \bibnamefont{et~al.} (\bibinfo{collaboration}{CDF}), \bibinfo{journal}{Phys. Rev. Lett.} \textbf{\bibinfo{volume}{81}}, \bibinfo{pages}{2432} (\bibinfo{year}{1998}), \eprint{hep-ex/9805034}.

\bibitem[{\citenamefont{Aad et~al.}(2014)}]{ATLAS:2014lga}
\bibinfo{author}{\bibfnamefont{G.}~\bibnamefont{Aad}} \bibnamefont{et~al.} (\bibinfo{collaboration}{ATLAS}), \bibinfo{journal}{Phys. Rev. Lett.} \textbf{\bibinfo{volume}{113}}, \bibinfo{pages}{212004} (\bibinfo{year}{2014}), \eprint{1407.1032}.

\bibitem[{\citenamefont{Roy and Sridhar}(1994)}]{Roy:1994ie}
\bibinfo{author}{\bibfnamefont{D.~P.} \bibnamefont{Roy}} \bibnamefont{and} \bibinfo{author}{\bibfnamefont{K.}~\bibnamefont{Sridhar}}, \bibinfo{journal}{Phys. Lett. B} \textbf{\bibinfo{volume}{339}}, \bibinfo{pages}{141} (\bibinfo{year}{1994}), \eprint{hep-ph/9406386}.

\bibitem[{\citenamefont{Cacciari et~al.}(1995)\citenamefont{Cacciari, Greco, Mangano, and Petrelli}}]{Cacciari:1995yt}
\bibinfo{author}{\bibfnamefont{M.}~\bibnamefont{Cacciari}}, \bibinfo{author}{\bibfnamefont{M.}~\bibnamefont{Greco}}, \bibinfo{author}{\bibfnamefont{M.~L.} \bibnamefont{Mangano}}, \bibnamefont{and} \bibinfo{author}{\bibfnamefont{A.}~\bibnamefont{Petrelli}}, \bibinfo{journal}{Phys. Lett. B} \textbf{\bibinfo{volume}{356}}, \bibinfo{pages}{553} (\bibinfo{year}{1995}), \eprint{hep-ph/9505379}.

\bibitem[{\citenamefont{Cacciari and Kr\"amer}(1996)}]{Cacciari:1996dg}
\bibinfo{author}{\bibfnamefont{M.}~\bibnamefont{Cacciari}} \bibnamefont{and} \bibinfo{author}{\bibfnamefont{M.}~\bibnamefont{Kr\"amer}}, \bibinfo{journal}{Phys. Rev. Lett.} \textbf{\bibinfo{volume}{76}}, \bibinfo{pages}{4128} (\bibinfo{year}{1996}), \eprint{hep-ph/9601276}.

\bibitem[{\citenamefont{Lansberg}(2020)}]{Lansberg:2019adr}
\bibinfo{author}{\bibfnamefont{J.-P.} \bibnamefont{Lansberg}}, \bibinfo{journal}{Phys. Rept.} \textbf{\bibinfo{volume}{889}}, \bibinfo{pages}{1} (\bibinfo{year}{2020}), \eprint{1903.09185}.

\bibitem[{\citenamefont{Kolodziej et~al.}(1995)\citenamefont{Kolodziej, Leike, and Ruckl}}]{Kolodziej:1995nv}
\bibinfo{author}{\bibfnamefont{K.}~\bibnamefont{Kolodziej}}, \bibinfo{author}{\bibfnamefont{A.}~\bibnamefont{Leike}}, \bibnamefont{and} \bibinfo{author}{\bibfnamefont{R.}~\bibnamefont{Ruckl}}, \bibinfo{journal}{Phys. Lett. B} \textbf{\bibinfo{volume}{355}}, \bibinfo{pages}{337} (\bibinfo{year}{1995}), \eprint{hep-ph/9505298}.

\bibitem[{\citenamefont{Artoisenet et~al.}(2007)\citenamefont{Artoisenet, Lansberg, and Maltoni}}]{Artoisenet:2007xi}
\bibinfo{author}{\bibfnamefont{P.}~\bibnamefont{Artoisenet}}, \bibinfo{author}{\bibfnamefont{J.~P.} \bibnamefont{Lansberg}}, \bibnamefont{and} \bibinfo{author}{\bibfnamefont{F.}~\bibnamefont{Maltoni}}, \bibinfo{journal}{Phys. Lett. B} \textbf{\bibinfo{volume}{653}}, \bibinfo{pages}{60} (\bibinfo{year}{2007}), \eprint{hep-ph/0703129}.

\bibitem[{\citenamefont{Kang et~al.}(2012)\citenamefont{Kang, Qiu, and Sterman}}]{Kang:2011mg}
\bibinfo{author}{\bibfnamefont{Z.-B.} \bibnamefont{Kang}}, \bibinfo{author}{\bibfnamefont{J.-W.} \bibnamefont{Qiu}}, \bibnamefont{and} \bibinfo{author}{\bibfnamefont{G.}~\bibnamefont{Sterman}}, \bibinfo{journal}{Phys. Rev. Lett.} \textbf{\bibinfo{volume}{108}}, \bibinfo{pages}{102002} (\bibinfo{year}{2012}), \eprint{1109.1520}.

\bibitem[{\citenamefont{Ma et~al.}(2014{\natexlab{a}})\citenamefont{Ma, Qiu, and Zhang}}]{Ma:2013yla}
\bibinfo{author}{\bibfnamefont{Y.-Q.} \bibnamefont{Ma}}, \bibinfo{author}{\bibfnamefont{J.-W.} \bibnamefont{Qiu}}, \bibnamefont{and} \bibinfo{author}{\bibfnamefont{H.}~\bibnamefont{Zhang}}, \bibinfo{journal}{Phys. Rev. D} \textbf{\bibinfo{volume}{89}}, \bibinfo{pages}{094029} (\bibinfo{year}{2014}{\natexlab{a}}), \eprint{1311.7078}.

\bibitem[{\citenamefont{Ma et~al.}(2014{\natexlab{b}})\citenamefont{Ma, Qiu, and Zhang}}]{Ma:2014eja}
\bibinfo{author}{\bibfnamefont{Y.-Q.} \bibnamefont{Ma}}, \bibinfo{author}{\bibfnamefont{J.-W.} \bibnamefont{Qiu}}, \bibnamefont{and} \bibinfo{author}{\bibfnamefont{H.}~\bibnamefont{Zhang}}, \bibinfo{journal}{Phys. Rev. D} \textbf{\bibinfo{volume}{89}}, \bibinfo{pages}{094030} (\bibinfo{year}{2014}{\natexlab{b}}), \eprint{1401.0524}.

\bibitem[{\citenamefont{Chang and Chen}(1992)}]{Chang:1992bb}
\bibinfo{author}{\bibfnamefont{C.-H.} \bibnamefont{Chang}} \bibnamefont{and} \bibinfo{author}{\bibfnamefont{Y.-Q.} \bibnamefont{Chen}}, \bibinfo{journal}{Phys. Rev. D} \textbf{\bibinfo{volume}{46}}, \bibinfo{pages}{3845} (\bibinfo{year}{1992}), \bibinfo{note}{[Erratum: Phys.Rev.D 50, 6013 (1994)]}.

\bibitem[{\citenamefont{Braaten et~al.}(1993{\natexlab{b}})\citenamefont{Braaten, Cheung, and Yuan}}]{Braaten:1993jn}
\bibinfo{author}{\bibfnamefont{E.}~\bibnamefont{Braaten}}, \bibinfo{author}{\bibfnamefont{K.-m.} \bibnamefont{Cheung}}, \bibnamefont{and} \bibinfo{author}{\bibfnamefont{T.~C.} \bibnamefont{Yuan}}, \bibinfo{journal}{Phys. Rev. D} \textbf{\bibinfo{volume}{48}}, \bibinfo{pages}{R5049} (\bibinfo{year}{1993}{\natexlab{b}}), \eprint{hep-ph/9305206}.

\bibitem[{\citenamefont{Ma}(1994)}]{Ma:1994zt}
\bibinfo{author}{\bibfnamefont{J.~P.} \bibnamefont{Ma}}, \bibinfo{journal}{Phys. Lett. B} \textbf{\bibinfo{volume}{332}}, \bibinfo{pages}{398} (\bibinfo{year}{1994}), \eprint{hep-ph/9401249}.

\bibitem[{\citenamefont{Zheng et~al.}(2019{\natexlab{a}})\citenamefont{Zheng, Chang, Feng, and Wu}}]{Zheng:2019gnb}
\bibinfo{author}{\bibfnamefont{X.-C.} \bibnamefont{Zheng}}, \bibinfo{author}{\bibfnamefont{C.-H.} \bibnamefont{Chang}}, \bibinfo{author}{\bibfnamefont{T.-F.} \bibnamefont{Feng}}, \bibnamefont{and} \bibinfo{author}{\bibfnamefont{X.-G.} \bibnamefont{Wu}}, \bibinfo{journal}{Phys. Rev. D} \textbf{\bibinfo{volume}{100}}, \bibinfo{pages}{034004} (\bibinfo{year}{2019}{\natexlab{a}}), \eprint{1901.03477}.

\bibitem[{\citenamefont{Zheng et~al.}(2022)\citenamefont{Zheng, Chang, and Wu}}]{Zheng:2021sdo}
\bibinfo{author}{\bibfnamefont{X.-C.} \bibnamefont{Zheng}}, \bibinfo{author}{\bibfnamefont{C.-H.} \bibnamefont{Chang}}, \bibnamefont{and} \bibinfo{author}{\bibfnamefont{X.-G.} \bibnamefont{Wu}}, \bibinfo{journal}{JHEP} \textbf{\bibinfo{volume}{05}}, \bibinfo{pages}{036} (\bibinfo{year}{2022}), \eprint{2112.10520}.

\bibitem[{\citenamefont{Feng et~al.}(2022{\natexlab{b}})\citenamefont{Feng, Jia, and Yang}}]{Feng:2021qjm}
\bibinfo{author}{\bibfnamefont{F.}~\bibnamefont{Feng}}, \bibinfo{author}{\bibfnamefont{Y.}~\bibnamefont{Jia}}, \bibnamefont{and} \bibinfo{author}{\bibfnamefont{D.}~\bibnamefont{Yang}}, \bibinfo{journal}{Phys. Rev. D} \textbf{\bibinfo{volume}{106}}, \bibinfo{pages}{054030} (\bibinfo{year}{2022}{\natexlab{b}}), \eprint{2112.15569}.

\bibitem[{\citenamefont{Feng and Jia}(2023)}]{Feng:2018ulg}
\bibinfo{author}{\bibfnamefont{F.}~\bibnamefont{Feng}} \bibnamefont{and} \bibinfo{author}{\bibfnamefont{Y.}~\bibnamefont{Jia}}, \bibinfo{journal}{Chin. Phys. C} \textbf{\bibinfo{volume}{47}}, \bibinfo{pages}{033103} (\bibinfo{year}{2023}), \eprint{1810.04138}.

\bibitem[{\citenamefont{Celiberto et~al.}(2016{\natexlab{a}})\citenamefont{Celiberto, Ivanov, Murdaca, and Papa}}]{Celiberto:2016hae}
\bibinfo{author}{\bibfnamefont{F.~G.} \bibnamefont{Celiberto}}, \bibinfo{author}{\bibfnamefont{D.~{\relax Yu}.} \bibnamefont{Ivanov}}, \bibinfo{author}{\bibfnamefont{B.}~\bibnamefont{Murdaca}}, \bibnamefont{and} \bibinfo{author}{\bibfnamefont{A.}~\bibnamefont{Papa}}, \bibinfo{journal}{Phys. Rev. D} \textbf{\bibinfo{volume}{94}}, \bibinfo{pages}{034013} (\bibinfo{year}{2016}{\natexlab{a}}), \eprint{1604.08013}.

\bibitem[{\citenamefont{Celiberto et~al.}(2017)\citenamefont{Celiberto, Ivanov, Murdaca, and Papa}}]{Celiberto:2017ptm}
\bibinfo{author}{\bibfnamefont{F.~G.} \bibnamefont{Celiberto}}, \bibinfo{author}{\bibfnamefont{D.~{\relax Yu}.} \bibnamefont{Ivanov}}, \bibinfo{author}{\bibfnamefont{B.}~\bibnamefont{Murdaca}}, \bibnamefont{and} \bibinfo{author}{\bibfnamefont{A.}~\bibnamefont{Papa}}, \bibinfo{journal}{Eur. Phys. J. C} \textbf{\bibinfo{volume}{77}}, \bibinfo{pages}{382} (\bibinfo{year}{2017}), \eprint{1701.05077}.

\bibitem[{\citenamefont{Celiberto et~al.}(2021{\natexlab{a}})\citenamefont{Celiberto, Fucilla, Ivanov, and Papa}}]{Celiberto:2021dzy}
\bibinfo{author}{\bibfnamefont{F.~G.} \bibnamefont{Celiberto}}, \bibinfo{author}{\bibfnamefont{M.}~\bibnamefont{Fucilla}}, \bibinfo{author}{\bibfnamefont{D.~{\relax Yu}.} \bibnamefont{Ivanov}}, \bibnamefont{and} \bibinfo{author}{\bibfnamefont{A.}~\bibnamefont{Papa}}, \bibinfo{journal}{Eur. Phys. J. C} \textbf{\bibinfo{volume}{81}}, \bibinfo{pages}{780} (\bibinfo{year}{2021}{\natexlab{a}}), \eprint{2105.06432}.

\bibitem[{\citenamefont{Celiberto et~al.}(2021{\natexlab{b}})\citenamefont{Celiberto, Fucilla, Ivanov, Mohammed, and Papa}}]{Celiberto:2021fdp}
\bibinfo{author}{\bibfnamefont{F.~G.} \bibnamefont{Celiberto}}, \bibinfo{author}{\bibfnamefont{M.}~\bibnamefont{Fucilla}}, \bibinfo{author}{\bibfnamefont{D.~{\relax Yu}.} \bibnamefont{Ivanov}}, \bibinfo{author}{\bibfnamefont{M.~M.~A.} \bibnamefont{Mohammed}}, \bibnamefont{and} \bibinfo{author}{\bibfnamefont{A.}~\bibnamefont{Papa}}, \bibinfo{journal}{Phys. Rev. D} \textbf{\bibinfo{volume}{104}}, \bibinfo{pages}{114007} (\bibinfo{year}{2021}{\natexlab{b}}), \eprint{2109.11875}.

\bibitem[{\citenamefont{Celiberto}(2023{\natexlab{b}})}]{Celiberto:2022kxx}
\bibinfo{author}{\bibfnamefont{F.~G.} \bibnamefont{Celiberto}}, \bibinfo{journal}{Eur. Phys. J. C} \textbf{\bibinfo{volume}{83}}, \bibinfo{pages}{332} (\bibinfo{year}{2023}{\natexlab{b}}), \eprint{2208.14577}.

\bibitem[{\citenamefont{Celiberto}(2023{\natexlab{c}})}]{Celiberto:2022grc}
\bibinfo{author}{\bibfnamefont{F.~G.} \bibnamefont{Celiberto}}, \bibinfo{journal}{Acta Phys. Polon. Supp.} \textbf{\bibinfo{volume}{16}}, \bibinfo{pages}{41} (\bibinfo{year}{2023}{\natexlab{c}}), \eprint{2211.11780}.

\bibitem[{\citenamefont{Celiberto et~al.}(2022)\citenamefont{Celiberto, Fucilla, Mohammed, and Papa}}]{Celiberto:2022zdg}
\bibinfo{author}{\bibfnamefont{F.~G.} \bibnamefont{Celiberto}}, \bibinfo{author}{\bibfnamefont{M.}~\bibnamefont{Fucilla}}, \bibinfo{author}{\bibfnamefont{M.~M.~A.} \bibnamefont{Mohammed}}, \bibnamefont{and} \bibinfo{author}{\bibfnamefont{A.}~\bibnamefont{Papa}}, \bibinfo{journal}{Phys. Rev. D} \textbf{\bibinfo{volume}{105}}, \bibinfo{pages}{114056} (\bibinfo{year}{2022}), \eprint{2205.13429}.

\bibitem[{\citenamefont{Aaij et~al.}(2015)}]{LHCb:2014iah}
\bibinfo{author}{\bibfnamefont{R.}~\bibnamefont{Aaij}} \bibnamefont{et~al.} (\bibinfo{collaboration}{LHCb}), \bibinfo{journal}{Phys. Rev. Lett.} \textbf{\bibinfo{volume}{114}}, \bibinfo{pages}{041801} (\bibinfo{year}{2015}), \eprint{1411.3104}.

\bibitem[{\citenamefont{Aaij et~al.}(2017)}]{LHCb:2016qpe}
\bibinfo{author}{\bibfnamefont{R.}~\bibnamefont{Aaij}} \bibnamefont{et~al.} (\bibinfo{collaboration}{LHCb}), \bibinfo{journal}{Phys. Rev. Lett.} \textbf{\bibinfo{volume}{118}}, \bibinfo{pages}{052002} (\bibinfo{year}{2017}), \bibinfo{note}{[Erratum: Phys.Rev.Lett. 119, 169901 (2017)]}, \eprint{1612.05140}.

\bibitem[{\citenamefont{Alekhin et~al.}(2010)\citenamefont{Alekhin, Blumlein, Klein, and Moch}}]{Alekhin:2009ni}
\bibinfo{author}{\bibfnamefont{S.}~\bibnamefont{Alekhin}}, \bibinfo{author}{\bibfnamefont{J.}~\bibnamefont{Blumlein}}, \bibinfo{author}{\bibfnamefont{S.}~\bibnamefont{Klein}}, \bibnamefont{and} \bibinfo{author}{\bibfnamefont{S.}~\bibnamefont{Moch}}, \bibinfo{journal}{Phys. Rev. D} \textbf{\bibinfo{volume}{81}}, \bibinfo{pages}{014032} (\bibinfo{year}{2010}), \eprint{0908.2766}.

\bibitem[{\citenamefont{Fleming et~al.}(2012)\citenamefont{Fleming, Leibovich, Mehen, and Rothstein}}]{Fleming:2012wy}
\bibinfo{author}{\bibfnamefont{S.}~\bibnamefont{Fleming}}, \bibinfo{author}{\bibfnamefont{A.~K.} \bibnamefont{Leibovich}}, \bibinfo{author}{\bibfnamefont{T.}~\bibnamefont{Mehen}}, \bibnamefont{and} \bibinfo{author}{\bibfnamefont{I.~Z.} \bibnamefont{Rothstein}}, \bibinfo{journal}{Phys. Rev. D} \textbf{\bibinfo{volume}{86}}, \bibinfo{pages}{094012} (\bibinfo{year}{2012}), \eprint{1207.2578}.

\bibitem[{\citenamefont{Echevarria}(2019)}]{Echevarria:2019ynx}
\bibinfo{author}{\bibfnamefont{M.~G.} \bibnamefont{Echevarria}}, \bibinfo{journal}{JHEP} \textbf{\bibinfo{volume}{10}}, \bibinfo{pages}{144} (\bibinfo{year}{2019}), \eprint{1907.06494}.

\bibitem[{\citenamefont{Boer et~al.}(2023)\citenamefont{Boer, Bor, Maxia, Pisano, and Yuan}}]{Boer:2023zit}
\bibinfo{author}{\bibfnamefont{D.}~\bibnamefont{Boer}}, \bibinfo{author}{\bibfnamefont{J.}~\bibnamefont{Bor}}, \bibinfo{author}{\bibfnamefont{L.}~\bibnamefont{Maxia}}, \bibinfo{author}{\bibfnamefont{C.}~\bibnamefont{Pisano}}, \bibnamefont{and} \bibinfo{author}{\bibfnamefont{F.}~\bibnamefont{Yuan}}, \bibinfo{journal}{JHEP} \textbf{\bibinfo{volume}{08}}, \bibinfo{pages}{105} (\bibinfo{year}{2023}), \eprint{2304.09473}.

\bibitem[{\citenamefont{Aad et~al.}(2023)}]{ATLAS:2023bft}
\bibinfo{author}{\bibfnamefont{G.}~\bibnamefont{Aad}} \bibnamefont{et~al.} (\bibinfo{collaboration}{ATLAS}), \bibinfo{journal}{Phys. Rev. Lett.} \textbf{\bibinfo{volume}{131}}, \bibinfo{pages}{151902} (\bibinfo{year}{2023}), \eprint{2304.08962}.

\bibitem[{\citenamefont{Hayrapetyan et~al.}(2024)}]{CMS:2023owd}
\bibinfo{author}{\bibfnamefont{A.}~\bibnamefont{Hayrapetyan}} \bibnamefont{et~al.} (\bibinfo{collaboration}{CMS}), \bibinfo{journal}{Phys. Rev. Lett.} \textbf{\bibinfo{volume}{132}}, \bibinfo{pages}{111901} (\bibinfo{year}{2024}), \eprint{2306.07164}.

\bibitem[{\citenamefont{Zhang and Ma}(2020)}]{Zhang:2020hoh}
\bibinfo{author}{\bibfnamefont{H.-F.} \bibnamefont{Zhang}} \bibnamefont{and} \bibinfo{author}{\bibfnamefont{Y.-Q.} \bibnamefont{Ma}} (\bibinfo{year}{2020}), \eprint{2009.08376}.

\bibitem[{\citenamefont{Zhu}(2021)}]{Zhu:2020xni}
\bibinfo{author}{\bibfnamefont{R.}~\bibnamefont{Zhu}}, \bibinfo{journal}{Nucl. Phys. B} \textbf{\bibinfo{volume}{966}}, \bibinfo{pages}{115393} (\bibinfo{year}{2021}), \eprint{2010.09082}.

\bibitem[{\citenamefont{Feng et~al.}(2023{\natexlab{a}})\citenamefont{Feng, Huang, Jia, Sang, Yang, and Zhang}}]{Feng:2023agq}
\bibinfo{author}{\bibfnamefont{F.}~\bibnamefont{Feng}}, \bibinfo{author}{\bibfnamefont{Y.}~\bibnamefont{Huang}}, \bibinfo{author}{\bibfnamefont{Y.}~\bibnamefont{Jia}}, \bibinfo{author}{\bibfnamefont{W.-L.} \bibnamefont{Sang}}, \bibinfo{author}{\bibfnamefont{D.-S.} \bibnamefont{Yang}}, \bibnamefont{and} \bibinfo{author}{\bibfnamefont{J.-Y.} \bibnamefont{Zhang}}, \bibinfo{journal}{Phys. Rev. D} \textbf{\bibinfo{volume}{108}}, \bibinfo{pages}{L051501} (\bibinfo{year}{2023}{\natexlab{a}}), \eprint{2304.11142}.

\bibitem[{\citenamefont{Lepage and Brodsky}(1980)}]{Lepage:1980fj}
\bibinfo{author}{\bibfnamefont{G.~P.} \bibnamefont{Lepage}} \bibnamefont{and} \bibinfo{author}{\bibfnamefont{S.~J.} \bibnamefont{Brodsky}}, \bibinfo{journal}{Phys. Rev. D} \textbf{\bibinfo{volume}{22}}, \bibinfo{pages}{2157} (\bibinfo{year}{1980}).

\bibitem[{\citenamefont{Brodsky and Ji}(1985)}]{Brodsky:1985cr}
\bibinfo{author}{\bibfnamefont{S.~J.} \bibnamefont{Brodsky}} \bibnamefont{and} \bibinfo{author}{\bibfnamefont{C.-R.} \bibnamefont{Ji}}, \bibinfo{journal}{Phys. Rev. Lett.} \textbf{\bibinfo{volume}{55}}, \bibinfo{pages}{2257} (\bibinfo{year}{1985}).

\bibitem[{\citenamefont{Zyla et~al.}(2020)}]{ParticleDataGroup:2020ssz}
\bibinfo{author}{\bibfnamefont{P.~A.} \bibnamefont{Zyla}} \bibnamefont{et~al.} (\bibinfo{collaboration}{Particle Data Group}), \bibinfo{journal}{PTEP} \textbf{\bibinfo{volume}{2020}}, \bibinfo{pages}{083C01} (\bibinfo{year}{2020}).

\bibitem[{\citenamefont{Gomshi~Nobary}(1994)}]{GomshiNobary:1994eq}
\bibinfo{author}{\bibfnamefont{M.~A.} \bibnamefont{Gomshi~Nobary}}, \bibinfo{journal}{J. Phys. G} \textbf{\bibinfo{volume}{20}}, \bibinfo{pages}{65} (\bibinfo{year}{1994}).

\bibitem[{\citenamefont{Bolognino et~al.}(2018{\natexlab{a}})\citenamefont{Bolognino, Celiberto, Ivanov, Mohammed, and Papa}}]{Bolognino:2018oth}
\bibinfo{author}{\bibfnamefont{A.~D.} \bibnamefont{Bolognino}}, \bibinfo{author}{\bibfnamefont{F.~G.} \bibnamefont{Celiberto}}, \bibinfo{author}{\bibfnamefont{D.~{\relax Yu}.} \bibnamefont{Ivanov}}, \bibinfo{author}{\bibfnamefont{M.~M.~A.} \bibnamefont{Mohammed}}, \bibnamefont{and} \bibinfo{author}{\bibfnamefont{A.}~\bibnamefont{Papa}}, \bibinfo{journal}{Eur. Phys. J. C} \textbf{\bibinfo{volume}{78}}, \bibinfo{pages}{772} (\bibinfo{year}{2018}{\natexlab{a}}), \eprint{1808.05483}.

\bibitem[{\citenamefont{Artoisenet and Braaten}(2015)}]{Artoisenet:2014lpa}
\bibinfo{author}{\bibfnamefont{P.}~\bibnamefont{Artoisenet}} \bibnamefont{and} \bibinfo{author}{\bibfnamefont{E.}~\bibnamefont{Braaten}}, \bibinfo{journal}{JHEP} \textbf{\bibinfo{volume}{04}}, \bibinfo{pages}{121} (\bibinfo{year}{2015}), \eprint{1412.3834}.

\bibitem[{\citenamefont{Zhang et~al.}(2019)\citenamefont{Zhang, Wang, Liu, Ma, Meng, and Chao}}]{Zhang:2018mlo}
\bibinfo{author}{\bibfnamefont{P.}~\bibnamefont{Zhang}}, \bibinfo{author}{\bibfnamefont{C.-Y.} \bibnamefont{Wang}}, \bibinfo{author}{\bibfnamefont{X.}~\bibnamefont{Liu}}, \bibinfo{author}{\bibfnamefont{Y.-Q.} \bibnamefont{Ma}}, \bibinfo{author}{\bibfnamefont{C.}~\bibnamefont{Meng}}, \bibnamefont{and} \bibinfo{author}{\bibfnamefont{K.-T.} \bibnamefont{Chao}}, \bibinfo{journal}{JHEP} \textbf{\bibinfo{volume}{04}}, \bibinfo{pages}{116} (\bibinfo{year}{2019}), \eprint{1810.07656}.

\bibitem[{\citenamefont{Zheng et~al.}(2021{\natexlab{a}})\citenamefont{Zheng, Wu, and Huang}}]{Zheng:2021ylc}
\bibinfo{author}{\bibfnamefont{X.-C.} \bibnamefont{Zheng}}, \bibinfo{author}{\bibfnamefont{X.-G.} \bibnamefont{Wu}}, \bibnamefont{and} \bibinfo{author}{\bibfnamefont{X.-D.} \bibnamefont{Huang}}, \bibinfo{journal}{JHEP} \textbf{\bibinfo{volume}{07}}, \bibinfo{pages}{014} (\bibinfo{year}{2021}{\natexlab{a}}), \eprint{2105.14580}.

\bibitem[{\citenamefont{Zheng et~al.}(2021{\natexlab{b}})\citenamefont{Zheng, Zhang, and Wu}}]{Zheng:2021mqr}
\bibinfo{author}{\bibfnamefont{X.-C.} \bibnamefont{Zheng}}, \bibinfo{author}{\bibfnamefont{Z.-Y.} \bibnamefont{Zhang}}, \bibnamefont{and} \bibinfo{author}{\bibfnamefont{X.-G.} \bibnamefont{Wu}}, \bibinfo{journal}{Phys. Rev. D} \textbf{\bibinfo{volume}{103}}, \bibinfo{pages}{074004} (\bibinfo{year}{2021}{\natexlab{b}}), \eprint{2101.01527}.

\bibitem[{\citenamefont{Bjorken}(1978)}]{Bjorken:1977md}
\bibinfo{author}{\bibfnamefont{J.~D.} \bibnamefont{Bjorken}}, \bibinfo{journal}{Phys. Rev. D} \textbf{\bibinfo{volume}{17}}, \bibinfo{pages}{171} (\bibinfo{year}{1978}).

\bibitem[{\citenamefont{Kinoshita}(1986)}]{Kinoshita:1985mh}
\bibinfo{author}{\bibfnamefont{K.}~\bibnamefont{Kinoshita}}, \bibinfo{journal}{Prog. Theor. Phys.} \textbf{\bibinfo{volume}{75}}, \bibinfo{pages}{84} (\bibinfo{year}{1986}).

\bibitem[{\citenamefont{Cacciari and Catani}(2001)}]{Cacciari:2001cw}
\bibinfo{author}{\bibfnamefont{M.}~\bibnamefont{Cacciari}} \bibnamefont{and} \bibinfo{author}{\bibfnamefont{S.}~\bibnamefont{Catani}}, \bibinfo{journal}{Nucl. Phys. B} \textbf{\bibinfo{volume}{617}}, \bibinfo{pages}{253} (\bibinfo{year}{2001}), \eprint{hep-ph/0107138}.

\bibitem[{\citenamefont{Fickinger et~al.}(2016)\citenamefont{Fickinger, Fleming, Kim, and Mereghetti}}]{Fickinger:2016rfd}
\bibinfo{author}{\bibfnamefont{M.}~\bibnamefont{Fickinger}}, \bibinfo{author}{\bibfnamefont{S.}~\bibnamefont{Fleming}}, \bibinfo{author}{\bibfnamefont{C.}~\bibnamefont{Kim}}, \bibnamefont{and} \bibinfo{author}{\bibfnamefont{E.}~\bibnamefont{Mereghetti}}, \bibinfo{journal}{JHEP} \textbf{\bibinfo{volume}{11}}, \bibinfo{pages}{095} (\bibinfo{year}{2016}), \eprint{1606.07737}.

\bibitem[{\citenamefont{Ridolfi et~al.}(2020)\citenamefont{Ridolfi, Ubiali, and Zaro}}]{Ridolfi:2019bch}
\bibinfo{author}{\bibfnamefont{G.}~\bibnamefont{Ridolfi}}, \bibinfo{author}{\bibfnamefont{M.}~\bibnamefont{Ubiali}}, \bibnamefont{and} \bibinfo{author}{\bibfnamefont{M.}~\bibnamefont{Zaro}}, \bibinfo{journal}{JHEP} \textbf{\bibinfo{volume}{01}}, \bibinfo{pages}{196} (\bibinfo{year}{2020}), \eprint{1911.01975}.

\bibitem[{\citenamefont{Maltoni et~al.}(2022)\citenamefont{Maltoni, Ridolfi, Ubiali, and Zaro}}]{Maltoni:2022bpy}
\bibinfo{author}{\bibfnamefont{F.}~\bibnamefont{Maltoni}}, \bibinfo{author}{\bibfnamefont{G.}~\bibnamefont{Ridolfi}}, \bibinfo{author}{\bibfnamefont{M.}~\bibnamefont{Ubiali}}, \bibnamefont{and} \bibinfo{author}{\bibfnamefont{M.}~\bibnamefont{Zaro}}, \bibinfo{journal}{JHEP} \textbf{\bibinfo{volume}{10}}, \bibinfo{pages}{027} (\bibinfo{year}{2022}), \eprint{2207.10038}.

\bibitem[{\citenamefont{Czakon et~al.}(2021)\citenamefont{Czakon, Generet, Mitov, and Poncelet}}]{Czakon:2021ohs}
\bibinfo{author}{\bibfnamefont{M.~L.} \bibnamefont{Czakon}}, \bibinfo{author}{\bibfnamefont{T.}~\bibnamefont{Generet}}, \bibinfo{author}{\bibfnamefont{A.}~\bibnamefont{Mitov}}, \bibnamefont{and} \bibinfo{author}{\bibfnamefont{R.}~\bibnamefont{Poncelet}}, \bibinfo{journal}{JHEP} \textbf{\bibinfo{volume}{10}}, \bibinfo{pages}{216} (\bibinfo{year}{2021}), \eprint{2102.08267}.

\bibitem[{\citenamefont{Czakon et~al.}(2023)\citenamefont{Czakon, Generet, Mitov, and Poncelet}}]{Czakon:2022pyz}
\bibinfo{author}{\bibfnamefont{M.}~\bibnamefont{Czakon}}, \bibinfo{author}{\bibfnamefont{T.}~\bibnamefont{Generet}}, \bibinfo{author}{\bibfnamefont{A.}~\bibnamefont{Mitov}}, \bibnamefont{and} \bibinfo{author}{\bibfnamefont{R.}~\bibnamefont{Poncelet}}, \bibinfo{journal}{JHEP} \textbf{\bibinfo{volume}{03}}, \bibinfo{pages}{251} (\bibinfo{year}{2023}), \eprint{2210.06078}.

\bibitem[{\citenamefont{Generet}(2023)}]{Generet:2023vte}
\bibinfo{author}{\bibfnamefont{T.}~\bibnamefont{Generet}}, Ph.D. thesis, \bibinfo{school}{RWTH Aachen University, RWTH Aachen U.} (\bibinfo{year}{2023}).

\bibitem[{\citenamefont{Aglietti et~al.}(2007)\citenamefont{Aglietti, Di~Giustino, Ferrera, Renzaglia, Ricciardi, and Trentadue}}]{Aglietti:2007bp}
\bibinfo{author}{\bibfnamefont{U.}~\bibnamefont{Aglietti}}, \bibinfo{author}{\bibfnamefont{L.}~\bibnamefont{Di~Giustino}}, \bibinfo{author}{\bibfnamefont{G.}~\bibnamefont{Ferrera}}, \bibinfo{author}{\bibfnamefont{A.}~\bibnamefont{Renzaglia}}, \bibinfo{author}{\bibfnamefont{G.}~\bibnamefont{Ricciardi}}, \bibnamefont{and} \bibinfo{author}{\bibfnamefont{L.}~\bibnamefont{Trentadue}}, \bibinfo{journal}{Phys. Lett. B} \textbf{\bibinfo{volume}{653}}, \bibinfo{pages}{38} (\bibinfo{year}{2007}), \eprint{0707.2010}.

\bibitem[{\citenamefont{Aglietti and Ferrera}(2023)}]{Aglietti:2022rcm}
\bibinfo{author}{\bibfnamefont{U.~G.} \bibnamefont{Aglietti}} \bibnamefont{and} \bibinfo{author}{\bibfnamefont{G.}~\bibnamefont{Ferrera}}, \bibinfo{journal}{Eur. Phys. J. C} \textbf{\bibinfo{volume}{83}}, \bibinfo{pages}{335} (\bibinfo{year}{2023}), \eprint{2211.14397}.

\bibitem[{\citenamefont{Gaggero et~al.}(2022)\citenamefont{Gaggero, Ghira, Marzani, and Ridolfi}}]{Gaggero:2022hmv}
\bibinfo{author}{\bibfnamefont{D.}~\bibnamefont{Gaggero}}, \bibinfo{author}{\bibfnamefont{A.}~\bibnamefont{Ghira}}, \bibinfo{author}{\bibfnamefont{S.}~\bibnamefont{Marzani}}, \bibnamefont{and} \bibinfo{author}{\bibfnamefont{G.}~\bibnamefont{Ridolfi}}, \bibinfo{journal}{JHEP} \textbf{\bibinfo{volume}{09}}, \bibinfo{pages}{058} (\bibinfo{year}{2022}), \eprint{2207.13567}.

\bibitem[{\citenamefont{Ghira et~al.}(2023)\citenamefont{Ghira, Marzani, and Ridolfi}}]{Ghira:2023bxr}
\bibinfo{author}{\bibfnamefont{A.}~\bibnamefont{Ghira}}, \bibinfo{author}{\bibfnamefont{S.}~\bibnamefont{Marzani}}, \bibnamefont{and} \bibinfo{author}{\bibfnamefont{G.}~\bibnamefont{Ridolfi}}, \bibinfo{journal}{JHEP} \textbf{\bibinfo{volume}{11}}, \bibinfo{pages}{120} (\bibinfo{year}{2023}), \eprint{2309.06139}.

\bibitem[{\citenamefont{Bonino et~al.}(2024)\citenamefont{Bonino, Cacciari, and Stagnitto}}]{Bonino:2023icn}
\bibinfo{author}{\bibfnamefont{L.}~\bibnamefont{Bonino}}, \bibinfo{author}{\bibfnamefont{M.}~\bibnamefont{Cacciari}}, \bibnamefont{and} \bibinfo{author}{\bibfnamefont{G.}~\bibnamefont{Stagnitto}}, \bibinfo{journal}{JHEP} \textbf{\bibinfo{volume}{06}}, \bibinfo{pages}{040} (\bibinfo{year}{2024}), \eprint{2312.12519}.

\bibitem[{\citenamefont{Cacciari et~al.}(2024)\citenamefont{Cacciari, Ghira, Marzani, and Ridolfi}}]{Cacciari:2024kaa}
\bibinfo{author}{\bibfnamefont{M.}~\bibnamefont{Cacciari}}, \bibinfo{author}{\bibfnamefont{A.}~\bibnamefont{Ghira}}, \bibinfo{author}{\bibfnamefont{S.}~\bibnamefont{Marzani}}, \bibnamefont{and} \bibinfo{author}{\bibfnamefont{G.}~\bibnamefont{Ridolfi}}, \bibinfo{journal}{Eur. Phys. J. C} \textbf{\bibinfo{volume}{84}}, \bibinfo{pages}{889} (\bibinfo{year}{2024}), \eprint{2406.04173}.

\bibitem[{\citenamefont{Zhao et~al.}(2020)\citenamefont{Zhao, Shi, and Zhuang}}]{Zhao:2020nwy}
\bibinfo{author}{\bibfnamefont{J.}~\bibnamefont{Zhao}}, \bibinfo{author}{\bibfnamefont{S.}~\bibnamefont{Shi}}, \bibnamefont{and} \bibinfo{author}{\bibfnamefont{P.}~\bibnamefont{Zhuang}}, \bibinfo{journal}{Phys. Rev. D} \textbf{\bibinfo{volume}{102}}, \bibinfo{pages}{114001} (\bibinfo{year}{2020}), \eprint{2009.10319}.

\bibitem[{\citenamefont{L\"u et~al.}(2020)\citenamefont{L\"u, Chen, and Dong}}]{Lu:2020cns}
\bibinfo{author}{\bibfnamefont{Q.-F.} \bibnamefont{L\"u}}, \bibinfo{author}{\bibfnamefont{D.-Y.} \bibnamefont{Chen}}, \bibnamefont{and} \bibinfo{author}{\bibfnamefont{Y.-B.} \bibnamefont{Dong}}, \bibinfo{journal}{Eur. Phys. J. C} \textbf{\bibinfo{volume}{80}}, \bibinfo{pages}{871} (\bibinfo{year}{2020}), \eprint{2006.14445}.

\bibitem[{\citenamefont{Liu et~al.}(2024)\citenamefont{Liu, Liu, Zhong, and Zhao}}]{liu:2020eha}
\bibinfo{author}{\bibfnamefont{M.-S.} \bibnamefont{Liu}}, \bibinfo{author}{\bibfnamefont{F.-X.} \bibnamefont{Liu}}, \bibinfo{author}{\bibfnamefont{X.-H.} \bibnamefont{Zhong}}, \bibnamefont{and} \bibinfo{author}{\bibfnamefont{Q.}~\bibnamefont{Zhao}}, \bibinfo{journal}{Phys. Rev. D} \textbf{\bibinfo{volume}{109}}, \bibinfo{pages}{076017} (\bibinfo{year}{2024}), \eprint{2006.11952}.

\bibitem[{\citenamefont{Eichten et~al.}(1975)\citenamefont{Eichten, Gottfried, Kinoshita, Kogut, Lane, and Yan}}]{Eichten:1974af}
\bibinfo{author}{\bibfnamefont{E.}~\bibnamefont{Eichten}}, \bibinfo{author}{\bibfnamefont{K.}~\bibnamefont{Gottfried}}, \bibinfo{author}{\bibfnamefont{T.}~\bibnamefont{Kinoshita}}, \bibinfo{author}{\bibfnamefont{J.~B.} \bibnamefont{Kogut}}, \bibinfo{author}{\bibfnamefont{K.~D.} \bibnamefont{Lane}}, \bibnamefont{and} \bibinfo{author}{\bibfnamefont{T.-M.} \bibnamefont{Yan}}, \bibinfo{journal}{Phys. Rev. Lett.} \textbf{\bibinfo{volume}{34}}, \bibinfo{pages}{369} (\bibinfo{year}{1975}), \bibinfo{note}{[Erratum: Phys.Rev.Lett. 36, 1276 (1976)]}.

\bibitem[{\citenamefont{Eichten et~al.}(1978)\citenamefont{Eichten, Gottfried, Kinoshita, Lane, and Yan}}]{Eichten:1978tg}
\bibinfo{author}{\bibfnamefont{E.}~\bibnamefont{Eichten}}, \bibinfo{author}{\bibfnamefont{K.}~\bibnamefont{Gottfried}}, \bibinfo{author}{\bibfnamefont{T.}~\bibnamefont{Kinoshita}}, \bibinfo{author}{\bibfnamefont{K.~D.} \bibnamefont{Lane}}, \bibnamefont{and} \bibinfo{author}{\bibfnamefont{T.-M.} \bibnamefont{Yan}}, \bibinfo{journal}{Phys. Rev. D} \textbf{\bibinfo{volume}{17}}, \bibinfo{pages}{3090} (\bibinfo{year}{1978}), \bibinfo{note}{[Erratum: Phys.Rev.D 21, 313 (1980)]}.

\bibitem[{\citenamefont{Sirunyan et~al.}(2018)}]{CMS:2017dju}
\bibinfo{author}{\bibfnamefont{A.~M.} \bibnamefont{Sirunyan}} \bibnamefont{et~al.} (\bibinfo{collaboration}{CMS}), \bibinfo{journal}{Phys. Lett. B} \textbf{\bibinfo{volume}{780}}, \bibinfo{pages}{251} (\bibinfo{year}{2018}), \eprint{1710.11002}.

\bibitem[{\citenamefont{Zheng et~al.}(2019{\natexlab{b}})\citenamefont{Zheng, Chang, and Wu}}]{Zheng:2019dfk}
\bibinfo{author}{\bibfnamefont{X.-C.} \bibnamefont{Zheng}}, \bibinfo{author}{\bibfnamefont{C.-H.} \bibnamefont{Chang}}, \bibnamefont{and} \bibinfo{author}{\bibfnamefont{X.-G.} \bibnamefont{Wu}}, \bibinfo{journal}{Phys. Rev. D} \textbf{\bibinfo{volume}{100}}, \bibinfo{pages}{014005} (\bibinfo{year}{2019}{\natexlab{b}}), \eprint{1905.09171}.

\bibitem[{\citenamefont{Bai et~al.}(2024{\natexlab{b}})\citenamefont{Bai, Huang, and Sang}}]{Bai:2024flh}
\bibinfo{author}{\bibfnamefont{X.-W.} \bibnamefont{Bai}}, \bibinfo{author}{\bibfnamefont{Y.}~\bibnamefont{Huang}}, \bibnamefont{and} \bibinfo{author}{\bibfnamefont{W.-L.} \bibnamefont{Sang}} (\bibinfo{year}{2024}{\natexlab{b}}), \eprint{2411.19296}.

\bibitem[{\citenamefont{Fadin et~al.}(1975)\citenamefont{Fadin, Kuraev, and Lipatov}}]{Fadin:1975cb}
\bibinfo{author}{\bibfnamefont{V.~S.} \bibnamefont{Fadin}}, \bibinfo{author}{\bibfnamefont{E.}~\bibnamefont{Kuraev}}, \bibnamefont{and} \bibinfo{author}{\bibfnamefont{L.}~\bibnamefont{Lipatov}}, \bibinfo{journal}{Phys. Lett. B} \textbf{\bibinfo{volume}{60}}, \bibinfo{pages}{50} (\bibinfo{year}{1975}).

\bibitem[{\citenamefont{Kuraev et~al.}(1977)\citenamefont{Kuraev, Lipatov, and Fadin}}]{Kuraev:1977fs}
\bibinfo{author}{\bibfnamefont{E.}~\bibnamefont{Kuraev}}, \bibinfo{author}{\bibfnamefont{L.}~\bibnamefont{Lipatov}}, \bibnamefont{and} \bibinfo{author}{\bibfnamefont{V.~S.} \bibnamefont{Fadin}}, \bibinfo{journal}{Sov.\ Phys.\ JETP} \textbf{\bibinfo{volume}{45}}, \bibinfo{pages}{199} (\bibinfo{year}{1977}).

\bibitem[{\citenamefont{Balitsky and Lipatov}(1978)}]{Balitsky:1978ic}
\bibinfo{author}{\bibfnamefont{I.}~\bibnamefont{Balitsky}} \bibnamefont{and} \bibinfo{author}{\bibfnamefont{L.}~\bibnamefont{Lipatov}}, \bibinfo{journal}{Sov.\ J.\ Nucl.\ Phys.} \textbf{\bibinfo{volume}{28}}, \bibinfo{pages}{822} (\bibinfo{year}{1978}).

\bibitem[{\citenamefont{Fadin and Lipatov}(1998)}]{Fadin:1998py}
\bibinfo{author}{\bibfnamefont{V.~S.} \bibnamefont{Fadin}} \bibnamefont{and} \bibinfo{author}{\bibfnamefont{L.~N.} \bibnamefont{Lipatov}}, \bibinfo{journal}{Phys. Lett. B} \textbf{\bibinfo{volume}{429}}, \bibinfo{pages}{127} (\bibinfo{year}{1998}), \eprint{hep-ph/9802290}.

\bibitem[{\citenamefont{Ciafaloni and Camici}(1998)}]{Ciafaloni:1998gs}
\bibinfo{author}{\bibfnamefont{M.}~\bibnamefont{Ciafaloni}} \bibnamefont{and} \bibinfo{author}{\bibfnamefont{G.}~\bibnamefont{Camici}}, \bibinfo{journal}{Phys. Lett. B} \textbf{\bibinfo{volume}{430}}, \bibinfo{pages}{349} (\bibinfo{year}{1998}), \eprint{hep-ph/9803389}.

\bibitem[{\citenamefont{Colferai et~al.}(2010)\citenamefont{Colferai, Schwennsen, Szymanowski, and Wallon}}]{Colferai:2010wu}
\bibinfo{author}{\bibfnamefont{D.}~\bibnamefont{Colferai}}, \bibinfo{author}{\bibfnamefont{F.}~\bibnamefont{Schwennsen}}, \bibinfo{author}{\bibfnamefont{L.}~\bibnamefont{Szymanowski}}, \bibnamefont{and} \bibinfo{author}{\bibfnamefont{S.}~\bibnamefont{Wallon}}, \bibinfo{journal}{JHEP} \textbf{\bibinfo{volume}{12}}, \bibinfo{pages}{026} (\bibinfo{year}{2010}), \eprint{1002.1365}.

\bibitem[{\citenamefont{Celiberto et~al.}(2021{\natexlab{c}})\citenamefont{Celiberto, Ivanov, Mohammed, and Papa}}]{Celiberto:2020tmb}
\bibinfo{author}{\bibfnamefont{F.~G.} \bibnamefont{Celiberto}}, \bibinfo{author}{\bibfnamefont{D.~{\relax Yu}.} \bibnamefont{Ivanov}}, \bibinfo{author}{\bibfnamefont{M.~M.~A.} \bibnamefont{Mohammed}}, \bibnamefont{and} \bibinfo{author}{\bibfnamefont{A.}~\bibnamefont{Papa}}, \bibinfo{journal}{Eur. Phys. J. C} \textbf{\bibinfo{volume}{81}}, \bibinfo{pages}{293} (\bibinfo{year}{2021}{\natexlab{c}}), \eprint{2008.00501}.

\bibitem[{\citenamefont{Bolognino et~al.}(2021{\natexlab{a}})\citenamefont{Bolognino, Celiberto, Fucilla, Ivanov, and Papa}}]{Bolognino:2021mrc}
\bibinfo{author}{\bibfnamefont{A.~D.} \bibnamefont{Bolognino}}, \bibinfo{author}{\bibfnamefont{F.~G.} \bibnamefont{Celiberto}}, \bibinfo{author}{\bibfnamefont{M.}~\bibnamefont{Fucilla}}, \bibinfo{author}{\bibfnamefont{D.~{\relax Yu}.} \bibnamefont{Ivanov}}, \bibnamefont{and} \bibinfo{author}{\bibfnamefont{A.}~\bibnamefont{Papa}}, \bibinfo{journal}{Phys. Rev. D} \textbf{\bibinfo{volume}{103}}, \bibinfo{pages}{094004} (\bibinfo{year}{2021}{\natexlab{a}}), \eprint{2103.07396}.

\bibitem[{\citenamefont{van Hameren et~al.}(2022)\citenamefont{van Hameren, Motyka, and Ziarko}}]{vanHameren:2022mtk}
\bibinfo{author}{\bibfnamefont{A.}~\bibnamefont{van Hameren}}, \bibinfo{author}{\bibfnamefont{L.}~\bibnamefont{Motyka}}, \bibnamefont{and} \bibinfo{author}{\bibfnamefont{G.}~\bibnamefont{Ziarko}}, \bibinfo{journal}{JHEP} \textbf{\bibinfo{volume}{11}}, \bibinfo{pages}{103} (\bibinfo{year}{2022}), \eprint{2205.09585}.

\bibitem[{\citenamefont{Giachino et~al.}(2024)\citenamefont{Giachino, van Hameren, and Ziarko}}]{Giachino:2023loc}
\bibinfo{author}{\bibfnamefont{A.}~\bibnamefont{Giachino}}, \bibinfo{author}{\bibfnamefont{A.}~\bibnamefont{van Hameren}}, \bibnamefont{and} \bibinfo{author}{\bibfnamefont{G.}~\bibnamefont{Ziarko}}, \bibinfo{journal}{JHEP} \textbf{\bibinfo{volume}{06}}, \bibinfo{pages}{167} (\bibinfo{year}{2024}), \eprint{2312.02808}.

\bibitem[{\citenamefont{Bonvini and Marzani}(2018)}]{Bonvini:2018ixe}
\bibinfo{author}{\bibfnamefont{M.}~\bibnamefont{Bonvini}} \bibnamefont{and} \bibinfo{author}{\bibfnamefont{S.}~\bibnamefont{Marzani}}, \bibinfo{journal}{Phys. Rev. Lett.} \textbf{\bibinfo{volume}{120}}, \bibinfo{pages}{202003} (\bibinfo{year}{2018}), \eprint{1802.07758}.

\bibitem[{\citenamefont{Silvetti and Bonvini}(2023)}]{Silvetti:2022hyc}
\bibinfo{author}{\bibfnamefont{F.}~\bibnamefont{Silvetti}} \bibnamefont{and} \bibinfo{author}{\bibfnamefont{M.}~\bibnamefont{Bonvini}}, \bibinfo{journal}{Eur. Phys. J. C} \textbf{\bibinfo{volume}{83}}, \bibinfo{pages}{267} (\bibinfo{year}{2023}), \eprint{2211.10142}.

\bibitem[{\citenamefont{Silvetti}(2023)}]{Silvetti:2023suu}
\bibinfo{author}{\bibfnamefont{F.}~\bibnamefont{Silvetti}}, \bibinfo{type}{Phd thesis}, \bibinfo{school}{Rome U.} (\bibinfo{year}{2023}), \eprint{2403.20315}.

\bibitem[{\citenamefont{Rinaudo}(2024)}]{Rinaudo:2024hdb}
\bibinfo{author}{\bibfnamefont{A.}~\bibnamefont{Rinaudo}}, \bibinfo{type}{Phd thesis}, \bibinfo{school}{Genoa U.} (\bibinfo{year}{2024}).

\bibitem[{\citenamefont{Mueller and Navelet}(1987)}]{Mueller:1986ey}
\bibinfo{author}{\bibfnamefont{A.~H.} \bibnamefont{Mueller}} \bibnamefont{and} \bibinfo{author}{\bibfnamefont{H.}~\bibnamefont{Navelet}}, \bibinfo{journal}{Nucl. Phys. B} \textbf{\bibinfo{volume}{282}}, \bibinfo{pages}{727} (\bibinfo{year}{1987}).

\bibitem[{\citenamefont{Duclou\'e et~al.}(2013)\citenamefont{Duclou\'e, Szymanowski, and Wallon}}]{Ducloue:2013hia}
\bibinfo{author}{\bibfnamefont{B.}~\bibnamefont{Duclou\'e}}, \bibinfo{author}{\bibfnamefont{L.}~\bibnamefont{Szymanowski}}, \bibnamefont{and} \bibinfo{author}{\bibfnamefont{S.}~\bibnamefont{Wallon}}, \bibinfo{journal}{JHEP} \textbf{\bibinfo{volume}{05}}, \bibinfo{pages}{096} (\bibinfo{year}{2013}), \eprint{1302.7012}.

\bibitem[{\citenamefont{Duclou\'e et~al.}(2014{\natexlab{a}})\citenamefont{Duclou\'e, Szymanowski, and Wallon}}]{Ducloue:2013bva}
\bibinfo{author}{\bibfnamefont{B.}~\bibnamefont{Duclou\'e}}, \bibinfo{author}{\bibfnamefont{L.}~\bibnamefont{Szymanowski}}, \bibnamefont{and} \bibinfo{author}{\bibfnamefont{S.}~\bibnamefont{Wallon}}, \bibinfo{journal}{Phys. Rev. Lett.} \textbf{\bibinfo{volume}{112}}, \bibinfo{pages}{082003} (\bibinfo{year}{2014}{\natexlab{a}}), \eprint{1309.3229}.

\bibitem[{\citenamefont{Caporale et~al.}(2014)\citenamefont{Caporale, Ivanov, Murdaca, and Papa}}]{Caporale:2014gpa}
\bibinfo{author}{\bibfnamefont{F.}~\bibnamefont{Caporale}}, \bibinfo{author}{\bibfnamefont{D.~{\relax Yu}.} \bibnamefont{Ivanov}}, \bibinfo{author}{\bibfnamefont{B.}~\bibnamefont{Murdaca}}, \bibnamefont{and} \bibinfo{author}{\bibfnamefont{A.}~\bibnamefont{Papa}}, \bibinfo{journal}{Eur. Phys. J. C} \textbf{\bibinfo{volume}{74}}, \bibinfo{pages}{3084} (\bibinfo{year}{2014}), \bibinfo{note}{[Erratum: Eur.Phys.J.C 75, 535 (2015)]}, \eprint{1407.8431}.

\bibitem[{\citenamefont{Celiberto et~al.}(2015{\natexlab{a}})\citenamefont{Celiberto, Ivanov, Murdaca, and Papa}}]{Celiberto:2015yba}
\bibinfo{author}{\bibfnamefont{F.~G.} \bibnamefont{Celiberto}}, \bibinfo{author}{\bibfnamefont{D.~{\relax Yu}.} \bibnamefont{Ivanov}}, \bibinfo{author}{\bibfnamefont{B.}~\bibnamefont{Murdaca}}, \bibnamefont{and} \bibinfo{author}{\bibfnamefont{A.}~\bibnamefont{Papa}}, \bibinfo{journal}{Eur. Phys. J. C} \textbf{\bibinfo{volume}{75}}, \bibinfo{pages}{292} (\bibinfo{year}{2015}{\natexlab{a}}), \eprint{1504.08233}.

\bibitem[{\citenamefont{Celiberto et~al.}(2015{\natexlab{b}})\citenamefont{Celiberto, Ivanov, Murdaca, and Papa}}]{Celiberto:2015mpa}
\bibinfo{author}{\bibfnamefont{F.~G.} \bibnamefont{Celiberto}}, \bibinfo{author}{\bibfnamefont{D.~{\relax Yu}.} \bibnamefont{Ivanov}}, \bibinfo{author}{\bibfnamefont{B.}~\bibnamefont{Murdaca}}, \bibnamefont{and} \bibinfo{author}{\bibfnamefont{A.}~\bibnamefont{Papa}}, \bibinfo{journal}{Acta Phys. Polon. Supp.} \textbf{\bibinfo{volume}{8}}, \bibinfo{pages}{935} (\bibinfo{year}{2015}{\natexlab{b}}), \eprint{1510.01626}.

\bibitem[{\citenamefont{Celiberto et~al.}(2016{\natexlab{b}})\citenamefont{Celiberto, Ivanov, Murdaca, and Papa}}]{Celiberto:2016ygs}
\bibinfo{author}{\bibfnamefont{F.~G.} \bibnamefont{Celiberto}}, \bibinfo{author}{\bibfnamefont{D.~{\relax Yu}.} \bibnamefont{Ivanov}}, \bibinfo{author}{\bibfnamefont{B.}~\bibnamefont{Murdaca}}, \bibnamefont{and} \bibinfo{author}{\bibfnamefont{A.}~\bibnamefont{Papa}}, \bibinfo{journal}{Eur. Phys. J. C} \textbf{\bibinfo{volume}{76}}, \bibinfo{pages}{224} (\bibinfo{year}{2016}{\natexlab{b}}), \eprint{1601.07847}.

\bibitem[{\citenamefont{Celiberto and Papa}(2022)}]{Celiberto:2022gji}
\bibinfo{author}{\bibfnamefont{F.~G.} \bibnamefont{Celiberto}} \bibnamefont{and} \bibinfo{author}{\bibfnamefont{A.}~\bibnamefont{Papa}}, \bibinfo{journal}{Phys. Rev. D} \textbf{\bibinfo{volume}{106}}, \bibinfo{pages}{114004} (\bibinfo{year}{2022}), \eprint{2207.05015}.

\bibitem[{\citenamefont{Caporale et~al.}(2018)\citenamefont{Caporale, Celiberto, Chachamis, Gordo~G\'omez, and Sabio~Vera}}]{Caporale:2018qnm}
\bibinfo{author}{\bibfnamefont{F.}~\bibnamefont{Caporale}}, \bibinfo{author}{\bibfnamefont{F.~G.} \bibnamefont{Celiberto}}, \bibinfo{author}{\bibfnamefont{G.}~\bibnamefont{Chachamis}}, \bibinfo{author}{\bibfnamefont{D.}~\bibnamefont{Gordo~G\'omez}}, \bibnamefont{and} \bibinfo{author}{\bibfnamefont{A.}~\bibnamefont{Sabio~Vera}}, \bibinfo{journal}{Nucl. Phys. B} \textbf{\bibinfo{volume}{935}}, \bibinfo{pages}{412} (\bibinfo{year}{2018}), \eprint{1806.06309}.

\bibitem[{\citenamefont{de~Le\'on et~al.}(2021)\citenamefont{de~Le\'on, Chachamis, and Sabio~Vera}}]{deLeon:2021ecb}
\bibinfo{author}{\bibfnamefont{N.~B.} \bibnamefont{de~Le\'on}}, \bibinfo{author}{\bibfnamefont{G.}~\bibnamefont{Chachamis}}, \bibnamefont{and} \bibinfo{author}{\bibfnamefont{A.}~\bibnamefont{Sabio~Vera}}, \bibinfo{journal}{Eur. Phys. J. C} \textbf{\bibinfo{volume}{81}}, \bibinfo{pages}{1019} (\bibinfo{year}{2021}), \eprint{2106.11255}.

\bibitem[{\citenamefont{Baldenegro et~al.}(2024)\citenamefont{Baldenegro, Chachamis, Kampshoff, Klasen, Milhano, Royon, and Sabio~Vera}}]{Baldenegro:2024ndr}
\bibinfo{author}{\bibfnamefont{C.}~\bibnamefont{Baldenegro}}, \bibinfo{author}{\bibfnamefont{G.}~\bibnamefont{Chachamis}}, \bibinfo{author}{\bibfnamefont{M.}~\bibnamefont{Kampshoff}}, \bibinfo{author}{\bibfnamefont{M.}~\bibnamefont{Klasen}}, \bibinfo{author}{\bibfnamefont{G.~J.} \bibnamefont{Milhano}}, \bibinfo{author}{\bibfnamefont{C.}~\bibnamefont{Royon}}, \bibnamefont{and} \bibinfo{author}{\bibfnamefont{A.}~\bibnamefont{Sabio~Vera}}, \bibinfo{journal}{Phys. Rev. D} \textbf{\bibinfo{volume}{110}}, \bibinfo{pages}{114027} (\bibinfo{year}{2024}), \eprint{2406.10681}.

\bibitem[{\citenamefont{Egorov and Kim}(2023)}]{Egorov:2023duz}
\bibinfo{author}{\bibfnamefont{A.~I.} \bibnamefont{Egorov}} \bibnamefont{and} \bibinfo{author}{\bibfnamefont{V.~T.} \bibnamefont{Kim}}, \bibinfo{journal}{Phys. Rev. D} \textbf{\bibinfo{volume}{108}}, \bibinfo{pages}{014010} (\bibinfo{year}{2023}), \eprint{2305.19854}.

\bibitem[{\citenamefont{Celiberto et~al.}(2020)\citenamefont{Celiberto, Ivanov, and Papa}}]{Celiberto:2020rxb}
\bibinfo{author}{\bibfnamefont{F.~G.} \bibnamefont{Celiberto}}, \bibinfo{author}{\bibfnamefont{D.~{\relax Yu}.} \bibnamefont{Ivanov}}, \bibnamefont{and} \bibinfo{author}{\bibfnamefont{A.}~\bibnamefont{Papa}}, \bibinfo{journal}{Phys. Rev. D} \textbf{\bibinfo{volume}{102}}, \bibinfo{pages}{094019} (\bibinfo{year}{2020}), \eprint{2008.10513}.

\bibitem[{\citenamefont{Bolognino et~al.}(2019{\natexlab{a}})\citenamefont{Bolognino, Celiberto, Ivanov, Mohammed, and Papa}}]{Bolognino:2019cac}
\bibinfo{author}{\bibfnamefont{A.~D.} \bibnamefont{Bolognino}}, \bibinfo{author}{\bibfnamefont{F.~G.} \bibnamefont{Celiberto}}, \bibinfo{author}{\bibfnamefont{D.~{\relax Yu}.} \bibnamefont{Ivanov}}, \bibinfo{author}{\bibfnamefont{M.~M.~A.} \bibnamefont{Mohammed}}, \bibnamefont{and} \bibinfo{author}{\bibfnamefont{A.}~\bibnamefont{Papa}}, \bibinfo{journal}{PoS} \textbf{\bibinfo{volume}{DIS2019}}, \bibinfo{pages}{049} (\bibinfo{year}{2019}{\natexlab{a}}), \eprint{1906.11800}.

\bibitem[{\citenamefont{Bolognino et~al.}(2019{\natexlab{b}})\citenamefont{Bolognino, Celiberto, Ivanov, Mohammed, and Papa}}]{Bolognino:2019yqj}
\bibinfo{author}{\bibfnamefont{A.~D.} \bibnamefont{Bolognino}}, \bibinfo{author}{\bibfnamefont{F.~G.} \bibnamefont{Celiberto}}, \bibinfo{author}{\bibfnamefont{D.~{\relax Yu}.} \bibnamefont{Ivanov}}, \bibinfo{author}{\bibfnamefont{M.~M.~A.} \bibnamefont{Mohammed}}, \bibnamefont{and} \bibinfo{author}{\bibfnamefont{A.}~\bibnamefont{Papa}}, \bibinfo{journal}{Acta Phys. Polon. Supp.} \textbf{\bibinfo{volume}{12}}, \bibinfo{pages}{773} (\bibinfo{year}{2019}{\natexlab{b}}), \eprint{1902.04511}.

\bibitem[{\citenamefont{Celiberto et~al.}(2018{\natexlab{a}})\citenamefont{Celiberto, Gordo~G\'omez, and Sabio~Vera}}]{Celiberto:2018muu}
\bibinfo{author}{\bibfnamefont{F.~G.} \bibnamefont{Celiberto}}, \bibinfo{author}{\bibfnamefont{D.}~\bibnamefont{Gordo~G\'omez}}, \bibnamefont{and} \bibinfo{author}{\bibfnamefont{A.}~\bibnamefont{Sabio~Vera}}, \bibinfo{journal}{Phys. Lett.} \textbf{\bibinfo{volume}{B786}}, \bibinfo{pages}{201} (\bibinfo{year}{2018}{\natexlab{a}}), \eprint{1808.09511}.

\bibitem[{\citenamefont{Golec-Biernat et~al.}(2018)\citenamefont{Golec-Biernat, Motyka, and Stebel}}]{Golec-Biernat:2018kem}
\bibinfo{author}{\bibfnamefont{K.}~\bibnamefont{Golec-Biernat}}, \bibinfo{author}{\bibfnamefont{L.}~\bibnamefont{Motyka}}, \bibnamefont{and} \bibinfo{author}{\bibfnamefont{T.}~\bibnamefont{Stebel}}, \bibinfo{journal}{JHEP} \textbf{\bibinfo{volume}{12}}, \bibinfo{pages}{091} (\bibinfo{year}{2018}), \eprint{1811.04361}.

\bibitem[{\citenamefont{Celiberto and Papa}(2023)}]{Celiberto:2023rtu}
\bibinfo{author}{\bibfnamefont{F.~G.} \bibnamefont{Celiberto}} \bibnamefont{and} \bibinfo{author}{\bibfnamefont{A.}~\bibnamefont{Papa}} (\bibinfo{year}{2023}), \eprint{2305.00962}.

\bibitem[{\citenamefont{Celiberto et~al.}(2023{\natexlab{a}})\citenamefont{Celiberto, Delle~Rose, Fucilla, Gatto, and Papa}}]{Celiberto:2023uuk}
\bibinfo{author}{\bibfnamefont{F.~G.} \bibnamefont{Celiberto}}, \bibinfo{author}{\bibfnamefont{L.}~\bibnamefont{Delle~Rose}}, \bibinfo{author}{\bibfnamefont{M.}~\bibnamefont{Fucilla}}, \bibinfo{author}{\bibfnamefont{G.}~\bibnamefont{Gatto}}, \bibnamefont{and} \bibinfo{author}{\bibfnamefont{A.}~\bibnamefont{Papa}}, in \emph{\bibinfo{booktitle}{{57th Rencontres de Moriond on QCD and High Energy Interactions}}} (\bibinfo{year}{2023}{\natexlab{a}}), \eprint{2305.05052}.

\bibitem[{\citenamefont{Celiberto et~al.}(2024{\natexlab{b}})\citenamefont{Celiberto, Delle~Rose, Fucilla, Gatto, and Papa}}]{Celiberto:2023eba}
\bibinfo{author}{\bibfnamefont{F.~G.} \bibnamefont{Celiberto}}, \bibinfo{author}{\bibfnamefont{L.}~\bibnamefont{Delle~Rose}}, \bibinfo{author}{\bibfnamefont{M.}~\bibnamefont{Fucilla}}, \bibinfo{author}{\bibfnamefont{G.}~\bibnamefont{Gatto}}, \bibnamefont{and} \bibinfo{author}{\bibfnamefont{A.}~\bibnamefont{Papa}}, \bibinfo{journal}{PoS} \textbf{\bibinfo{volume}{RADCOR2023}}, \bibinfo{pages}{069} (\bibinfo{year}{2024}{\natexlab{b}}), \eprint{2309.11573}.

\bibitem[{\citenamefont{Celiberto et~al.}(2024{\natexlab{c}})\citenamefont{Celiberto, Delle~Rose, Fucilla, Gatto, and Papa}}]{Celiberto:2023nym}
\bibinfo{author}{\bibfnamefont{F.~G.} \bibnamefont{Celiberto}}, \bibinfo{author}{\bibfnamefont{L.}~\bibnamefont{Delle~Rose}}, \bibinfo{author}{\bibfnamefont{M.}~\bibnamefont{Fucilla}}, \bibinfo{author}{\bibfnamefont{G.}~\bibnamefont{Gatto}}, \bibnamefont{and} \bibinfo{author}{\bibfnamefont{A.}~\bibnamefont{Papa}}, \bibinfo{journal}{PoS} \textbf{\bibinfo{volume}{EPS-HEP2023}}, \bibinfo{pages}{390} (\bibinfo{year}{2024}{\natexlab{c}}), \eprint{2310.16967}.

\bibitem[{\citenamefont{Celiberto et~al.}(2024{\natexlab{d}})\citenamefont{Celiberto, Fucilla, Mohammed, Ivanov, and Papa}}]{Celiberto:2023rqp}
\bibinfo{author}{\bibfnamefont{F.~G.} \bibnamefont{Celiberto}}, \bibinfo{author}{\bibfnamefont{M.}~\bibnamefont{Fucilla}}, \bibinfo{author}{\bibfnamefont{M.~M.~A.} \bibnamefont{Mohammed}}, \bibinfo{author}{\bibfnamefont{D.~{\relax Yu}.} \bibnamefont{Ivanov}}, \bibnamefont{and} \bibinfo{author}{\bibfnamefont{A.}~\bibnamefont{Papa}}, \bibinfo{journal}{PoS} \textbf{\bibinfo{volume}{RADCOR2023}}, \bibinfo{pages}{091} (\bibinfo{year}{2024}{\natexlab{d}}), \eprint{2309.07570}.

\bibitem[{\citenamefont{Nefedov}(2019)}]{Nefedov:2019mrg}
\bibinfo{author}{\bibfnamefont{M.~A.} \bibnamefont{Nefedov}}, \bibinfo{journal}{Nucl. Phys. B} \textbf{\bibinfo{volume}{946}}, \bibinfo{pages}{114715} (\bibinfo{year}{2019}), \eprint{1902.11030}.

\bibitem[{\citenamefont{Celiberto et~al.}(2018{\natexlab{b}})\citenamefont{Celiberto, Ivanov, Murdaca, and Papa}}]{Celiberto:2017nyx}
\bibinfo{author}{\bibfnamefont{F.~G.} \bibnamefont{Celiberto}}, \bibinfo{author}{\bibfnamefont{D.~{\relax Yu}.} \bibnamefont{Ivanov}}, \bibinfo{author}{\bibfnamefont{B.}~\bibnamefont{Murdaca}}, \bibnamefont{and} \bibinfo{author}{\bibfnamefont{A.}~\bibnamefont{Papa}}, \bibinfo{journal}{Phys. Lett. B} \textbf{\bibinfo{volume}{777}}, \bibinfo{pages}{141} (\bibinfo{year}{2018}{\natexlab{b}}), \eprint{1709.10032}.

\bibitem[{\citenamefont{Boussarie et~al.}(2018)\citenamefont{Boussarie, Duclou\'e, Szymanowski, and Wallon}}]{Boussarie:2017oae}
\bibinfo{author}{\bibfnamefont{R.}~\bibnamefont{Boussarie}}, \bibinfo{author}{\bibfnamefont{B.}~\bibnamefont{Duclou\'e}}, \bibinfo{author}{\bibfnamefont{L.}~\bibnamefont{Szymanowski}}, \bibnamefont{and} \bibinfo{author}{\bibfnamefont{S.}~\bibnamefont{Wallon}}, \bibinfo{journal}{Phys. Rev. D} \textbf{\bibinfo{volume}{97}}, \bibinfo{pages}{014008} (\bibinfo{year}{2018}), \eprint{1709.01380}.

\bibitem[{\citenamefont{Bolognino et~al.}(2019{\natexlab{c}})\citenamefont{Bolognino, Celiberto, Fucilla, Ivanov, Murdaca, and Papa}}]{Bolognino:2019ouc}
\bibinfo{author}{\bibfnamefont{A.~D.} \bibnamefont{Bolognino}}, \bibinfo{author}{\bibfnamefont{F.~G.} \bibnamefont{Celiberto}}, \bibinfo{author}{\bibfnamefont{M.}~\bibnamefont{Fucilla}}, \bibinfo{author}{\bibfnamefont{D.~{\relax Yu}.} \bibnamefont{Ivanov}}, \bibinfo{author}{\bibfnamefont{B.}~\bibnamefont{Murdaca}}, \bibnamefont{and} \bibinfo{author}{\bibfnamefont{A.}~\bibnamefont{Papa}}, \bibinfo{journal}{PoS} \textbf{\bibinfo{volume}{DIS2019}}, \bibinfo{pages}{067} (\bibinfo{year}{2019}{\natexlab{c}}), \eprint{1906.05940}.

\bibitem[{\citenamefont{Bolognino et~al.}(2019{\natexlab{d}})\citenamefont{Bolognino, Celiberto, Fucilla, Ivanov, and Papa}}]{Bolognino:2019yls}
\bibinfo{author}{\bibfnamefont{A.~D.} \bibnamefont{Bolognino}}, \bibinfo{author}{\bibfnamefont{F.~G.} \bibnamefont{Celiberto}}, \bibinfo{author}{\bibfnamefont{M.}~\bibnamefont{Fucilla}}, \bibinfo{author}{\bibfnamefont{D.~{\relax Yu}.} \bibnamefont{Ivanov}}, \bibnamefont{and} \bibinfo{author}{\bibfnamefont{A.}~\bibnamefont{Papa}}, \bibinfo{journal}{Eur. Phys. J. C} \textbf{\bibinfo{volume}{79}}, \bibinfo{pages}{939} (\bibinfo{year}{2019}{\natexlab{d}}), \eprint{1909.03068}.

\bibitem[{\citenamefont{Bolognino et~al.}(2023)\citenamefont{Bolognino, Celiberto, Fucilla, Ivanov, Mohammed, and Papa}}]{Bolognino:2022paj}
\bibinfo{author}{\bibfnamefont{A.~D.} \bibnamefont{Bolognino}}, \bibinfo{author}{\bibfnamefont{F.~G.} \bibnamefont{Celiberto}}, \bibinfo{author}{\bibfnamefont{M.}~\bibnamefont{Fucilla}}, \bibinfo{author}{\bibfnamefont{D.~{\relax Yu}.} \bibnamefont{Ivanov}}, \bibinfo{author}{\bibfnamefont{M.~M.~A.} \bibnamefont{Mohammed}}, \bibnamefont{and} \bibinfo{author}{\bibfnamefont{A.}~\bibnamefont{Papa}}, \bibinfo{journal}{Acta Phys. Polon. Supp.} \textbf{\bibinfo{volume}{16}}, \bibinfo{pages}{17} (\bibinfo{year}{2023}), \eprint{2211.16818}.

\bibitem[{\citenamefont{Celiberto and Fucilla}(2022{\natexlab{b}})}]{Celiberto:2022kza}
\bibinfo{author}{\bibfnamefont{F.~G.} \bibnamefont{Celiberto}} \bibnamefont{and} \bibinfo{author}{\bibfnamefont{M.}~\bibnamefont{Fucilla}}, in \emph{\bibinfo{booktitle}{{29th International Workshop on Deep-Inelastic Scattering and Related Subjects}}} (\bibinfo{year}{2022}{\natexlab{b}}), \eprint{2208.07206}.

\bibitem[{\citenamefont{Anikin et~al.}(2011)\citenamefont{Anikin, Besse, Ivanov, Pire, Szymanowski, and Wallon}}]{Anikin:2011sa}
\bibinfo{author}{\bibfnamefont{I.}~\bibnamefont{Anikin}}, \bibinfo{author}{\bibfnamefont{A.}~\bibnamefont{Besse}}, \bibinfo{author}{\bibfnamefont{D.~{\relax Yu}.} \bibnamefont{Ivanov}}, \bibinfo{author}{\bibfnamefont{B.}~\bibnamefont{Pire}}, \bibinfo{author}{\bibfnamefont{L.}~\bibnamefont{Szymanowski}}, \bibnamefont{and} \bibinfo{author}{\bibfnamefont{S.}~\bibnamefont{Wallon}}, \bibinfo{journal}{Phys. Rev. D} \textbf{\bibinfo{volume}{84}}, \bibinfo{pages}{054004} (\bibinfo{year}{2011}), \eprint{1105.1761}.

\bibitem[{\citenamefont{Bolognino et~al.}(2018{\natexlab{b}})\citenamefont{Bolognino, Celiberto, Ivanov, and Papa}}]{Bolognino:2018rhb}
\bibinfo{author}{\bibfnamefont{A.~D.} \bibnamefont{Bolognino}}, \bibinfo{author}{\bibfnamefont{F.~G.} \bibnamefont{Celiberto}}, \bibinfo{author}{\bibfnamefont{D.~{\relax Yu}.} \bibnamefont{Ivanov}}, \bibnamefont{and} \bibinfo{author}{\bibfnamefont{A.}~\bibnamefont{Papa}}, \bibinfo{journal}{Eur. Phys. J.} \textbf{\bibinfo{volume}{C78}}, \bibinfo{pages}{1023} (\bibinfo{year}{2018}{\natexlab{b}}), \eprint{1808.02395}.

\bibitem[{\citenamefont{Bolognino et~al.}(2018{\natexlab{c}})\citenamefont{Bolognino, Celiberto, Ivanov, and Papa}}]{Bolognino:2018mlw}
\bibinfo{author}{\bibfnamefont{A.~D.} \bibnamefont{Bolognino}}, \bibinfo{author}{\bibfnamefont{F.~G.} \bibnamefont{Celiberto}}, \bibinfo{author}{\bibfnamefont{D.~{\relax Yu}.} \bibnamefont{Ivanov}}, \bibnamefont{and} \bibinfo{author}{\bibfnamefont{A.}~\bibnamefont{Papa}}, \bibinfo{journal}{Frascati Phys. Ser.} \textbf{\bibinfo{volume}{67}}, \bibinfo{pages}{76} (\bibinfo{year}{2018}{\natexlab{c}}), \eprint{1808.02958}.

\bibitem[{\citenamefont{Bolognino et~al.}(2019{\natexlab{e}})\citenamefont{Bolognino, Celiberto, Ivanov, and Papa}}]{Bolognino:2019bko}
\bibinfo{author}{\bibfnamefont{A.~D.} \bibnamefont{Bolognino}}, \bibinfo{author}{\bibfnamefont{F.~G.} \bibnamefont{Celiberto}}, \bibinfo{author}{\bibfnamefont{D.~{\relax Yu}.} \bibnamefont{Ivanov}}, \bibnamefont{and} \bibinfo{author}{\bibfnamefont{A.}~\bibnamefont{Papa}}, \bibinfo{journal}{Acta Phys. Polon. Supp.} \textbf{\bibinfo{volume}{12}}, \bibinfo{pages}{891} (\bibinfo{year}{2019}{\natexlab{e}}), \eprint{1902.04520}.

\bibitem[{\citenamefont{Bolognino et~al.}(2020)\citenamefont{Bolognino, Szczurek, and Schaefer}}]{Bolognino:2019pba}
\bibinfo{author}{\bibfnamefont{A.~D.} \bibnamefont{Bolognino}}, \bibinfo{author}{\bibfnamefont{A.}~\bibnamefont{Szczurek}}, \bibnamefont{and} \bibinfo{author}{\bibfnamefont{W.}~\bibnamefont{Schaefer}}, \bibinfo{journal}{Phys. Rev. D} \textbf{\bibinfo{volume}{101}}, \bibinfo{pages}{054041} (\bibinfo{year}{2020}), \eprint{1912.06507}.

\bibitem[{\citenamefont{Celiberto}(2019)}]{Celiberto:2019slj}
\bibinfo{author}{\bibfnamefont{F.~G.} \bibnamefont{Celiberto}}, \bibinfo{journal}{Nuovo Cim.} \textbf{\bibinfo{volume}{C42}}, \bibinfo{pages}{220} (\bibinfo{year}{2019}), \eprint{1912.11313}.

\bibitem[{\citenamefont{\L{}uszczak et~al.}(2022)\citenamefont{\L{}uszczak, \L{}uszczak, and Sch\"afer}}]{Luszczak:2022fkf}
\bibinfo{author}{\bibfnamefont{A.}~\bibnamefont{\L{}uszczak}}, \bibinfo{author}{\bibfnamefont{M.}~\bibnamefont{\L{}uszczak}}, \bibnamefont{and} \bibinfo{author}{\bibfnamefont{W.}~\bibnamefont{Sch\"afer}}, \bibinfo{journal}{Phys. Lett. B} \textbf{\bibinfo{volume}{835}}, \bibinfo{pages}{137582} (\bibinfo{year}{2022}), \eprint{2210.02877}.

\bibitem[{\citenamefont{Bolognino et~al.}(2021{\natexlab{b}})\citenamefont{Bolognino, Celiberto, Ivanov, Papa, Sch\"afer, and Szczurek}}]{Bolognino:2021niq}
\bibinfo{author}{\bibfnamefont{A.~D.} \bibnamefont{Bolognino}}, \bibinfo{author}{\bibfnamefont{F.~G.} \bibnamefont{Celiberto}}, \bibinfo{author}{\bibfnamefont{D.~{\relax Yu}.} \bibnamefont{Ivanov}}, \bibinfo{author}{\bibfnamefont{A.}~\bibnamefont{Papa}}, \bibinfo{author}{\bibfnamefont{W.}~\bibnamefont{Sch\"afer}}, \bibnamefont{and} \bibinfo{author}{\bibfnamefont{A.}~\bibnamefont{Szczurek}}, \bibinfo{journal}{Eur. Phys. J. C} \textbf{\bibinfo{volume}{81}}, \bibinfo{pages}{846} (\bibinfo{year}{2021}{\natexlab{b}}), \eprint{2107.13415}.

\bibitem[{\citenamefont{Bolognino et~al.}(2022{\natexlab{a}})\citenamefont{Bolognino, Celiberto, Ivanov, and Papa}}]{Bolognino:2021gjm}
\bibinfo{author}{\bibfnamefont{A.~D.} \bibnamefont{Bolognino}}, \bibinfo{author}{\bibfnamefont{F.~G.} \bibnamefont{Celiberto}}, \bibinfo{author}{\bibfnamefont{D.~{\relax Yu}.} \bibnamefont{Ivanov}}, \bibnamefont{and} \bibinfo{author}{\bibfnamefont{A.}~\bibnamefont{Papa}}, \bibinfo{journal}{SciPost Phys. Proc.} \textbf{\bibinfo{volume}{8}}, \bibinfo{pages}{089} (\bibinfo{year}{2022}{\natexlab{a}}), \eprint{2107.12725}.

\bibitem[{\citenamefont{Bolognino et~al.}(2022{\natexlab{b}})\citenamefont{Bolognino, Celiberto, Fucilla, Ivanov, Papa, Sch\"afer, and Szczurek}}]{Bolognino:2022uty}
\bibinfo{author}{\bibfnamefont{A.~D.} \bibnamefont{Bolognino}}, \bibinfo{author}{\bibfnamefont{F.~G.} \bibnamefont{Celiberto}}, \bibinfo{author}{\bibfnamefont{M.}~\bibnamefont{Fucilla}}, \bibinfo{author}{\bibfnamefont{D.~{\relax Yu}.} \bibnamefont{Ivanov}}, \bibinfo{author}{\bibfnamefont{A.}~\bibnamefont{Papa}}, \bibinfo{author}{\bibfnamefont{W.}~\bibnamefont{Sch\"afer}}, \bibnamefont{and} \bibinfo{author}{\bibfnamefont{A.}~\bibnamefont{Szczurek}}, \bibinfo{journal}{Rev. Mex. Fis. Suppl.} \textbf{\bibinfo{volume}{3}}, \bibinfo{pages}{0308109} (\bibinfo{year}{2022}{\natexlab{b}}), \eprint{2202.02513}.

\bibitem[{\citenamefont{Bolognino et~al.}(2022{\natexlab{c}})\citenamefont{Bolognino, Celiberto, Ivanov, Papa, Sch\"afer, and Szczurek}}]{Bolognino:2022ndh}
\bibinfo{author}{\bibfnamefont{A.~D.} \bibnamefont{Bolognino}}, \bibinfo{author}{\bibfnamefont{F.~G.} \bibnamefont{Celiberto}}, \bibinfo{author}{\bibfnamefont{D.~{\relax Yu.}.} \bibnamefont{Ivanov}}, \bibinfo{author}{\bibfnamefont{A.}~\bibnamefont{Papa}}, \bibinfo{author}{\bibfnamefont{W.}~\bibnamefont{Sch\"afer}}, \bibnamefont{and} \bibinfo{author}{\bibfnamefont{A.}~\bibnamefont{Szczurek}}, in \emph{\bibinfo{booktitle}{{29th International Workshop on Deep-Inelastic Scattering and Related Subjects}}} (\bibinfo{year}{2022}{\natexlab{c}}), \eprint{2207.05726}.

\bibitem[{\citenamefont{Arroyo~Garcia et~al.}(2019)\citenamefont{Arroyo~Garcia, Hentschinski, and Kutak}}]{Garcia:2019tne}
\bibinfo{author}{\bibfnamefont{A.}~\bibnamefont{Arroyo~Garcia}}, \bibinfo{author}{\bibfnamefont{M.}~\bibnamefont{Hentschinski}}, \bibnamefont{and} \bibinfo{author}{\bibfnamefont{K.}~\bibnamefont{Kutak}}, \bibinfo{journal}{Phys. Lett. B} \textbf{\bibinfo{volume}{795}}, \bibinfo{pages}{569} (\bibinfo{year}{2019}), \eprint{1904.04394}.

\bibitem[{\citenamefont{Hentschinski and Padr\'on~Molina}(2021)}]{Hentschinski:2020yfm}
\bibinfo{author}{\bibfnamefont{M.}~\bibnamefont{Hentschinski}} \bibnamefont{and} \bibinfo{author}{\bibfnamefont{E.}~\bibnamefont{Padr\'on~Molina}}, \bibinfo{journal}{Phys. Rev. D} \textbf{\bibinfo{volume}{103}}, \bibinfo{pages}{074008} (\bibinfo{year}{2021}), \eprint{2011.02640}.

\bibitem[{\citenamefont{Peredo and Hentschinski}(2024)}]{Peredo:2023oym}
\bibinfo{author}{\bibfnamefont{M.~A.} \bibnamefont{Peredo}} \bibnamefont{and} \bibinfo{author}{\bibfnamefont{M.}~\bibnamefont{Hentschinski}}, \bibinfo{journal}{Phys. Rev. D} \textbf{\bibinfo{volume}{109}}, \bibinfo{pages}{014032} (\bibinfo{year}{2024}), \eprint{2308.15430}.

\bibitem[{\citenamefont{Ball et~al.}(2018)\citenamefont{Ball, Bertone, Bonvini, Marzani, Rojo, and Rottoli}}]{Ball:2017otu}
\bibinfo{author}{\bibfnamefont{R.~D.} \bibnamefont{Ball}}, \bibinfo{author}{\bibfnamefont{V.}~\bibnamefont{Bertone}}, \bibinfo{author}{\bibfnamefont{M.}~\bibnamefont{Bonvini}}, \bibinfo{author}{\bibfnamefont{S.}~\bibnamefont{Marzani}}, \bibinfo{author}{\bibfnamefont{J.}~\bibnamefont{Rojo}}, \bibnamefont{and} \bibinfo{author}{\bibfnamefont{L.}~\bibnamefont{Rottoli}}, \bibinfo{journal}{Eur. Phys. J.} \textbf{\bibinfo{volume}{C78}}, \bibinfo{pages}{321} (\bibinfo{year}{2018}), \eprint{1710.05935}.

\bibitem[{\citenamefont{Abdolmaleki et~al.}(2018)}]{Abdolmaleki:2018jln}
\bibinfo{author}{\bibfnamefont{H.}~\bibnamefont{Abdolmaleki}} \bibnamefont{et~al.} (\bibinfo{collaboration}{xFitter Developers' Team}), \bibinfo{journal}{Eur. Phys. J. C} \textbf{\bibinfo{volume}{78}}, \bibinfo{pages}{621} (\bibinfo{year}{2018}), \eprint{1802.00064}.

\bibitem[{\citenamefont{Bonvini and Giuli}(2019)}]{Bonvini:2019wxf}
\bibinfo{author}{\bibfnamefont{M.}~\bibnamefont{Bonvini}} \bibnamefont{and} \bibinfo{author}{\bibfnamefont{F.}~\bibnamefont{Giuli}}, \bibinfo{journal}{Eur. Phys. J. Plus} \textbf{\bibinfo{volume}{134}}, \bibinfo{pages}{531} (\bibinfo{year}{2019}), \eprint{1902.11125}.

\bibitem[{\citenamefont{Bacchetta et~al.}(2020)\citenamefont{Bacchetta, Celiberto, Radici, and Taels}}]{Bacchetta:2020vty}
\bibinfo{author}{\bibfnamefont{A.}~\bibnamefont{Bacchetta}}, \bibinfo{author}{\bibfnamefont{F.~G.} \bibnamefont{Celiberto}}, \bibinfo{author}{\bibfnamefont{M.}~\bibnamefont{Radici}}, \bibnamefont{and} \bibinfo{author}{\bibfnamefont{P.}~\bibnamefont{Taels}}, \bibinfo{journal}{Eur. Phys. J. C} \textbf{\bibinfo{volume}{80}}, \bibinfo{pages}{733} (\bibinfo{year}{2020}), \eprint{2005.02288}.

\bibitem[{\citenamefont{Bacchetta et~al.}(2024{\natexlab{a}})\citenamefont{Bacchetta, Celiberto, and Radici}}]{Bacchetta:2024fci}
\bibinfo{author}{\bibfnamefont{A.}~\bibnamefont{Bacchetta}}, \bibinfo{author}{\bibfnamefont{F.~G.} \bibnamefont{Celiberto}}, \bibnamefont{and} \bibinfo{author}{\bibfnamefont{M.}~\bibnamefont{Radici}}, \bibinfo{journal}{Eur. Phys. J. C} \textbf{\bibinfo{volume}{84}}, \bibinfo{pages}{576} (\bibinfo{year}{2024}{\natexlab{a}}), \eprint{2402.17556}.

\bibitem[{\citenamefont{Celiberto}(2021{\natexlab{b}})}]{Celiberto:2021zww}
\bibinfo{author}{\bibfnamefont{F.~G.} \bibnamefont{Celiberto}}, \bibinfo{journal}{Nuovo Cim.} \textbf{\bibinfo{volume}{C44}}, \bibinfo{pages}{36} (\bibinfo{year}{2021}{\natexlab{b}}), \eprint{2101.04630}.

\bibitem[{\citenamefont{Bacchetta et~al.}(2022{\natexlab{a}})\citenamefont{Bacchetta, Celiberto, Radici, and Taels}}]{Bacchetta:2021oht}
\bibinfo{author}{\bibfnamefont{A.}~\bibnamefont{Bacchetta}}, \bibinfo{author}{\bibfnamefont{F.~G.} \bibnamefont{Celiberto}}, \bibinfo{author}{\bibfnamefont{M.}~\bibnamefont{Radici}}, \bibnamefont{and} \bibinfo{author}{\bibfnamefont{P.}~\bibnamefont{Taels}}, \bibinfo{journal}{SciPost Phys. Proc.} \textbf{\bibinfo{volume}{8}}, \bibinfo{pages}{040} (\bibinfo{year}{2022}{\natexlab{a}}), \eprint{2107.13446}.

\bibitem[{\citenamefont{Bacchetta et~al.}(2022{\natexlab{b}})\citenamefont{Bacchetta, Celiberto, and Radici}}]{Bacchetta:2021lvw}
\bibinfo{author}{\bibfnamefont{A.}~\bibnamefont{Bacchetta}}, \bibinfo{author}{\bibfnamefont{F.~G.} \bibnamefont{Celiberto}}, \bibnamefont{and} \bibinfo{author}{\bibfnamefont{M.}~\bibnamefont{Radici}}, \bibinfo{journal}{PoS} \textbf{\bibinfo{volume}{EPS-HEP2021}}, \bibinfo{pages}{376} (\bibinfo{year}{2022}{\natexlab{b}}), \eprint{2111.01686}.

\bibitem[{\citenamefont{Bacchetta et~al.}(2022{\natexlab{c}})\citenamefont{Bacchetta, Celiberto, and Radici}}]{Bacchetta:2021twk}
\bibinfo{author}{\bibfnamefont{A.}~\bibnamefont{Bacchetta}}, \bibinfo{author}{\bibfnamefont{F.~G.} \bibnamefont{Celiberto}}, \bibnamefont{and} \bibinfo{author}{\bibfnamefont{M.}~\bibnamefont{Radici}}, \bibinfo{journal}{PoS} \textbf{\bibinfo{volume}{PANIC2021}}, \bibinfo{pages}{378} (\bibinfo{year}{2022}{\natexlab{c}}), \eprint{2111.03567}.

\bibitem[{\citenamefont{Bacchetta et~al.}(2022{\natexlab{d}})\citenamefont{Bacchetta, Celiberto, and Radici}}]{Bacchetta:2022esb}
\bibinfo{author}{\bibfnamefont{A.}~\bibnamefont{Bacchetta}}, \bibinfo{author}{\bibfnamefont{F.~G.} \bibnamefont{Celiberto}}, \bibnamefont{and} \bibinfo{author}{\bibfnamefont{M.}~\bibnamefont{Radici}}, \bibinfo{journal}{JPS Conf. Proc.} \textbf{\bibinfo{volume}{37}}, \bibinfo{pages}{020124} (\bibinfo{year}{2022}{\natexlab{d}}), \eprint{2201.10508}.

\bibitem[{\citenamefont{Bacchetta et~al.}(2022{\natexlab{e}})\citenamefont{Bacchetta, Celiberto, and Radici}}]{Bacchetta:2022crh}
\bibinfo{author}{\bibfnamefont{A.}~\bibnamefont{Bacchetta}}, \bibinfo{author}{\bibfnamefont{F.~G.} \bibnamefont{Celiberto}}, \bibnamefont{and} \bibinfo{author}{\bibfnamefont{M.}~\bibnamefont{Radici}}, \bibinfo{journal}{Rev. Mex. Fis. Suppl.} \textbf{\bibinfo{volume}{3}}, \bibinfo{pages}{0308108} (\bibinfo{year}{2022}{\natexlab{e}}), \eprint{2206.07815}.

\bibitem[{\citenamefont{Bacchetta et~al.}(2022{\natexlab{f}})\citenamefont{Bacchetta, Celiberto, Radici, and Signori}}]{Bacchetta:2022nyv}
\bibinfo{author}{\bibfnamefont{A.}~\bibnamefont{Bacchetta}}, \bibinfo{author}{\bibfnamefont{F.~G.} \bibnamefont{Celiberto}}, \bibinfo{author}{\bibfnamefont{M.}~\bibnamefont{Radici}}, \bibnamefont{and} \bibinfo{author}{\bibfnamefont{A.}~\bibnamefont{Signori}}, in \emph{\bibinfo{booktitle}{{29th International Workshop on Deep-Inelastic Scattering and Related Subjects}}} (\bibinfo{year}{2022}{\natexlab{f}}), \eprint{2208.06252}.

\bibitem[{\citenamefont{Celiberto}(2022{\natexlab{c}})}]{Celiberto:2022omz}
\bibinfo{author}{\bibfnamefont{F.~G.} \bibnamefont{Celiberto}}, \bibinfo{journal}{Universe} \textbf{\bibinfo{volume}{8}}, \bibinfo{pages}{661} (\bibinfo{year}{2022}{\natexlab{c}}), \eprint{2210.08322}.

\bibitem[{\citenamefont{Bacchetta et~al.}(2024{\natexlab{b}})\citenamefont{Bacchetta, Celiberto, and Radici}}]{Bacchetta:2023zir}
\bibinfo{author}{\bibfnamefont{A.}~\bibnamefont{Bacchetta}}, \bibinfo{author}{\bibfnamefont{F.~G.} \bibnamefont{Celiberto}}, \bibnamefont{and} \bibinfo{author}{\bibfnamefont{M.}~\bibnamefont{Radici}}, \bibinfo{journal}{PoS} \textbf{\bibinfo{volume}{EPS-HEP2023}}, \bibinfo{pages}{247} (\bibinfo{year}{2024}{\natexlab{b}}), \eprint{2310.19916}.

\bibitem[{\citenamefont{Bacchetta et~al.}(2024{\natexlab{c}})\citenamefont{Bacchetta, Celiberto, and Radici}}]{Bacchetta:2024uxb}
\bibinfo{author}{\bibfnamefont{A.}~\bibnamefont{Bacchetta}}, \bibinfo{author}{\bibfnamefont{F.~G.} \bibnamefont{Celiberto}}, \bibnamefont{and} \bibinfo{author}{\bibfnamefont{M.}~\bibnamefont{Radici}}, \bibinfo{journal}{PoS} \textbf{\bibinfo{volume}{SPIN2023}}, \bibinfo{pages}{049} (\bibinfo{year}{2024}{\natexlab{c}}), \eprint{2406.04893}.

\bibitem[{\citenamefont{Hentschinski}(2021)}]{Hentschinski:2021lsh}
\bibinfo{author}{\bibfnamefont{M.}~\bibnamefont{Hentschinski}}, \bibinfo{journal}{Phys. Rev. D} \textbf{\bibinfo{volume}{104}}, \bibinfo{pages}{054014} (\bibinfo{year}{2021}), \eprint{2107.06203}.

\bibitem[{\citenamefont{Mukherjee et~al.}(2024)\citenamefont{Mukherjee, Skokov, Tarasov, and Tiwari}}]{Mukherjee:2023snp}
\bibinfo{author}{\bibfnamefont{S.}~\bibnamefont{Mukherjee}}, \bibinfo{author}{\bibfnamefont{V.~V.} \bibnamefont{Skokov}}, \bibinfo{author}{\bibfnamefont{A.}~\bibnamefont{Tarasov}}, \bibnamefont{and} \bibinfo{author}{\bibfnamefont{S.}~\bibnamefont{Tiwari}}, \bibinfo{journal}{Phys. Rev. D} \textbf{\bibinfo{volume}{109}}, \bibinfo{pages}{034035} (\bibinfo{year}{2024}), \eprint{2311.16402}.

\bibitem[{\citenamefont{Boroun}(2023)}]{Boroun:2023goy}
\bibinfo{author}{\bibfnamefont{G.~R.} \bibnamefont{Boroun}}, \bibinfo{journal}{Phys. Rev. D} \textbf{\bibinfo{volume}{108}}, \bibinfo{pages}{034025} (\bibinfo{year}{2023}), \eprint{2301.01083}.

\bibitem[{\citenamefont{Boroun}(2024)}]{Boroun:2023ldq}
\bibinfo{author}{\bibfnamefont{G.~R.} \bibnamefont{Boroun}}, \bibinfo{journal}{Eur. Phys. J. A} \textbf{\bibinfo{volume}{60}}, \bibinfo{pages}{48} (\bibinfo{year}{2024}), \eprint{2309.04832}.

\bibitem[{\citenamefont{Caporale et~al.}(2013{\natexlab{a}})\citenamefont{Caporale, Ivanov, Murdaca, and Papa}}]{Caporale:2012ih}
\bibinfo{author}{\bibfnamefont{F.}~\bibnamefont{Caporale}}, \bibinfo{author}{\bibfnamefont{D.~{\relax Yu}.} \bibnamefont{Ivanov}}, \bibinfo{author}{\bibfnamefont{B.}~\bibnamefont{Murdaca}}, \bibnamefont{and} \bibinfo{author}{\bibfnamefont{A.}~\bibnamefont{Papa}}, \bibinfo{journal}{Nucl. Phys. B} \textbf{\bibinfo{volume}{877}}, \bibinfo{pages}{73} (\bibinfo{year}{2013}{\natexlab{a}}), \eprint{1211.7225}.

\bibitem[{\citenamefont{Kotikov and Lipatov}(2000)}]{Kotikov:2000pm}
\bibinfo{author}{\bibfnamefont{A.~V.} \bibnamefont{Kotikov}} \bibnamefont{and} \bibinfo{author}{\bibfnamefont{L.~N.} \bibnamefont{Lipatov}}, \bibinfo{journal}{Nucl. Phys. B} \textbf{\bibinfo{volume}{582}}, \bibinfo{pages}{19} (\bibinfo{year}{2000}), \eprint{hep-ph/0004008}.

\bibitem[{\citenamefont{Ivanov and Papa}(2012{\natexlab{a}})}]{Ivanov:2012iv}
\bibinfo{author}{\bibfnamefont{D.~{\relax Yu}.} \bibnamefont{Ivanov}} \bibnamefont{and} \bibinfo{author}{\bibfnamefont{A.}~\bibnamefont{Papa}}, \bibinfo{journal}{JHEP} \textbf{\bibinfo{volume}{07}}, \bibinfo{pages}{045} (\bibinfo{year}{2012}{\natexlab{a}}), \eprint{1205.6068}.

\bibitem[{\citenamefont{Colferai and Niccoli}(2015)}]{Colferai:2015zfa}
\bibinfo{author}{\bibfnamefont{D.}~\bibnamefont{Colferai}} \bibnamefont{and} \bibinfo{author}{\bibfnamefont{A.}~\bibnamefont{Niccoli}}, \bibinfo{journal}{JHEP} \textbf{\bibinfo{volume}{04}}, \bibinfo{pages}{071} (\bibinfo{year}{2015}), \eprint{1501.07442}.

\bibitem[{\citenamefont{Ivanov and Papa}(2012{\natexlab{b}})}]{Ivanov:2012ms}
\bibinfo{author}{\bibfnamefont{D.~{\relax Yu}.} \bibnamefont{Ivanov}} \bibnamefont{and} \bibinfo{author}{\bibfnamefont{A.}~\bibnamefont{Papa}}, \bibinfo{journal}{JHEP} \textbf{\bibinfo{volume}{05}}, \bibinfo{pages}{086} (\bibinfo{year}{2012}{\natexlab{b}}), \eprint{1202.1082}.

\bibitem[{\citenamefont{Furman}(1982)}]{Furman:1981kf}
\bibinfo{author}{\bibfnamefont{M.}~\bibnamefont{Furman}}, \bibinfo{journal}{Nucl. Phys. B} \textbf{\bibinfo{volume}{197}}, \bibinfo{pages}{413} (\bibinfo{year}{1982}).

\bibitem[{\citenamefont{Aversa et~al.}(1989)\citenamefont{Aversa, Chiappetta, Greco, and Guillet}}]{Aversa:1988vb}
\bibinfo{author}{\bibfnamefont{F.}~\bibnamefont{Aversa}}, \bibinfo{author}{\bibfnamefont{P.}~\bibnamefont{Chiappetta}}, \bibinfo{author}{\bibfnamefont{M.}~\bibnamefont{Greco}}, \bibnamefont{and} \bibinfo{author}{\bibfnamefont{J.~P.} \bibnamefont{Guillet}}, \bibinfo{journal}{Nucl. Phys. B} \textbf{\bibinfo{volume}{327}}, \bibinfo{pages}{105} (\bibinfo{year}{1989}).

\bibitem[{\citenamefont{Alioli et~al.}(2010)\citenamefont{Alioli, Nason, Oleari, and Re}}]{Alioli:2010xd}
\bibinfo{author}{\bibfnamefont{S.}~\bibnamefont{Alioli}}, \bibinfo{author}{\bibfnamefont{P.}~\bibnamefont{Nason}}, \bibinfo{author}{\bibfnamefont{C.}~\bibnamefont{Oleari}}, \bibnamefont{and} \bibinfo{author}{\bibfnamefont{E.}~\bibnamefont{Re}}, \bibinfo{journal}{JHEP} \textbf{\bibinfo{volume}{06}}, \bibinfo{pages}{043} (\bibinfo{year}{2010}), \eprint{1002.2581}.

\bibitem[{\citenamefont{Campbell et~al.}(2012)\citenamefont{Campbell, Ellis, Frederix, Nason, Oleari, and Williams}}]{Campbell:2012am}
\bibinfo{author}{\bibfnamefont{J.~M.} \bibnamefont{Campbell}}, \bibinfo{author}{\bibfnamefont{R.~K.} \bibnamefont{Ellis}}, \bibinfo{author}{\bibfnamefont{R.}~\bibnamefont{Frederix}}, \bibinfo{author}{\bibfnamefont{P.}~\bibnamefont{Nason}}, \bibinfo{author}{\bibfnamefont{C.}~\bibnamefont{Oleari}}, \bibnamefont{and} \bibinfo{author}{\bibfnamefont{C.}~\bibnamefont{Williams}}, \bibinfo{journal}{JHEP} \textbf{\bibinfo{volume}{07}}, \bibinfo{pages}{092} (\bibinfo{year}{2012}), \eprint{1202.5475}.

\bibitem[{\citenamefont{Hamilton et~al.}(2013)\citenamefont{Hamilton, Nason, Oleari, and Zanderighi}}]{Hamilton:2012rf}
\bibinfo{author}{\bibfnamefont{K.}~\bibnamefont{Hamilton}}, \bibinfo{author}{\bibfnamefont{P.}~\bibnamefont{Nason}}, \bibinfo{author}{\bibfnamefont{C.}~\bibnamefont{Oleari}}, \bibnamefont{and} \bibinfo{author}{\bibfnamefont{G.}~\bibnamefont{Zanderighi}}, \bibinfo{journal}{JHEP} \textbf{\bibinfo{volume}{05}}, \bibinfo{pages}{082} (\bibinfo{year}{2013}), \eprint{1212.4504}.

\bibitem[{\citenamefont{Ball et~al.}(2021)}]{NNPDF:2021uiq}
\bibinfo{author}{\bibfnamefont{R.~D.} \bibnamefont{Ball}} \bibnamefont{et~al.} (\bibinfo{collaboration}{NNPDF}), \bibinfo{journal}{Eur. Phys. J. C} \textbf{\bibinfo{volume}{81}}, \bibinfo{pages}{958} (\bibinfo{year}{2021}), \eprint{2109.02671}.

\bibitem[{\citenamefont{Ball et~al.}(2022{\natexlab{a}})}]{NNPDF:2021njg}
\bibinfo{author}{\bibfnamefont{R.~D.} \bibnamefont{Ball}} \bibnamefont{et~al.} (\bibinfo{collaboration}{NNPDF}), \bibinfo{journal}{Eur. Phys. J. C} \textbf{\bibinfo{volume}{82}}, \bibinfo{pages}{428} (\bibinfo{year}{2022}{\natexlab{a}}), \eprint{2109.02653}.

\bibitem[{\citenamefont{Buckley et~al.}(2015)\citenamefont{Buckley, Ferrando, Lloyd, Nordstr\"om, Page, R\"ufenacht, Sch\"onherr, and Watt}}]{Buckley:2014ana}
\bibinfo{author}{\bibfnamefont{A.}~\bibnamefont{Buckley}}, \bibinfo{author}{\bibfnamefont{J.}~\bibnamefont{Ferrando}}, \bibinfo{author}{\bibfnamefont{S.}~\bibnamefont{Lloyd}}, \bibinfo{author}{\bibfnamefont{K.}~\bibnamefont{Nordstr\"om}}, \bibinfo{author}{\bibfnamefont{B.}~\bibnamefont{Page}}, \bibinfo{author}{\bibfnamefont{M.}~\bibnamefont{R\"ufenacht}}, \bibinfo{author}{\bibfnamefont{M.}~\bibnamefont{Sch\"onherr}}, \bibnamefont{and} \bibinfo{author}{\bibfnamefont{G.}~\bibnamefont{Watt}}, \bibinfo{journal}{Eur. Phys. J. C} \textbf{\bibinfo{volume}{75}}, \bibinfo{pages}{132} (\bibinfo{year}{2015}), \eprint{1412.7420}.

\bibitem[{\citenamefont{Khachatryan et~al.}(2016)}]{Khachatryan:2016udy}
\bibinfo{author}{\bibfnamefont{V.}~\bibnamefont{Khachatryan}} \bibnamefont{et~al.} (\bibinfo{collaboration}{CMS}), \bibinfo{journal}{JHEP} \textbf{\bibinfo{volume}{08}}, \bibinfo{pages}{139} (\bibinfo{year}{2016}), \eprint{1601.06713}.

\bibitem[{\citenamefont{Khachatryan et~al.}(2021{\natexlab{a}})}]{CMS:2020ldm}
\bibinfo{author}{\bibfnamefont{V.}~\bibnamefont{Khachatryan}} \bibnamefont{et~al.} (\bibinfo{collaboration}{CMS}), \bibinfo{journal}{JINST} \textbf{\bibinfo{volume}{16}}, \bibinfo{pages}{P02010} (\bibinfo{year}{2021}{\natexlab{a}}), \eprint{2011.01185}.

\bibitem[{\citenamefont{Khachatryan et~al.}(2021{\natexlab{b}})}]{Khachatryan:2020mpd}
\bibinfo{author}{\bibfnamefont{V.}~\bibnamefont{Khachatryan}} \bibnamefont{et~al.} (\bibinfo{collaboration}{CMS}), \bibinfo{journal}{JINST} \textbf{\bibinfo{volume}{16}}, \bibinfo{pages}{P02010} (\bibinfo{year}{2021}{\natexlab{b}}), \eprint{2011.01185}.

\bibitem[{\citenamefont{Mueller et~al.}(2013)\citenamefont{Mueller, Xiao, and Yuan}}]{Mueller:2013wwa}
\bibinfo{author}{\bibfnamefont{A.}~\bibnamefont{Mueller}}, \bibinfo{author}{\bibfnamefont{B.-W.} \bibnamefont{Xiao}}, \bibnamefont{and} \bibinfo{author}{\bibfnamefont{F.}~\bibnamefont{Yuan}}, \bibinfo{journal}{Phys. Rev. D} \textbf{\bibinfo{volume}{88}}, \bibinfo{pages}{114010} (\bibinfo{year}{2013}), \eprint{1308.2993}.

\bibitem[{\citenamefont{Marzani}(2016)}]{Marzani:2015oyb}
\bibinfo{author}{\bibfnamefont{S.}~\bibnamefont{Marzani}}, \bibinfo{journal}{Phys. Rev. D} \textbf{\bibinfo{volume}{93}}, \bibinfo{pages}{054047} (\bibinfo{year}{2016}), \eprint{1511.06039}.

\bibitem[{\citenamefont{Mueller et~al.}(2016)\citenamefont{Mueller, Szymanowski, Wallon, Xiao, and Yuan}}]{Mueller:2015ael}
\bibinfo{author}{\bibfnamefont{A.}~\bibnamefont{Mueller}}, \bibinfo{author}{\bibfnamefont{L.}~\bibnamefont{Szymanowski}}, \bibinfo{author}{\bibfnamefont{S.}~\bibnamefont{Wallon}}, \bibinfo{author}{\bibfnamefont{B.-W.} \bibnamefont{Xiao}}, \bibnamefont{and} \bibinfo{author}{\bibfnamefont{F.}~\bibnamefont{Yuan}}, \bibinfo{journal}{JHEP} \textbf{\bibinfo{volume}{03}}, \bibinfo{pages}{096} (\bibinfo{year}{2016}), \eprint{1512.07127}.

\bibitem[{\citenamefont{Xiao and Yuan}(2018)}]{Xiao:2018esv}
\bibinfo{author}{\bibfnamefont{B.-W.} \bibnamefont{Xiao}} \bibnamefont{and} \bibinfo{author}{\bibfnamefont{F.}~\bibnamefont{Yuan}}, \bibinfo{journal}{Phys. Lett. B} \textbf{\bibinfo{volume}{782}}, \bibinfo{pages}{28} (\bibinfo{year}{2018}), \eprint{1801.05478}.

\bibitem[{\citenamefont{Hatta et~al.}(2021{\natexlab{a}})\citenamefont{Hatta, Xiao, Yuan, and Zhou}}]{Hatta:2020bgy}
\bibinfo{author}{\bibfnamefont{Y.}~\bibnamefont{Hatta}}, \bibinfo{author}{\bibfnamefont{B.-W.} \bibnamefont{Xiao}}, \bibinfo{author}{\bibfnamefont{F.}~\bibnamefont{Yuan}}, \bibnamefont{and} \bibinfo{author}{\bibfnamefont{J.}~\bibnamefont{Zhou}}, \bibinfo{journal}{Phys. Rev. Lett.} \textbf{\bibinfo{volume}{126}}, \bibinfo{pages}{142001} (\bibinfo{year}{2021}{\natexlab{a}}), \eprint{2010.10774}.

\bibitem[{\citenamefont{Hatta et~al.}(2021{\natexlab{b}})\citenamefont{Hatta, Xiao, Yuan, and Zhou}}]{Hatta:2021jcd}
\bibinfo{author}{\bibfnamefont{Y.}~\bibnamefont{Hatta}}, \bibinfo{author}{\bibfnamefont{B.-W.} \bibnamefont{Xiao}}, \bibinfo{author}{\bibfnamefont{F.}~\bibnamefont{Yuan}}, \bibnamefont{and} \bibinfo{author}{\bibfnamefont{J.}~\bibnamefont{Zhou}}, \bibinfo{journal}{Phys. Rev. D} \textbf{\bibinfo{volume}{104}}, \bibinfo{pages}{054037} (\bibinfo{year}{2021}{\natexlab{b}}), \eprint{2106.05307}.

\bibitem[{\citenamefont{Andersen et~al.}(2001)\citenamefont{Andersen, Del~Duca, Frixione, Schmidt, and Stirling}}]{Andersen:2001kta}
\bibinfo{author}{\bibfnamefont{J.~R.} \bibnamefont{Andersen}}, \bibinfo{author}{\bibfnamefont{V.}~\bibnamefont{Del~Duca}}, \bibinfo{author}{\bibfnamefont{S.}~\bibnamefont{Frixione}}, \bibinfo{author}{\bibfnamefont{C.~R.} \bibnamefont{Schmidt}}, \bibnamefont{and} \bibinfo{author}{\bibfnamefont{W.~J.} \bibnamefont{Stirling}}, \bibinfo{journal}{JHEP} \textbf{\bibinfo{volume}{02}}, \bibinfo{pages}{007} (\bibinfo{year}{2001}), \eprint{hep-ph/0101180}.

\bibitem[{\citenamefont{Fontannaz et~al.}(2001)\citenamefont{Fontannaz, Guillet, and Heinrich}}]{Fontannaz:2001nq}
\bibinfo{author}{\bibfnamefont{M.}~\bibnamefont{Fontannaz}}, \bibinfo{author}{\bibfnamefont{J.~P.} \bibnamefont{Guillet}}, \bibnamefont{and} \bibinfo{author}{\bibfnamefont{G.}~\bibnamefont{Heinrich}}, \bibinfo{journal}{Eur. Phys. J. C} \textbf{\bibinfo{volume}{22}}, \bibinfo{pages}{303} (\bibinfo{year}{2001}), \eprint{hep-ph/0107262}.

\bibitem[{\citenamefont{Duclou\'e et~al.}(2014{\natexlab{b}})\citenamefont{Duclou\'e, Szymanowski, and Wallon}}]{Ducloue:2014koa}
\bibinfo{author}{\bibfnamefont{B.}~\bibnamefont{Duclou\'e}}, \bibinfo{author}{\bibfnamefont{L.}~\bibnamefont{Szymanowski}}, \bibnamefont{and} \bibinfo{author}{\bibfnamefont{S.}~\bibnamefont{Wallon}}, \bibinfo{journal}{Phys. Lett. B} \textbf{\bibinfo{volume}{738}}, \bibinfo{pages}{311} (\bibinfo{year}{2014}{\natexlab{b}}), \eprint{1407.6593}.

\bibitem[{\citenamefont{Chatrchyan et~al.}(2012)}]{Chatrchyan:2012xg}
\bibinfo{author}{\bibfnamefont{S.}~\bibnamefont{Chatrchyan}} \bibnamefont{et~al.} (\bibinfo{collaboration}{CMS}), \bibinfo{journal}{Phys. Lett. B} \textbf{\bibinfo{volume}{714}}, \bibinfo{pages}{136} (\bibinfo{year}{2012}), \eprint{1205.0594}.

\bibitem[{\citenamefont{Sterman}(1987)}]{Sterman:1986aj}
\bibinfo{author}{\bibfnamefont{G.~F.} \bibnamefont{Sterman}}, \bibinfo{journal}{Nucl. Phys. B} \textbf{\bibinfo{volume}{281}}, \bibinfo{pages}{310} (\bibinfo{year}{1987}).

\bibitem[{\citenamefont{Catani and Trentadue}(1989)}]{Catani:1989ne}
\bibinfo{author}{\bibfnamefont{S.}~\bibnamefont{Catani}} \bibnamefont{and} \bibinfo{author}{\bibfnamefont{L.}~\bibnamefont{Trentadue}}, \bibinfo{journal}{Nucl. Phys. B} \textbf{\bibinfo{volume}{327}}, \bibinfo{pages}{323} (\bibinfo{year}{1989}).

\bibitem[{\citenamefont{Catani et~al.}(1996)\citenamefont{Catani, Mangano, Nason, and Trentadue}}]{Catani:1996yz}
\bibinfo{author}{\bibfnamefont{S.}~\bibnamefont{Catani}}, \bibinfo{author}{\bibfnamefont{M.~L.} \bibnamefont{Mangano}}, \bibinfo{author}{\bibfnamefont{P.}~\bibnamefont{Nason}}, \bibnamefont{and} \bibinfo{author}{\bibfnamefont{L.}~\bibnamefont{Trentadue}}, \bibinfo{journal}{Nucl. Phys. B} \textbf{\bibinfo{volume}{478}}, \bibinfo{pages}{273} (\bibinfo{year}{1996}), \eprint{hep-ph/9604351}.

\bibitem[{\citenamefont{Bonciani et~al.}(2003)\citenamefont{Bonciani, Catani, Mangano, and Nason}}]{Bonciani:2003nt}
\bibinfo{author}{\bibfnamefont{R.}~\bibnamefont{Bonciani}}, \bibinfo{author}{\bibfnamefont{S.}~\bibnamefont{Catani}}, \bibinfo{author}{\bibfnamefont{M.~L.} \bibnamefont{Mangano}}, \bibnamefont{and} \bibinfo{author}{\bibfnamefont{P.}~\bibnamefont{Nason}}, \bibinfo{journal}{Phys. Lett. B} \textbf{\bibinfo{volume}{575}}, \bibinfo{pages}{268} (\bibinfo{year}{2003}), \eprint{hep-ph/0307035}.

\bibitem[{\citenamefont{de~Florian et~al.}(2006)\citenamefont{de~Florian, Kulesza, and Vogelsang}}]{deFlorian:2005fzc}
\bibinfo{author}{\bibfnamefont{D.}~\bibnamefont{de~Florian}}, \bibinfo{author}{\bibfnamefont{A.}~\bibnamefont{Kulesza}}, \bibnamefont{and} \bibinfo{author}{\bibfnamefont{W.}~\bibnamefont{Vogelsang}}, \bibinfo{journal}{JHEP} \textbf{\bibinfo{volume}{02}}, \bibinfo{pages}{047} (\bibinfo{year}{2006}), \eprint{hep-ph/0511205}.

\bibitem[{\citenamefont{Ahrens et~al.}(2009)\citenamefont{Ahrens, Becher, Neubert, and Yang}}]{Ahrens:2009cxz}
\bibinfo{author}{\bibfnamefont{V.}~\bibnamefont{Ahrens}}, \bibinfo{author}{\bibfnamefont{T.}~\bibnamefont{Becher}}, \bibinfo{author}{\bibfnamefont{M.}~\bibnamefont{Neubert}}, \bibnamefont{and} \bibinfo{author}{\bibfnamefont{L.~L.} \bibnamefont{Yang}}, \bibinfo{journal}{Eur. Phys. J. C} \textbf{\bibinfo{volume}{62}}, \bibinfo{pages}{333} (\bibinfo{year}{2009}), \eprint{0809.4283}.

\bibitem[{\citenamefont{de~Florian and Grazzini}(2012)}]{deFlorian:2012yg}
\bibinfo{author}{\bibfnamefont{D.}~\bibnamefont{de~Florian}} \bibnamefont{and} \bibinfo{author}{\bibfnamefont{M.}~\bibnamefont{Grazzini}}, \bibinfo{journal}{Phys. Lett. B} \textbf{\bibinfo{volume}{718}}, \bibinfo{pages}{117} (\bibinfo{year}{2012}), \eprint{1206.4133}.

\bibitem[{\citenamefont{Forte et~al.}(2021)\citenamefont{Forte, Ridolfi, and Rota}}]{Forte:2021wxe}
\bibinfo{author}{\bibfnamefont{S.}~\bibnamefont{Forte}}, \bibinfo{author}{\bibfnamefont{G.}~\bibnamefont{Ridolfi}}, \bibnamefont{and} \bibinfo{author}{\bibfnamefont{S.}~\bibnamefont{Rota}}, \bibinfo{journal}{JHEP} \textbf{\bibinfo{volume}{08}}, \bibinfo{pages}{110} (\bibinfo{year}{2021}), \eprint{2106.11321}.

\bibitem[{\citenamefont{Mukherjee and Vogelsang}(2006)}]{Mukherjee:2006uu}
\bibinfo{author}{\bibfnamefont{A.}~\bibnamefont{Mukherjee}} \bibnamefont{and} \bibinfo{author}{\bibfnamefont{W.}~\bibnamefont{Vogelsang}}, \bibinfo{journal}{Phys. Rev. D} \textbf{\bibinfo{volume}{73}}, \bibinfo{pages}{074005} (\bibinfo{year}{2006}), \eprint{hep-ph/0601162}.

\bibitem[{\citenamefont{Bolzoni}(2006)}]{Bolzoni:2006ky}
\bibinfo{author}{\bibfnamefont{P.}~\bibnamefont{Bolzoni}}, \bibinfo{journal}{Phys. Lett. B} \textbf{\bibinfo{volume}{643}}, \bibinfo{pages}{325} (\bibinfo{year}{2006}), \eprint{hep-ph/0609073}.

\bibitem[{\citenamefont{Becher and Neubert}(2006)}]{Becher:2006nr}
\bibinfo{author}{\bibfnamefont{T.}~\bibnamefont{Becher}} \bibnamefont{and} \bibinfo{author}{\bibfnamefont{M.}~\bibnamefont{Neubert}}, \bibinfo{journal}{Phys. Rev. Lett.} \textbf{\bibinfo{volume}{97}}, \bibinfo{pages}{082001} (\bibinfo{year}{2006}), \eprint{hep-ph/0605050}.

\bibitem[{\citenamefont{Becher et~al.}(2008)\citenamefont{Becher, Neubert, and Xu}}]{Becher:2007ty}
\bibinfo{author}{\bibfnamefont{T.}~\bibnamefont{Becher}}, \bibinfo{author}{\bibfnamefont{M.}~\bibnamefont{Neubert}}, \bibnamefont{and} \bibinfo{author}{\bibfnamefont{G.}~\bibnamefont{Xu}}, \bibinfo{journal}{JHEP} \textbf{\bibinfo{volume}{07}}, \bibinfo{pages}{030} (\bibinfo{year}{2008}), \eprint{0710.0680}.

\bibitem[{\citenamefont{Bonvini et~al.}(2011)\citenamefont{Bonvini, Forte, and Ridolfi}}]{Bonvini:2010tp}
\bibinfo{author}{\bibfnamefont{M.}~\bibnamefont{Bonvini}}, \bibinfo{author}{\bibfnamefont{S.}~\bibnamefont{Forte}}, \bibnamefont{and} \bibinfo{author}{\bibfnamefont{G.}~\bibnamefont{Ridolfi}}, \bibinfo{journal}{Nucl. Phys. B} \textbf{\bibinfo{volume}{847}}, \bibinfo{pages}{93} (\bibinfo{year}{2011}), \eprint{1009.5691}.

\bibitem[{\citenamefont{Ahmed et~al.}(2015)\citenamefont{Ahmed, Mandal, Rana, and Ravindran}}]{Ahmed:2014era}
\bibinfo{author}{\bibfnamefont{T.}~\bibnamefont{Ahmed}}, \bibinfo{author}{\bibfnamefont{M.~K.} \bibnamefont{Mandal}}, \bibinfo{author}{\bibfnamefont{N.}~\bibnamefont{Rana}}, \bibnamefont{and} \bibinfo{author}{\bibfnamefont{V.}~\bibnamefont{Ravindran}}, \bibinfo{journal}{JHEP} \textbf{\bibinfo{volume}{02}}, \bibinfo{pages}{131} (\bibinfo{year}{2015}), \eprint{1411.5301}.

\bibitem[{\citenamefont{Muselli et~al.}(2017)\citenamefont{Muselli, Forte, and Ridolfi}}]{Muselli:2017bad}
\bibinfo{author}{\bibfnamefont{C.}~\bibnamefont{Muselli}}, \bibinfo{author}{\bibfnamefont{S.}~\bibnamefont{Forte}}, \bibnamefont{and} \bibinfo{author}{\bibfnamefont{G.}~\bibnamefont{Ridolfi}}, \bibinfo{journal}{JHEP} \textbf{\bibinfo{volume}{03}}, \bibinfo{pages}{106} (\bibinfo{year}{2017}), \eprint{1701.01464}.

\bibitem[{\citenamefont{Banerjee et~al.}(2018)\citenamefont{Banerjee, Das, Dhani, and Ravindran}}]{Banerjee:2018vvb}
\bibinfo{author}{\bibfnamefont{P.}~\bibnamefont{Banerjee}}, \bibinfo{author}{\bibfnamefont{G.}~\bibnamefont{Das}}, \bibinfo{author}{\bibfnamefont{P.~K.} \bibnamefont{Dhani}}, \bibnamefont{and} \bibinfo{author}{\bibfnamefont{V.}~\bibnamefont{Ravindran}}, \bibinfo{journal}{Phys. Rev. D} \textbf{\bibinfo{volume}{98}}, \bibinfo{pages}{054018} (\bibinfo{year}{2018}), \eprint{1805.01186}.

\bibitem[{\citenamefont{Duhr et~al.}(2022{\natexlab{a}})\citenamefont{Duhr, Mistlberger, and Vita}}]{Duhr:2022cob}
\bibinfo{author}{\bibfnamefont{C.}~\bibnamefont{Duhr}}, \bibinfo{author}{\bibfnamefont{B.}~\bibnamefont{Mistlberger}}, \bibnamefont{and} \bibinfo{author}{\bibfnamefont{G.}~\bibnamefont{Vita}}, \bibinfo{journal}{JHEP} \textbf{\bibinfo{volume}{09}}, \bibinfo{pages}{155} (\bibinfo{year}{2022}{\natexlab{a}}), \eprint{2205.04493}.

\bibitem[{\citenamefont{Shi et~al.}(2022)\citenamefont{Shi, Wang, Wei, and Xiao}}]{Shi:2021hwx}
\bibinfo{author}{\bibfnamefont{Y.}~\bibnamefont{Shi}}, \bibinfo{author}{\bibfnamefont{L.}~\bibnamefont{Wang}}, \bibinfo{author}{\bibfnamefont{S.-Y.} \bibnamefont{Wei}}, \bibnamefont{and} \bibinfo{author}{\bibfnamefont{B.-W.} \bibnamefont{Xiao}}, \bibinfo{journal}{Phys. Rev. Lett.} \textbf{\bibinfo{volume}{128}}, \bibinfo{pages}{202302} (\bibinfo{year}{2022}), \eprint{2112.06975}.

\bibitem[{\citenamefont{Wang et~al.}(2023)\citenamefont{Wang, Chen, Gao, Shi, Wei, and Xiao}}]{Wang:2022zdu}
\bibinfo{author}{\bibfnamefont{L.}~\bibnamefont{Wang}}, \bibinfo{author}{\bibfnamefont{L.}~\bibnamefont{Chen}}, \bibinfo{author}{\bibfnamefont{Z.}~\bibnamefont{Gao}}, \bibinfo{author}{\bibfnamefont{Y.}~\bibnamefont{Shi}}, \bibinfo{author}{\bibfnamefont{S.-Y.} \bibnamefont{Wei}}, \bibnamefont{and} \bibinfo{author}{\bibfnamefont{B.-W.} \bibnamefont{Xiao}}, \bibinfo{journal}{Phys. Rev. D} \textbf{\bibinfo{volume}{107}}, \bibinfo{pages}{016016} (\bibinfo{year}{2023}), \eprint{2211.08322}.

\bibitem[{\citenamefont{Bonvini and Marinelli}(2023)}]{Bonvini:2023mfj}
\bibinfo{author}{\bibfnamefont{M.}~\bibnamefont{Bonvini}} \bibnamefont{and} \bibinfo{author}{\bibfnamefont{G.}~\bibnamefont{Marinelli}}, \bibinfo{journal}{Eur. Phys. J. C} \textbf{\bibinfo{volume}{83}}, \bibinfo{pages}{931} (\bibinfo{year}{2023}), \eprint{2306.03568}.

\bibitem[{\citenamefont{Ball et~al.}(2013)\citenamefont{Ball, Bonvini, Forte, Marzani, and Ridolfi}}]{Ball:2013bra}
\bibinfo{author}{\bibfnamefont{R.~D.} \bibnamefont{Ball}}, \bibinfo{author}{\bibfnamefont{M.}~\bibnamefont{Bonvini}}, \bibinfo{author}{\bibfnamefont{S.}~\bibnamefont{Forte}}, \bibinfo{author}{\bibfnamefont{S.}~\bibnamefont{Marzani}}, \bibnamefont{and} \bibinfo{author}{\bibfnamefont{G.}~\bibnamefont{Ridolfi}}, \bibinfo{journal}{Nucl. Phys. B} \textbf{\bibinfo{volume}{874}}, \bibinfo{pages}{746} (\bibinfo{year}{2013}), \eprint{1303.3590}.

\bibitem[{\citenamefont{Bonvini and Marzani}(2014)}]{Bonvini:2014joa}
\bibinfo{author}{\bibfnamefont{M.}~\bibnamefont{Bonvini}} \bibnamefont{and} \bibinfo{author}{\bibfnamefont{S.}~\bibnamefont{Marzani}}, \bibinfo{journal}{JHEP} \textbf{\bibinfo{volume}{09}}, \bibinfo{pages}{007} (\bibinfo{year}{2014}), \eprint{1405.3654}.

\bibitem[{\citenamefont{Bartels and Lotter}(1993)}]{Bartels:1993du}
\bibinfo{author}{\bibfnamefont{J.}~\bibnamefont{Bartels}} \bibnamefont{and} \bibinfo{author}{\bibfnamefont{H.}~\bibnamefont{Lotter}}, \bibinfo{journal}{Phys. Lett. B} \textbf{\bibinfo{volume}{309}}, \bibinfo{pages}{400} (\bibinfo{year}{1993}).

\bibitem[{\citenamefont{Caporale et~al.}(2013{\natexlab{b}})\citenamefont{Caporale, Chachamis, Madrigal, Murdaca, and Sabio~Vera}}]{Caporale:2013bva}
\bibinfo{author}{\bibfnamefont{F.}~\bibnamefont{Caporale}}, \bibinfo{author}{\bibfnamefont{G.}~\bibnamefont{Chachamis}}, \bibinfo{author}{\bibfnamefont{J.~D.} \bibnamefont{Madrigal}}, \bibinfo{author}{\bibfnamefont{B.}~\bibnamefont{Murdaca}}, \bibnamefont{and} \bibinfo{author}{\bibfnamefont{A.}~\bibnamefont{Sabio~Vera}}, \bibinfo{journal}{Phys. Lett. B} \textbf{\bibinfo{volume}{724}}, \bibinfo{pages}{127} (\bibinfo{year}{2013}{\natexlab{b}}), \eprint{1305.1474}.

\bibitem[{\citenamefont{Ross and Sabio~Vera}(2016)}]{Ross:2016zwl}
\bibinfo{author}{\bibfnamefont{D.~A.} \bibnamefont{Ross}} \bibnamefont{and} \bibinfo{author}{\bibfnamefont{A.}~\bibnamefont{Sabio~Vera}}, \bibinfo{journal}{JHEP} \textbf{\bibinfo{volume}{08}}, \bibinfo{pages}{071} (\bibinfo{year}{2016}), \eprint{1605.08265}.

\bibitem[{\citenamefont{Catani et~al.}(2001)\citenamefont{Catani, de~Florian, and Grazzini}}]{Catani:2000vq}
\bibinfo{author}{\bibfnamefont{S.}~\bibnamefont{Catani}}, \bibinfo{author}{\bibfnamefont{D.}~\bibnamefont{de~Florian}}, \bibnamefont{and} \bibinfo{author}{\bibfnamefont{M.}~\bibnamefont{Grazzini}}, \bibinfo{journal}{Nucl. Phys. B} \textbf{\bibinfo{volume}{596}}, \bibinfo{pages}{299} (\bibinfo{year}{2001}), \eprint{hep-ph/0008184}.

\bibitem[{\citenamefont{Bozzi et~al.}(2006)\citenamefont{Bozzi, Catani, de~Florian, and Grazzini}}]{Bozzi:2005wk}
\bibinfo{author}{\bibfnamefont{G.}~\bibnamefont{Bozzi}}, \bibinfo{author}{\bibfnamefont{S.}~\bibnamefont{Catani}}, \bibinfo{author}{\bibfnamefont{D.}~\bibnamefont{de~Florian}}, \bibnamefont{and} \bibinfo{author}{\bibfnamefont{M.}~\bibnamefont{Grazzini}}, \bibinfo{journal}{Nucl. Phys. B} \textbf{\bibinfo{volume}{737}}, \bibinfo{pages}{73} (\bibinfo{year}{2006}), \eprint{hep-ph/0508068}.

\bibitem[{\citenamefont{Bozzi et~al.}(2009)\citenamefont{Bozzi, Catani, Ferrera, de~Florian, and Grazzini}}]{Bozzi:2008bb}
\bibinfo{author}{\bibfnamefont{G.}~\bibnamefont{Bozzi}}, \bibinfo{author}{\bibfnamefont{S.}~\bibnamefont{Catani}}, \bibinfo{author}{\bibfnamefont{G.}~\bibnamefont{Ferrera}}, \bibinfo{author}{\bibfnamefont{D.}~\bibnamefont{de~Florian}}, \bibnamefont{and} \bibinfo{author}{\bibfnamefont{M.}~\bibnamefont{Grazzini}}, \bibinfo{journal}{Nucl. Phys. B} \textbf{\bibinfo{volume}{815}}, \bibinfo{pages}{174} (\bibinfo{year}{2009}), \eprint{0812.2862}.

\bibitem[{\citenamefont{Catani and Grazzini}(2011)}]{Catani:2010pd}
\bibinfo{author}{\bibfnamefont{S.}~\bibnamefont{Catani}} \bibnamefont{and} \bibinfo{author}{\bibfnamefont{M.}~\bibnamefont{Grazzini}}, \bibinfo{journal}{Nucl. Phys. B} \textbf{\bibinfo{volume}{845}}, \bibinfo{pages}{297} (\bibinfo{year}{2011}), \eprint{1011.3918}.

\bibitem[{\citenamefont{Catani and Grazzini}(2012)}]{Catani:2011kr}
\bibinfo{author}{\bibfnamefont{S.}~\bibnamefont{Catani}} \bibnamefont{and} \bibinfo{author}{\bibfnamefont{M.}~\bibnamefont{Grazzini}}, \bibinfo{journal}{Eur. Phys. J. C} \textbf{\bibinfo{volume}{72}}, \bibinfo{pages}{2013} (\bibinfo{year}{2012}), \bibinfo{note}{[Erratum: Eur.Phys.J.C 72, 2132 (2012)]}, \eprint{1106.4652}.

\bibitem[{\citenamefont{Catani et~al.}(2014)\citenamefont{Catani, Cieri, de~Florian, Ferrera, and Grazzini}}]{Catani:2013tia}
\bibinfo{author}{\bibfnamefont{S.}~\bibnamefont{Catani}}, \bibinfo{author}{\bibfnamefont{L.}~\bibnamefont{Cieri}}, \bibinfo{author}{\bibfnamefont{D.}~\bibnamefont{de~Florian}}, \bibinfo{author}{\bibfnamefont{G.}~\bibnamefont{Ferrera}}, \bibnamefont{and} \bibinfo{author}{\bibfnamefont{M.}~\bibnamefont{Grazzini}}, \bibinfo{journal}{Nucl. Phys. B} \textbf{\bibinfo{volume}{881}}, \bibinfo{pages}{414} (\bibinfo{year}{2014}), \eprint{1311.1654}.

\bibitem[{\citenamefont{Catani et~al.}(2015)\citenamefont{Catani, de~Florian, Ferrera, and Grazzini}}]{Catani:2015vma}
\bibinfo{author}{\bibfnamefont{S.}~\bibnamefont{Catani}}, \bibinfo{author}{\bibfnamefont{D.}~\bibnamefont{de~Florian}}, \bibinfo{author}{\bibfnamefont{G.}~\bibnamefont{Ferrera}}, \bibnamefont{and} \bibinfo{author}{\bibfnamefont{M.}~\bibnamefont{Grazzini}}, \bibinfo{journal}{JHEP} \textbf{\bibinfo{volume}{12}}, \bibinfo{pages}{047} (\bibinfo{year}{2015}), \eprint{1507.06937}.

\bibitem[{\citenamefont{Duhr et~al.}(2022{\natexlab{b}})\citenamefont{Duhr, Mistlberger, and Vita}}]{Duhr:2022yyp}
\bibinfo{author}{\bibfnamefont{C.}~\bibnamefont{Duhr}}, \bibinfo{author}{\bibfnamefont{B.}~\bibnamefont{Mistlberger}}, \bibnamefont{and} \bibinfo{author}{\bibfnamefont{G.}~\bibnamefont{Vita}}, \bibinfo{journal}{Phys. Rev. Lett.} \textbf{\bibinfo{volume}{129}}, \bibinfo{pages}{162001} (\bibinfo{year}{2022}{\natexlab{b}}), \eprint{2205.02242}.

\bibitem[{\citenamefont{Cieri et~al.}(2015)\citenamefont{Cieri, Coradeschi, and de~Florian}}]{Cieri:2015rqa}
\bibinfo{author}{\bibfnamefont{L.}~\bibnamefont{Cieri}}, \bibinfo{author}{\bibfnamefont{F.}~\bibnamefont{Coradeschi}}, \bibnamefont{and} \bibinfo{author}{\bibfnamefont{D.}~\bibnamefont{de~Florian}}, \bibinfo{journal}{JHEP} \textbf{\bibinfo{volume}{06}}, \bibinfo{pages}{185} (\bibinfo{year}{2015}), \eprint{1505.03162}.

\bibitem[{\citenamefont{Alioli et~al.}(2021)\citenamefont{Alioli, Broggio, Gavardi, Kallweit, Lim, Nagar, Napoletano, and Rottoli}}]{Alioli:2020qrd}
\bibinfo{author}{\bibfnamefont{S.}~\bibnamefont{Alioli}}, \bibinfo{author}{\bibfnamefont{A.}~\bibnamefont{Broggio}}, \bibinfo{author}{\bibfnamefont{A.}~\bibnamefont{Gavardi}}, \bibinfo{author}{\bibfnamefont{S.}~\bibnamefont{Kallweit}}, \bibinfo{author}{\bibfnamefont{M.~A.} \bibnamefont{Lim}}, \bibinfo{author}{\bibfnamefont{R.}~\bibnamefont{Nagar}}, \bibinfo{author}{\bibfnamefont{D.}~\bibnamefont{Napoletano}}, \bibnamefont{and} \bibinfo{author}{\bibfnamefont{L.}~\bibnamefont{Rottoli}}, \bibinfo{journal}{JHEP} \textbf{\bibinfo{volume}{04}}, \bibinfo{pages}{041} (\bibinfo{year}{2021}), \eprint{2010.10498}.

\bibitem[{\citenamefont{Becher and Neumann}(2021)}]{Becher:2020ugp}
\bibinfo{author}{\bibfnamefont{T.}~\bibnamefont{Becher}} \bibnamefont{and} \bibinfo{author}{\bibfnamefont{T.}~\bibnamefont{Neumann}}, \bibinfo{journal}{JHEP} \textbf{\bibinfo{volume}{03}}, \bibinfo{pages}{199} (\bibinfo{year}{2021}), \eprint{2009.11437}.

\bibitem[{\citenamefont{Neumann}(2021)}]{Neumann:2021zkb}
\bibinfo{author}{\bibfnamefont{T.}~\bibnamefont{Neumann}}, \bibinfo{journal}{Eur. Phys. J. C} \textbf{\bibinfo{volume}{81}}, \bibinfo{pages}{905} (\bibinfo{year}{2021}), \eprint{2107.12478}.

\bibitem[{\citenamefont{Ferrera and Pires}(2017)}]{Ferrera:2016prr}
\bibinfo{author}{\bibfnamefont{G.}~\bibnamefont{Ferrera}} \bibnamefont{and} \bibinfo{author}{\bibfnamefont{J.}~\bibnamefont{Pires}}, \bibinfo{journal}{JHEP} \textbf{\bibinfo{volume}{02}}, \bibinfo{pages}{139} (\bibinfo{year}{2017}), \eprint{1609.01691}.

\bibitem[{\citenamefont{Ju and Sch\"onherr}(2021)}]{Ju:2021lah}
\bibinfo{author}{\bibfnamefont{W.-L.} \bibnamefont{Ju}} \bibnamefont{and} \bibinfo{author}{\bibfnamefont{M.}~\bibnamefont{Sch\"onherr}}, \bibinfo{journal}{JHEP} \textbf{\bibinfo{volume}{10}}, \bibinfo{pages}{088} (\bibinfo{year}{2021}), \eprint{2106.11260}.

\bibitem[{\citenamefont{Monni et~al.}(2020)\citenamefont{Monni, Rottoli, and Torrielli}}]{Monni:2019yyr}
\bibinfo{author}{\bibfnamefont{P.~F.} \bibnamefont{Monni}}, \bibinfo{author}{\bibfnamefont{L.}~\bibnamefont{Rottoli}}, \bibnamefont{and} \bibinfo{author}{\bibfnamefont{P.}~\bibnamefont{Torrielli}}, \bibinfo{journal}{Phys. Rev. Lett.} \textbf{\bibinfo{volume}{124}}, \bibinfo{pages}{252001} (\bibinfo{year}{2020}), \eprint{1909.04704}.

\bibitem[{\citenamefont{Buonocore et~al.}(2022)\citenamefont{Buonocore, Grazzini, Haag, and Rottoli}}]{Buonocore:2021akg}
\bibinfo{author}{\bibfnamefont{L.}~\bibnamefont{Buonocore}}, \bibinfo{author}{\bibfnamefont{M.}~\bibnamefont{Grazzini}}, \bibinfo{author}{\bibfnamefont{J.}~\bibnamefont{Haag}}, \bibnamefont{and} \bibinfo{author}{\bibfnamefont{L.}~\bibnamefont{Rottoli}}, \bibinfo{journal}{Eur. Phys. J. C} \textbf{\bibinfo{volume}{82}}, \bibinfo{pages}{27} (\bibinfo{year}{2022}), \eprint{2110.06913}.

\bibitem[{\citenamefont{Wiesemann et~al.}(2020)\citenamefont{Wiesemann, Rottoli, and Torrielli}}]{Wiesemann:2020gbm}
\bibinfo{author}{\bibfnamefont{M.}~\bibnamefont{Wiesemann}}, \bibinfo{author}{\bibfnamefont{L.}~\bibnamefont{Rottoli}}, \bibnamefont{and} \bibinfo{author}{\bibfnamefont{P.}~\bibnamefont{Torrielli}}, \bibinfo{journal}{Phys. Lett. B} \textbf{\bibinfo{volume}{809}}, \bibinfo{pages}{135718} (\bibinfo{year}{2020}), \eprint{2006.09338}.

\bibitem[{\citenamefont{Ebert et~al.}(2021)\citenamefont{Ebert, Michel, Stewart, and Tackmann}}]{Ebert:2020dfc}
\bibinfo{author}{\bibfnamefont{M.~A.} \bibnamefont{Ebert}}, \bibinfo{author}{\bibfnamefont{J.~K.~L.} \bibnamefont{Michel}}, \bibinfo{author}{\bibfnamefont{I.~W.} \bibnamefont{Stewart}}, \bibnamefont{and} \bibinfo{author}{\bibfnamefont{F.~J.} \bibnamefont{Tackmann}}, \bibinfo{journal}{JHEP} \textbf{\bibinfo{volume}{04}}, \bibinfo{pages}{102} (\bibinfo{year}{2021}), \eprint{2006.11382}.

\bibitem[{\citenamefont{Re et~al.}(2021)\citenamefont{Re, Rottoli, and Torrielli}}]{Re:2021con}
\bibinfo{author}{\bibfnamefont{E.}~\bibnamefont{Re}}, \bibinfo{author}{\bibfnamefont{L.}~\bibnamefont{Rottoli}}, \bibnamefont{and} \bibinfo{author}{\bibfnamefont{P.}~\bibnamefont{Torrielli}}, \bibinfo{journal}{JHEP} \textbf{\bibinfo{volume}{09}}, \bibinfo{pages}{108} (\bibinfo{year}{2021}), \eprint{2104.07509}.

\bibitem[{\citenamefont{Chen et~al.}(2022)\citenamefont{Chen, Gehrmann, Glover, Huss, Monni, Re, Rottoli, and Torrielli}}]{Chen:2022cgv}
\bibinfo{author}{\bibfnamefont{X.}~\bibnamefont{Chen}}, \bibinfo{author}{\bibfnamefont{T.}~\bibnamefont{Gehrmann}}, \bibinfo{author}{\bibfnamefont{E.~W.~N.} \bibnamefont{Glover}}, \bibinfo{author}{\bibfnamefont{A.}~\bibnamefont{Huss}}, \bibinfo{author}{\bibfnamefont{P.~F.} \bibnamefont{Monni}}, \bibinfo{author}{\bibfnamefont{E.}~\bibnamefont{Re}}, \bibinfo{author}{\bibfnamefont{L.}~\bibnamefont{Rottoli}}, \bibnamefont{and} \bibinfo{author}{\bibfnamefont{P.}~\bibnamefont{Torrielli}}, \bibinfo{journal}{Phys. Rev. Lett.} \textbf{\bibinfo{volume}{128}}, \bibinfo{pages}{252001} (\bibinfo{year}{2022}), \eprint{2203.01565}.

\bibitem[{\citenamefont{Neumann and Campbell}(2023)}]{Neumann:2022lft}
\bibinfo{author}{\bibfnamefont{T.}~\bibnamefont{Neumann}} \bibnamefont{and} \bibinfo{author}{\bibfnamefont{J.}~\bibnamefont{Campbell}}, \bibinfo{journal}{Phys. Rev. D} \textbf{\bibinfo{volume}{107}}, \bibinfo{pages}{L011506} (\bibinfo{year}{2023}), \eprint{2207.07056}.

\bibitem[{\citenamefont{Bizon et~al.}(2018)\citenamefont{Bizon, Monni, Re, Rottoli, and Torrielli}}]{Bizon:2017rah}
\bibinfo{author}{\bibfnamefont{W.}~\bibnamefont{Bizon}}, \bibinfo{author}{\bibfnamefont{P.~F.} \bibnamefont{Monni}}, \bibinfo{author}{\bibfnamefont{E.}~\bibnamefont{Re}}, \bibinfo{author}{\bibfnamefont{L.}~\bibnamefont{Rottoli}}, \bibnamefont{and} \bibinfo{author}{\bibfnamefont{P.}~\bibnamefont{Torrielli}}, \bibinfo{journal}{JHEP} \textbf{\bibinfo{volume}{02}}, \bibinfo{pages}{108} (\bibinfo{year}{2018}), \eprint{1705.09127}.

\bibitem[{\citenamefont{Billis et~al.}(2021)\citenamefont{Billis, Dehnadi, Ebert, Michel, and Tackmann}}]{Billis:2021ecs}
\bibinfo{author}{\bibfnamefont{G.}~\bibnamefont{Billis}}, \bibinfo{author}{\bibfnamefont{B.}~\bibnamefont{Dehnadi}}, \bibinfo{author}{\bibfnamefont{M.~A.} \bibnamefont{Ebert}}, \bibinfo{author}{\bibfnamefont{J.~K.~L.} \bibnamefont{Michel}}, \bibnamefont{and} \bibinfo{author}{\bibfnamefont{F.~J.} \bibnamefont{Tackmann}}, \bibinfo{journal}{Phys. Rev. Lett.} \textbf{\bibinfo{volume}{127}}, \bibinfo{pages}{072001} (\bibinfo{year}{2021}), \eprint{2102.08039}.

\bibitem[{\citenamefont{Caola et~al.}(2022)\citenamefont{Caola, Chen, Duhr, Liu, Mistlberger, Petriello, Vita, and Weinzierl}}]{Caola:2022ayt}
\bibinfo{author}{\bibfnamefont{F.}~\bibnamefont{Caola}}, \bibinfo{author}{\bibfnamefont{W.}~\bibnamefont{Chen}}, \bibinfo{author}{\bibfnamefont{C.}~\bibnamefont{Duhr}}, \bibinfo{author}{\bibfnamefont{X.}~\bibnamefont{Liu}}, \bibinfo{author}{\bibfnamefont{B.}~\bibnamefont{Mistlberger}}, \bibinfo{author}{\bibfnamefont{F.}~\bibnamefont{Petriello}}, \bibinfo{author}{\bibfnamefont{G.}~\bibnamefont{Vita}}, \bibnamefont{and} \bibinfo{author}{\bibfnamefont{S.}~\bibnamefont{Weinzierl}}, in \emph{\bibinfo{booktitle}{{2022 Snowmass Summer Study}}} (\bibinfo{year}{2022}), \eprint{2203.06730}.

\bibitem[{\citenamefont{Kallweit et~al.}(2020)\citenamefont{Kallweit, Re, Rottoli, and Wiesemann}}]{Kallweit:2020gva}
\bibinfo{author}{\bibfnamefont{S.}~\bibnamefont{Kallweit}}, \bibinfo{author}{\bibfnamefont{E.}~\bibnamefont{Re}}, \bibinfo{author}{\bibfnamefont{L.}~\bibnamefont{Rottoli}}, \bibnamefont{and} \bibinfo{author}{\bibfnamefont{M.}~\bibnamefont{Wiesemann}}, \bibinfo{journal}{JHEP} \textbf{\bibinfo{volume}{12}}, \bibinfo{pages}{147} (\bibinfo{year}{2020}), \eprint{2004.07720}.

\bibitem[{\citenamefont{Celiberto et~al.}(2023{\natexlab{b}})\citenamefont{Celiberto, Fucilla, Ivanov, Mohammed, and Papa}}]{Celiberto:2023dkr}
\bibinfo{author}{\bibfnamefont{F.~G.} \bibnamefont{Celiberto}}, \bibinfo{author}{\bibfnamefont{M.}~\bibnamefont{Fucilla}}, \bibinfo{author}{\bibfnamefont{D.~{\relax Yu}.} \bibnamefont{Ivanov}}, \bibinfo{author}{\bibfnamefont{M.~M.~A.} \bibnamefont{Mohammed}}, \bibnamefont{and} \bibinfo{author}{\bibfnamefont{A.}~\bibnamefont{Papa}}, in \emph{\bibinfo{booktitle}{{57th Rencontres de Moriond on QCD and High Energy Interactions}}} (\bibinfo{year}{2023}{\natexlab{b}}), \eprint{2305.11760}.

\bibitem[{\citenamefont{Bartels et~al.}(2002)\citenamefont{Bartels, Colferai, and Vacca}}]{Bartels:2001ge}
\bibinfo{author}{\bibfnamefont{J.}~\bibnamefont{Bartels}}, \bibinfo{author}{\bibfnamefont{D.}~\bibnamefont{Colferai}}, \bibnamefont{and} \bibinfo{author}{\bibfnamefont{G.~P.} \bibnamefont{Vacca}}, \bibinfo{journal}{Eur. Phys. J. C} \textbf{\bibinfo{volume}{24}}, \bibinfo{pages}{83} (\bibinfo{year}{2002}), \eprint{hep-ph/0112283}.

\bibitem[{\citenamefont{Caucal et~al.}(2022)\citenamefont{Caucal, Salazar, Schenke, and Venugopalan}}]{Caucal:2022ulg}
\bibinfo{author}{\bibfnamefont{P.}~\bibnamefont{Caucal}}, \bibinfo{author}{\bibfnamefont{F.}~\bibnamefont{Salazar}}, \bibinfo{author}{\bibfnamefont{B.}~\bibnamefont{Schenke}}, \bibnamefont{and} \bibinfo{author}{\bibfnamefont{R.}~\bibnamefont{Venugopalan}}, \bibinfo{journal}{JHEP} \textbf{\bibinfo{volume}{11}}, \bibinfo{pages}{169} (\bibinfo{year}{2022}), \eprint{2208.13872}.

\bibitem[{\citenamefont{Taels et~al.}(2022)\citenamefont{Taels, Altinoluk, Beuf, and Marquet}}]{Taels:2022tza}
\bibinfo{author}{\bibfnamefont{P.}~\bibnamefont{Taels}}, \bibinfo{author}{\bibfnamefont{T.}~\bibnamefont{Altinoluk}}, \bibinfo{author}{\bibfnamefont{G.}~\bibnamefont{Beuf}}, \bibnamefont{and} \bibinfo{author}{\bibfnamefont{C.}~\bibnamefont{Marquet}}, \bibinfo{journal}{JHEP} \textbf{\bibinfo{volume}{10}}, \bibinfo{pages}{184} (\bibinfo{year}{2022}), \eprint{2204.11650}.

\bibitem[{\citenamefont{Dasgupta et~al.}(2015)\citenamefont{Dasgupta, Dreyer, Salam, and Soyez}}]{Dasgupta:2014yra}
\bibinfo{author}{\bibfnamefont{M.}~\bibnamefont{Dasgupta}}, \bibinfo{author}{\bibfnamefont{F.}~\bibnamefont{Dreyer}}, \bibinfo{author}{\bibfnamefont{G.~P.} \bibnamefont{Salam}}, \bibnamefont{and} \bibinfo{author}{\bibfnamefont{G.}~\bibnamefont{Soyez}}, \bibinfo{journal}{JHEP} \textbf{\bibinfo{volume}{04}}, \bibinfo{pages}{039} (\bibinfo{year}{2015}), \eprint{1411.5182}.

\bibitem[{\citenamefont{Dasgupta et~al.}(2016)\citenamefont{Dasgupta, Dreyer, Salam, and Soyez}}]{Dasgupta:2016bnd}
\bibinfo{author}{\bibfnamefont{M.}~\bibnamefont{Dasgupta}}, \bibinfo{author}{\bibfnamefont{F.~A.} \bibnamefont{Dreyer}}, \bibinfo{author}{\bibfnamefont{G.~P.} \bibnamefont{Salam}}, \bibnamefont{and} \bibinfo{author}{\bibfnamefont{G.}~\bibnamefont{Soyez}}, \bibinfo{journal}{JHEP} \textbf{\bibinfo{volume}{06}}, \bibinfo{pages}{057} (\bibinfo{year}{2016}), \eprint{1602.01110}.

\bibitem[{\citenamefont{Banfi et~al.}(2012)\citenamefont{Banfi, Monni, Salam, and Zanderighi}}]{Banfi:2012jm}
\bibinfo{author}{\bibfnamefont{A.}~\bibnamefont{Banfi}}, \bibinfo{author}{\bibfnamefont{P.~F.} \bibnamefont{Monni}}, \bibinfo{author}{\bibfnamefont{G.~P.} \bibnamefont{Salam}}, \bibnamefont{and} \bibinfo{author}{\bibfnamefont{G.}~\bibnamefont{Zanderighi}}, \bibinfo{journal}{Phys. Rev. Lett.} \textbf{\bibinfo{volume}{109}}, \bibinfo{pages}{202001} (\bibinfo{year}{2012}), \eprint{1206.4998}.

\bibitem[{\citenamefont{Banfi et~al.}(2016)\citenamefont{Banfi, Caola, Dreyer, Monni, Salam, Zanderighi, and Dulat}}]{Banfi:2015pju}
\bibinfo{author}{\bibfnamefont{A.}~\bibnamefont{Banfi}}, \bibinfo{author}{\bibfnamefont{F.}~\bibnamefont{Caola}}, \bibinfo{author}{\bibfnamefont{F.~A.} \bibnamefont{Dreyer}}, \bibinfo{author}{\bibfnamefont{P.~F.} \bibnamefont{Monni}}, \bibinfo{author}{\bibfnamefont{G.~P.} \bibnamefont{Salam}}, \bibinfo{author}{\bibfnamefont{G.}~\bibnamefont{Zanderighi}}, \bibnamefont{and} \bibinfo{author}{\bibfnamefont{F.}~\bibnamefont{Dulat}}, \bibinfo{journal}{JHEP} \textbf{\bibinfo{volume}{04}}, \bibinfo{pages}{049} (\bibinfo{year}{2016}), \eprint{1511.02886}.

\bibitem[{\citenamefont{Liu et~al.}(2017)\citenamefont{Liu, Moch, and Ringer}}]{Liu:2017pbb}
\bibinfo{author}{\bibfnamefont{X.}~\bibnamefont{Liu}}, \bibinfo{author}{\bibfnamefont{S.-O.} \bibnamefont{Moch}}, \bibnamefont{and} \bibinfo{author}{\bibfnamefont{F.}~\bibnamefont{Ringer}}, \bibinfo{journal}{Phys. Rev. Lett.} \textbf{\bibinfo{volume}{119}}, \bibinfo{pages}{212001} (\bibinfo{year}{2017}), \eprint{1708.04641}.

\bibitem[{\citenamefont{Luisoni and Marzani}(2015)}]{Luisoni:2015xha}
\bibinfo{author}{\bibfnamefont{G.}~\bibnamefont{Luisoni}} \bibnamefont{and} \bibinfo{author}{\bibfnamefont{S.}~\bibnamefont{Marzani}}, \bibinfo{journal}{J. Phys. G} \textbf{\bibinfo{volume}{42}}, \bibinfo{pages}{103101} (\bibinfo{year}{2015}), \eprint{1505.04084}.

\bibitem[{\citenamefont{Caletti et~al.}(2021)\citenamefont{Caletti, Fedkevych, Marzani, Reichelt, Schumann, Soyez, and Theeuwes}}]{Caletti:2021oor}
\bibinfo{author}{\bibfnamefont{S.}~\bibnamefont{Caletti}}, \bibinfo{author}{\bibfnamefont{O.}~\bibnamefont{Fedkevych}}, \bibinfo{author}{\bibfnamefont{S.}~\bibnamefont{Marzani}}, \bibinfo{author}{\bibfnamefont{D.}~\bibnamefont{Reichelt}}, \bibinfo{author}{\bibfnamefont{S.}~\bibnamefont{Schumann}}, \bibinfo{author}{\bibfnamefont{G.}~\bibnamefont{Soyez}}, \bibnamefont{and} \bibinfo{author}{\bibfnamefont{V.}~\bibnamefont{Theeuwes}}, \bibinfo{journal}{JHEP} \textbf{\bibinfo{volume}{07}}, \bibinfo{pages}{076} (\bibinfo{year}{2021}), \eprint{2104.06920}.

\bibitem[{\citenamefont{Reichelt et~al.}(2022)\citenamefont{Reichelt, Caletti, Fedkevych, Marzani, Schumann, and Soyez}}]{Reichelt:2021svh}
\bibinfo{author}{\bibfnamefont{D.}~\bibnamefont{Reichelt}}, \bibinfo{author}{\bibfnamefont{S.}~\bibnamefont{Caletti}}, \bibinfo{author}{\bibfnamefont{O.}~\bibnamefont{Fedkevych}}, \bibinfo{author}{\bibfnamefont{S.}~\bibnamefont{Marzani}}, \bibinfo{author}{\bibfnamefont{S.}~\bibnamefont{Schumann}}, \bibnamefont{and} \bibinfo{author}{\bibfnamefont{G.}~\bibnamefont{Soyez}}, \bibinfo{journal}{JHEP} \textbf{\bibinfo{volume}{03}}, \bibinfo{pages}{131} (\bibinfo{year}{2022}), \eprint{2112.09545}.

\bibitem[{\citenamefont{Chapon et~al.}(2022)}]{Chapon:2020heu}
\bibinfo{author}{\bibfnamefont{E.}~\bibnamefont{Chapon}} \bibnamefont{et~al.}, \bibinfo{journal}{Prog. Part. Nucl. Phys.} \textbf{\bibinfo{volume}{122}}, \bibinfo{pages}{103906} (\bibinfo{year}{2022}), \eprint{2012.14161}.

\bibitem[{\citenamefont{Anchordoqui et~al.}(2022)}]{Anchordoqui:2021ghd}
\bibinfo{author}{\bibfnamefont{L.~A.} \bibnamefont{Anchordoqui}} \bibnamefont{et~al.}, \bibinfo{journal}{Phys. Rept.} \textbf{\bibinfo{volume}{968}}, \bibinfo{pages}{1} (\bibinfo{year}{2022}), \eprint{2109.10905}.

\bibitem[{\citenamefont{Feng et~al.}(2023{\natexlab{b}})}]{Feng:2022inv}
\bibinfo{author}{\bibfnamefont{J.~L.} \bibnamefont{Feng}} \bibnamefont{et~al.}, \bibinfo{journal}{J. Phys. G} \textbf{\bibinfo{volume}{50}}, \bibinfo{pages}{030501} (\bibinfo{year}{2023}{\natexlab{b}}), \eprint{2203.05090}.

\bibitem[{\citenamefont{Adachi et~al.}(2022)}]{AlexanderAryshev:2022pkx}
\bibinfo{author}{\bibfnamefont{I.}~\bibnamefont{Adachi}} \bibnamefont{et~al.} (\bibinfo{collaboration}{ILC International Development Team and ILC Community}) (\bibinfo{year}{2022}), \eprint{2203.07622}.

\bibitem[{\citenamefont{Arbuzov et~al.}(2021)}]{Arbuzov:2020cqg}
\bibinfo{author}{\bibfnamefont{A.}~\bibnamefont{Arbuzov}} \bibnamefont{et~al.}, \bibinfo{journal}{Prog. Part. Nucl. Phys.} \textbf{\bibinfo{volume}{119}}, \bibinfo{pages}{103858} (\bibinfo{year}{2021}), \eprint{2011.15005}.

\bibitem[{\citenamefont{Accettura et~al.}(2023)}]{Accettura:2023ked}
\bibinfo{author}{\bibfnamefont{C.}~\bibnamefont{Accettura}} \bibnamefont{et~al.}, \bibinfo{journal}{Eur. Phys. J. C} \textbf{\bibinfo{volume}{83}}, \bibinfo{pages}{864} (\bibinfo{year}{2023}), \bibinfo{note}{[Erratum: Eur.Phys.J.C 84, 36 (2024)]}, \eprint{2303.08533}.

\bibitem[{\citenamefont{Accettura et~al.}(2024{\natexlab{a}})}]{InternationalMuonCollider:2024jyv}
\bibinfo{author}{\bibfnamefont{C.}~\bibnamefont{Accettura}} \bibnamefont{et~al.} (\bibinfo{collaboration}{International Muon Collider}), \textbf{\bibinfo{volume}{2/2024}} (\bibinfo{year}{2024}{\natexlab{a}}), \eprint{2407.12450}.

\bibitem[{\citenamefont{Accettura et~al.}(2024{\natexlab{b}})}]{MuCoL:2024oxj}
\bibinfo{author}{\bibfnamefont{C.}~\bibnamefont{Accettura}} \bibnamefont{et~al.} (\bibinfo{collaboration}{MuCoL}) (\bibinfo{year}{2024}{\natexlab{b}}), \eprint{2411.02966}.

\bibitem[{\citenamefont{Black et~al.}(2024)}]{Black:2022cth}
\bibinfo{author}{\bibfnamefont{K.~M.} \bibnamefont{Black}} \bibnamefont{et~al.}, \bibinfo{journal}{JINST} \textbf{\bibinfo{volume}{19}}, \bibinfo{pages}{T02015} (\bibinfo{year}{2024}), \eprint{2209.01318}.

\bibitem[{\citenamefont{Accardi et~al.}(2024)}]{Accardi:2023chb}
\bibinfo{author}{\bibfnamefont{A.}~\bibnamefont{Accardi}} \bibnamefont{et~al.}, \bibinfo{journal}{Eur. Phys. J. A} \textbf{\bibinfo{volume}{60}}, \bibinfo{pages}{173} (\bibinfo{year}{2024}), \eprint{2306.09360}.

\bibitem[{\citenamefont{Kassabov et~al.}(2023)\citenamefont{Kassabov, Ubiali, and Voisey}}]{Kassabov:2022orn}
\bibinfo{author}{\bibfnamefont{Z.}~\bibnamefont{Kassabov}}, \bibinfo{author}{\bibfnamefont{M.}~\bibnamefont{Ubiali}}, \bibnamefont{and} \bibinfo{author}{\bibfnamefont{C.}~\bibnamefont{Voisey}}, \bibinfo{journal}{JHEP} \textbf{\bibinfo{volume}{03}}, \bibinfo{pages}{148} (\bibinfo{year}{2023}), \eprint{2207.07616}.

\bibitem[{\citenamefont{Harland-Lang and Thorne}(2019)}]{Harland-Lang:2018bxd}
\bibinfo{author}{\bibfnamefont{L.~A.} \bibnamefont{Harland-Lang}} \bibnamefont{and} \bibinfo{author}{\bibfnamefont{R.~S.} \bibnamefont{Thorne}}, \bibinfo{journal}{Eur. Phys. J. C} \textbf{\bibinfo{volume}{79}}, \bibinfo{pages}{225} (\bibinfo{year}{2019}), \eprint{1811.08434}.

\bibitem[{\citenamefont{Ball and Pearson}(2021)}]{Ball:2021icz}
\bibinfo{author}{\bibfnamefont{R.~D.} \bibnamefont{Ball}} \bibnamefont{and} \bibinfo{author}{\bibfnamefont{R.~L.} \bibnamefont{Pearson}}, \bibinfo{journal}{Eur. Phys. J. C} \textbf{\bibinfo{volume}{81}}, \bibinfo{pages}{830} (\bibinfo{year}{2021}), \eprint{2105.05114}.

\bibitem[{\citenamefont{McGowan et~al.}(2023)\citenamefont{McGowan, Cridge, Harland-Lang, and Thorne}}]{McGowan:2022nag}
\bibinfo{author}{\bibfnamefont{J.}~\bibnamefont{McGowan}}, \bibinfo{author}{\bibfnamefont{T.}~\bibnamefont{Cridge}}, \bibinfo{author}{\bibfnamefont{L.~A.} \bibnamefont{Harland-Lang}}, \bibnamefont{and} \bibinfo{author}{\bibfnamefont{R.~S.} \bibnamefont{Thorne}}, \bibinfo{journal}{Eur. Phys. J. C} \textbf{\bibinfo{volume}{83}}, \bibinfo{pages}{185} (\bibinfo{year}{2023}), \bibinfo{note}{[Erratum: Eur.Phys.J.C 83, 302 (2023)]}, \eprint{2207.04739}.

\bibitem[{\citenamefont{Ball et~al.}(2024{\natexlab{a}})}]{NNPDF:2024dpb}
\bibinfo{author}{\bibfnamefont{R.~D.} \bibnamefont{Ball}} \bibnamefont{et~al.} (\bibinfo{collaboration}{NNPDF}), \bibinfo{journal}{Eur. Phys. J. C} \textbf{\bibinfo{volume}{84}}, \bibinfo{pages}{517} (\bibinfo{year}{2024}{\natexlab{a}}), \eprint{2401.10319}.

\bibitem[{\citenamefont{Pasquini et~al.}(2023)\citenamefont{Pasquini, Rodini, and Venturini}}]{Pasquini:2023aaf}
\bibinfo{author}{\bibfnamefont{B.}~\bibnamefont{Pasquini}}, \bibinfo{author}{\bibfnamefont{S.}~\bibnamefont{Rodini}}, \bibnamefont{and} \bibinfo{author}{\bibfnamefont{S.}~\bibnamefont{Venturini}} (\bibinfo{collaboration}{MAP (Multi-dimensional Analyses of Partonic distributions)}), \bibinfo{journal}{Phys. Rev. D} \textbf{\bibinfo{volume}{107}}, \bibinfo{pages}{114023} (\bibinfo{year}{2023}), \eprint{2303.01789}.

\bibitem[{\citenamefont{Ball et~al.}(2022{\natexlab{b}})\citenamefont{Ball, Candido, Cruz-Martinez, Forte, Giani, Hekhorn, Kudashkin, Magni, and Rojo}}]{Ball:2022qks}
\bibinfo{author}{\bibfnamefont{R.~D.} \bibnamefont{Ball}}, \bibinfo{author}{\bibfnamefont{A.}~\bibnamefont{Candido}}, \bibinfo{author}{\bibfnamefont{J.}~\bibnamefont{Cruz-Martinez}}, \bibinfo{author}{\bibfnamefont{S.}~\bibnamefont{Forte}}, \bibinfo{author}{\bibfnamefont{T.}~\bibnamefont{Giani}}, \bibinfo{author}{\bibfnamefont{F.}~\bibnamefont{Hekhorn}}, \bibinfo{author}{\bibfnamefont{K.}~\bibnamefont{Kudashkin}}, \bibinfo{author}{\bibfnamefont{G.}~\bibnamefont{Magni}}, \bibnamefont{and} \bibinfo{author}{\bibfnamefont{J.}~\bibnamefont{Rojo}} (\bibinfo{collaboration}{NNPDF}), \bibinfo{journal}{Nature} \textbf{\bibinfo{volume}{608}}, \bibinfo{pages}{483} (\bibinfo{year}{2022}{\natexlab{b}}), \eprint{2208.08372}.

\bibitem[{\citenamefont{Brodsky et~al.}(1980)\citenamefont{Brodsky, Hoyer, Peterson, and Sakai}}]{Brodsky:1980pb}
\bibinfo{author}{\bibfnamefont{S.~J.} \bibnamefont{Brodsky}}, \bibinfo{author}{\bibfnamefont{P.}~\bibnamefont{Hoyer}}, \bibinfo{author}{\bibfnamefont{C.}~\bibnamefont{Peterson}}, \bibnamefont{and} \bibinfo{author}{\bibfnamefont{N.}~\bibnamefont{Sakai}}, \bibinfo{journal}{Phys. Lett. B} \textbf{\bibinfo{volume}{93}}, \bibinfo{pages}{451} (\bibinfo{year}{1980}).

\bibitem[{\citenamefont{Brodsky et~al.}(2015)\citenamefont{Brodsky, Kusina, Lyonnet, Schienbein, Spiesberger, and Vogt}}]{Brodsky:2015fna}
\bibinfo{author}{\bibfnamefont{S.~J.} \bibnamefont{Brodsky}}, \bibinfo{author}{\bibfnamefont{A.}~\bibnamefont{Kusina}}, \bibinfo{author}{\bibfnamefont{F.}~\bibnamefont{Lyonnet}}, \bibinfo{author}{\bibfnamefont{I.}~\bibnamefont{Schienbein}}, \bibinfo{author}{\bibfnamefont{H.}~\bibnamefont{Spiesberger}}, \bibnamefont{and} \bibinfo{author}{\bibfnamefont{R.}~\bibnamefont{Vogt}}, \bibinfo{journal}{Adv. High Energy Phys.} \textbf{\bibinfo{volume}{2015}}, \bibinfo{pages}{231547} (\bibinfo{year}{2015}), \eprint{1504.06287}.

\bibitem[{\citenamefont{Jimenez-Delgado et~al.}(2015)\citenamefont{Jimenez-Delgado, Hobbs, Londergan, and Melnitchouk}}]{Jimenez-Delgado:2014zga}
\bibinfo{author}{\bibfnamefont{P.}~\bibnamefont{Jimenez-Delgado}}, \bibinfo{author}{\bibfnamefont{T.~J.} \bibnamefont{Hobbs}}, \bibinfo{author}{\bibfnamefont{J.~T.} \bibnamefont{Londergan}}, \bibnamefont{and} \bibinfo{author}{\bibfnamefont{W.}~\bibnamefont{Melnitchouk}}, \bibinfo{journal}{Phys. Rev. Lett.} \textbf{\bibinfo{volume}{114}}, \bibinfo{pages}{082002} (\bibinfo{year}{2015}), \eprint{1408.1708}.

\bibitem[{\citenamefont{Ball et~al.}(2016)\citenamefont{Ball, Bertone, Bonvini, Carrazza, Forte, Guffanti, Hartland, Rojo, and Rottoli}}]{Ball:2016neh}
\bibinfo{author}{\bibfnamefont{R.~D.} \bibnamefont{Ball}}, \bibinfo{author}{\bibfnamefont{V.}~\bibnamefont{Bertone}}, \bibinfo{author}{\bibfnamefont{M.}~\bibnamefont{Bonvini}}, \bibinfo{author}{\bibfnamefont{S.}~\bibnamefont{Carrazza}}, \bibinfo{author}{\bibfnamefont{S.}~\bibnamefont{Forte}}, \bibinfo{author}{\bibfnamefont{A.}~\bibnamefont{Guffanti}}, \bibinfo{author}{\bibfnamefont{N.~P.} \bibnamefont{Hartland}}, \bibinfo{author}{\bibfnamefont{J.}~\bibnamefont{Rojo}}, \bibnamefont{and} \bibinfo{author}{\bibfnamefont{L.}~\bibnamefont{Rottoli}} (\bibinfo{collaboration}{NNPDF}), \bibinfo{journal}{Eur. Phys. J. C} \textbf{\bibinfo{volume}{76}}, \bibinfo{pages}{647} (\bibinfo{year}{2016}), \eprint{1605.06515}.

\bibitem[{\citenamefont{Hou et~al.}(2018)\citenamefont{Hou, Dulat, Gao, Guzzi, Huston, Nadolsky, Schmidt, Winter, Xie, and Yuan}}]{Hou:2017khm}
\bibinfo{author}{\bibfnamefont{T.-J.} \bibnamefont{Hou}}, \bibinfo{author}{\bibfnamefont{S.}~\bibnamefont{Dulat}}, \bibinfo{author}{\bibfnamefont{J.}~\bibnamefont{Gao}}, \bibinfo{author}{\bibfnamefont{M.}~\bibnamefont{Guzzi}}, \bibinfo{author}{\bibfnamefont{J.}~\bibnamefont{Huston}}, \bibinfo{author}{\bibfnamefont{P.}~\bibnamefont{Nadolsky}}, \bibinfo{author}{\bibfnamefont{C.}~\bibnamefont{Schmidt}}, \bibinfo{author}{\bibfnamefont{J.}~\bibnamefont{Winter}}, \bibinfo{author}{\bibfnamefont{K.}~\bibnamefont{Xie}}, \bibnamefont{and} \bibinfo{author}{\bibfnamefont{C.~P.} \bibnamefont{Yuan}}, \bibinfo{journal}{JHEP} \textbf{\bibinfo{volume}{02}}, \bibinfo{pages}{059} (\bibinfo{year}{2018}), \eprint{1707.00657}.

\bibitem[{\citenamefont{Guzzi et~al.}(2023)\citenamefont{Guzzi, Hobbs, Xie, Huston, Nadolsky, and Yuan}}]{Guzzi:2022rca}
\bibinfo{author}{\bibfnamefont{M.}~\bibnamefont{Guzzi}}, \bibinfo{author}{\bibfnamefont{T.~J.} \bibnamefont{Hobbs}}, \bibinfo{author}{\bibfnamefont{K.}~\bibnamefont{Xie}}, \bibinfo{author}{\bibfnamefont{J.}~\bibnamefont{Huston}}, \bibinfo{author}{\bibfnamefont{P.}~\bibnamefont{Nadolsky}}, \bibnamefont{and} \bibinfo{author}{\bibfnamefont{C.~P.} \bibnamefont{Yuan}}, \bibinfo{journal}{Phys. Lett. B} \textbf{\bibinfo{volume}{843}}, \bibinfo{pages}{137975} (\bibinfo{year}{2023}), \eprint{2211.01387}.

\bibitem[{\citenamefont{Ball et~al.}(2024{\natexlab{b}})\citenamefont{Ball, Candido, Cruz-Martinez, Forte, Giani, Hekhorn, Magni, Nocera, Rojo, and Stegeman}}]{NNPDF:2023tyk}
\bibinfo{author}{\bibfnamefont{R.~D.} \bibnamefont{Ball}}, \bibinfo{author}{\bibfnamefont{A.}~\bibnamefont{Candido}}, \bibinfo{author}{\bibfnamefont{J.}~\bibnamefont{Cruz-Martinez}}, \bibinfo{author}{\bibfnamefont{S.}~\bibnamefont{Forte}}, \bibinfo{author}{\bibfnamefont{T.}~\bibnamefont{Giani}}, \bibinfo{author}{\bibfnamefont{F.}~\bibnamefont{Hekhorn}}, \bibinfo{author}{\bibfnamefont{G.}~\bibnamefont{Magni}}, \bibinfo{author}{\bibfnamefont{E.~R.} \bibnamefont{Nocera}}, \bibinfo{author}{\bibfnamefont{J.}~\bibnamefont{Rojo}}, \bibnamefont{and} \bibinfo{author}{\bibfnamefont{R.}~\bibnamefont{Stegeman}} (\bibinfo{collaboration}{NNPDF}), \bibinfo{journal}{Phys. Rev. D} \textbf{\bibinfo{volume}{109}}, \bibinfo{pages}{L091501} (\bibinfo{year}{2024}{\natexlab{b}}), \eprint{2311.00743}.

\bibitem[{\citenamefont{Vogt}(2024)}]{Vogt:2024fky}
\bibinfo{author}{\bibfnamefont{R.}~\bibnamefont{Vogt}}, \bibinfo{journal}{Phys. Rev. D} \textbf{\bibinfo{volume}{110}}, \bibinfo{pages}{074036} (\bibinfo{year}{2024}), \eprint{2405.09018}.

\bibitem[{\citenamefont{Celiberto and Gatto}(2024{\natexlab{a}})}]{Celiberto:2024_TQHL11}
\bibinfo{author}{\bibfnamefont{F.~G.} \bibnamefont{Celiberto}} \bibnamefont{and} \bibinfo{author}{\bibfnamefont{G.}~\bibnamefont{Gatto}}, \emph{\bibinfo{title}{{TQHL1.1: TetraQuarks with Heavy and Light flavor collinear VFNS FFs}}} (\bibinfo{year}{2024}{\natexlab{a}}), \urlprefix\url{{https://github.com/FGCeliberto/Collinear_FFs/}}.

\bibitem[{\citenamefont{Celiberto and Gatto}(2024{\natexlab{b}})}]{Celiberto:2024_TQ4Q11}
\bibinfo{author}{\bibfnamefont{F.~G.} \bibnamefont{Celiberto}} \bibnamefont{and} \bibinfo{author}{\bibfnamefont{G.}~\bibnamefont{Gatto}}, \emph{\bibinfo{title}{{TQ4Q1.1: TetraQuarks with 4 heavy Quarks VFNS FFs}}} (\bibinfo{year}{2024}{\natexlab{b}}), \urlprefix\url{{https://github.com/FGCeliberto/Collinear_FFs/}}.

\bibitem[{\citenamefont{Binosi et~al.}(2009)\citenamefont{Binosi, Collins, Kaufhold, and Theussl}}]{Binosi:2008ig}
\bibinfo{author}{\bibfnamefont{D.}~\bibnamefont{Binosi}}, \bibinfo{author}{\bibfnamefont{J.}~\bibnamefont{Collins}}, \bibinfo{author}{\bibfnamefont{C.}~\bibnamefont{Kaufhold}}, \bibnamefont{and} \bibinfo{author}{\bibfnamefont{L.}~\bibnamefont{Theussl}}, \bibinfo{journal}{Comput. Phys. Commun.} \textbf{\bibinfo{volume}{180}}, \bibinfo{pages}{1709} (\bibinfo{year}{2009}), \eprint{0811.4113}.

\end{thebibliography}

\end{document}